\documentclass[
  twocolumn,english,aps,pra,
  superscriptaddress,amsmath,amssymb,floatfix,nofootinbib,longbibliography
]{revtex4-2}

\usepackage{amsthm}
\usepackage{amsfonts}
\usepackage{amsmath}
\usepackage{xcolor}
\usepackage{physics}
\usepackage{amssymb}
\usepackage{graphicx}
\usepackage{dcolumn}
\usepackage{bm}
\usepackage{mathtools}
\usepackage{hyperref}
\usepackage{mathrsfs}

\begin{document}

\title{Boltzmann Sampling of Frustrated J1 - J2 Ising Models with Programmable Quantum Annealers}

\author{Elijah Pelofske}
\email[]{epelofske@lanl.gov}
\affiliation{Information Systems \& Modeling, Los Alamos National Laboratory}
\affiliation{Theoretical Division, Quantum \& Condensed Matter Physics, Los Alamos National Laboratory}
\affiliation{Center for Quantum Computing, Los Alamos National Laboratory}

\begin{abstract}

One of the surprising, and potentially very useful, capabilities of analog quantum computers, such as D-Wave quantum annealers, is sampling from the Boltzmann, or Gibbs, distribution defined by a classical Hamiltonian. In this study, we thoroughly examine the ability of D-Wave quantum annealers to sample from the Boltzmann distribution defined by a canonical type of competing magnetic frustration $J_1$-$J_2$ model; the 1-dimensional ANNNI (axial next-nearest-neighbor Ising) model. Boltzmann sampling error rate is quantified for standard linear-ramp anneals ranging from $5$ nanosecond annealing times up to $2000$ microseconds on two different D-Wave quantum annealing processors. Interestingly, we find some analog hardware parameters which result in a very high accuracy (down to a TVD of $0.0003$) and low temperature sampling (down to $\beta=32.2$) in a frustrated region of the ANNNI model magnetic phase diagram. This bolsters the viability of current analog quantum computers for thermodynamic sampling applications of highly frustrated magnetic spin systems.

\end{abstract}

\maketitle

\section{Introduction}
\label{section:Introduction}

The ANNNI (axial next-nearest-neighbor Ising) model was first introduced as a model for the complex magnetic ordering of certain rare earth compounds~\cite{PhysRev.124.346, PhysRevLett.44.1502}, and subsequently has been extensively studied due to it exhibiting a variety of rich magnetic phenomena including various phase transitions, Devils staircase of commensurate and incommensurate phases, and in general providing a simple-to-define Ising model with controllable frustration~\cite{selke1988annni, fisher1981low}. The ANNNI model is the prototypical $J_1$-$J_2$ frustrated magnetic model. The general class of interactions characterized by short-range particle attraction and long-range particle repulsion yield a rich class of phenomenology~\cite{PhysRevLett.93.055701}, and the 1-dimensional ANNNI model is arguably the simplest version of this type of interaction. These types of frustrated competing interaction models have been subsequently generalized not only to higher dimensions, but also to include both thermal and quantum fluctuations to drive state transitions and time dynamics of these models~\cite{Rieger_1996, PhysRevB.81.094425, PhysRevB.45.2876, dutta2015quantumphasetransitionstransverse, Chandra_2007, selke1988annni}. Here, we will consider the original, classical, Ising model form of the 1D ANNNI model, given by

\begin{equation*}
{\mathcal H}_{\text{ANNNI}} = -J_1 \sum_{i} S_j S_{i+1} + J_2 \sum_{i} S_{i} S_{i+2}, 
\label{equation:ANNNI}
\end{equation*}

where $-J_1 > 0$ is the nearest neighbor ferromagnetic coupling, and $J_2 > 0$ is the antiferromagnetic coupling on the next-nearest-neighbors, and $S_i$ are the spins which in this case will be simulated by qubits. No local fields, $h_i$, are present in this model.

This study addresses how well current analog quantum computers, specifically D-Wave quantum annealers, can sample from the Boltzmann distribution defined by classical ANNNI models at various points in the frustrated magnetic phase diagram. Quantum annealing is a quantum computational algorithm which aims to find good solutions of combinatorial optimization problems using the principles of quantum adiabatic evolution~\cite{Kadowaki_1998, santoro2006optimization, Santoro_2002, Morita_2008, farhi2000quantumcomputationadiabaticevolution}. Noisy quantum annealers have been physically created, and one of the surprising findings is that they are quite effective at being programmable magnetic system simulators~\cite{harris2018phase, king2025beyond, PRXQuantum.2.030317, qubit_spin_ice, PRXQuantum.1.020320, pelofske2024simulatingheavyhextransversefield, Narasimhan_2024}, and moreover can, surprisingly, be efficient Boltzmann (thermal) samplers of classical, and potentially even quantum, Hamiltonians~\cite{PhysRevApplied.11.044083, Izquierdo_2021, Marshall_2017, nelson2021single, PRXQuantum.3.020317, PhysRevApplied.17.044046, buffoni2020thermodynamics, sathe2025classicalcriticalityquantumannealing, Raymond_2016, sandt2023efficient, PhysRevResearch.6.043050, Chancellor_2016}. The core computational capability that is of interest is being able to efficiently approximate the Boltzmann distribution defined by a classical Ising model;

\begin{equation}
    p(z) \propto e^{-\beta H(z)}, 
\end{equation}

\begin{figure*}[ht!]
    \centering
    \includegraphics[width=0.999\linewidth]{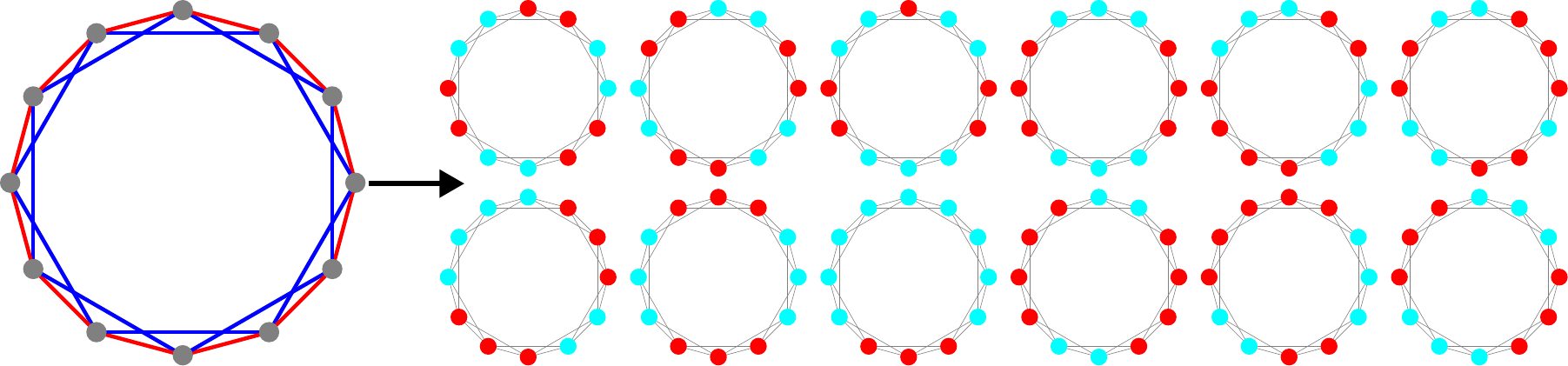}
    \caption{$12$-spin ANNNI model rendering (left), where red edges represent ferromagnetic coupling and blue edges represent antiferromagnetic coupling. This model is then sampled on the D-Wave QPU hardware; representative sampled spin configurations are shown on the right. Cyan nodes are spin down $\downarrow$ and red nodes are spin up $\uparrow$. These samples are from $5$ nanosecond annealing times, with frustration parameter coupling of $J_2=0.5$ (bottom row) and $J_2=1$ (top row). All of these spin configurations were measured on \texttt{Advantage2\_prototype1.4}.  }
    \label{fig:ANNNI_model}
\end{figure*}

where $p(z)$ is the probability distribution, over $z$ spin configurations. $H(z)$ is the classical Hamiltonian evaluation of the energy of a particular configuration. $\beta$ is the inverse thermodynamic temperature $\beta = 1/k_{B}T$, and we use natural units of the Boltzmann constant $k_B=1$ for numerical computations and visual plots. Note that we will generally refer to this sampling task as Boltzmann sampling, however, this can be equivalently referred to as either Gibbs sampling or thermal sampling. Boltzmann sampling, equivalently Gibbs sampling, is an important computational capability for many domains in information processing~\cite{5495896, metropolis1953equation, PhysRevE.79.041127, huang2024hardnesssamplingantiferromagneticising, spiridon2017hamiltonian, bodini2010multi}, and finding novel ways of speeding up this computation is an active area of study~\cite{noe2019boltzmann, goto2018boltzmann, npp4-b1xb, PhysRevResearch.5.L012029}.

The overall open question is how well the analog quantum hardware performs as a Boltzmann sampler of the ANNNI model in the different regions of the magnetic phase diagram. For example, it could be reasonable to assume that near or at the maximum frustration point, the analog hardware struggles with unbiased sampling, and therefore perhaps would be a worse Boltzmann sampler in that regime. However, maybe that is not the case, we need to test it to see what the results are -- and if the analog hardware is a good Boltzmann sampler in very highly frustrated regimes, that could be interesting because those regimes are also where more sophisticated classical Monte Carlo update schemes are needed.

The ANNNI model has been recently been studied in the context of sampling based quantum algorithms, including quantum annealing. For example, ref.~\cite{hasegawa2025residual} numerically studied quantum annealing simulation of the 1D ANNNI model using infinite time evolving block decimation (iTEBD). Ref.~\cite{Pexe_2024} used a digital non-variational feedback-based quantum algorithm to study properties of the 1D quantum ANNNI model. Ref.~\cite{cea2024exploringphasediagramquantum} studied the 1D quantum ANNNI model using a variety of techniques including tensor network approximations and quantum neural networks. Ref.~\cite{marin2024modelingfrustratedisingsquare} studied equilibrium simulations of the 2D classical $J_1-J_2$ (ANNNI) model using D-Wave quantum annealers.

\section{Methods}
\label{section:methods}

The specific type of analog quantum computers we use in this study are superconducting flux qubit quantum annealers which operate at approximately $15$ millikelvin, manufactured by the commercial company D-Wave~\cite{johnson2011quantum, Bunyk_2014, dickson2013thermally, PhysRevB.82.024511, Johnson_2010}. The physical Hamiltonian that the analog D-Wave processors implement is 

\begin{equation*}
    {\mathcal H} = - \frac{A(s)}{2} \Big( \sum_i \sigma_{x}^{(i)} \Big) + \frac{B(s)} {2} \Big( \sum_i h_i \sigma_z^{(i)} + \sum_{i>j} J_{i, j} \sigma_z^{(i)} \sigma_z^{(j)} \Big), 
    \label{equation:QA_Hamiltonian}
\end{equation*}

where $A(s)$ defines the transverse field strength over time, controlled by the normalized parameter $s \in [0, 1]$, and $B(s)$ similarly controls the energy scale of the classical Hamiltonian that we wish to sample. $h_i$ and $J_{i,j}$ are the user-programmable coefficients that define the Ising model we want to sample configurations of. The transverse field term $\sum_i \sigma_{x}^{(i)}$ does not commute with the diagonal Z basis terms, and thereby is the driving mechanism in the quantum annealing processor that facilitates state transitions. At the end of each anneal, the state of each (active) qubit is measured in the computational (Pauli Z) basis. 

Only two hardware control parameters are adjusted for these Boltzmann distribution sampling simulations. The first is the total annealing time, ranging from $2000$ microseconds to $5$ nanoseconds. The second is the analog coupler energy scale, which we artificially set to values weaker than the maximum that the hardware allows, by turning off \texttt{auto-scale}. The tuning of both of these parameters is motivated by prior studies that have shown that both of these mechanisms change the quality and the effective temperature of the Ising model sampling~\cite{PRXQuantum.3.020317, PhysRevApplied.17.044046, nelson2021single}. In particular, degeneracy lifting due to the transverse field for particular Ising models can cause thermodynamic sampling issues where configurations within the same energy level, in particular the ground-state energy, can be non-uniformly sampled~\cite{matsuda2009quantum, 9605329, PhysRevA.100.030303, PhysRevLett.118.070502, PhysRevE.99.063314, job2018test, PhysRevA.91.042314, Albash_2015_decoherence, boixo2013experimental}. In general, it has been shown that these issues can be mitigated by reducing the programmed energy scale of the Ising model on the analog hardware~\cite{PRXQuantum.3.020317, PhysRevApplied.17.044046, nelson2021single}. All other D-Wave QPU settings are left to default values, in particular only the standard linear-ramp anneals are used.

\begin{table*}[ht!]
    \begin{center}
        \begin{tabular}{|l||l|l|l|p{2.3cm}|}
            \hline
            D-Wave QPU Chip & Graph name & Qubits & Couplers & Disjoint native embedding count  \\
            \hline
            \hline
            \texttt{Advantage\_system4.1} & Pegasus $P_{16}$ & 5627 & 40279 & 337 \\
            \hline
            \texttt{Advantage2\_system1.4} & Zephyr $Z_{12}$ & 4593 & 41796 & 204 \\
            \hline
        \end{tabular}
    \end{center}
    \caption{Summary of the D-Wave QPUs used in this study, including the number of independent disjoint native embeddings of the ANNNI model used for each processor graph.  }
    \label{table:hardware_summary}
\end{table*}

The specific model we use is a $12$-spin ANNNI model, with periodic boundary conditions. The periodic boundary conditions remove edge effects and allow us to approximate the thermodynamic limit better than with open boundaries. The relatively small system size makes the numeric computations for fitting the thermodynamic sampling properties more tractable compared to larger spin systems, where some approximations of the true Boltzmann distribution would likely be required. Figure~\ref{fig:ANNNI_model} shows a rendering of the ANNNI model, along with some representative spin configurations measured on the D-Wave QPUs. The phase diagram of the ANNNI model has been extensively studied~\cite{selke1988annni, redner1981one, PhysRevB.103.094441, PhysRevB.1.4405, PhysRevB.104.144429, PhysRevB.90.144410, Dhar_2000, yeomans1995splittingmultiphasepoint, Torre_2025}, but for the purposes of this study the goal is to approximately sample from the Boltzmann distribution defined by the classical 1D ANNNI model at various frustration parameters defined by the next-nearest-neighbor antiferromagnetic coupling $J_2$. $J_2 < 0.5$ is the ferromagnetic portion of the phase diagram, where the spins tend to be aligned. $J_2=0.5$ is a critical frustration point (also known as a multiphase point), and denotes the boundary between the ferromagnetic phase and the antiferromagnetic phase. At $J_2>0.5$ the spin ordering becomes antiferromagnetic, with various terms being used for the particular spin grouping ranging from super-antiferromagnetic, antiphase, and even stripe phase in the case of 2D ANNNI. Concretely, the $J_2>0.5$ ground-state spin pattern is $\uparrow \uparrow \downarrow \downarrow \uparrow \uparrow \dots$, which we can see some examples of in the spin configurations shown in Figure~\ref{fig:ANNNI_model}. These main ordered phases are specifically for the low temperature regions of the phase diagram; at high temperatures the model becomes more disordered. One of the primary questions that the ANNNI model lets us investigate with regard to quantum annealer sampling capabilities is to evaluate whether there is a significant difference within regions of high frustration, and at the critical frustration point of $J_2=0.5$ as well as just before and after the critical frustration point. Therefore, we sample the ANNNI model at the frustration parameters of $J_2=0.01, 0.25, 0.49, 0.5, 0.51, 0.75, 1.0$. Classical Monte Carlo sampling methods are extremely efficient for this model, because the system size is small. This system size is ideal for extensive physical analog hardware parameter study, where ground truth is efficient to compute. Future studies should examine sampling properties at larger system sizes with analog quantum computing hardware in comparison to classical methods.

We use a direct spin-to-qubit mapping of the Ising model onto the QPU hardware graph. This is accomplished specifically using the Glasgow subgraph isomorphism finder~\cite{mccreesh2020glasgow}, part of the \texttt{minorminer} package~\cite{Chern_2023, cai2014practicalheuristicfindinggraph}. This isomorphism finder is applied iteratively to the hardware graphs in order to find many disjoint embeddings of the Ising model onto the QPU graph. This then enables many independent configurations of the Ising model to be sampled within the same anneal-readout cycle, in parallel~\cite{PhysRevA.91.042314, parallel_QA, Pelofske_2022_boolean}. Table~\ref{table:hardware_summary} details the D-Wave QPU hardware graph specifications, along with how many of these parallel embeddings are used. The QPU hardware graphs are called Pegasus~\cite{dattani2019pegasussecondconnectivitygraph, boothby2020nextgenerationtopologydwavequantum} and Zephyr~\cite{zephyr}, which have slightly different connectivities. One of the primary advantages of using a direct spin-to-qubit mapping on the hardware, as opposed to using minor embeddings, is that we remove various thermodynamic sampling issues, not to mention additional sources of error, caused by minor embedding~\cite{PhysRevResearch.2.023020}. No other types of error mitigation, besides independently embedded instances within the same anneal-readout cycle, were used (no readout bias mitigation, and no gauge transforms were applied). Appendix~\ref{section:appendix_hardware_embeddings} shows the exact embeddings on the hardware graphs.

The total number of samples obtained for each hardware parameter combination of annealing time, energy scale, ANNNI frustration parameter ($J_2$), and D-Wave device, is exactly $1{,}000$ anneal-readout cycles, multiplied by the disjoint embedding count shown in Table~\ref{table:hardware_summary} ($337{,}000$ samples on \texttt{Advantage\_system4.1} and $204{,}000$ samples on \texttt{Advantage2\_system1.4}). This means that the results from the \texttt{Advantage\_system4.1} device had a larger total sample count, which could make the results for that device better than the other QPU. For annealing times greater than $100$ microseconds, the maximum QPU time limit of the D-Wave software prevents jobs with a full $1{,}000$ anneals, and therefore for the longer annealing times four device-jobs are used, each using $250$ anneals. 

The experimental settings we apply are as follows. The energy scales that are evaluated are in the range of $0.01$ to $1.0$ in steps of $0.01$, along with $0.0001, 0.0005, 0.001, 0.005$. These energy scales are in normalized hardware programmable units, with more details given in Appendix~\ref{section:appendix_DWave_QPU_energy_scales}. These energy scale normalizations are applied to the entire ANNNI model; meaning, the magnetic interactions in the ANNNI model are the same relative to each other, but on the analog quantum hardware the corresponding sampling properties are different, due to a variety of mechanisms. To this end, when reporting results we will report both the overall $J$ energy scale, between 0 and 1, as well as the ANNNI model frustration parameter $J_2$. 
The annealing times used are $5$ through $100$ nanoseconds, in steps of $1$ nanosecond, then $200, 300, 400, 500, 600, 700, 800, 900$ nanoseconds, then $1$ through $100$ microseconds in steps of $1$ microsecond, then $200$ through $2000$ microseconds in steps of $100$ microseconds. The reason for these analog parameter ranges is to cover a reasonably fine resolution search from the minimum to the maximum possible values (where the minimum coupling energy scale is simply a sufficiently small value close to the precision limit of the hardware).

\begin{figure*}[ht!]
    \centering
    \includegraphics[width=0.999\linewidth]{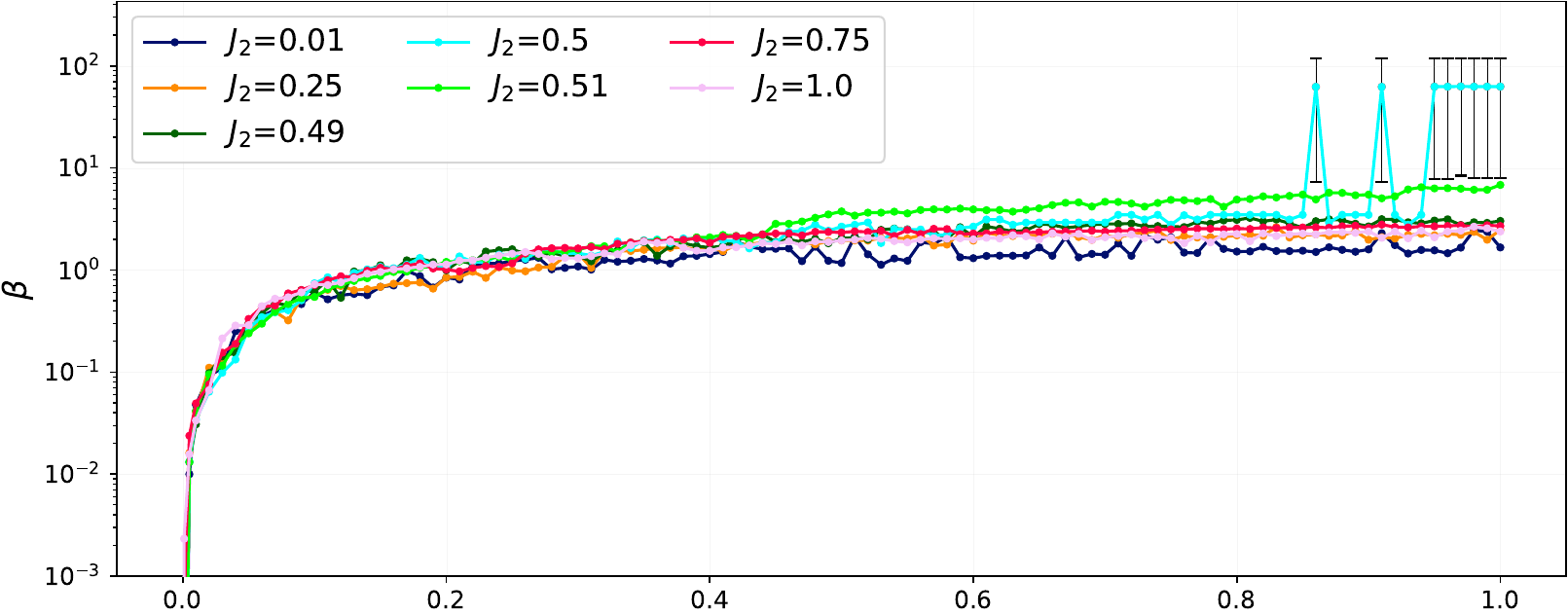}
    \includegraphics[width=0.999\linewidth]{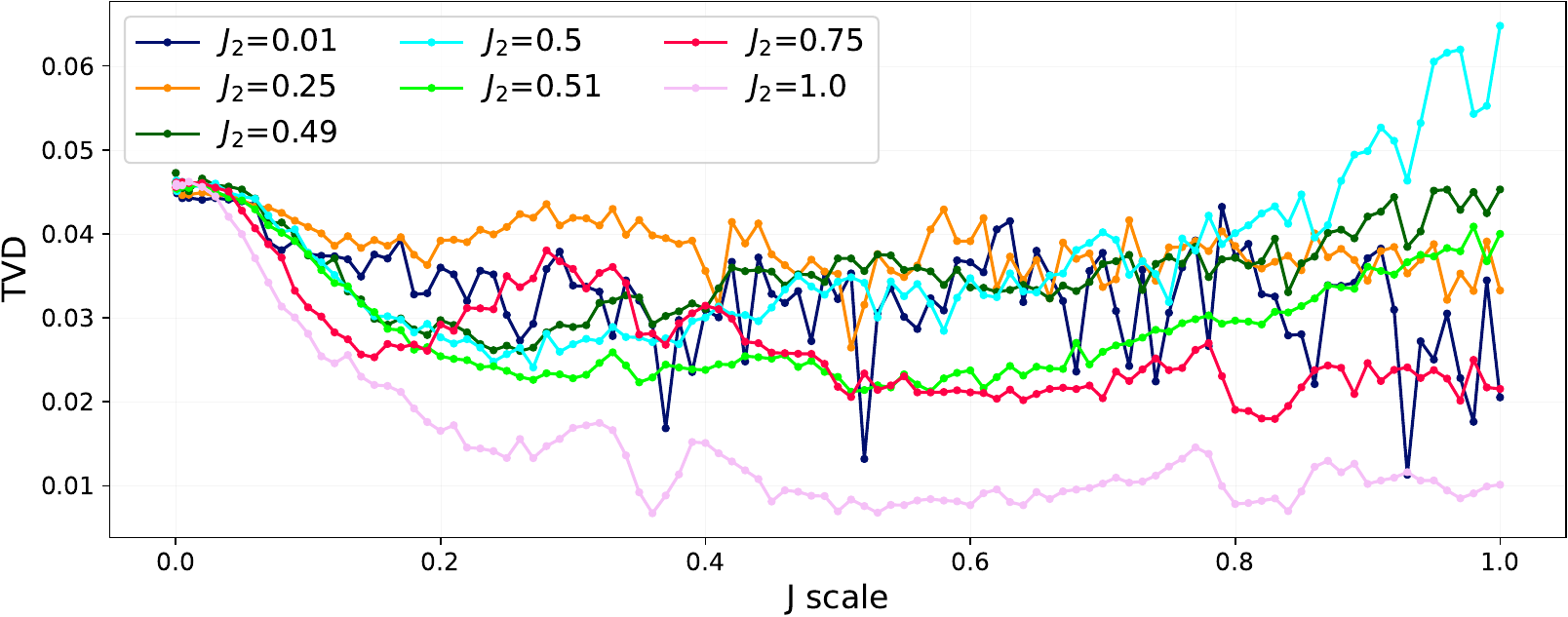}
    \caption{Minimum error rate (TVD), from all evaluated annealing times, Boltzmann sampling as a function of the J coupling energy scale programmed on the analog hardware (x-axis), for the \texttt{Advantage\_system4.1} D-Wave quantum annealing processor. The bottom plot shows the value of the minimum error rate, TVD, found across all evaluated annealing times, and the top plot shows the corresponding $\beta$ value at which the QPU samples a Boltzmann distribution at that given error rate. Each separate line denotes a different ANNNI model frustration parameter $J_2$.  }
    \label{fig:Pegasus_function_of_J}
\end{figure*}

There is evidence from Kibble-Zurek mechanism~\cite{kibble1976topology, zurek1985cosmological} scaling that the dynamics of the QPU in the very fast annealing times\footnote{Although, these very fast anneals are really quenches given the timescales involved} are closed-quantum system coherent dynamics~\cite{bando2020probing, king2023quantum, King_2022}. Then at much longer annealing times, the system becomes a quasistatic open quantum system~\cite{Amin_2015, buffoni2020thermodynamics}. In general, these processors have a variety of sources of error and noise contributing to the observed dynamics, including spurious qubit coupling, spin bath polarization, and noise drift over time~\cite{Pelofske_2023_noise, PhysRevApplied.19.034053, Zaborniak_2021, PhysRevApplied.8.064025, tüysüz2025learningresponsefunctionsanalog, PRXQuantum.3.020317, lanting2020probingenvironmentalspinpolarization}. All of these sources of error can be non-ideal for Boltzmann sampling, for instance, spin bath polarization causes anneal-readout sequences measured near to each other in time (e.g., a single device ``job'') to be correlated.

Note that during the execution of these experiments, there were hardware changes of the \texttt{Advantage2\_prototype1.4} graph where several qubits and edges were unexpectedly de-activated; this affected some of the disjoint minor embeddings, which were discarded in the post-processing stage. The disjoint embedding count of 204 accurately reflects the data reported in this study, and the qubit and coupler counts are for the more current hardware, and for consistency the chip id of \texttt{Advantage2\_prototype1.4} is used throughout the text.

\begin{figure*}[ht!]
    \centering
    \includegraphics[width=0.999\linewidth]{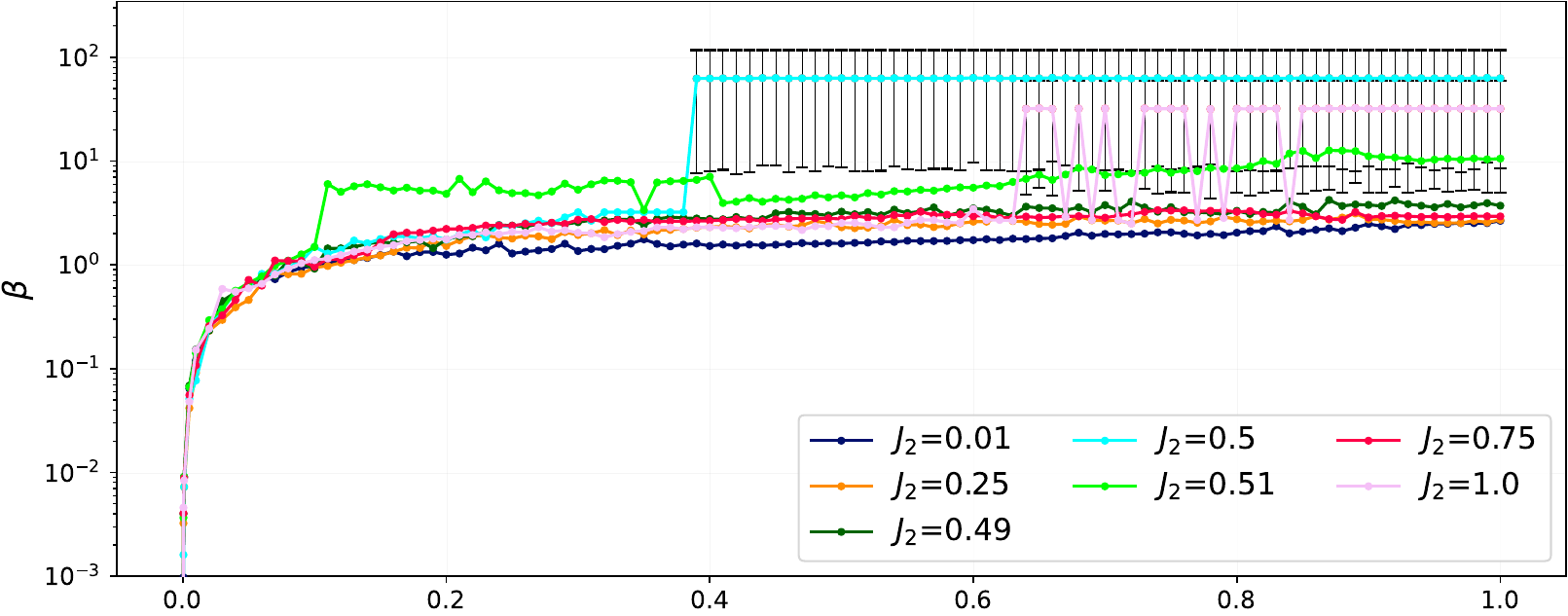}\\
    \includegraphics[width=0.999\linewidth]{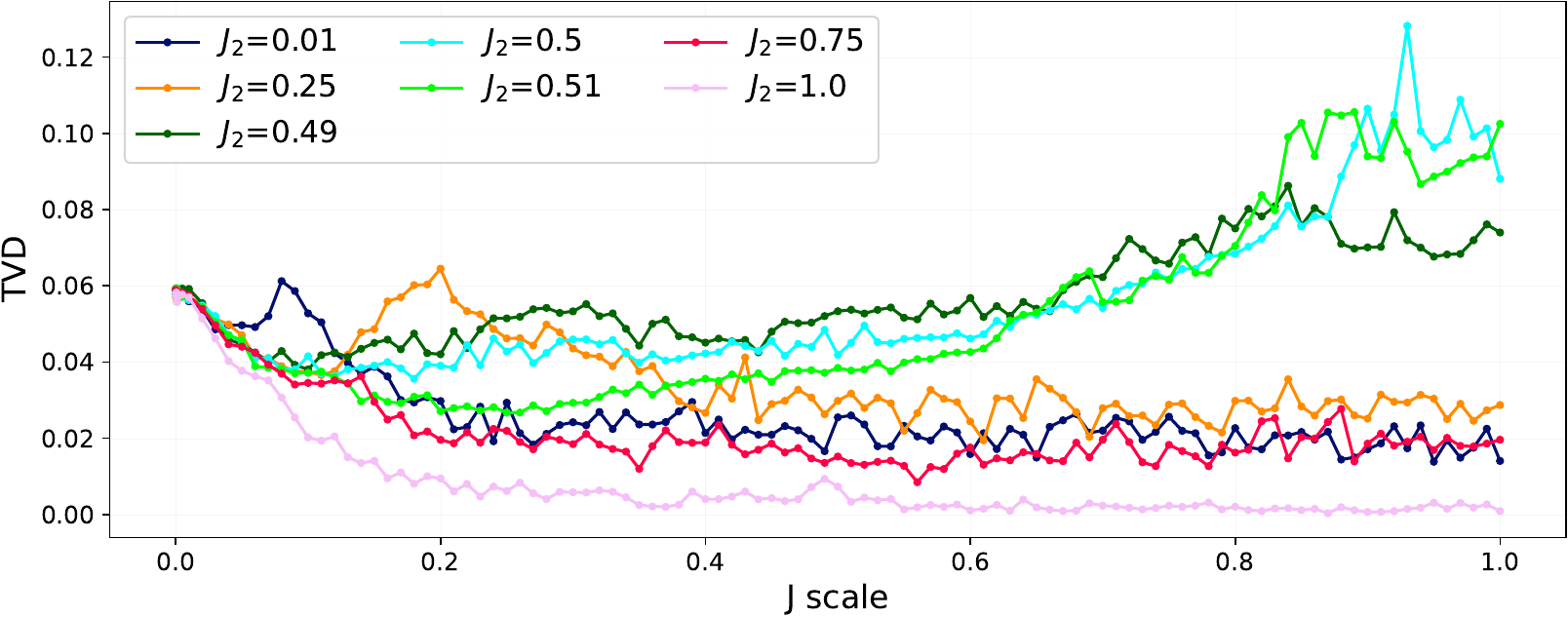}
    \caption{Minimum error rate (TVD), from all evaluated annealing times, Boltzmann sampling as a function of the J coupling energy scale programmed on the analog hardware (x-axis), for the \texttt{Advantage2\_system1.4} D-Wave quantum annealing processor. The bottom plot shows the value of the minimum error rate, TVD, found across all evaluated annealing times, and the top plot shows the corresponding $\beta$ value at which the QPU samples a Boltzmann distribution at that given error rate. Each separate line denotes a different ANNNI model frustration parameter $J_2$. }
    \label{fig:Zephyr2_function_of_J}
\end{figure*}

Next, we need an error rate measure between two probability distributions in order to quantify how well the quantum annealing hardware approximates a Gibbs distribution at a particular temperature. To this end, we use total variation distance (TVD), defined for two probability distributions $P(x)$ and $Q(x)$ as  

\begin{equation}
\text{TVD} = 0.5 \sum_x |P(x) - Q(x)|. 
\label{equation:TVD}
\end{equation}

\begin{table*}[ht]
    \begin{center}
        \begin{tabular}{|l||p{6.7cm}|p{6.7cm}|}
            \hline
            Frustration parameter & \texttt{Advantage2\_system1.4} & \texttt{Advantage\_system4.1}   \\
            \hline
            $J_2 = 0.01$ & TVD$=0.0138$, $\beta=2.484$, AT=$0.8$, $J=0.95$ & TVD=$0.011$, $\beta=1.451$, AT=$0.1$, $J=0.93$ \\
            \hline
            $J_2 = 0.25$ & TVD$=0.0195$, $\beta=2.613$, AT=$0.7$, $J=0.61$ & TVD$=0.0318$, $\beta=1.953$, AT=$2000$ $J=0.51$ \\
            \hline
            $J_2 = 0.49$ & TVD=$0.0381$, $\beta=0.921$, AT=$0.7$, $J=0.1$ & TVD=$0.0261$, $\beta=1.44$, AT=$500$, $J=0.26$ \\
            \hline
            $J_2 = 0.5$ & TVD=$0.026$, $\beta=1.43$, AT=$0.8$, $J=0.18$ & TVD=$0.0241$, $\beta=1.521$, AT=$2000$, $J=0.27$ \\
            \hline
            $J_2 = 0.51$ & TVD$=0.0236$, $\beta=4.913$, AT$=0.8$, $J=0.25$ & TVD=$0.0212$, $\beta=3.414$, AT=$1400$, $J=0.51$ \\
            \hline
            $J_2 = 0.75$ & TVD=$0.0085$, $\beta=3.257$, AT=$1600$, $J=0.56$ & TVD=$0.018$, $\beta=2.579$, AT=$1500$, $J=0.83$ \\
            \hline
            $J_2 = 1.0$ & TVD=$0.0003$, $\beta=32.166$, AT$=26$, $J=0.87$ & TVD=$0.0067$, $\beta=1.84$, AT=$800$, $J=0.36$ \\
            \hline
            \hline
        \end{tabular}
    \end{center}
    \caption{Lowest error rate (TVD) D-Wave quantum processor results, for different ANNNI model frustration parameters $J_2$. All annealing times are in units of microseconds, and $J$ denotes the overall hardware coupler scale factor.  }
    \label{table:lowest_error_rate_params}
\end{table*}

This error rate measure is minimized when it is zero, meaning the two distributions are identical, and if the TVD is $1$ then the distributions are entirely disjoint. Typically, the maximum possible error rate we will see is $0.5$, which corresponds to the approximate distribution being disordered with respect to the correct distribution. The full Boltzmann distribution fitting process is performed using a combination of black-box optimization solvers in scipy~\cite{2020SciPy-NMeth}, and a rigorous gridsearch. The fitting is performed independently for each analog hardware parameter range. Because of the finite sampling effect, in many cases the best fitted distribution has many different values of $\beta$ that result in the same error rate -- in this case in plots we report the error bar of the full $\beta$ range if the range exceeds $0.1$. And when reporting single $\beta$ values in the case where the $\beta$ error surface is flat, we report the average $\beta$ value that gives that minimum TVD error measure. The following blackbox optimizers were used, each being independently initialized at $28$ different values of $\beta$ ranging from $10^{5}$ to $10^{-8}$; \texttt{nelder-mead}~\cite{10.1007/s10589-010-9329-3}, \texttt{cobyla}, \texttt{l-bfgs-b}, \texttt{powell}, \texttt{slsqp}~\cite{nocedal2006numerical}, \texttt{trust-constr}~\cite{conn2000trust}, and \texttt{tnc}. The grid search is then performed over $100$ $\beta$ values in logarithmically spaced intervals from $10^{-3}$ to $10^{-15}$. An additional gridsearch over $\beta$ starting at $10^{-4}$ in steps of $10^{-4}$ is performed until sufficiently large $\beta$ values result in division by zero errors. All TVD values and $\beta$ values are rounded to reasonable numerical precision of at least 7 decimal places. At the end of the process, the relevant quantity we extract is simply what the absolute minimum TVD that was found, and then what the corresponding $\beta$ is of that distribution. The computation of the thermal distribution fitting was greatly accelerated by the Python~3 libraries Numba and Numpy~\cite{10.1145/2833157.2833162, harris2020array}. Appendix~\ref{section:appendix_beta_vs_TVD_examples} shows some examples of this temperature fitting procedure, including instances of flat response surfaces where different $\beta$ values result in the same TVD. The temperature fitting process is independently performed for the distribution (each distribution consisting of either $337{,}000$ or $204{,}000$ ANNNI model measurements) defined by each unique hardware parameter combination defined by the 4-tuple of D-Wave QPU, annealing time, overall J energy scale, and $J_2$ frustration parameter.

\section{Results}
\label{section:results}

Figure~\ref{fig:ANNNI_model} details some representative spin configurations sampled on one of the D-Wave QPUs, showing some of the differences in the magnetic ordering seen when the frustration parameter of the ANNNI model changes.

The experimental question is, given we are able to tune the analog hardware control properties of the coupler energy scale and the total annealing time, what is the absolute lowest error rate Boltzmann sampling that can be achieved?

To this end, Figure~\ref{fig:Pegasus_function_of_J} and Figure~\ref{fig:Zephyr2_function_of_J} plot the $\beta$ corresponding to the minimum error rate found across all evaluated annealing times, as a function of the $J$ energy scale. The seven different lines show the results for different ANNNI model frustration parameters, ranging from the $J_2=0.01$ (ferromagnetic ordering region of the model phase diagram) to $J_2=1.0$ (the more frustrated region). These plots show what the absolute lowest fitted error rate is, for each version of the ANNNI model at different frustration parameters, and then what the corresponding sampling temperature is in the top sub-plots. Notably, the model with the lowest error rate overall, on both QPUs is unexpectedly the highly frustrated $J_2=1=J_1$. Table~\ref{table:lowest_error_rate_params} lists the absolute lowest error rates, and what the corresponding physical analog parameters were that produced that distribution. Appendix~\ref{section:appendix_sub_sampling} analyzes sample convergence for some representative D-Wave device distributions, showing that the lowest reported error rates are stable. The newer of the two processors, \texttt{Advantage2\_system1.4}, which has a Zephyr connectivity graph, is able to achieve lower error rates compared to the \texttt{Advantage\_system4.1} processor, at some value of $\beta$, for all ANNNI frustration parameters except $J_2=0.49$. Note that the y-axis scales in Figure~\ref{fig:Pegasus_function_of_J} and Figure~\ref{fig:Zephyr2_function_of_J} for $\beta$ (top sub-plots) were cut off at very small $\beta$ for improved visual interpretability, those points corresponding to very weak hardware $J$ coupling, and therefore close to random sampling and therefore very high temperature sampling.

\begin{figure*}[p!]
    \centering
    \includegraphics[width=0.495\linewidth]{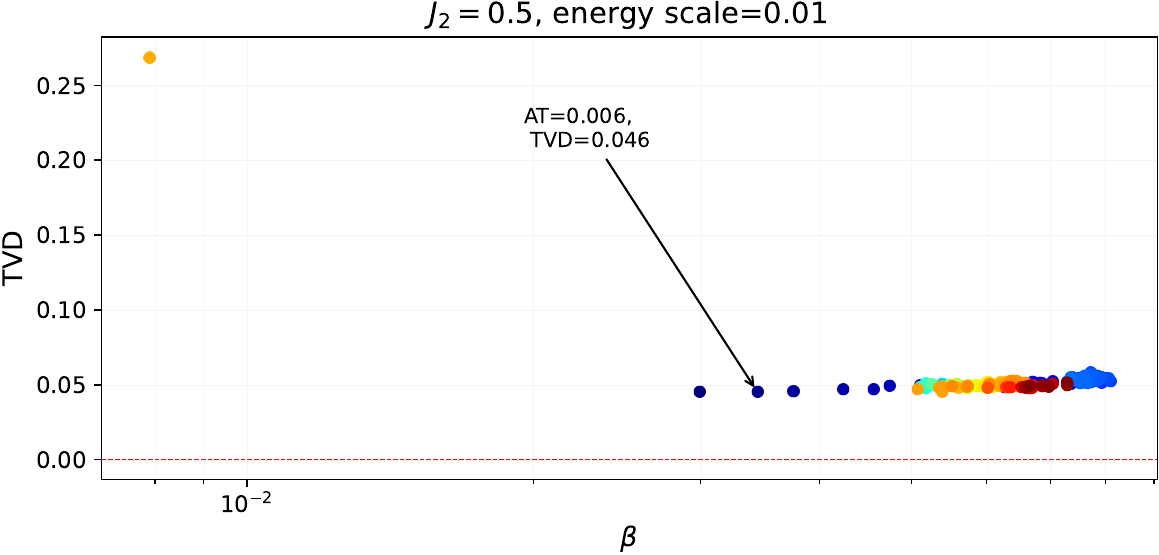}
    \includegraphics[width=0.495\linewidth]{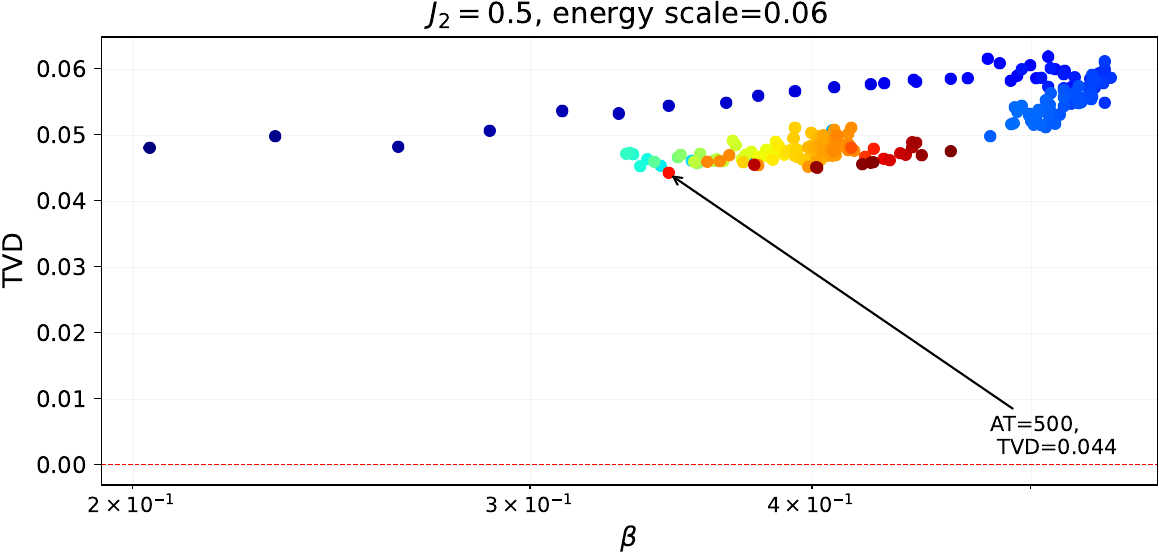}
    \includegraphics[width=0.495\linewidth]{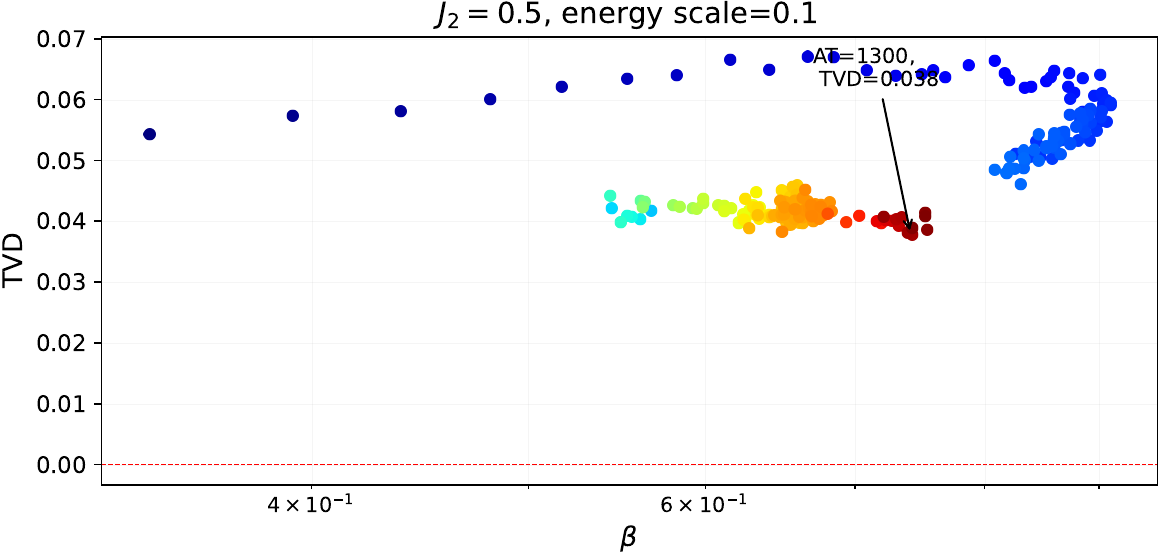}
    \includegraphics[width=0.495\linewidth]{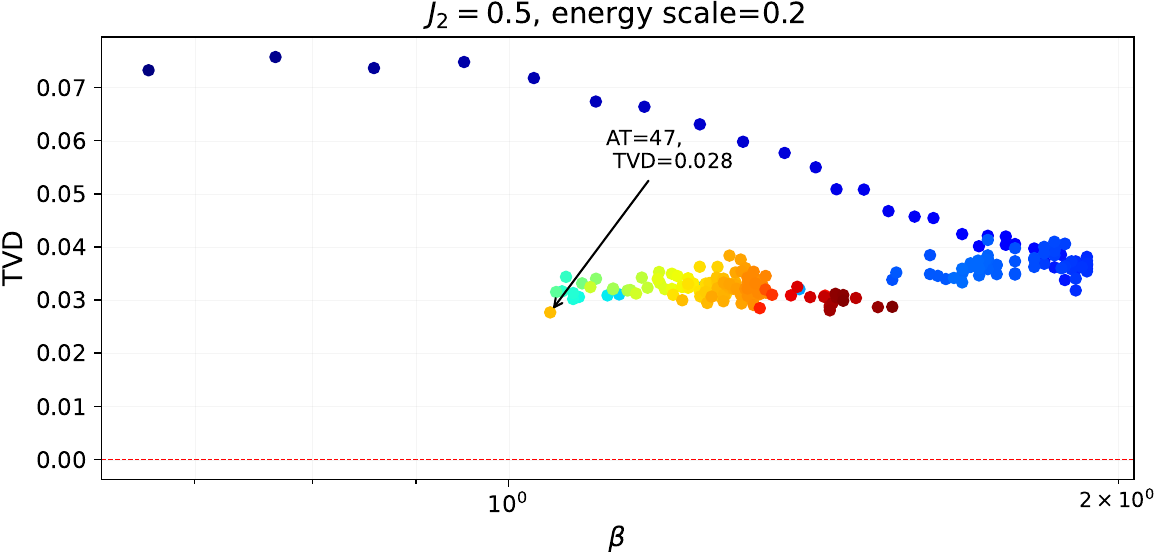}
    \includegraphics[width=0.495\linewidth]{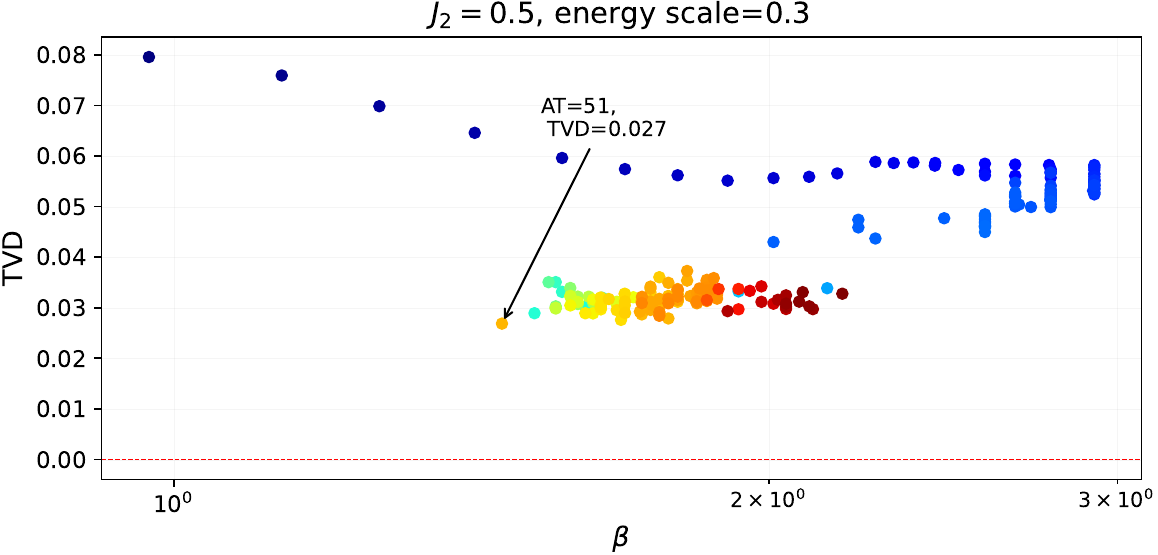}
    \includegraphics[width=0.495\linewidth]{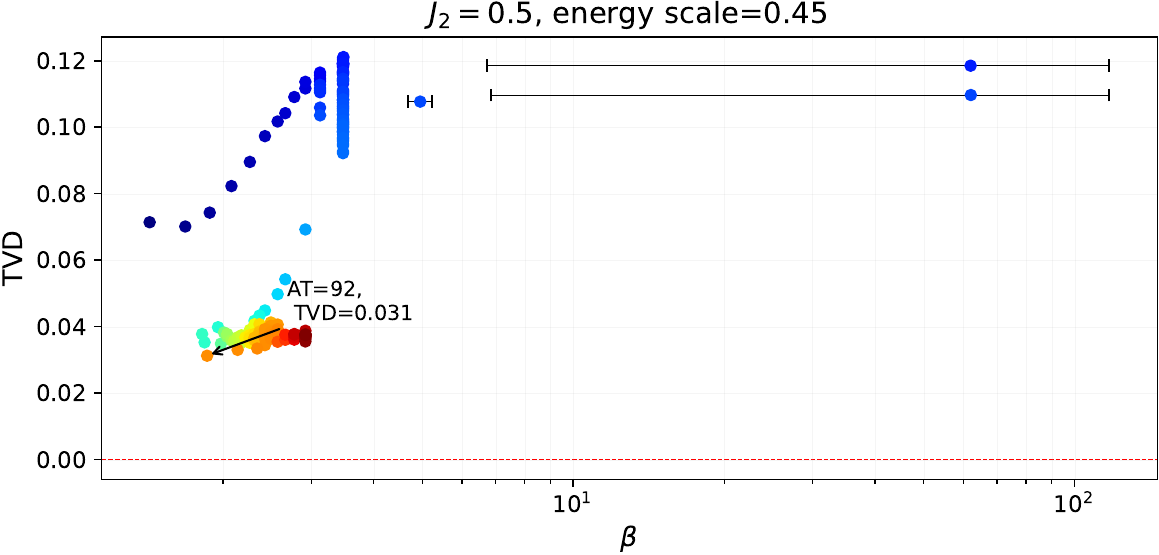}
    \includegraphics[width=0.495\linewidth]{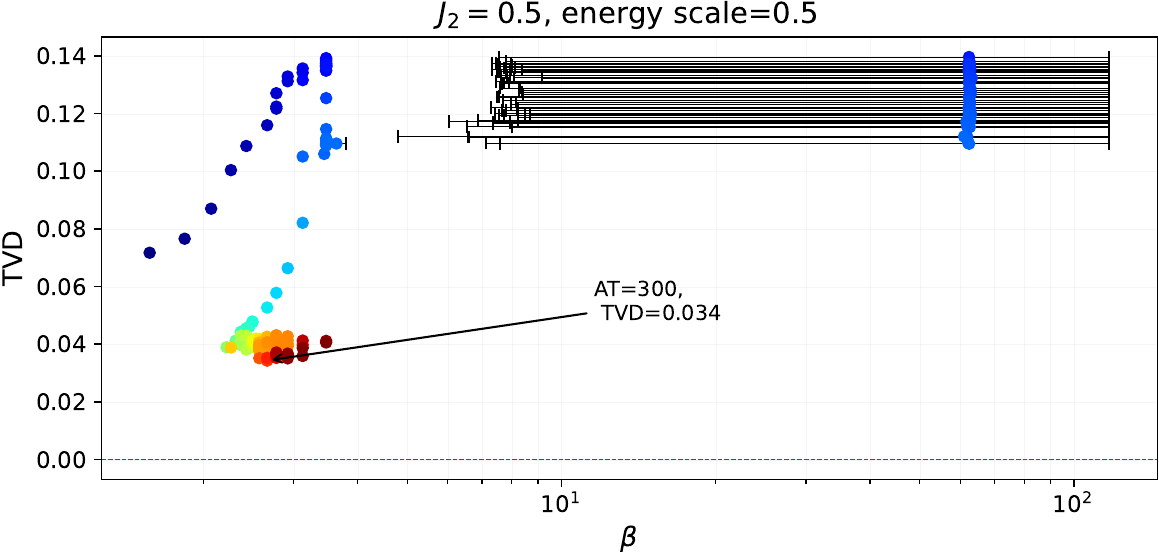}
    \includegraphics[width=0.495\linewidth]{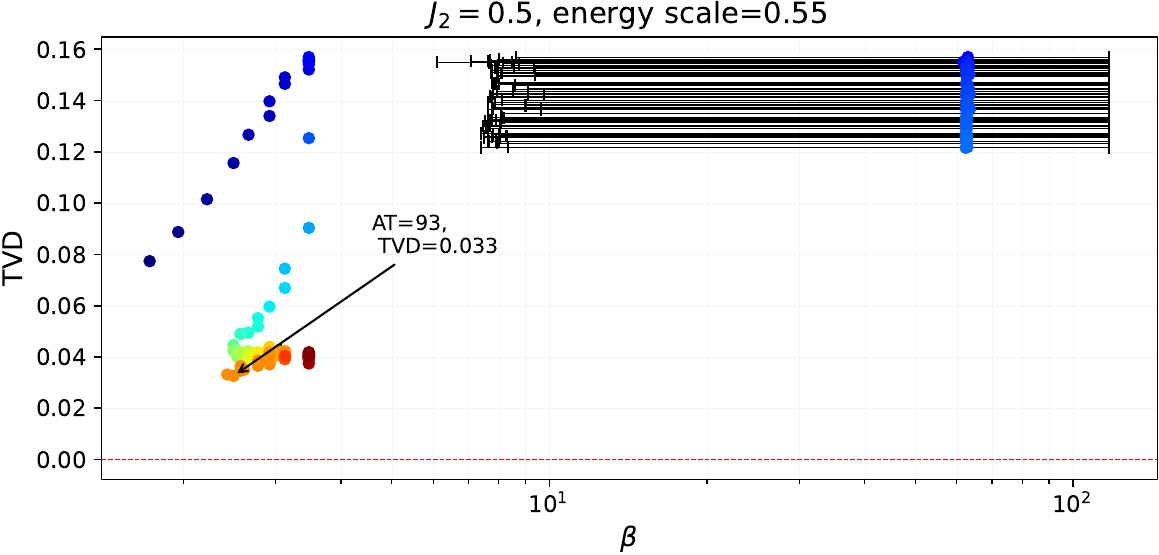}
    \includegraphics[width=0.495\linewidth]{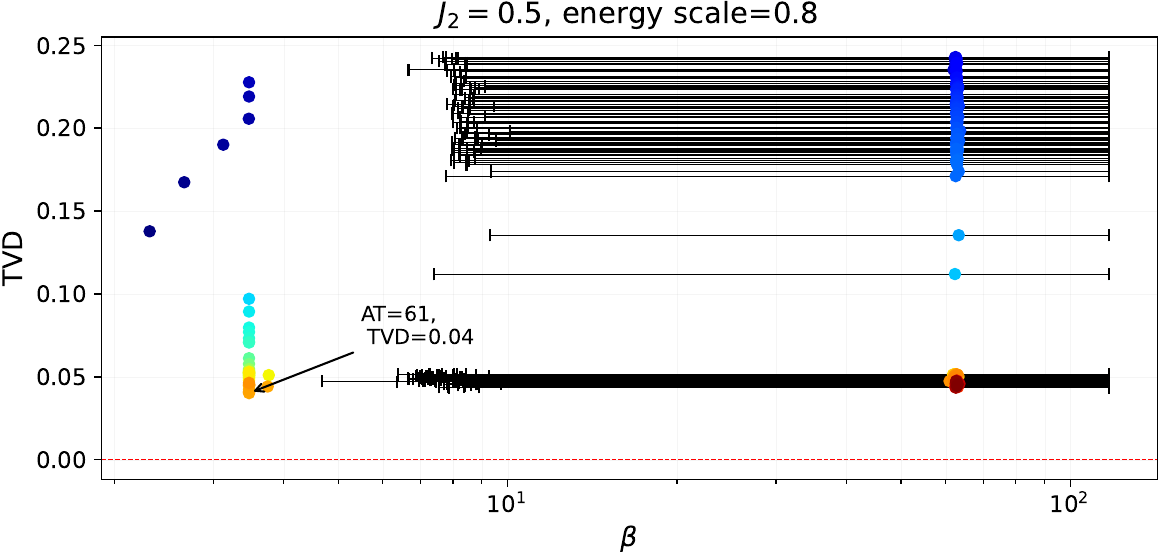}
    \includegraphics[width=0.495\linewidth]{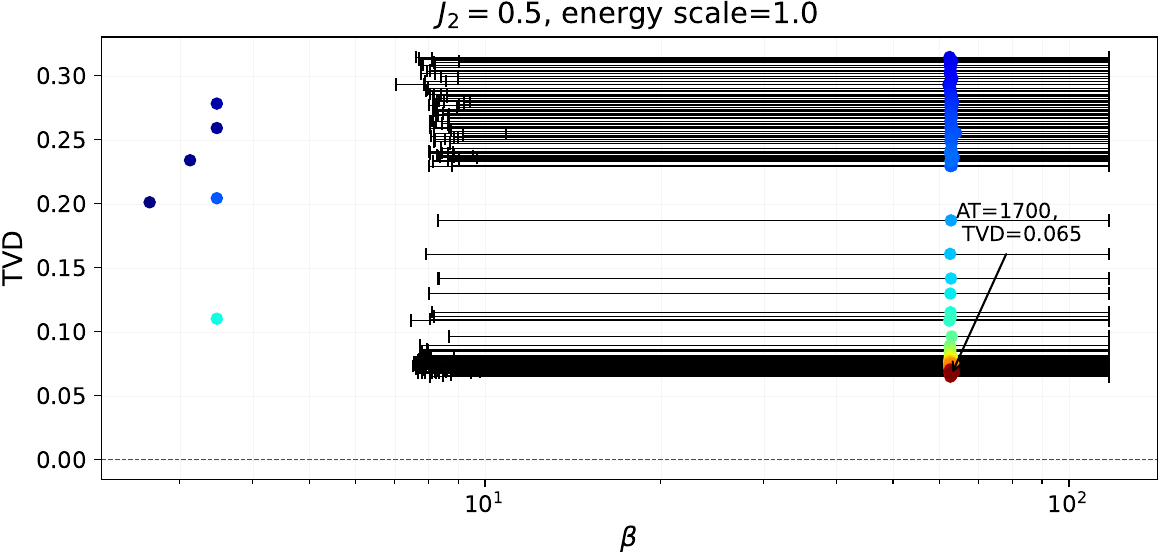}
    \includegraphics[width=0.6\linewidth]{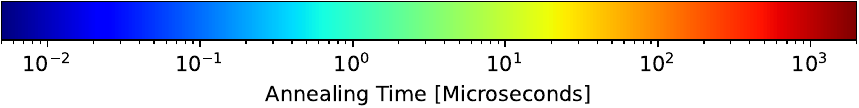}
    \caption{ Error rate (TVD) as a function of inverse temperature $\beta$, across the entire spectrum of evaluated annealing times (color coded by the log-scale heatmap below the sub-plots), for the ANNNI frustration parameter $J_2=0.5$ run on the \texttt{Advantage\_system4.1} processor. Each sub-plot corresponds to a different analog hardware energy scale, denoted in the title of each sub-plot. Error bars are shown on datapoints where there were many best-fitted $\beta$ values that spanned at least a range of $0.1$. The annealing time which had the absolute lowest error rate within each sub-plot is annotated on the plot with the exact annealing time, and the resulting TVD.  }
    \label{fig:beta_vs_TVD_J20.5_Pegasus}
\end{figure*}

\begin{figure*}[p!]
    \centering
    \includegraphics[width=0.495\linewidth]{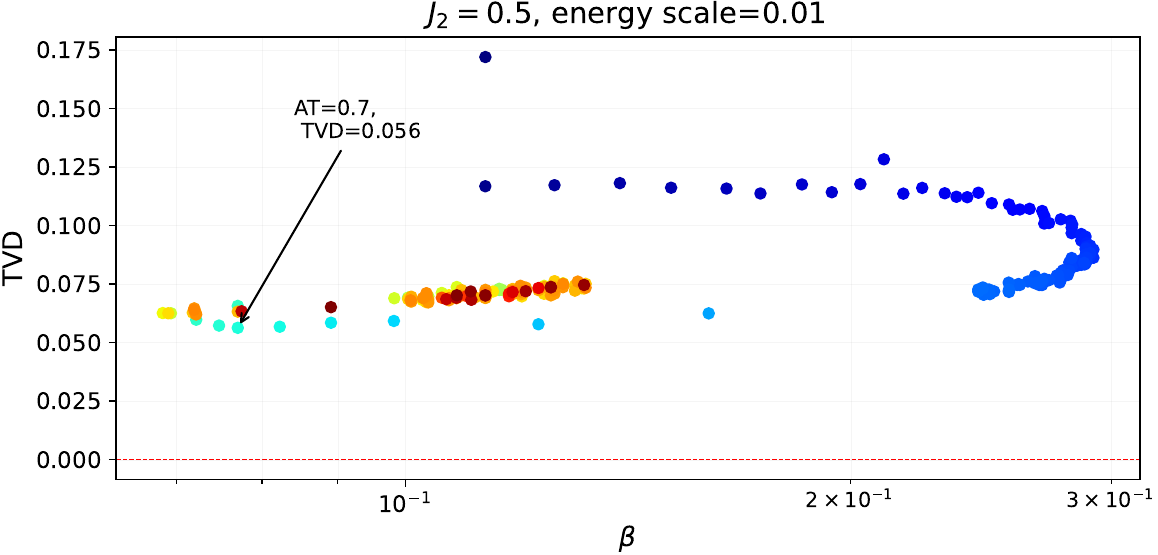}
    \includegraphics[width=0.495\linewidth]{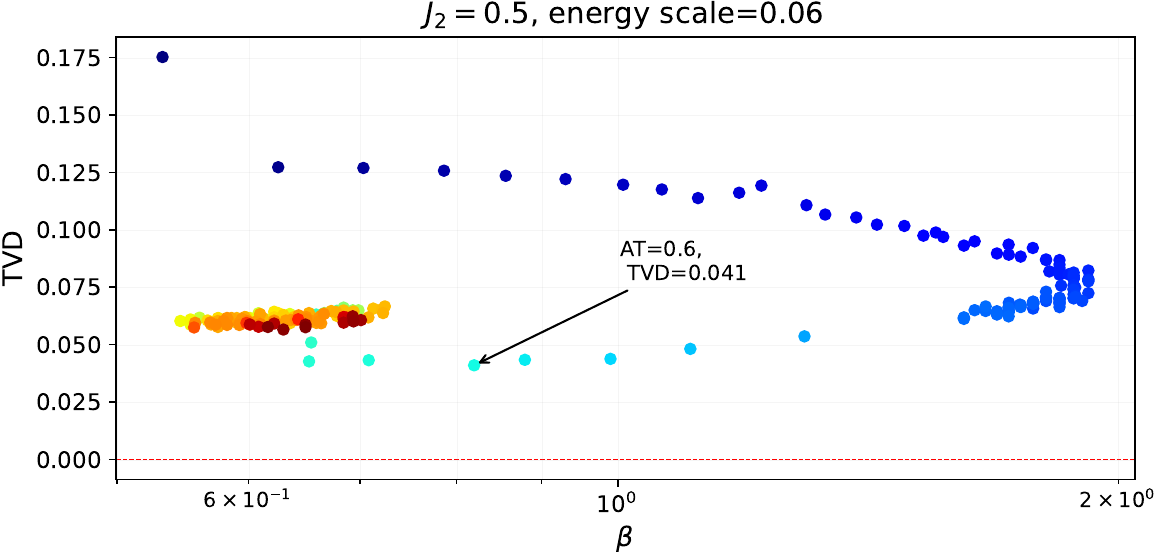}
    \includegraphics[width=0.495\linewidth]{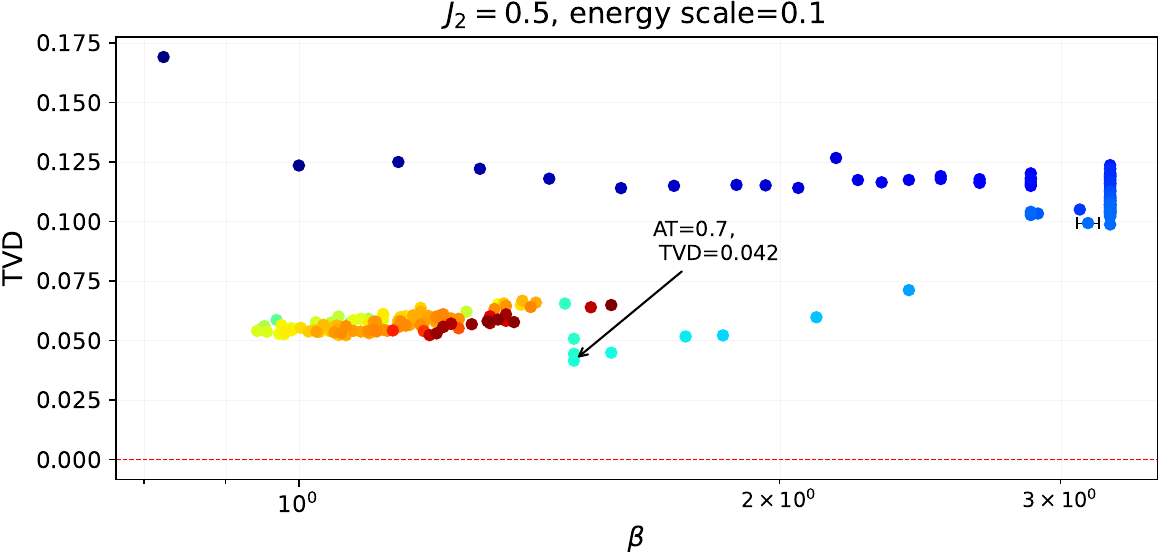}
    \includegraphics[width=0.495\linewidth]{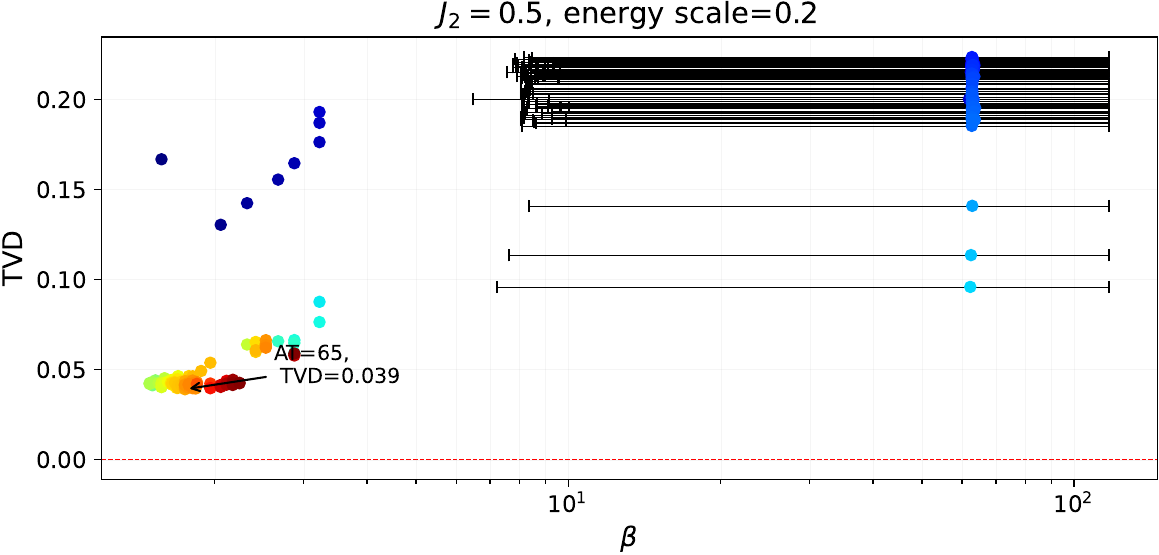}
    \includegraphics[width=0.495\linewidth]{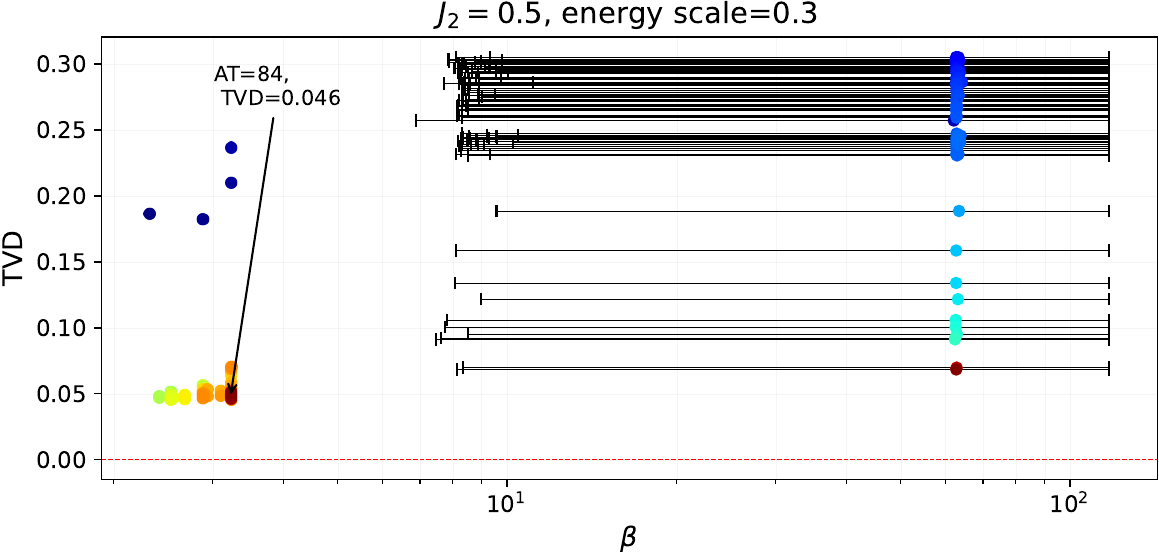}
    \includegraphics[width=0.495\linewidth]{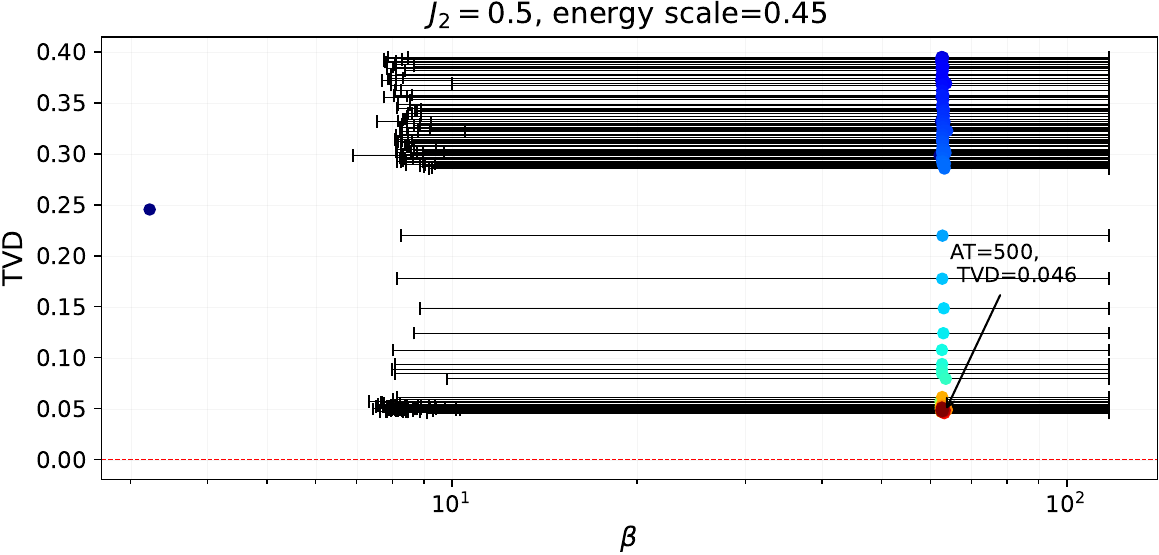}
    \includegraphics[width=0.495\linewidth]{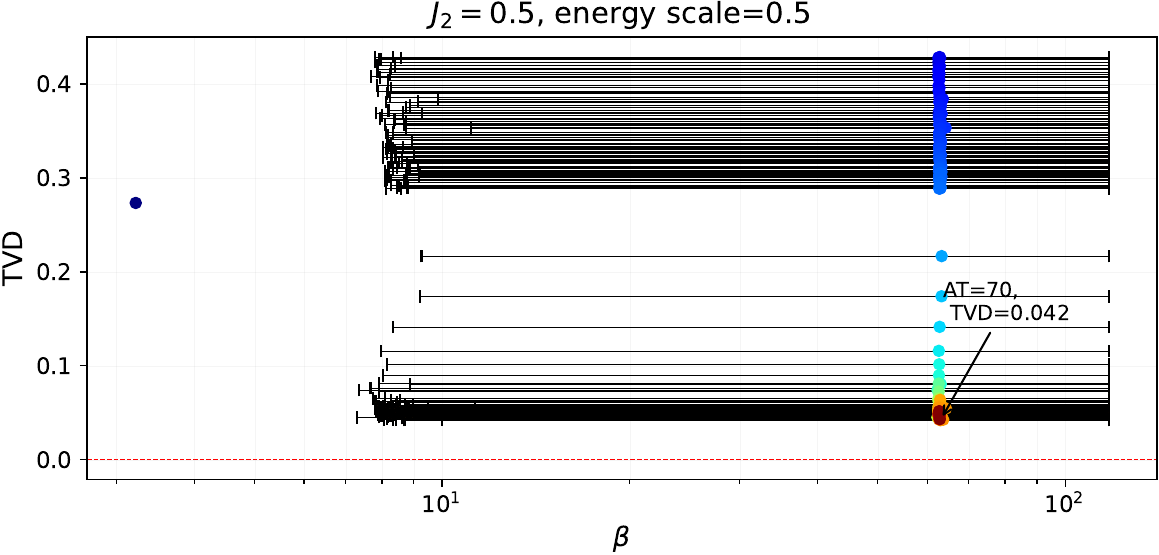}
    \includegraphics[width=0.495\linewidth]{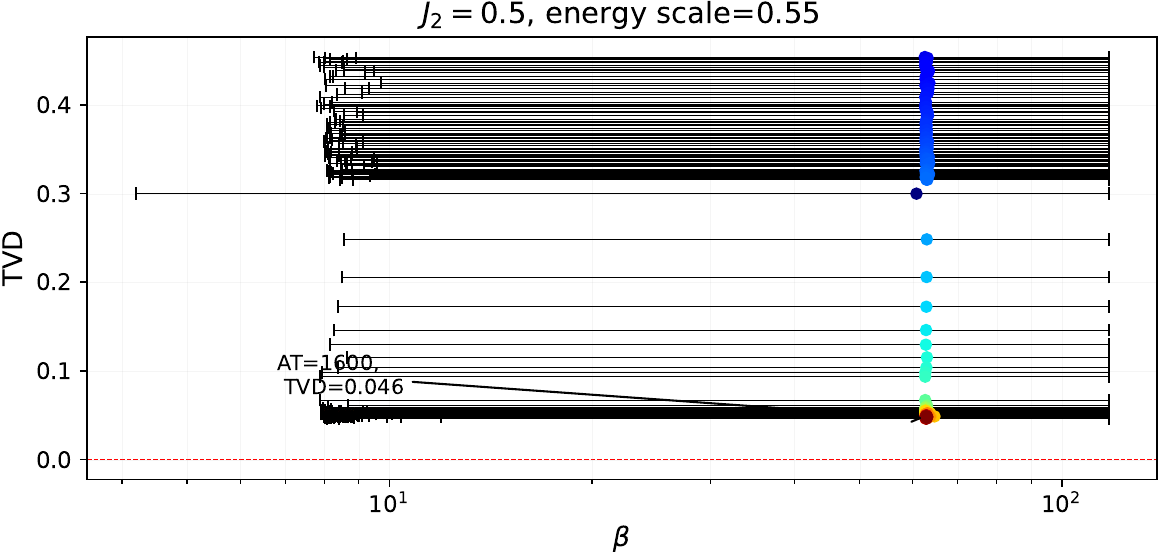}
    \includegraphics[width=0.495\linewidth]{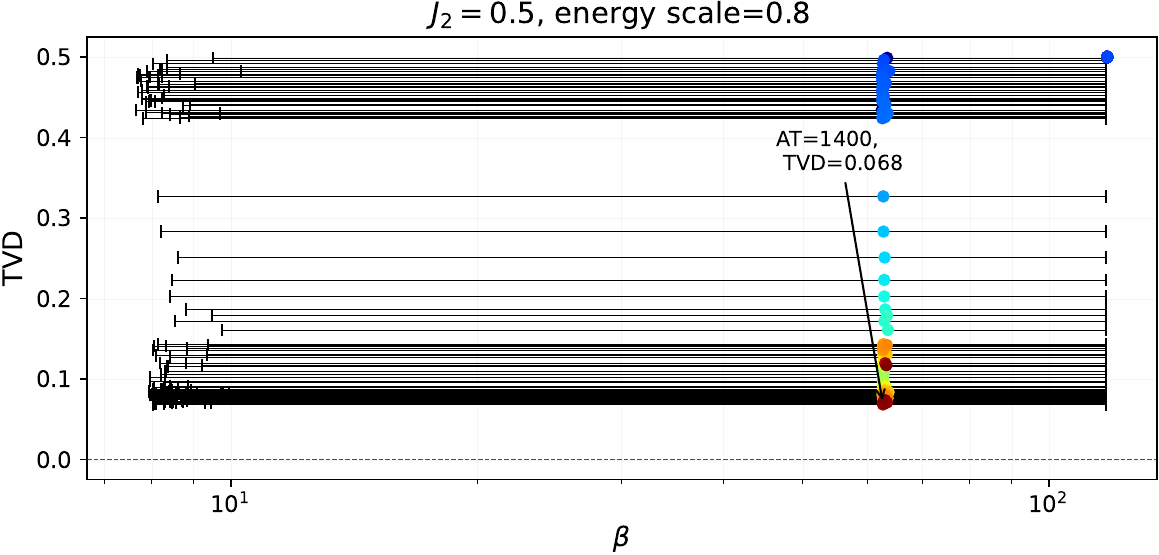}
    \includegraphics[width=0.495\linewidth]{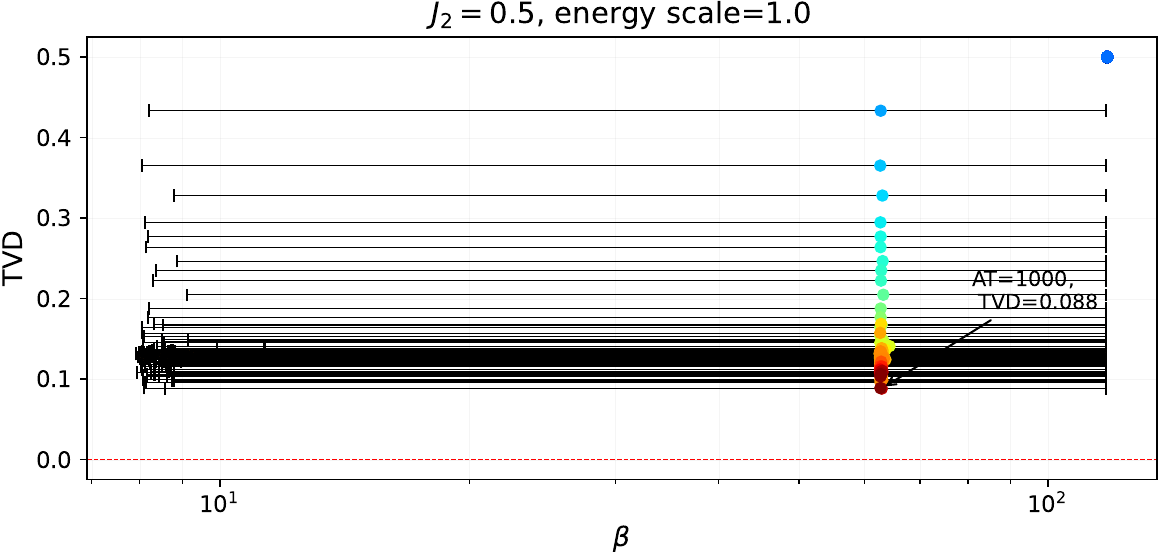}
    \includegraphics[width=0.6\linewidth]{figures/scatter_beta_vs_TVD/AT_colorbar.pdf}
    \caption{ Error rate (TVD) as a function of inverse temperature $\beta$, across the entire spectrum of evaluated annealing times (color coded by the log-scale heatmap below the sub-plots), for the ANNNI frustration parameter $J_2=0.5$ run on the \texttt{Advantage2\_system1.4} processor. Each sub-plot corresponds to a different analog hardware energy scale, denoted in the title of each sub-plot. Error bars are shown on datapoints where there were many best-fitted $\beta$ values that spanned at least a range of $0.1$. The annealing time which had the absolute lowest error rate within each sub-plot is annotated on the plot with the exact annealing time, and the resulting TVD.  }
    \label{fig:beta_vs_TVD_J20.5_Zephyr}
\end{figure*}

Interestingly, we see that in general the highest error rates are seen for the critical frustration point at $J_2=0.5$, or the models with frustration near the critical point with $J_2=0.49, 0.51$. 

\begin{figure*}[p!]
    \centering
    \includegraphics[width=0.495\linewidth]{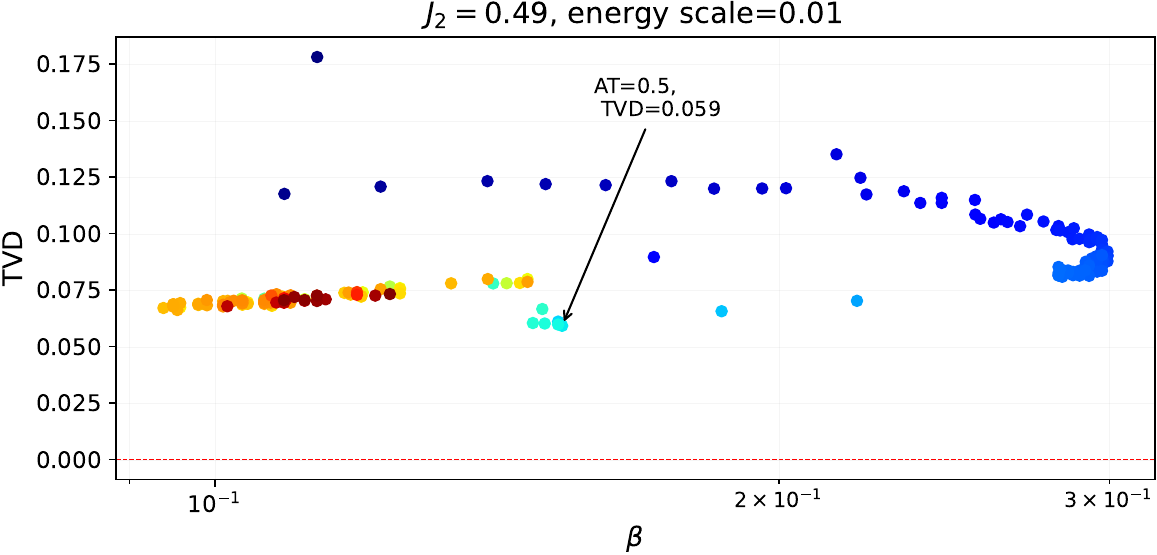}
    \includegraphics[width=0.495\linewidth]{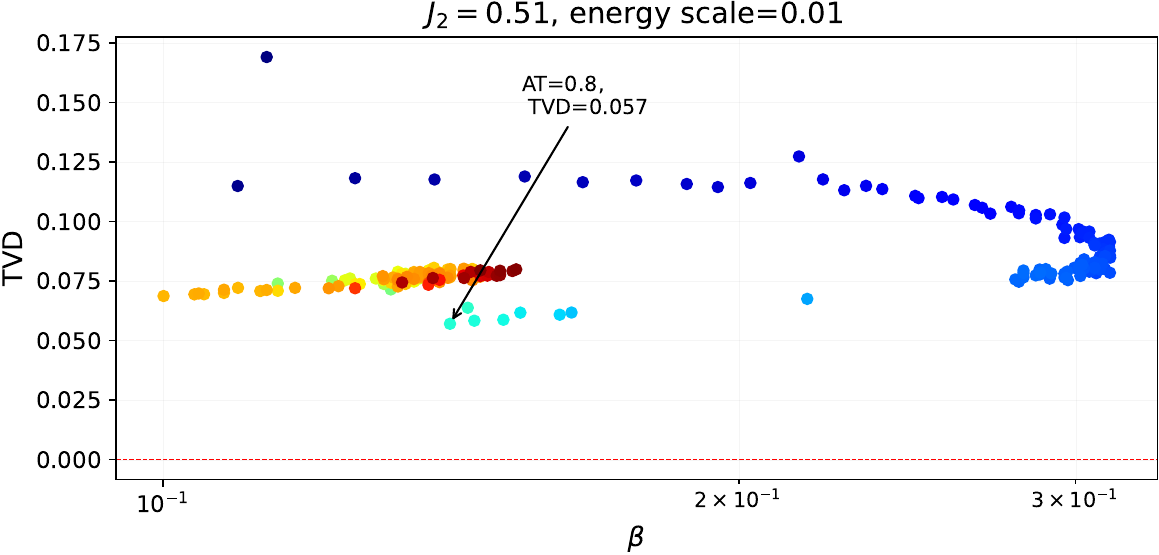}
    \includegraphics[width=0.495\linewidth]{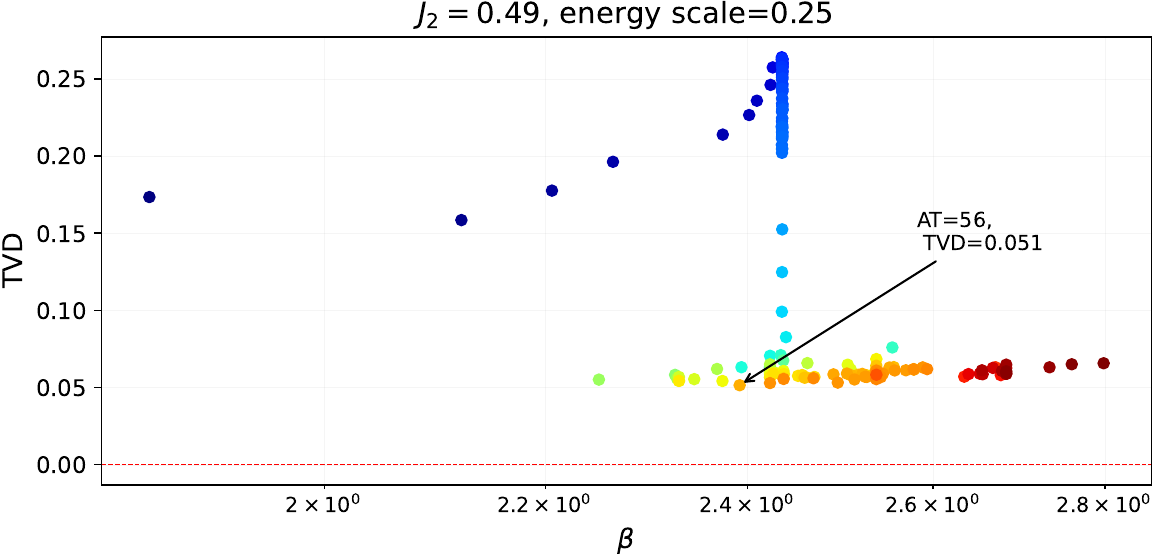}
    \includegraphics[width=0.495\linewidth]{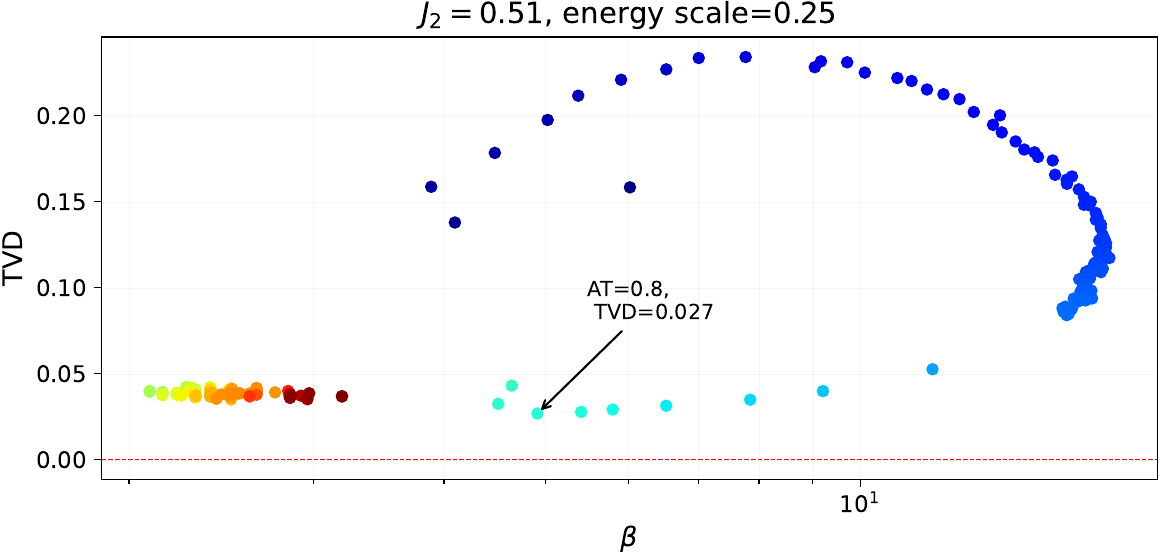}
    \includegraphics[width=0.495\linewidth]{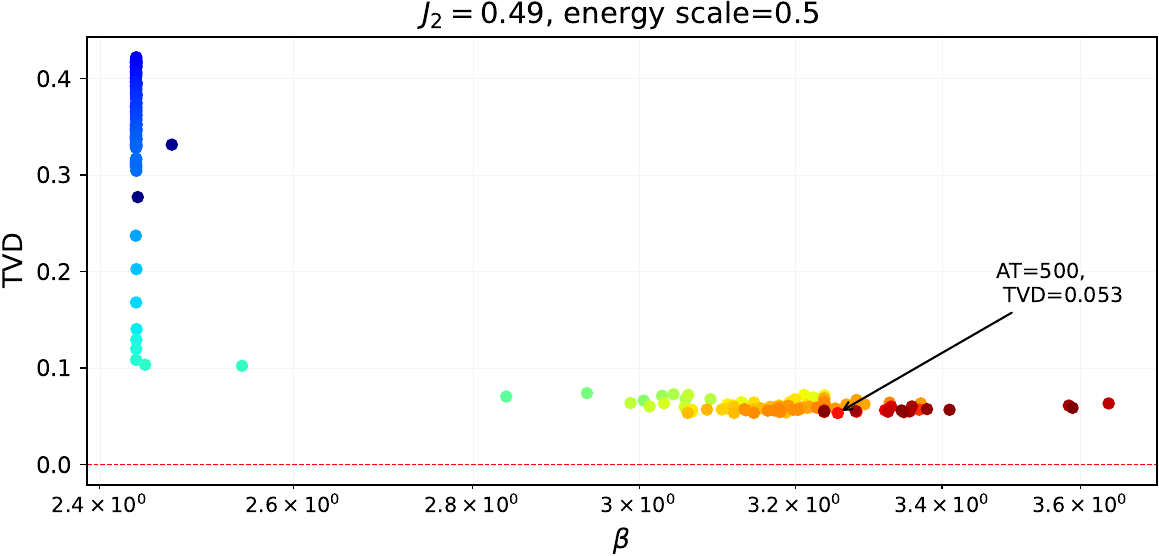}
    \includegraphics[width=0.495\linewidth]{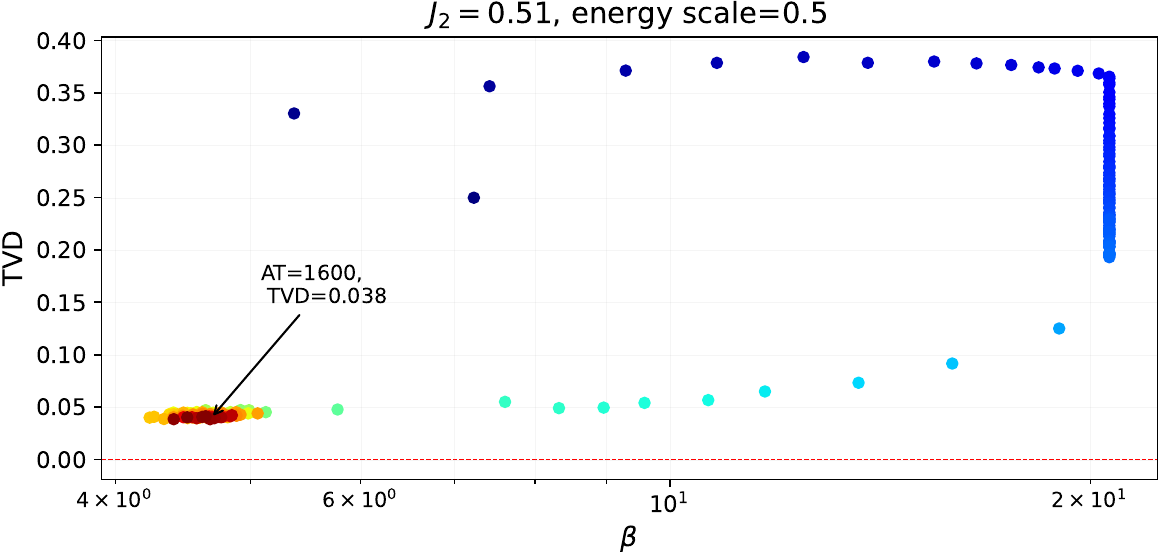}
    \includegraphics[width=0.495\linewidth]{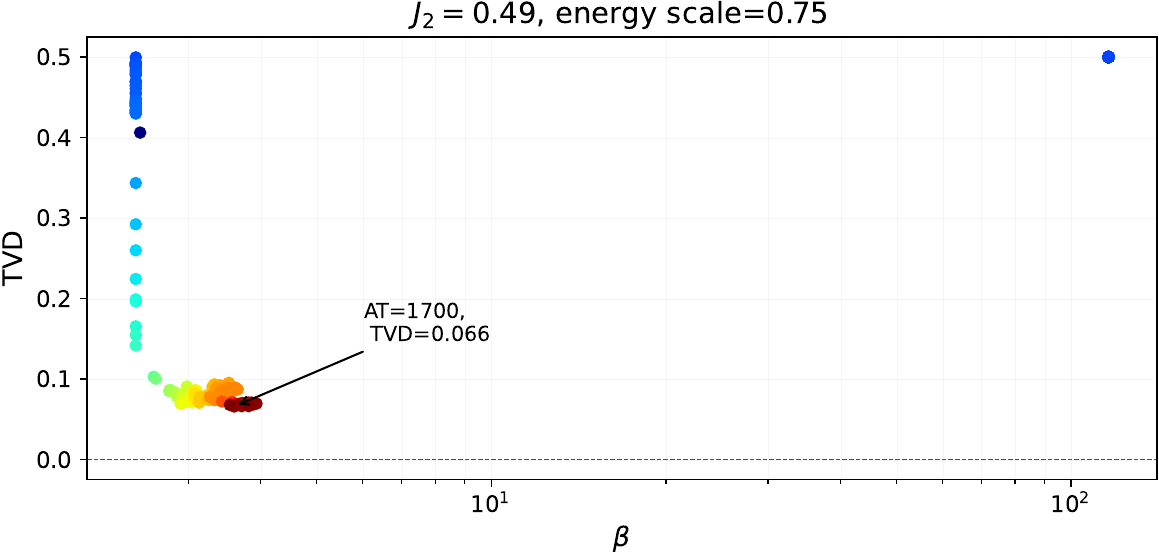}
    \includegraphics[width=0.495\linewidth]{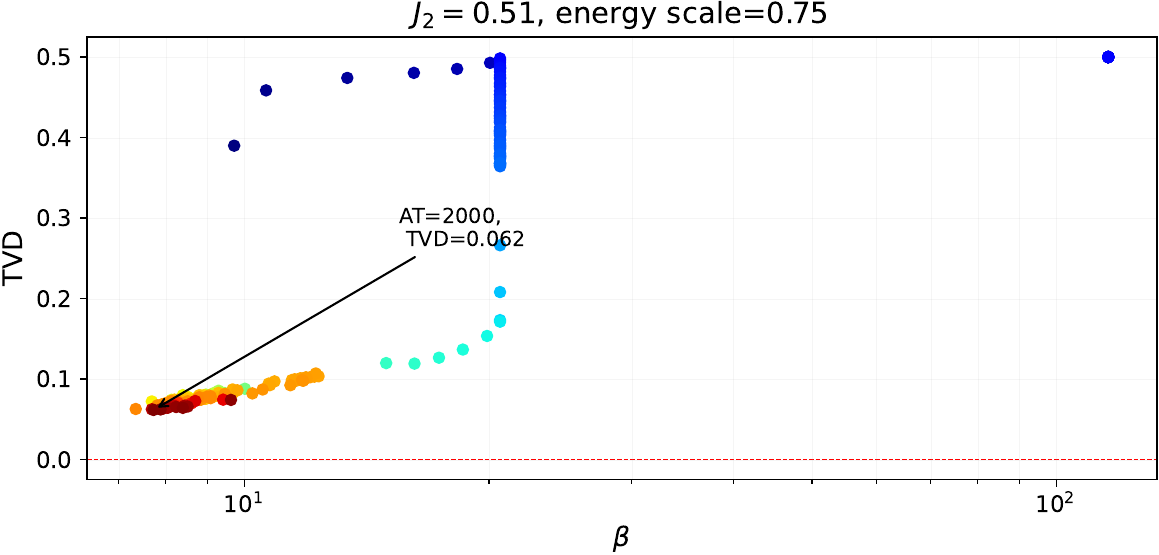}
    \includegraphics[width=0.495\linewidth]{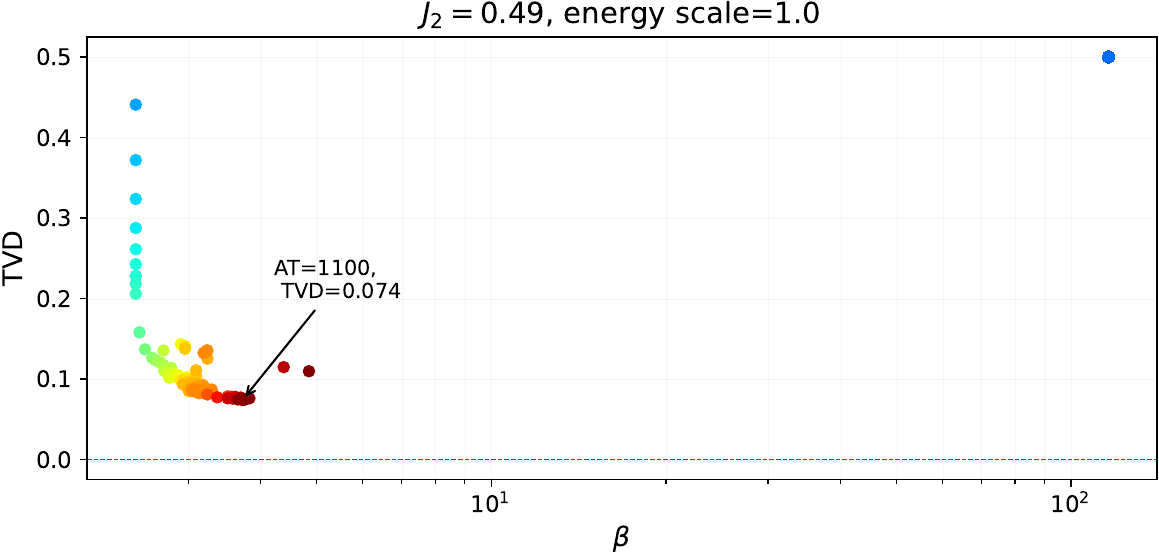}
    \includegraphics[width=0.495\linewidth]{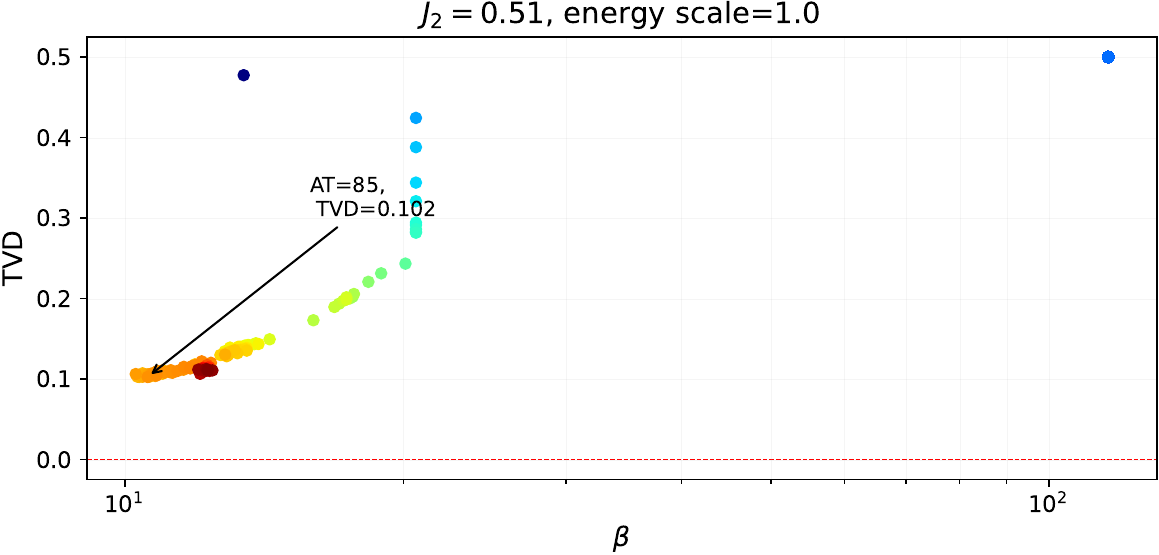}
    \includegraphics[width=0.6\linewidth]{figures/scatter_beta_vs_TVD/AT_colorbar.pdf}
    \caption{ \texttt{Advantage2\_system1.4}, comparing results for the ANNNI frustration parameters $J=0.49$ (left column) compared to $J=0.51$ (right column) at different overall coupler energy scales (rows).  }
    \label{fig:beta_vs_TVD_Zephyr_0.49_0.51}
\end{figure*}

\begin{figure*}[p!]
    \centering
    \includegraphics[width=0.495\linewidth]{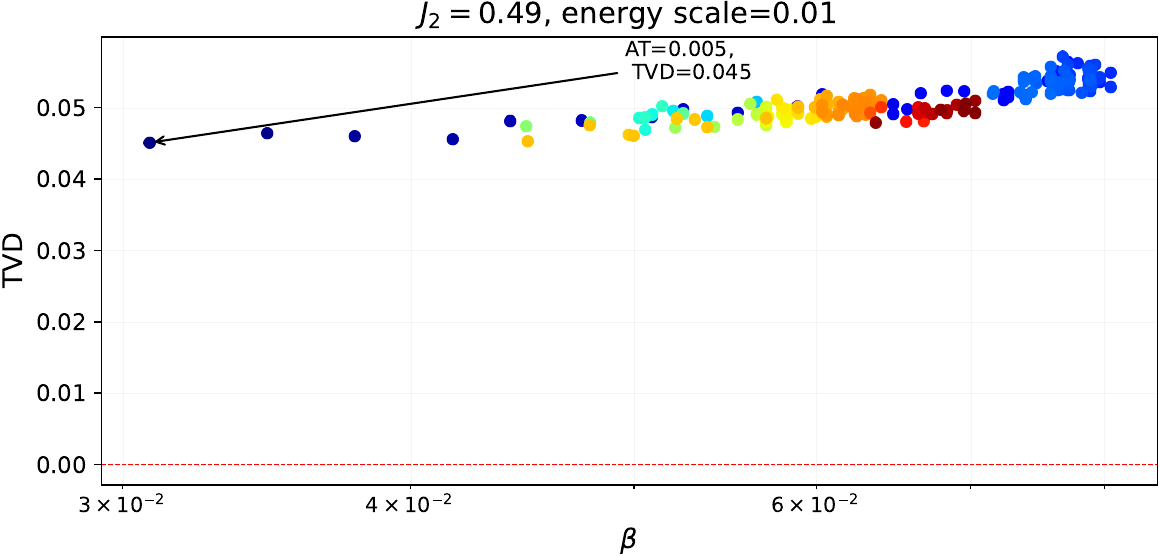}
    \includegraphics[width=0.495\linewidth]{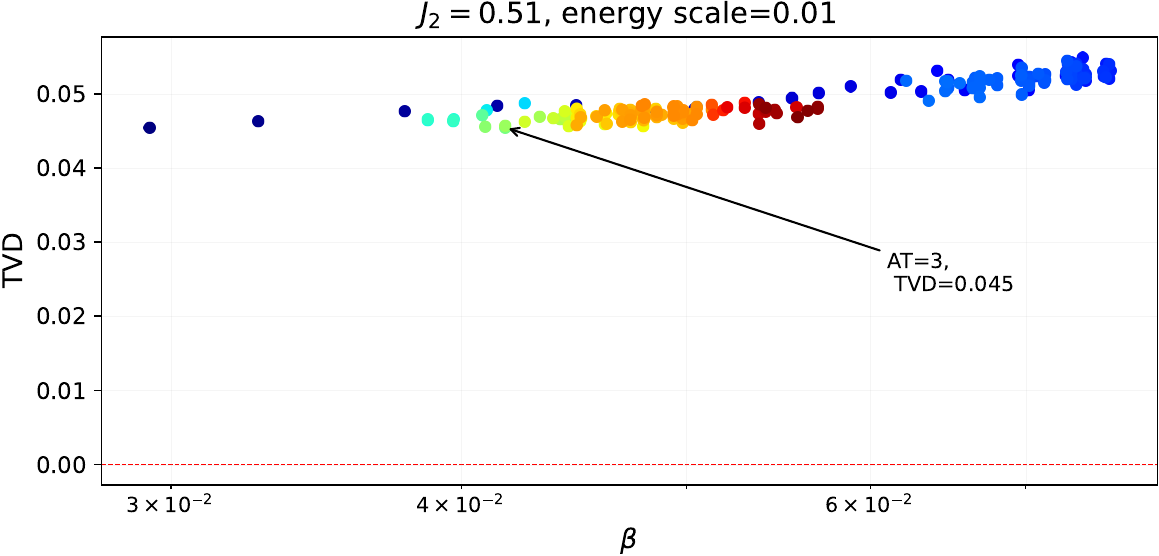}
    \includegraphics[width=0.495\linewidth]{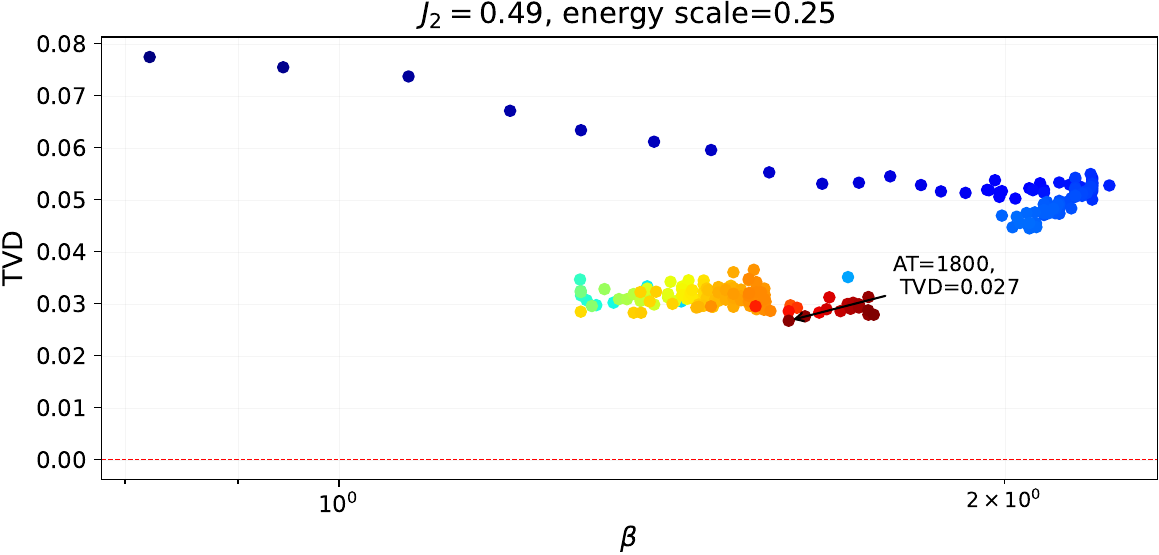}
    \includegraphics[width=0.495\linewidth]{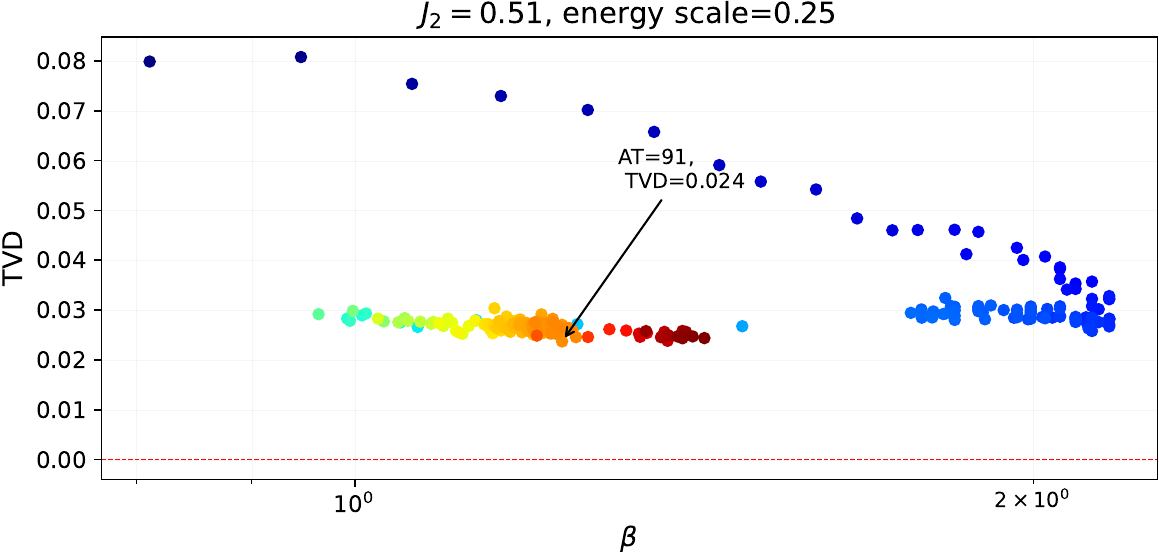}
    \includegraphics[width=0.495\linewidth]{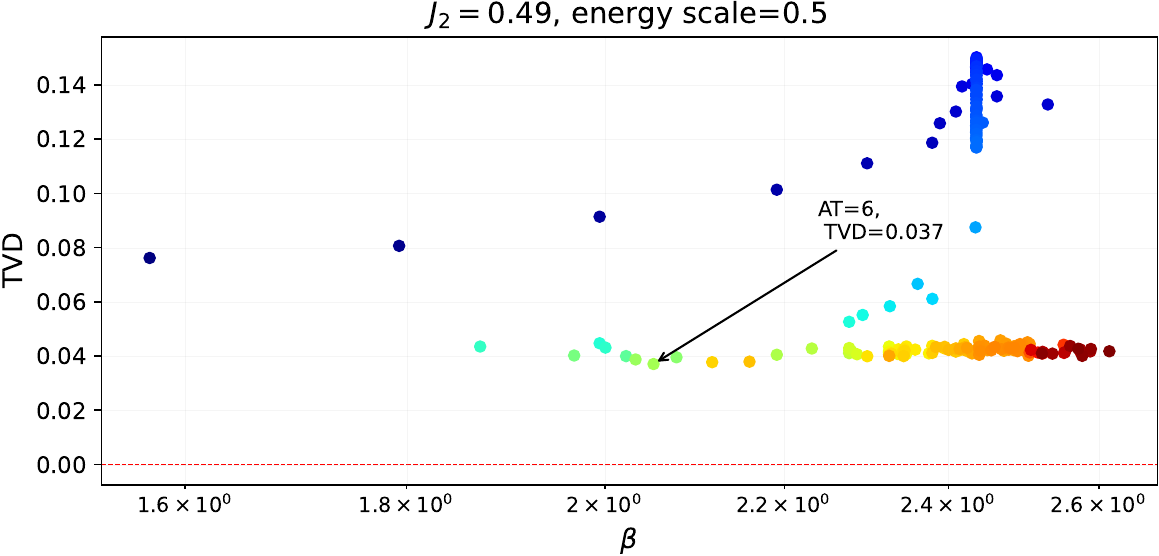}
    \includegraphics[width=0.495\linewidth]{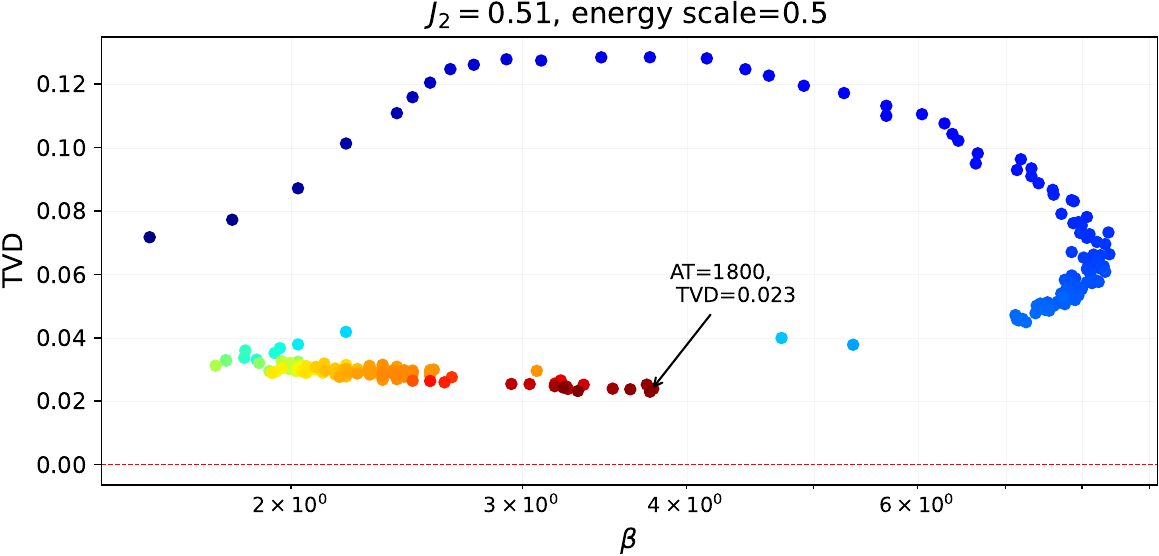}
    \includegraphics[width=0.495\linewidth]{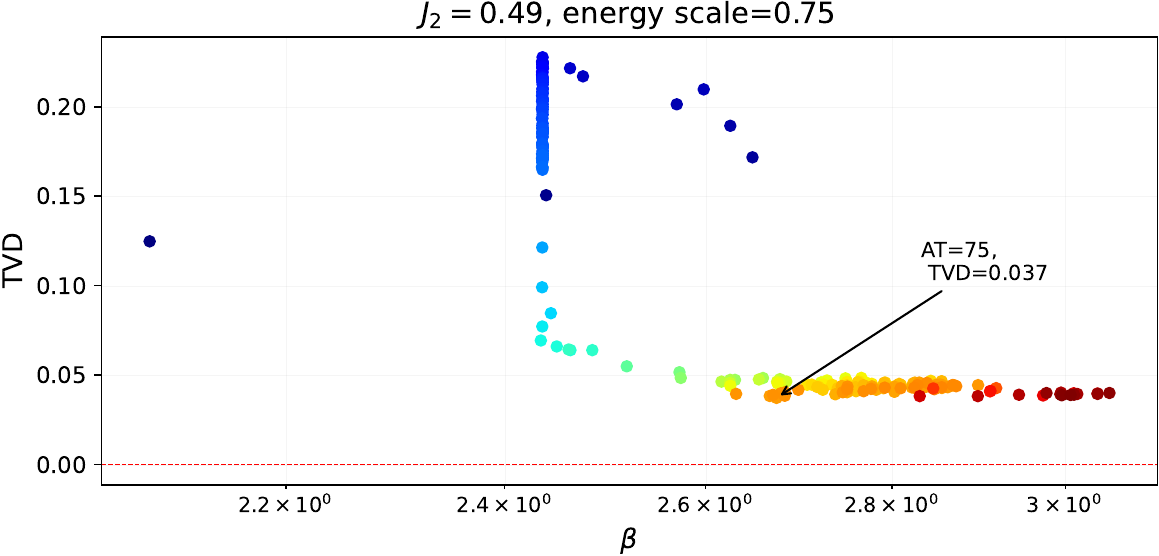}
    \includegraphics[width=0.495\linewidth]{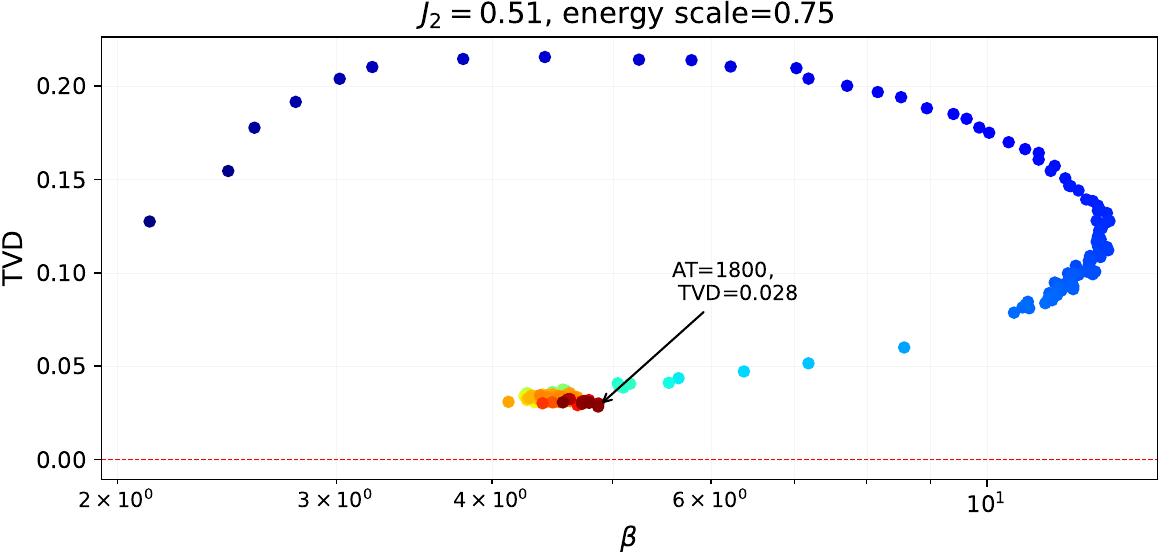}
    \includegraphics[width=0.495\linewidth]{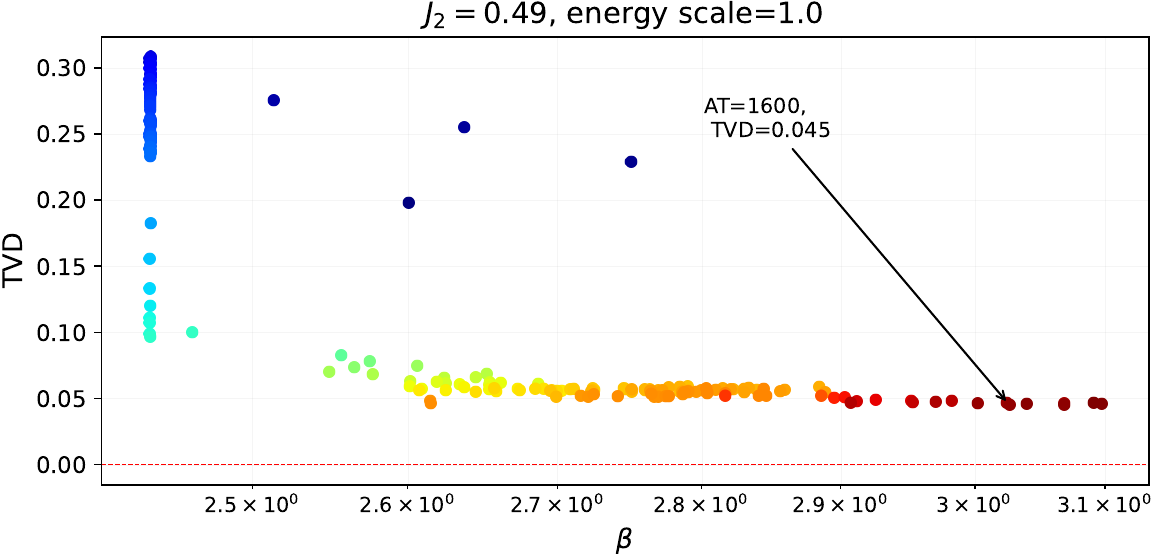}
    \includegraphics[width=0.495\linewidth]{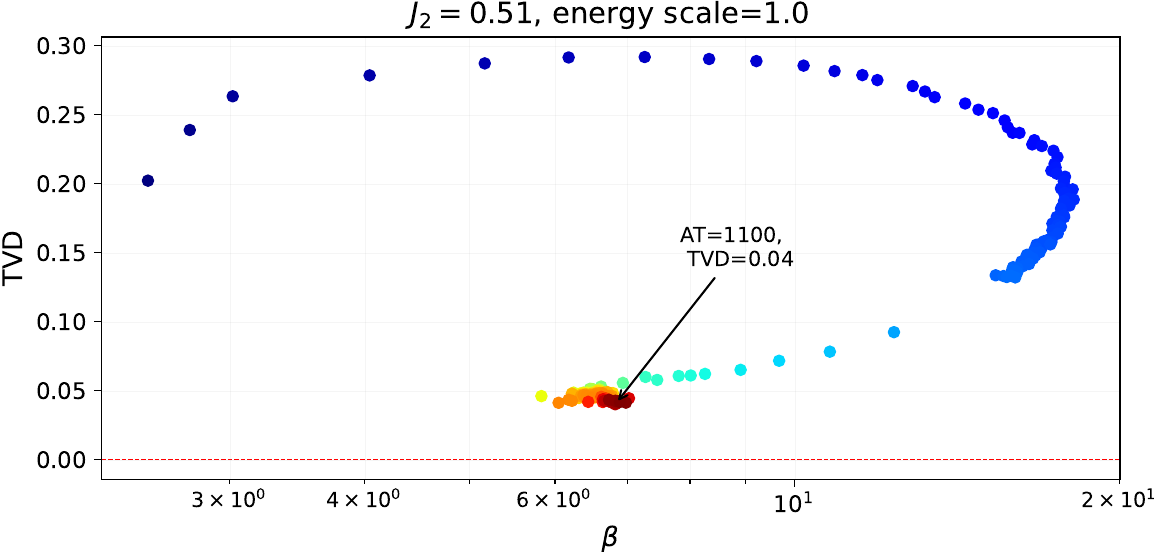}
    \includegraphics[width=0.6\linewidth]{figures/scatter_beta_vs_TVD/AT_colorbar.pdf}
    \caption{ \texttt{Advantage\_system4.1}, comparing results for the ANNNI frustration parameters $J=0.49$ (left column) compared to $J=0.51$ (right column) at different overall coupler energy scales (rows).  }
    \label{fig:beta_vs_TVD_Pegasus_0.49_0.51}
\end{figure*}

Figure~\ref{fig:beta_vs_TVD_J20.5_Pegasus} and Figure~\ref{fig:beta_vs_TVD_J20.5_Zephyr} provide more detail on the sampling characteristics over different annealing times, in the form of scatterplots between the best-fitted TVD versus $\beta$, where each point on the sub-plots is a different annealing time. These show that the sample distribution changes dramatically as both annealing time and the entire $J$ energy scale is changed. Figure~\ref{fig:beta_vs_TVD_J20.5_Pegasus} and Figure~\ref{fig:beta_vs_TVD_J20.5_Zephyr} show specifically simulations from the ANNNI model at the critical frustration point, and the general quality of these simulations is that the TVD error rate is quite high. Notably, these plots show that the estimated $\beta$ spans a wide range of possible values (mostly at higher temperatures) -- overall showing that for most of these energy scales (except at very small values, less than $\approx 0.2$) there is significant uncertainty of the effective temperature of the sampled distributions. This holds true for a majority of annealing times on the \texttt{Advantage2\_system1.4} processor, and these estimated temperature ranges when the overall energy scale is greater than $\approx 0.5$ typically span more than an order of magnitude. In the temperature fitting process, what this corresponds to is that there is a wide range of inverse temperatures which all result in the same TVD, up to numerical precision. This result is a fairly unique characteristic of sampling at this high-degeneracy critical frustration point at $J_2=0.5$ -- looking at slightly different $J_2$ frustration parameters such as Figure~\ref{fig:beta_vs_TVD_Zephyr_0.49_0.51} and Figure~\ref{fig:beta_vs_TVD_Pegasus_0.49_0.51} there is not this prevalence of temperature distribution fitting uncertainty. This suggests that this specific inverse temperature fitting uncertainty resulting from high energy scale quantum annealer sampling is a sensitive probe of critical frustration points of spin models, and could be especially prominent because of finite sampling. For context, the degeneracy at $J_2=0.5$ is $(\frac{1 + \sqrt{5}}{2})^{12} \approx 322$ (the golden ratio to the power of $N$)~\cite{selke1988annni, suzuki2012quantum}. 

Next, Figure~\ref{fig:beta_vs_TVD_Zephyr_0.49_0.51} directly compares ANNNI frustration parameters slightly into the ferromagnetic phase ($J_2=0.49$) and slightly into the antiphase (antiferromagnetic-like) region ($J_2=0.51$), at different analog hardware energy scales run on the Zephyr graph QPU. Figure~\ref{fig:beta_vs_TVD_Pegasus_0.49_0.51} does the same, but on the Pegasus graph QPU. The primary notable observation is that these two different ANNNI models do exhibit dramatically different characteristics. This suggests that, as approximate Boltzmann samplers, analog quantum annealers are sensitive to frustrated models at different regions in a magnetic phase diagram -- in particular, separated by a critical frustration point. Generally, when the energy scale on the hardware is larger, the error rates become much larger for faster annealing times and lower for longer annealing times.

Because the very fast anneals used here, e.g., down to $5$ nanoseconds, are within the coherence time of the processor, these computations are coherent quantum quenches. It is therefore interesting to note that some of the very fast quenches produce distributions that approximate a Boltzmann distribution of the classical ANNNI model reasonably well -- see for example the low-energy scale plots in Figure~\ref{fig:beta_vs_TVD_Pegasus_0.49_0.51} for the frustration parameter of $J_2=0.49$. The important caveat to this observation is that at the full energy scale of the J couplers, the distributions that are produced are very much not thermal -- in order for these fast anneals to approximately sample thermal distributions, the coupling energy scale must be reduced. Because of the quench dynamics of these fast anneals, it is plausible that the distributions that are produced happen to be close to relatively high temperature distributions. Importantly, the D-Wave quantum annealers can not measure in the X basis, or any basis except the computational Z basis. Therefore, we are not able to examine what sorts of dynamics are occurring here to lead to relatively low-error rate approximations of classical Boltzmann distributions. Nonetheless, this is an interesting property in particular because of the extremely fast simulation times, and therefore the overall low amount of QPU time used to obtain these approximate thermal distributions. One way of interpreting these very fast anneal-quenches is that they are effectively quenching through the quantum magnetic phase diagram of the quantum ANNNI model\footnote{perhaps analogous to similar types of quenches performed in ref.~\cite{PhysRevX.11.031062}, the difference being in that study the quench was performed across frustration parameters, such as from the ferromagnetic phase into the antiphase}, where the transverse field starts out dominating the system and then is very quickly quenched to zero transverse field.

\section{Discussion and Conclusion}
\label{section:conclusion}

This study has shown that analog quantum computers, specifically noisy superconducting qubit quantum annealers, are good Boltzmann samplers of highly frustrated Ising models such as the ANNNI model at various magnetic frustration parameters. Remarkably, this includes being able to sample at very low error rate (TVD $0.0003$) the ANNNI model at $J_2=1$ at a low temperature ($\beta = 32.166$). 

Interestingly, the sampling did become worse at and near the critical frustration point of $J_2=0.5$. This suggests that very high degeneracy and frustration does pose a problem for the analog hardware to accurately sample from an unbiased Boltzmann distribution. Moreover, the sampling characteristics are quite sensitive to where the ANNNI model is in its magnetic phase diagram with respect to the frustration parameter parameter $J_2$; namely, clear differences are seen at, and on either side of, the critical frustration point $J_2=0.5$. This suggests that sampling on the D-Wave hardware, regardless of how good of a thermal sampler it is, could be a probe of critical frustration points in magnetic models. An interesting topic of future study is the role of degeneracy lifting in the transverse field driven D-Wave QPU sampling of the ANNNI model. For example, at the critical frustration point if there is substantial degeneracy lifting of some of the ground-state configurations, that could explain why the sampling is worse at that point. Moreover, a lack of degeneracy lifting within the antiphase region ($J_2=1$) could explain why the noisy thermal sampling is very low error rate in that region of the ANNNI model. This is speculative; the role of degeneracy lifting was not quantified in the reported results.

The clearest open question is whether this Boltzmann sampling technique on D-Wave quantum annealers can be significantly scaled up in system size, and whether those simulations can compete with state of the art classical heuristic methods, namely the many monte carlo variants. The primary open question here is whether there are diagnostics that can be applied to the D-Wave QPU simulations which are proxies for the quality of the thermal sampling being performed. For sufficiently large system sizes, classical monte carlo methods may begin to struggle, and so an interesting open question is whether there are techniques that can be applied in order to validate equivalent D-Wave QPU sampling. There is no reason to expect that D-Wave quantum annealers will begin to fail at thermal sampling at larger system sizes; prior studies have demonstrated Gibbs sampling on reasonably large Ising models~\cite{sathe2025classicalcriticalityquantumannealing}, and there is no indication of systematic errors that would hinder the D-Wave hardware. It would be very interesting to show that D-Wave analog quantum computers can outperform state of the art classical methods for sufficiently large system sizes at the task of low-temperature sampling the Boltzmann distribution of a frustrated $J_1-J_2$ Ising model, in particular at known frustration points in the models phase diagram. Related to this, it would be useful to validate whether the best-performing D-Wave analog hardware parameters, shown in Table~\ref{table:lowest_error_rate_params}, are size-independent. If this is the case, then that would show that D-Wave hardware is a very valuable Boltzmann sampler -- being able to sample from the Boltzmann distribution of the frustrated ANNNI model at significantly larger system sizes.

\section*{Acknowledgments}
\label{sec:acknowledgments}
This work was supported by the U.S. Department of Energy through the Los Alamos National Laboratory. Los Alamos National Laboratory is operated by Triad National Security, LLC, for the National Nuclear Security Administration of U.S. Department of Energy (Contract No. 89233218CNA000001). The research presented in this article was supported by the Laboratory Directed Research and Development program of Los Alamos National Laboratory under project number 20240032DR. This research used resources provided by the Los Alamos National Laboratory Institutional Computing Program. The author would also like to thank the New Mexico Consortium, under subcontract C2778, the Quantum Cloud Access Project (QCAP), for providing quantum computing resources. LA-UR-25-30042.


\appendix

\begin{figure}[ht!]
    \centering
    \includegraphics[width=1.0\linewidth]{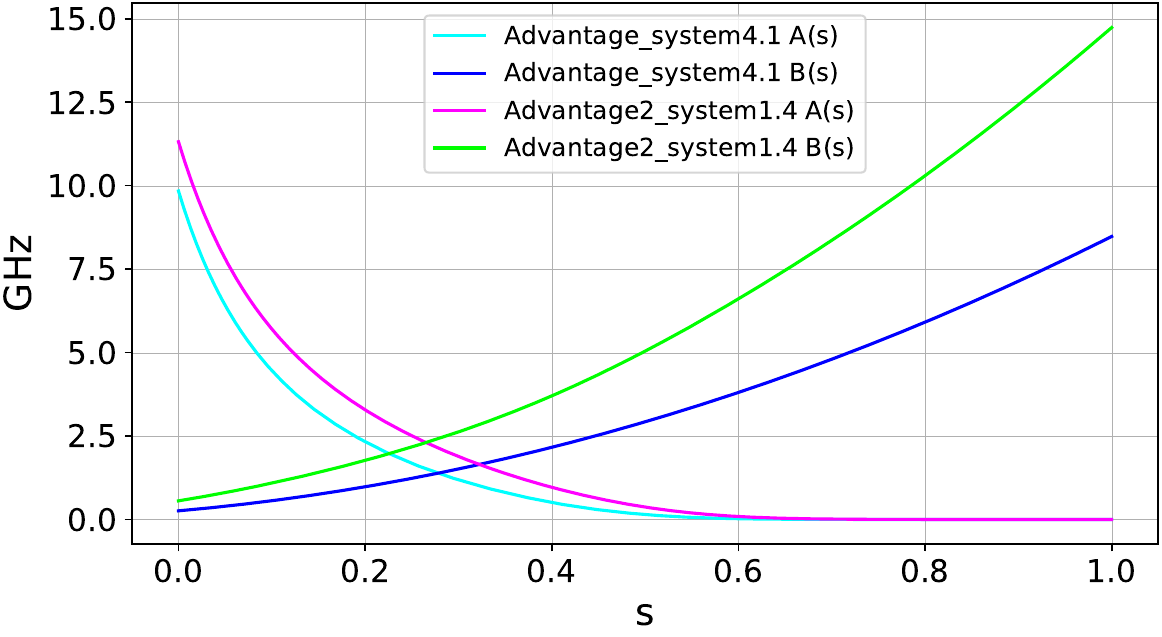}
    \caption{D-Wave QPU hardware time-dependent Hamiltonian energy scale functions $A(s)$ and $B(s)$. }
    \label{fig:DWave_hardware_energy_scales}
\end{figure}

\begin{figure*}[ht!]
    \centering
    \includegraphics[width=0.49\linewidth]{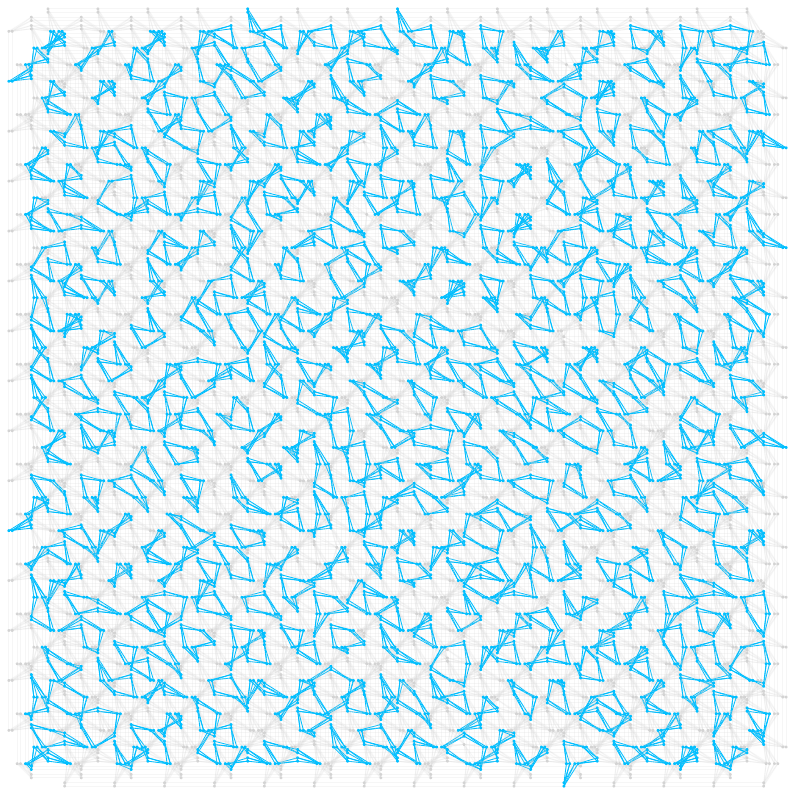}
    \includegraphics[width=0.49\linewidth]{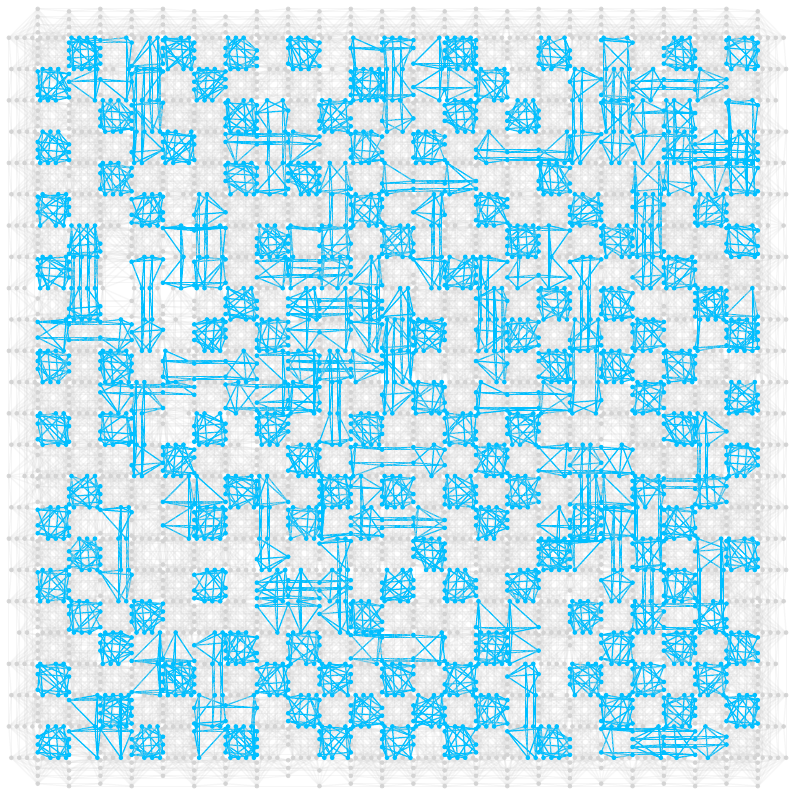}
    \caption{ Disjoint native (spin to qubit mappings) of multiple $N=12$ ANNNI models onto the hardware graphs of \texttt{Advantage\_system4.1} (left) and \texttt{Advantage2\_system1.5} (right). Blue qubits (nodes) and blue couplers (edges) are used in the embeddings, and gray nodes/edges indicate unused hardware.  }
    \label{fig:hardware_embeddings}
\end{figure*}

\section{D-Wave QPU Anneal Schedule Energy Scales}
\label{section:appendix_DWave_QPU_energy_scales}

Because all of the programmed units are in hardware specific and normalized parameters, the physical units are not clear. Therefore, in Figure~\ref{fig:DWave_hardware_energy_scales} we show the full anneal schedules of the functions $A(s)$ and $B(s)$, which, for example, give the physical energy units for the $J$ coupler energy scales.

\section{D-Wave Hardware Graph Embeddings}
\label{section:appendix_hardware_embeddings}

Figure~\ref{fig:hardware_embeddings} shows the exact $N=12$ ANNNI model disjoint embeddings, with periodic boundaries, on the two D-Wave hardware graphs.

\section{$\beta$ versus TVD Example Data Fitting}
\label{section:appendix_beta_vs_TVD_examples}

\begin{figure*}[ht!]
    \centering
    \includegraphics[width=0.245\linewidth]{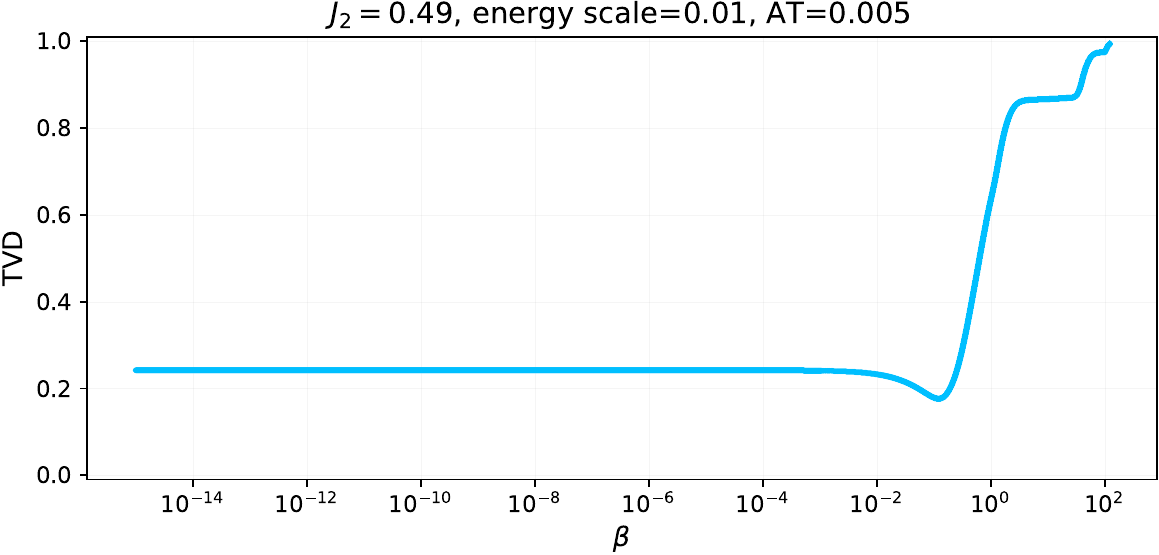}
    \includegraphics[width=0.245\linewidth]{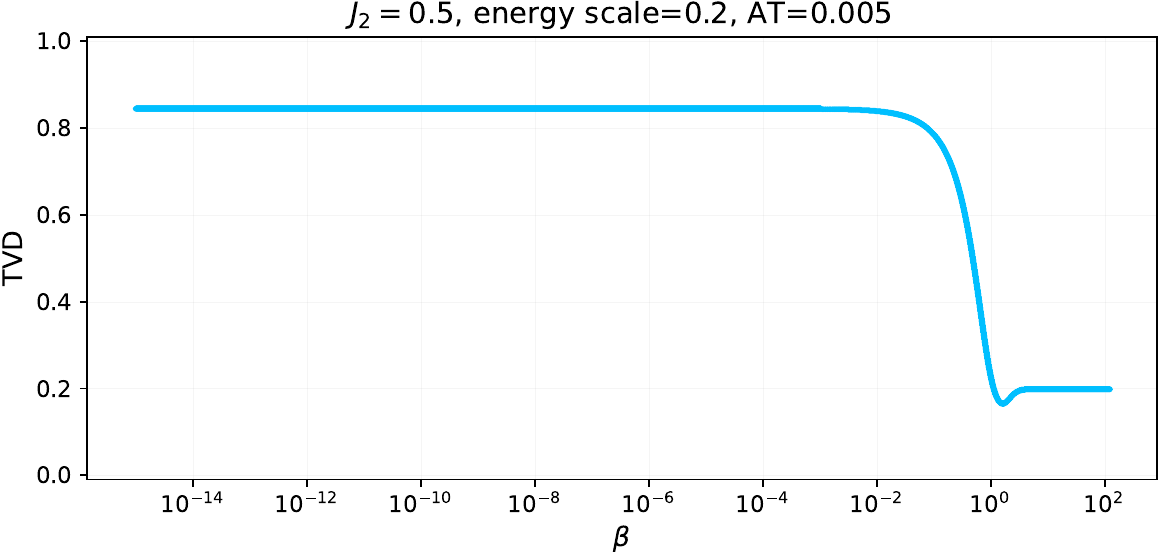}
    \includegraphics[width=0.245\linewidth]{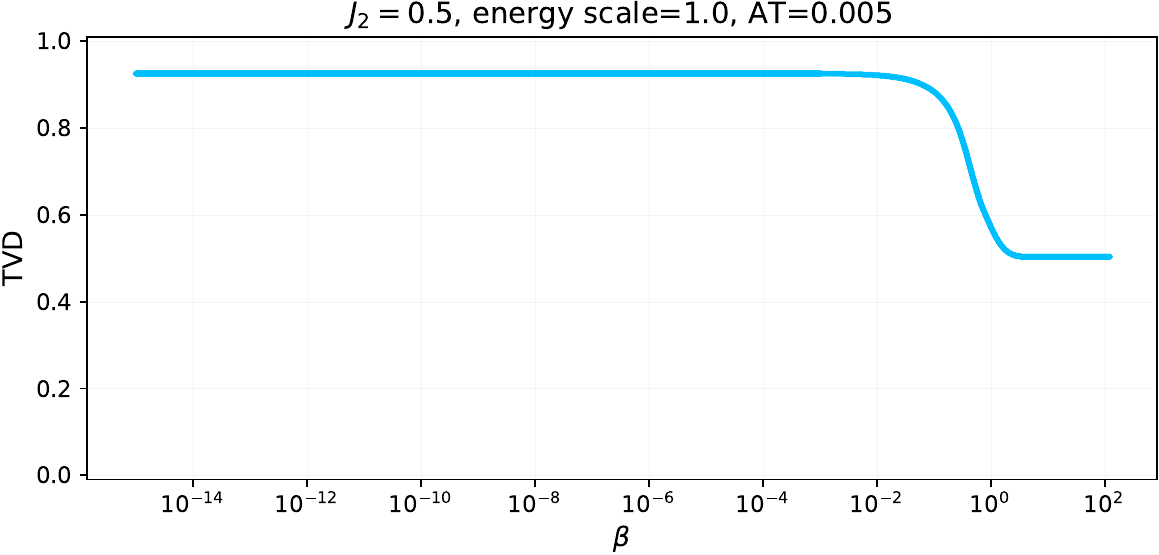}
    \includegraphics[width=0.245\linewidth]{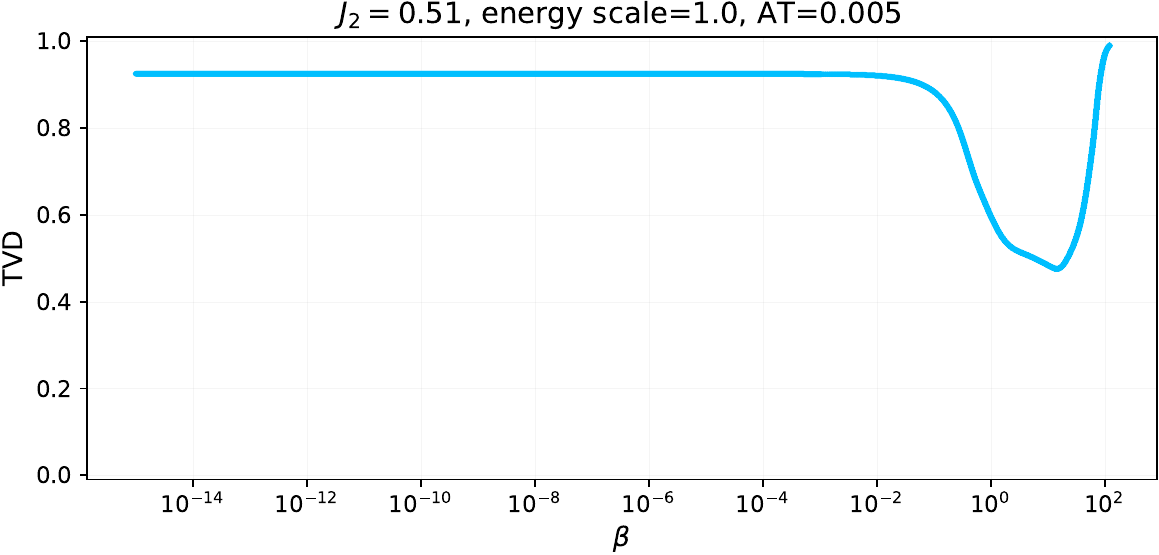}
    \includegraphics[width=0.245\linewidth]{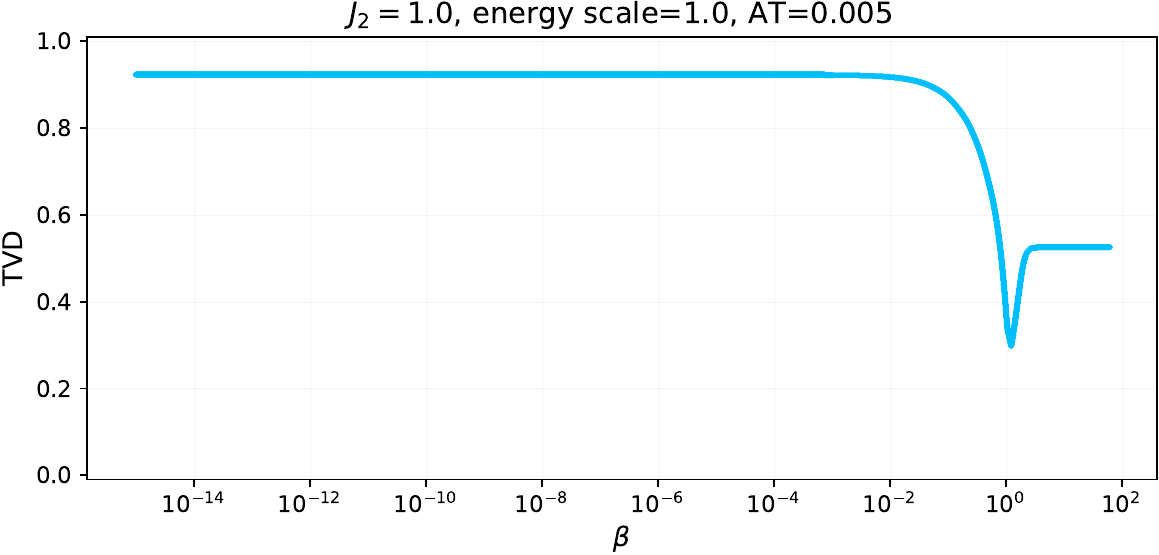}
    \includegraphics[width=0.245\linewidth]{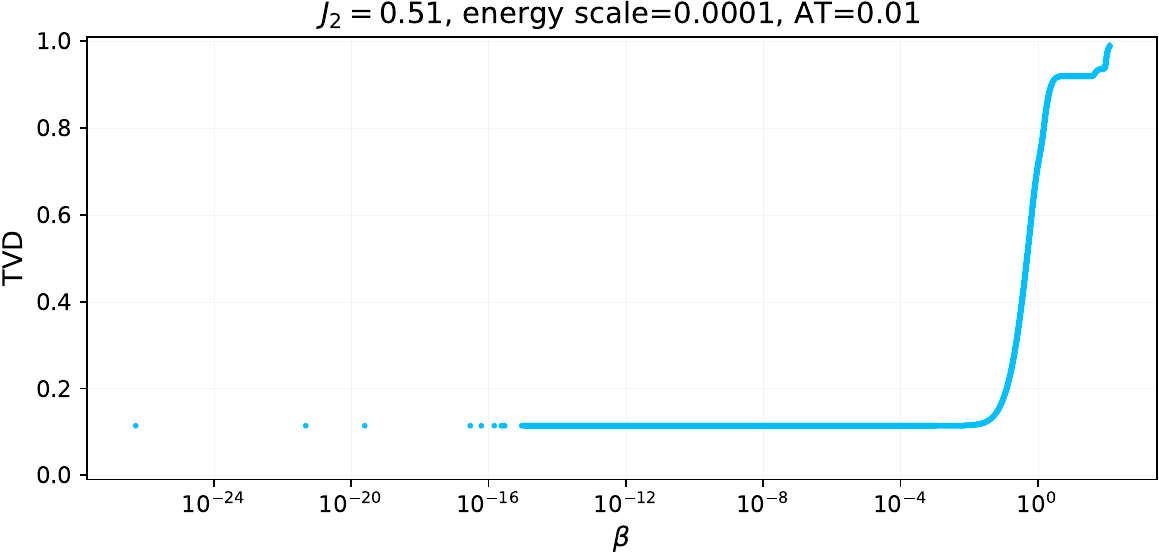}
    \includegraphics[width=0.245\linewidth]{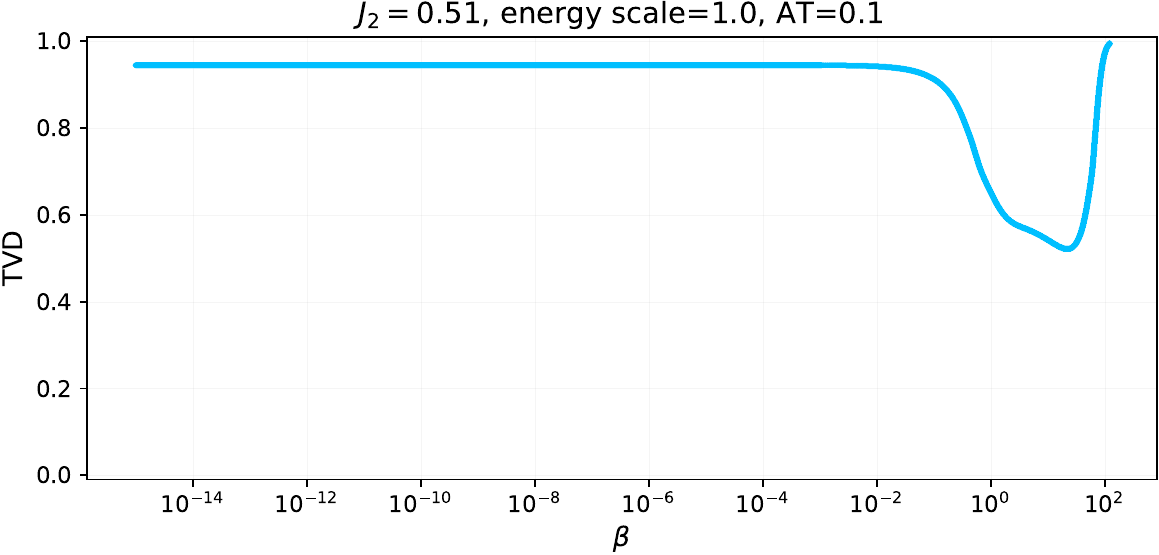}
    \includegraphics[width=0.245\linewidth]{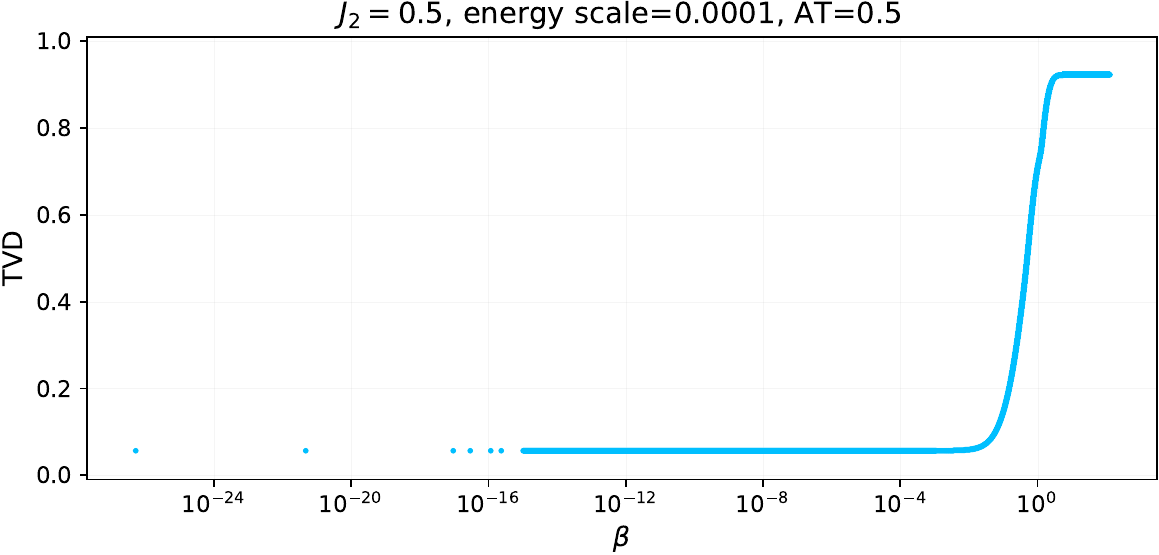}
    \includegraphics[width=0.245\linewidth]{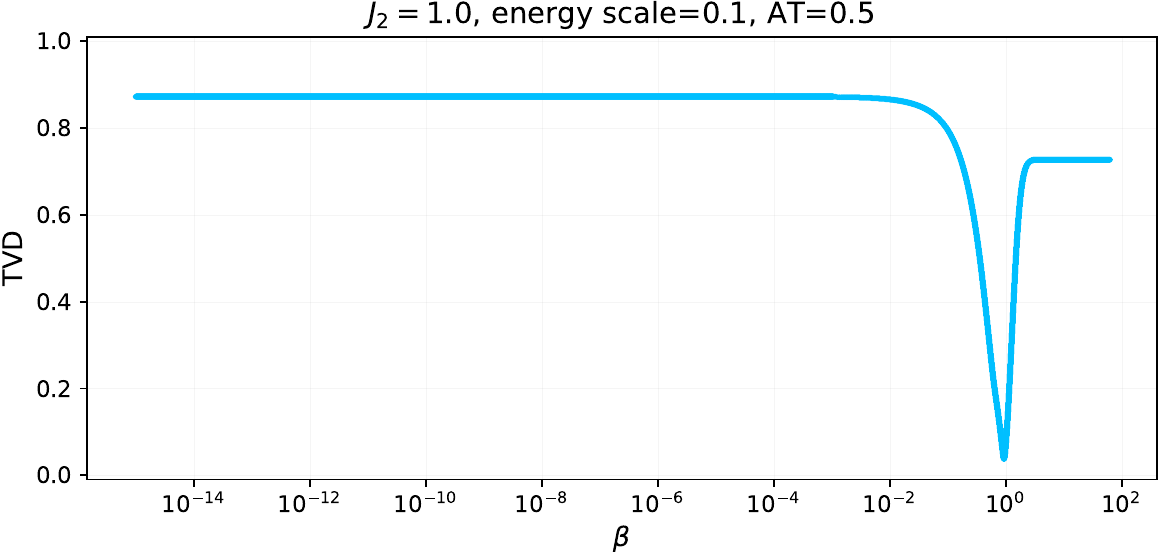}
    \includegraphics[width=0.245\linewidth]{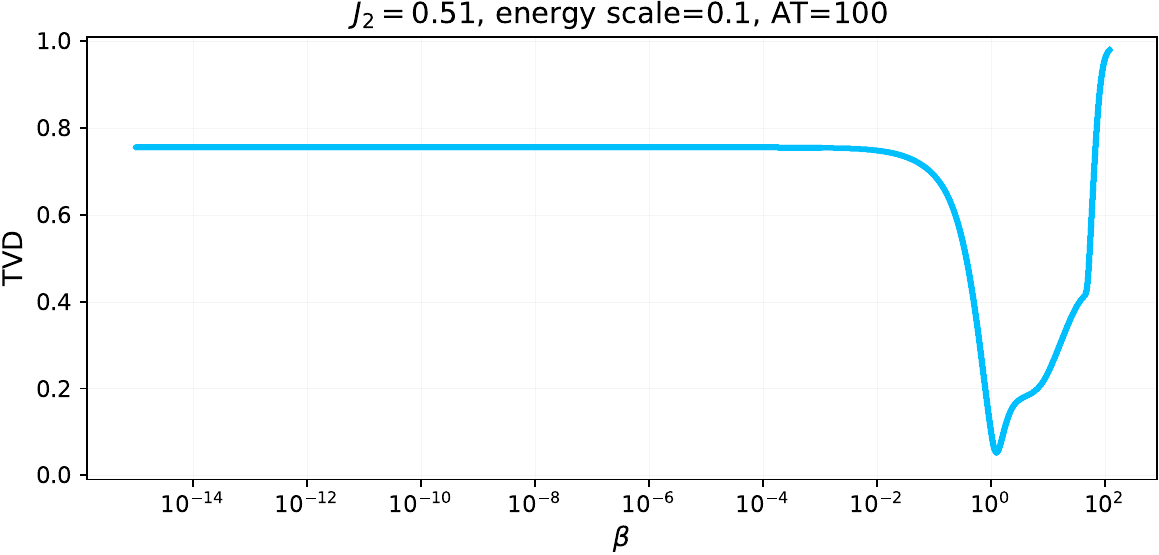}
    \includegraphics[width=0.245\linewidth]{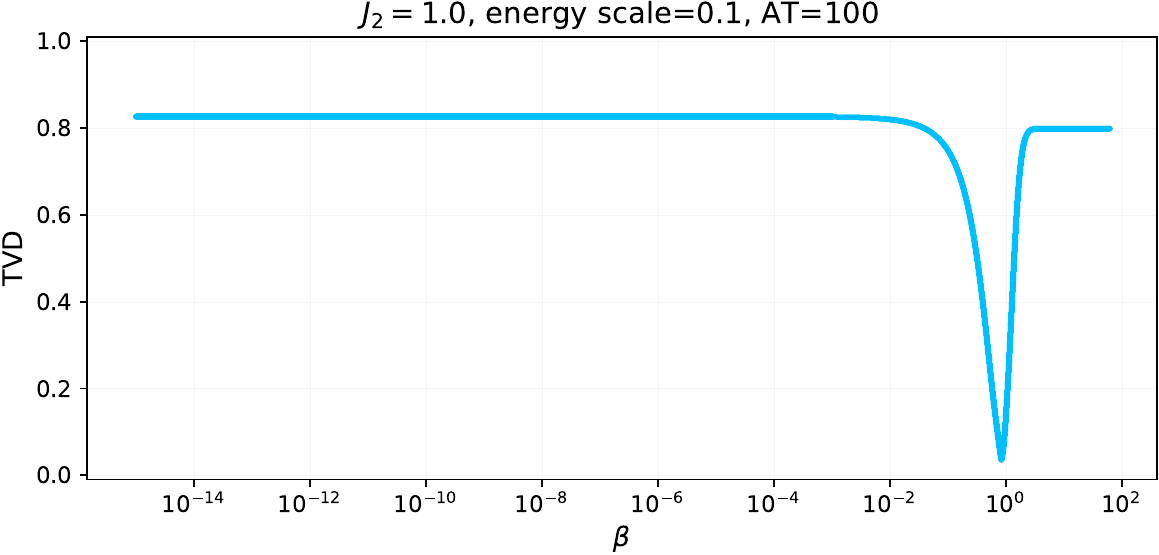}
    \includegraphics[width=0.245\linewidth]{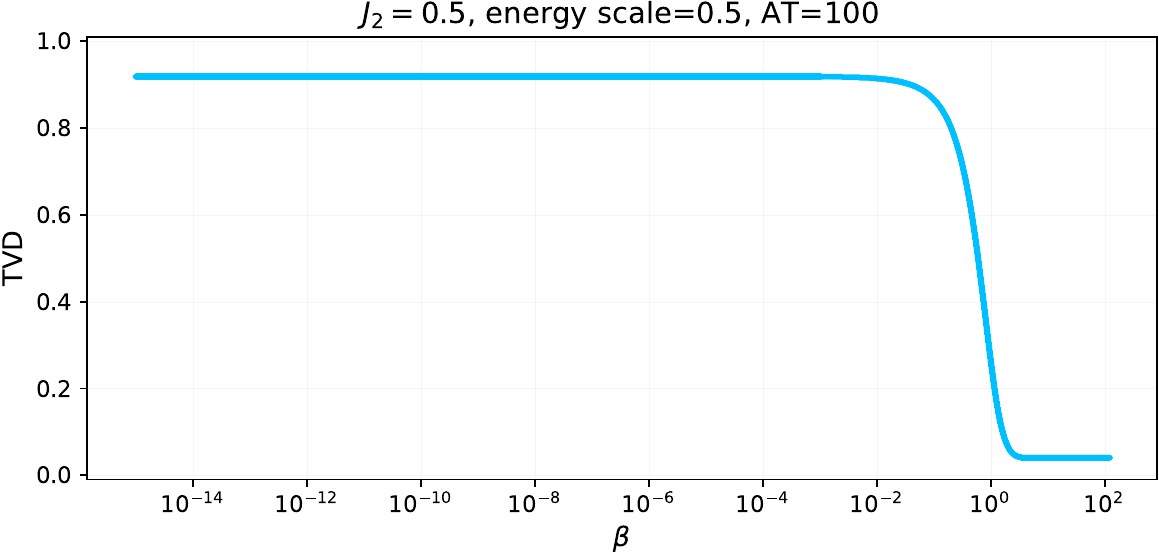}
    \includegraphics[width=0.245\linewidth]{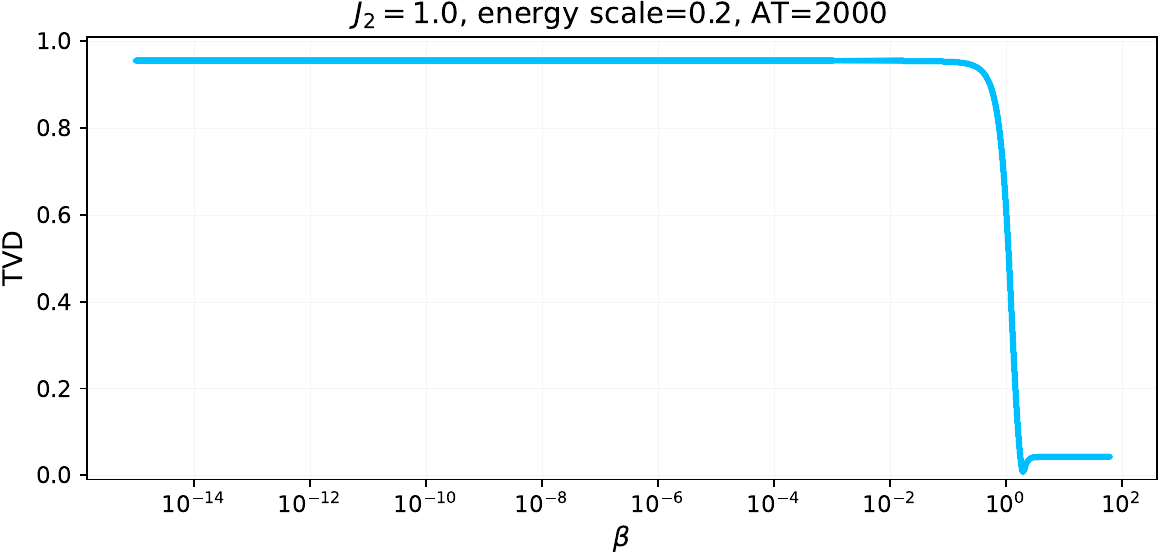}
    \includegraphics[width=0.245\linewidth]{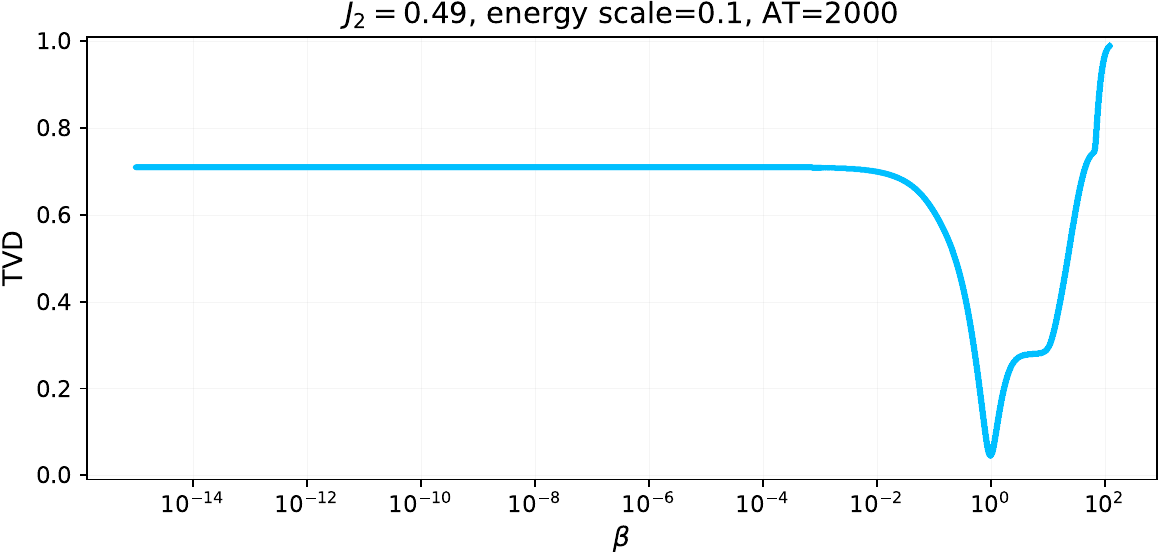}
    \includegraphics[width=0.245\linewidth]{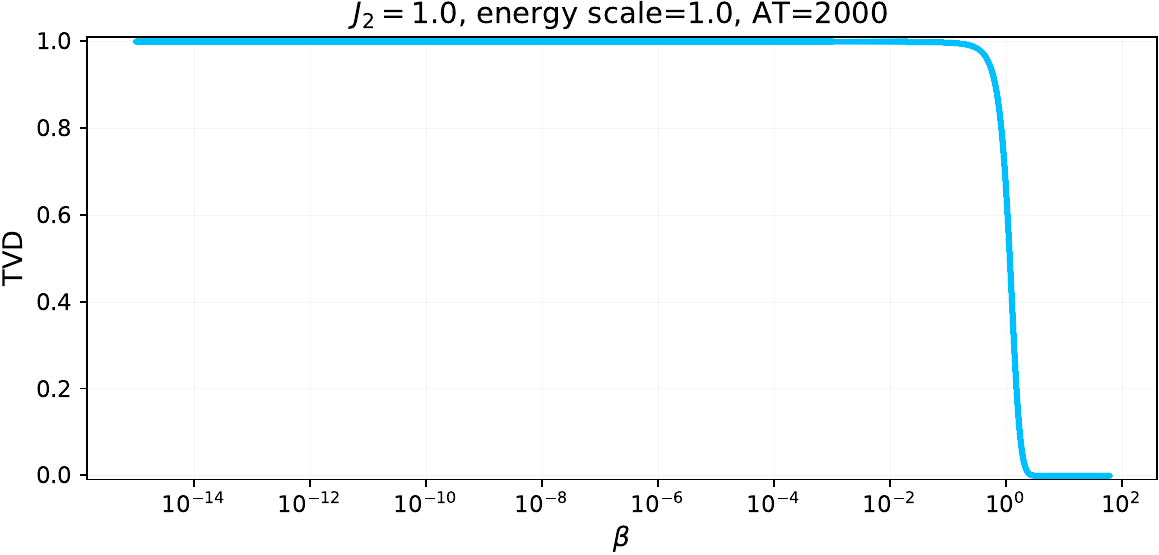}
    \includegraphics[width=0.245\linewidth]{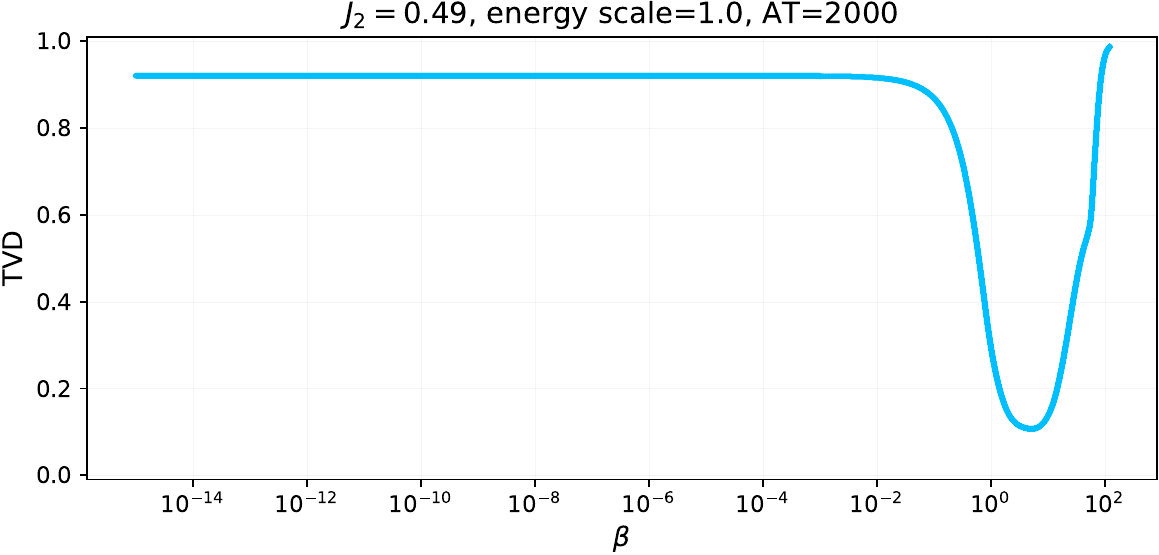}
    \caption{ Representative examples of the $\beta$ fitting search landscape where the TVD is plotted as a function of a large search over many orders of magnitude of $\beta$ values. Each plot corresponds to the measurements collected using a particular QPU analog hardware parameter set, which is specified in the title of each sub-plot (AT is an abbreviation of annealing time), and all data is from the \texttt{Advantage2\_system1.4} processor. This shows that in some instances there is a clearly defined unique minimum, whereas in other instances there are many $\beta$ values that result in the same error rate (a case where this occurs frequently is at higher temperature sampling, such as when the coupler energy scale is set to near or effectively zero due to precision limits on the hardware).  }
    \label{fig:appendix_beta_vs_TVD_fitting_examples}
\end{figure*}

Figure~\ref{fig:appendix_beta_vs_TVD_fitting_examples} reports some example $\beta$ versus TVD fitting curves. These illustrate the process of finding the overall lowest error (TVD) that a given quantum annealer sampling distribution has with respect to the ground-truth of the full Boltzmann distribution. Importantly, in some cases the lowest TVD $\beta$ is uniquely determined, in other cases it is not uniquely determined. So as to counteract potential numerical instability by reporting either the largest or the smallest $\beta$, in cases of a non-unique minimum error $\beta$, here we report the mean $\beta$. This process, for each physical analog set of parameters, involves searching over many orders of magnitude of $\beta$ values, and selecting the lowest found TVD.

\section{Sub-Sampling Bootstrap Error Convergence Analysis}
\label{section:appendix_sub_sampling}

\begin{figure*}[ht!]
    \centering
    \includegraphics[width=0.49\linewidth]{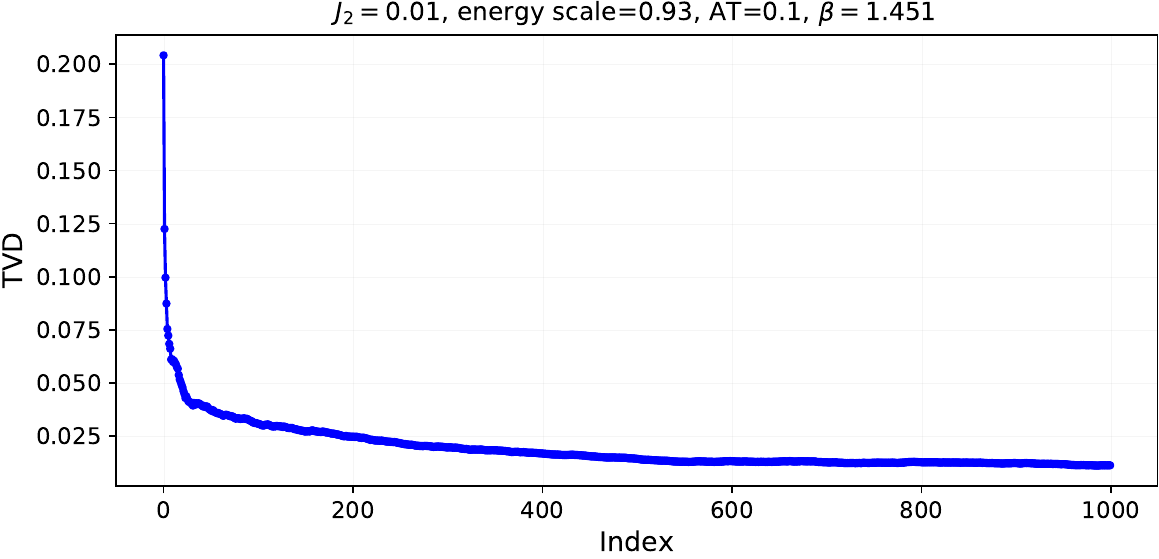}
    \includegraphics[width=0.49\linewidth]{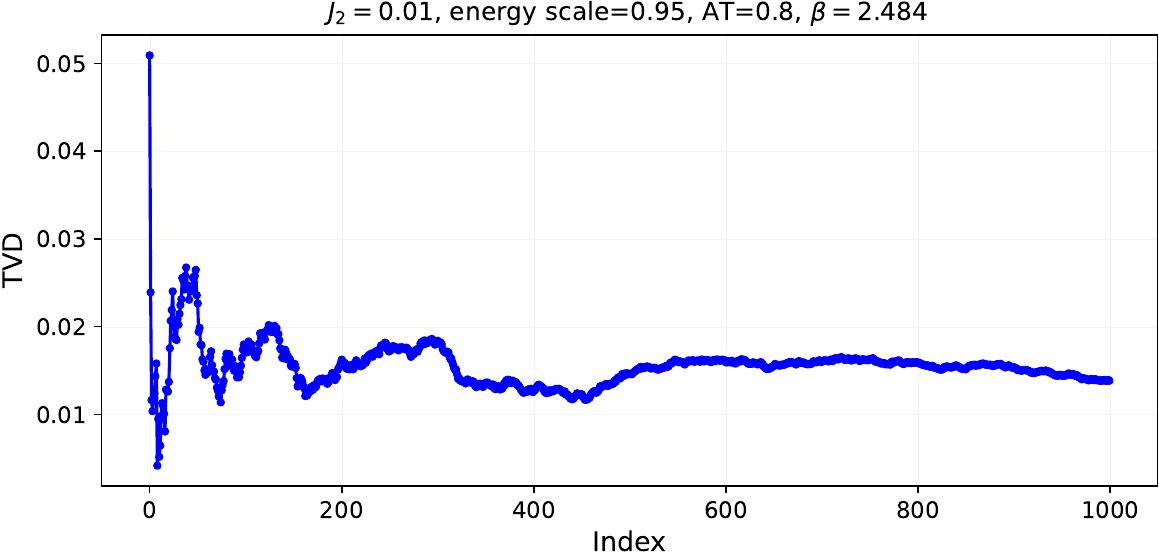}
    \includegraphics[width=0.49\linewidth]{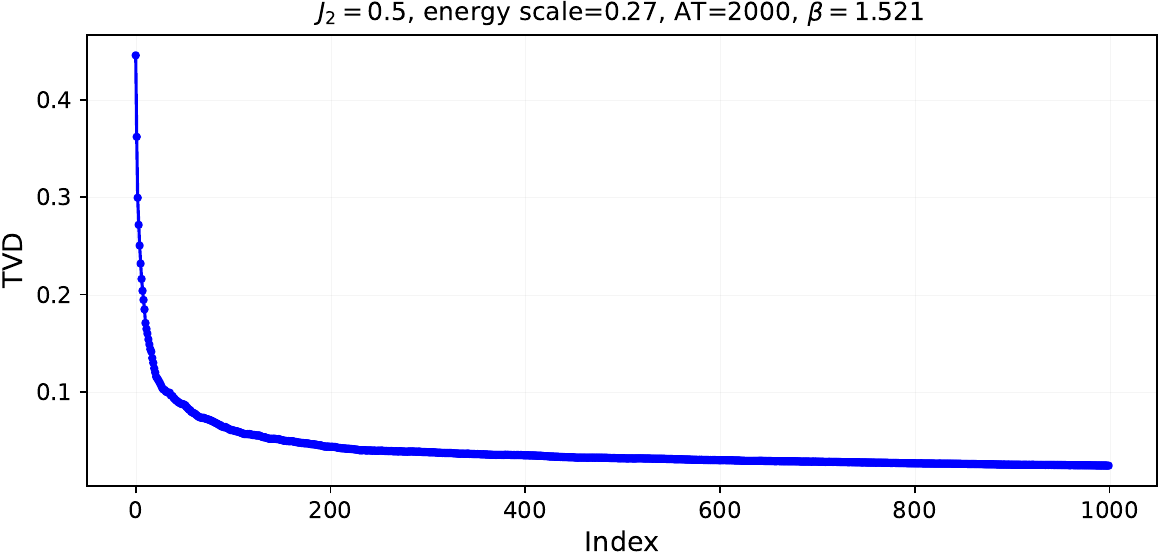}
    \includegraphics[width=0.49\linewidth]{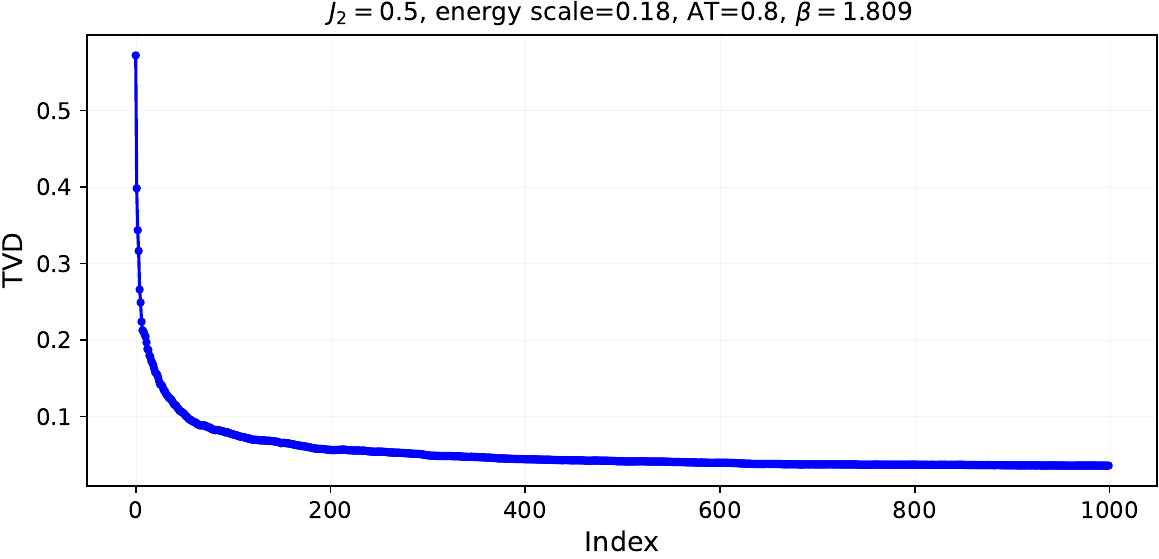}
    \includegraphics[width=0.49\linewidth]{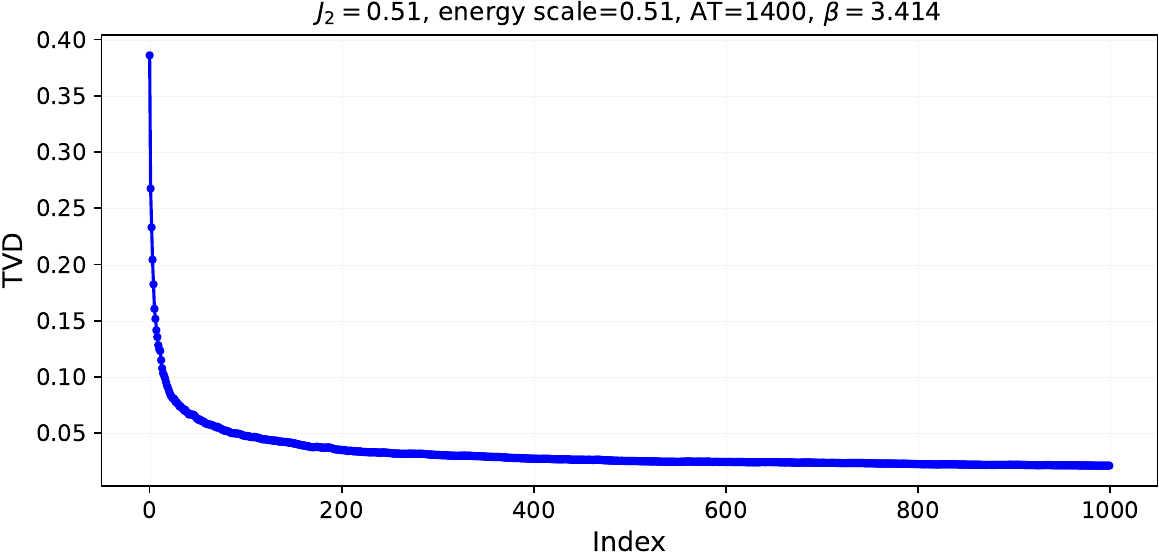}
    \includegraphics[width=0.49\linewidth]{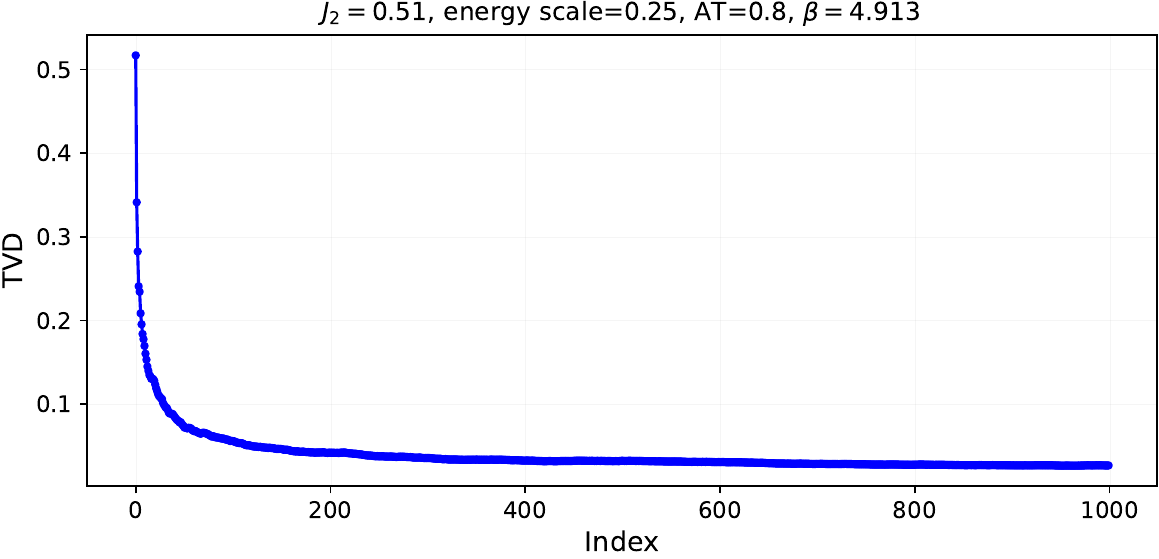}
    \includegraphics[width=0.49\linewidth]{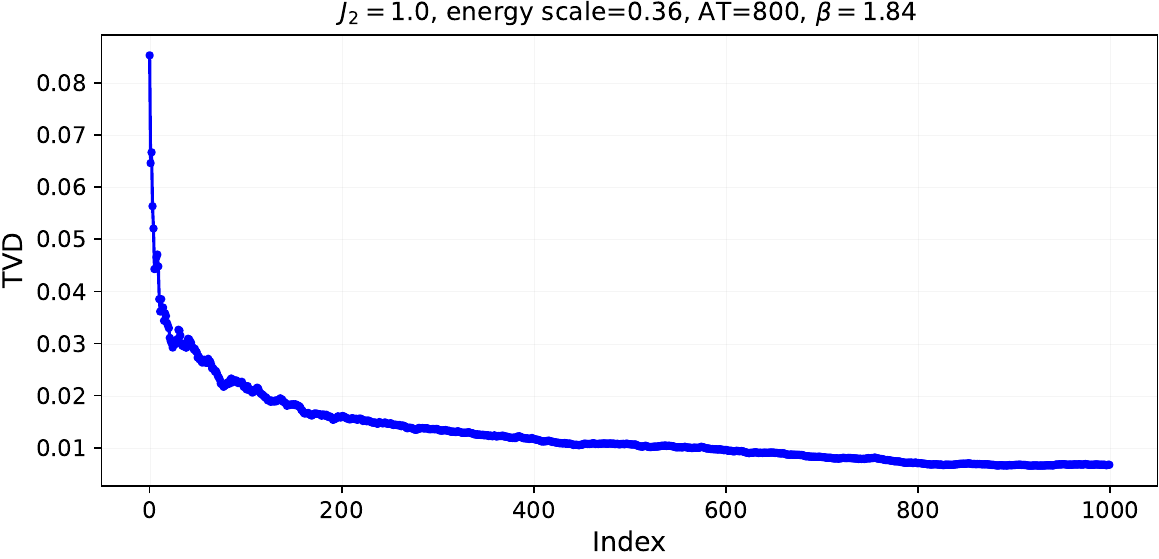}
    \includegraphics[width=0.49\linewidth]{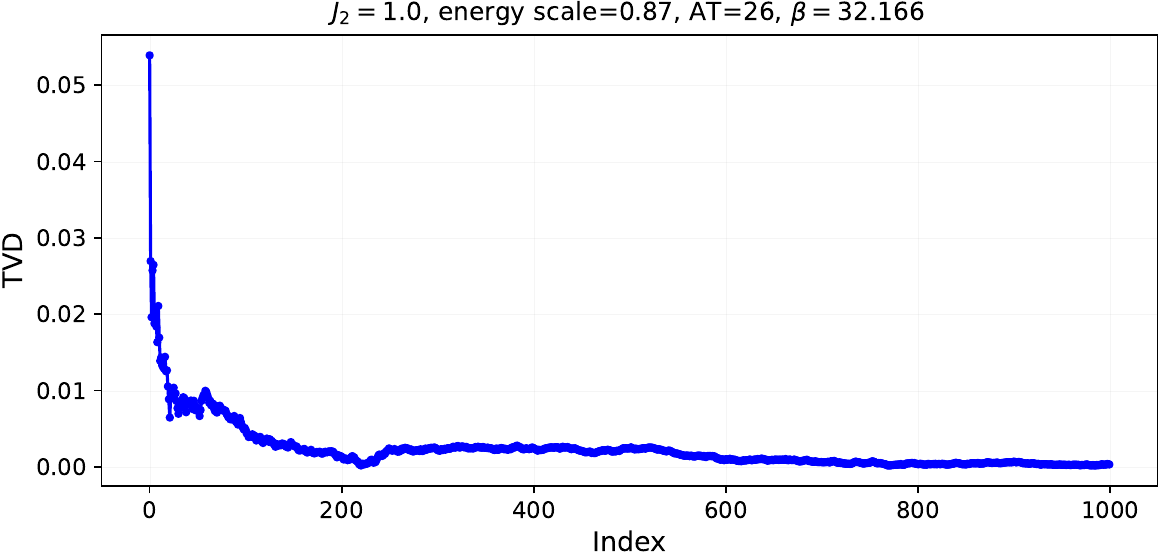}
    \caption{Convergence analysis with respect to total sample count (x-axis) for a representative set of the lowest error rate distributions. The x-axis is an anneal index, where each anneal is comprised of many independent tiled copies of the Ising model. Data from the \texttt{Advantage\_system4.1} processor (left) and the \texttt{Advantage2\_system1.4} processor (right).  }
    \label{fig:sub_sampling_error}
\end{figure*}

Figure~\ref{fig:sub_sampling_error} shows the sub-sampling analysis of some of the lowest error rate distributions. This shows that overall the error rates are reasonably stable once a sufficient number of samples is obtained.

\section{Additional Error versus Temperature Sampling Tradeoff Scatterplots}
\label{section:appendix_more_scatterplots}

\begin{figure*}[ht!]
    \centering
    \includegraphics[width=0.495\linewidth]{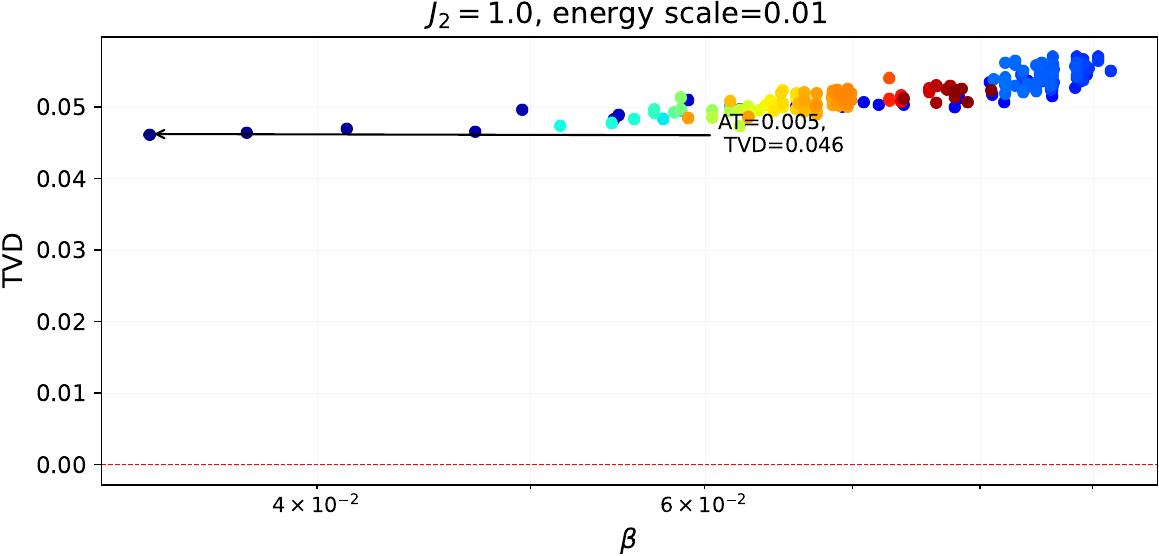}
    \includegraphics[width=0.495\linewidth]{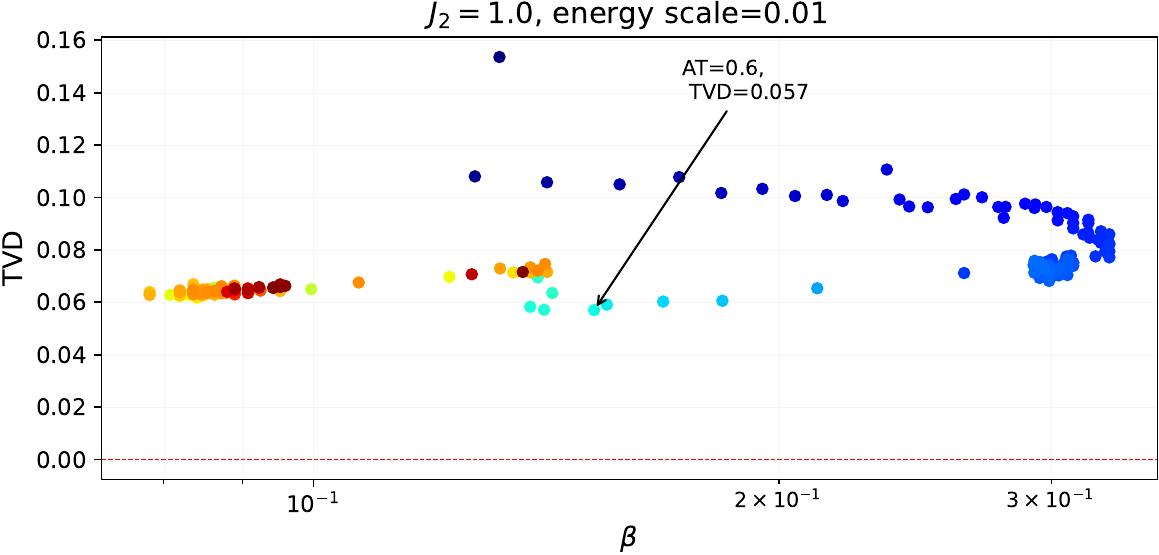}
    \includegraphics[width=0.495\linewidth]{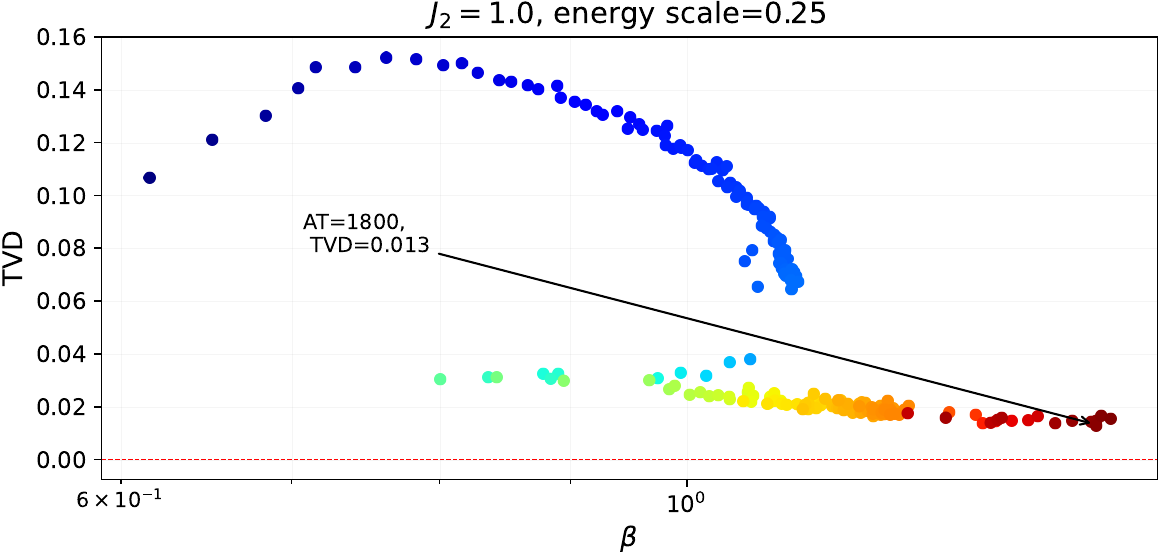}
    \includegraphics[width=0.495\linewidth]{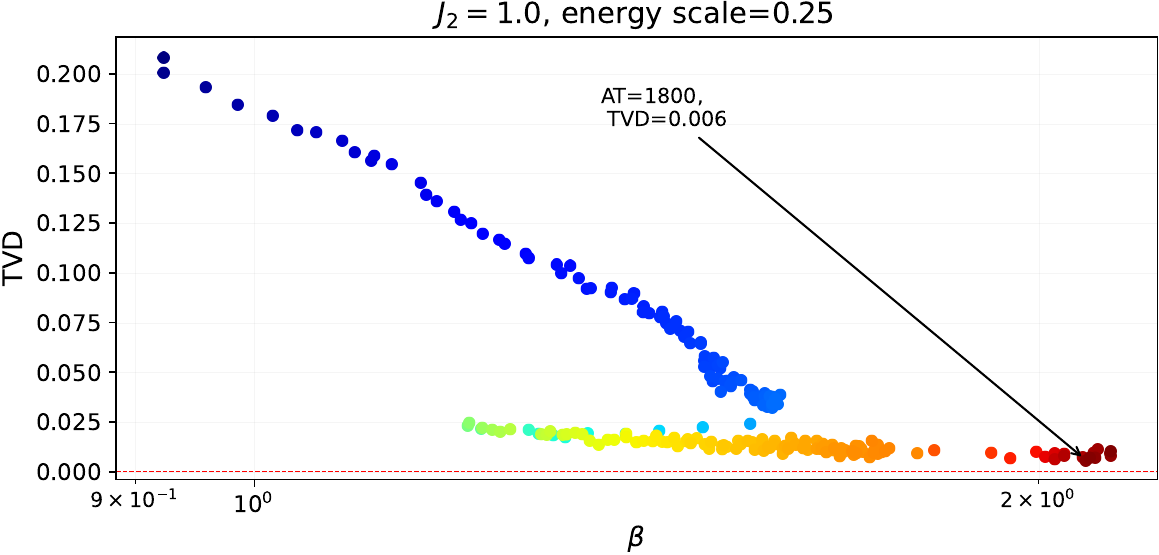}
    \includegraphics[width=0.495\linewidth]{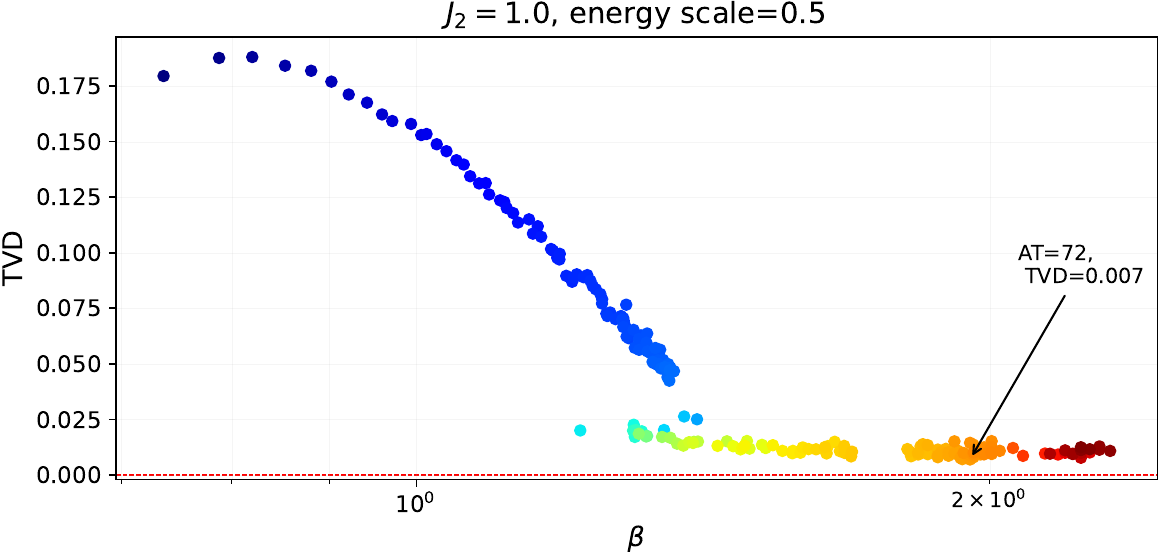}
    \includegraphics[width=0.495\linewidth]{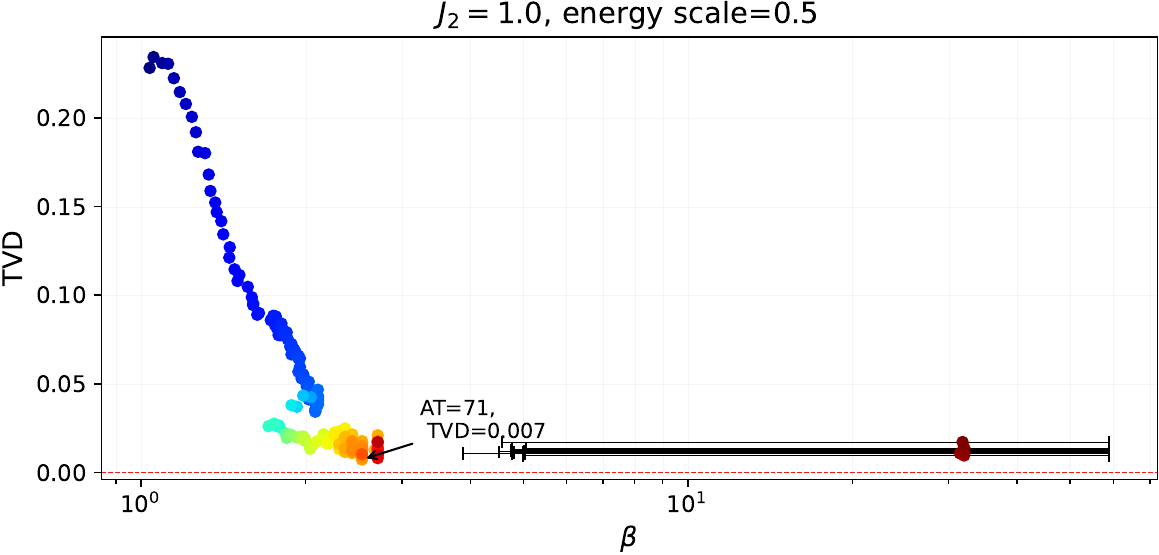}
    \includegraphics[width=0.495\linewidth]{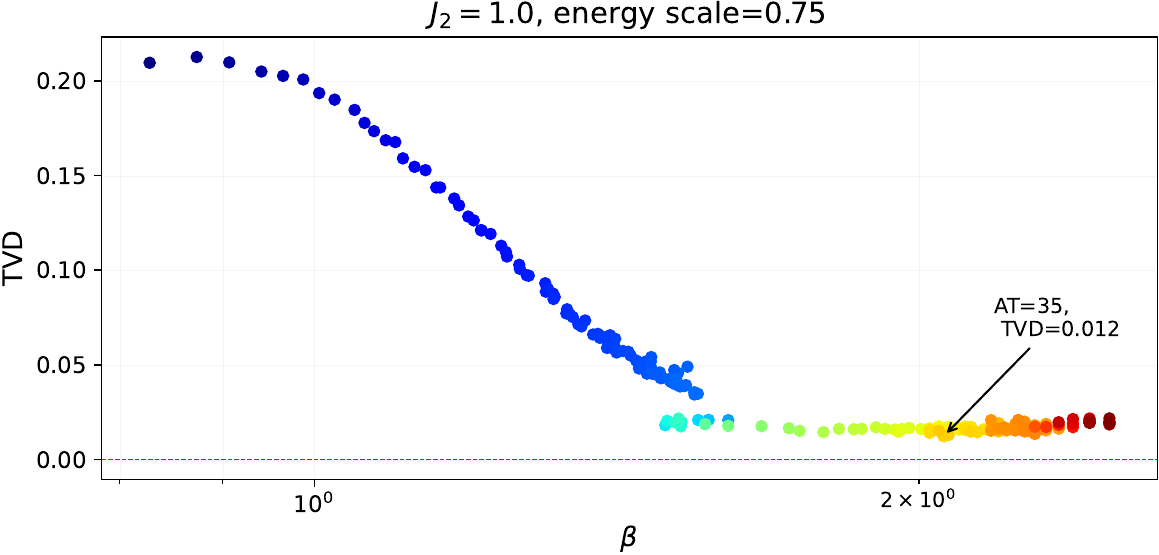}
    \includegraphics[width=0.495\linewidth]{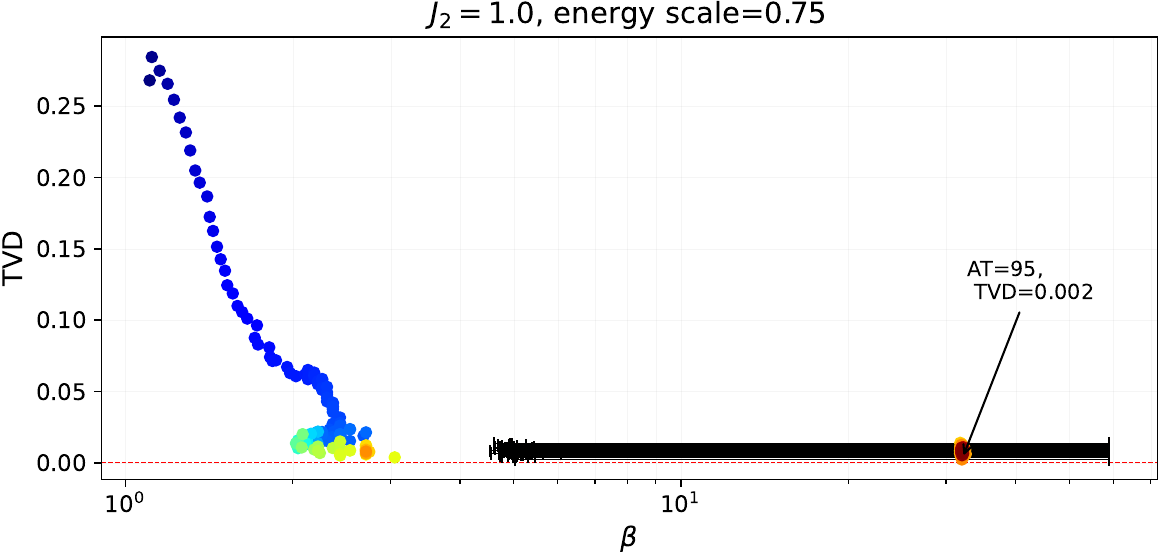}
    \includegraphics[width=0.495\linewidth]{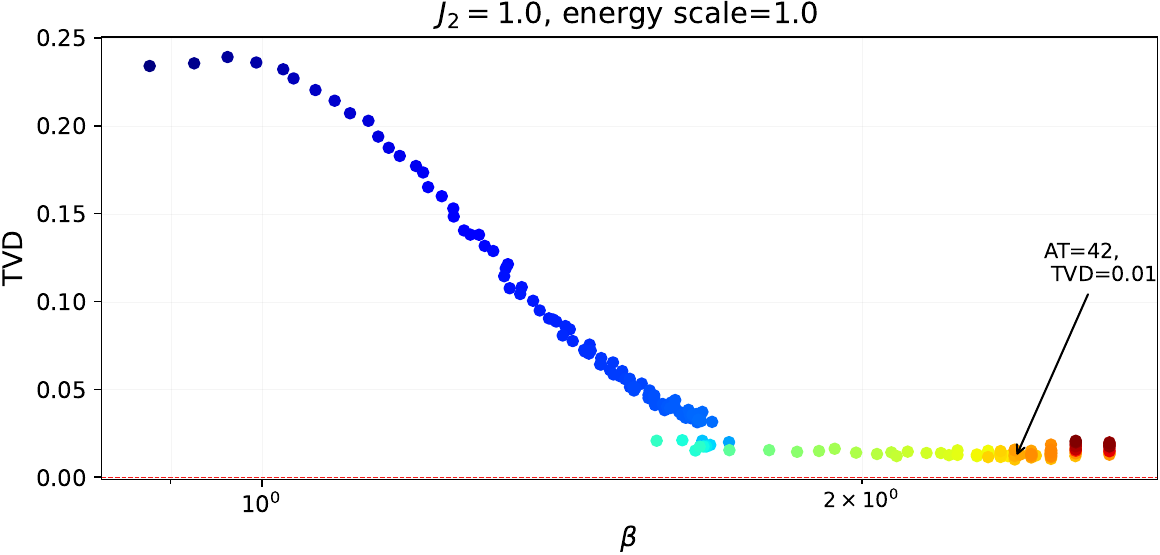}
    \includegraphics[width=0.495\linewidth]{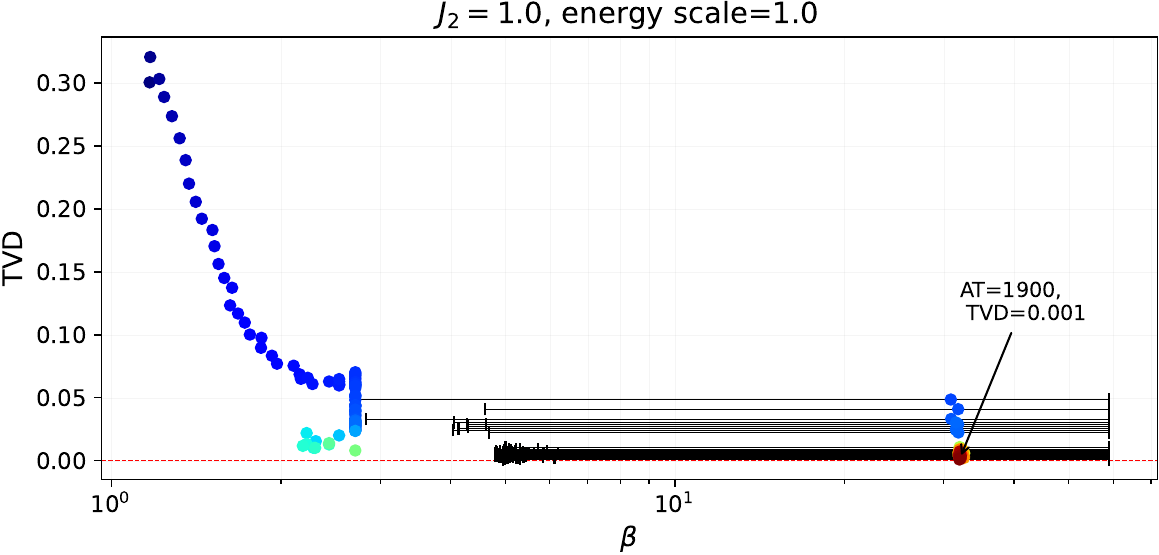}
    \includegraphics[width=0.6\linewidth]{figures/scatter_beta_vs_TVD/AT_colorbar.pdf}
    \caption{ $J_2=1$ ANNNI frustration parameter, run on \texttt{Advantage2\_system1.4} (right column) and \texttt{Advantage\_system4.1} (left column).  }
    \label{fig:J2_1_QPU_comparison}
\end{figure*}

\begin{figure*}[ht!]
    \centering
    \includegraphics[width=0.495\linewidth]{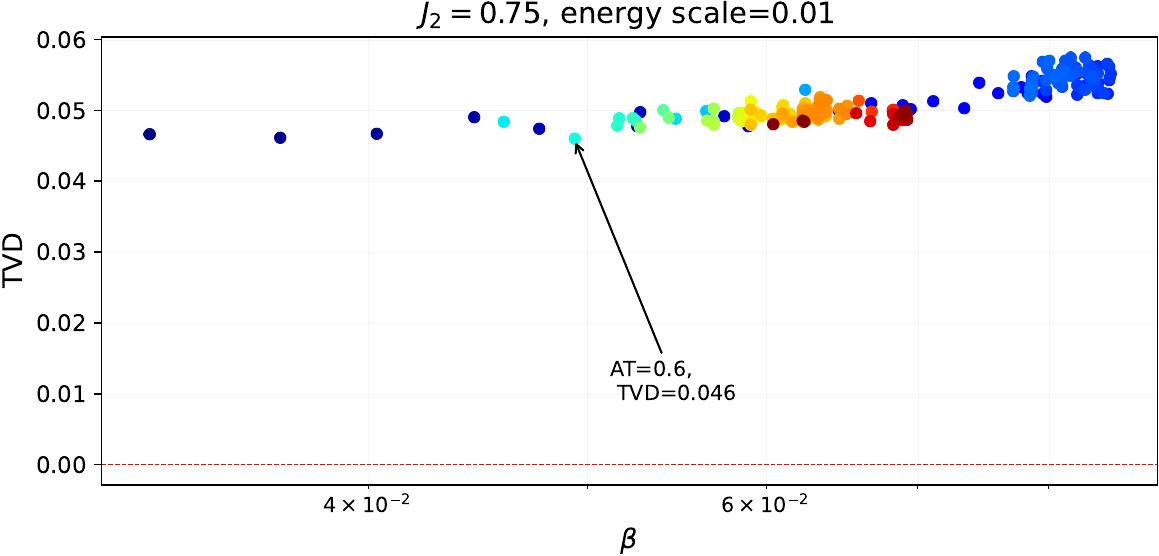}
    \includegraphics[width=0.495\linewidth]{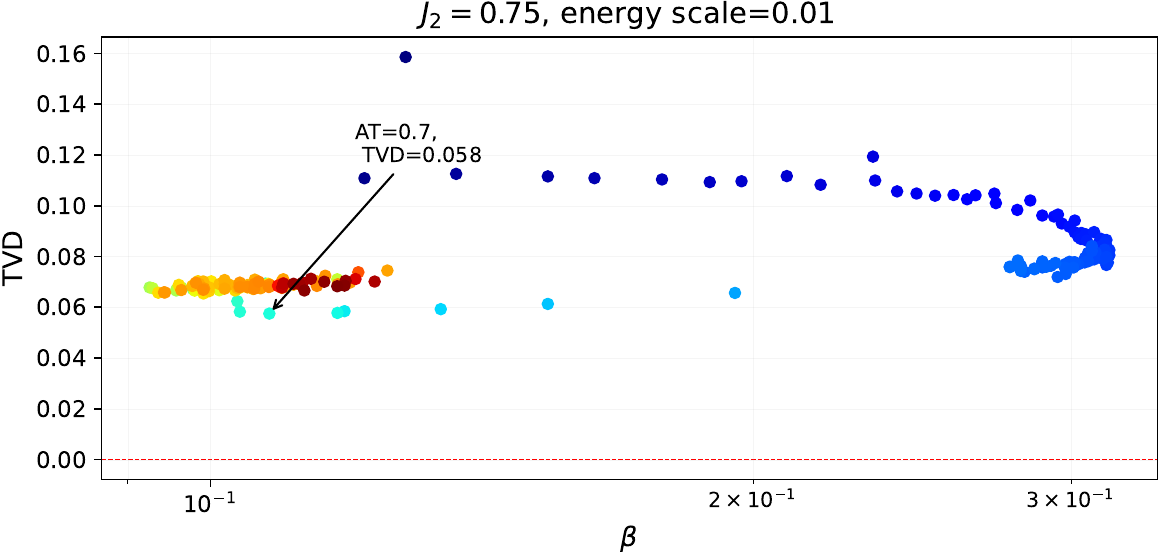}
    \includegraphics[width=0.495\linewidth]{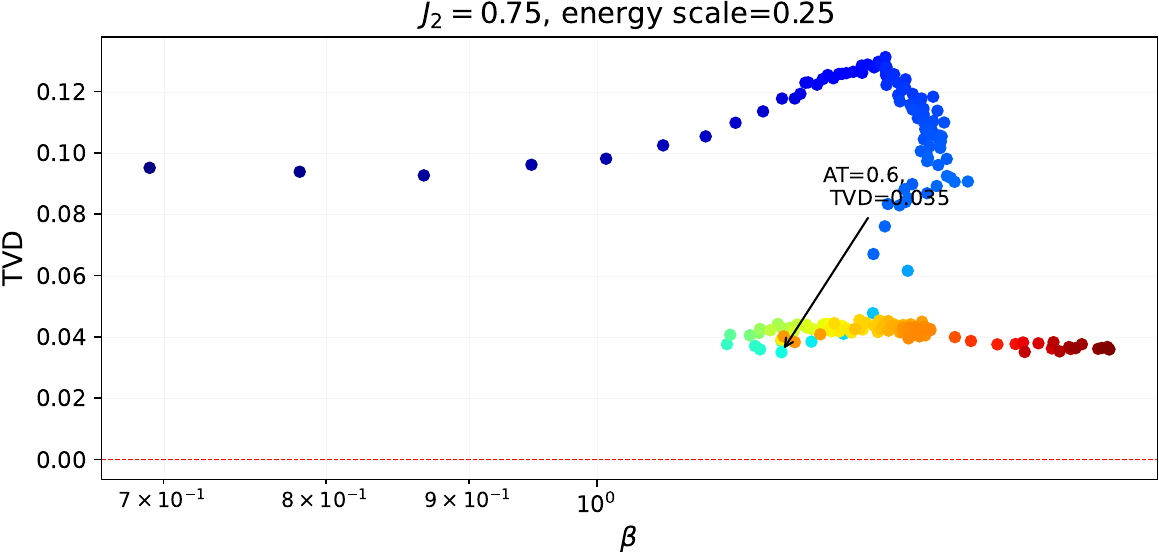}
    \includegraphics[width=0.495\linewidth]{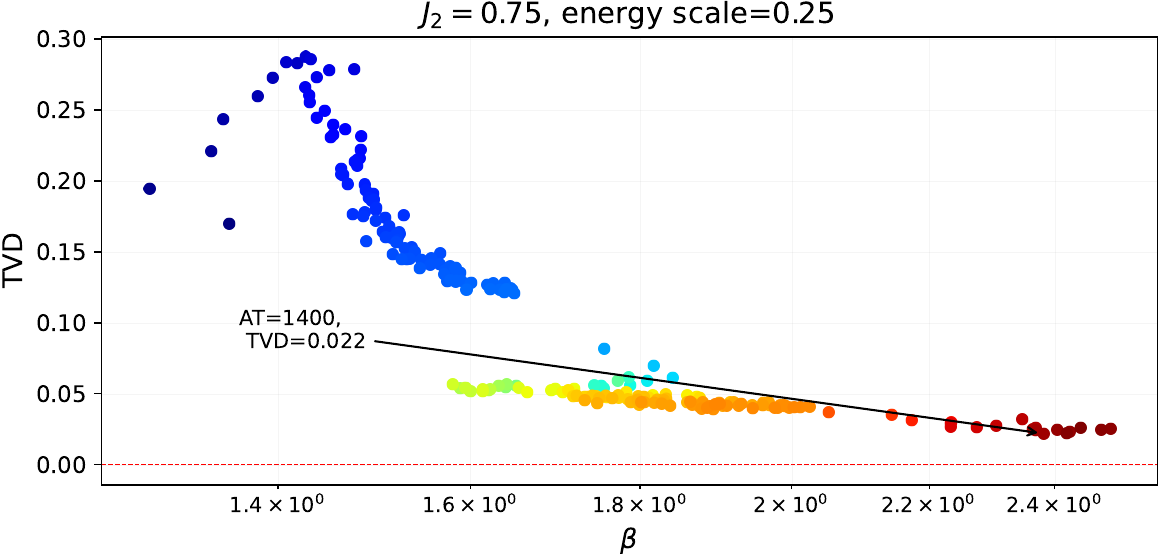}
    \includegraphics[width=0.495\linewidth]{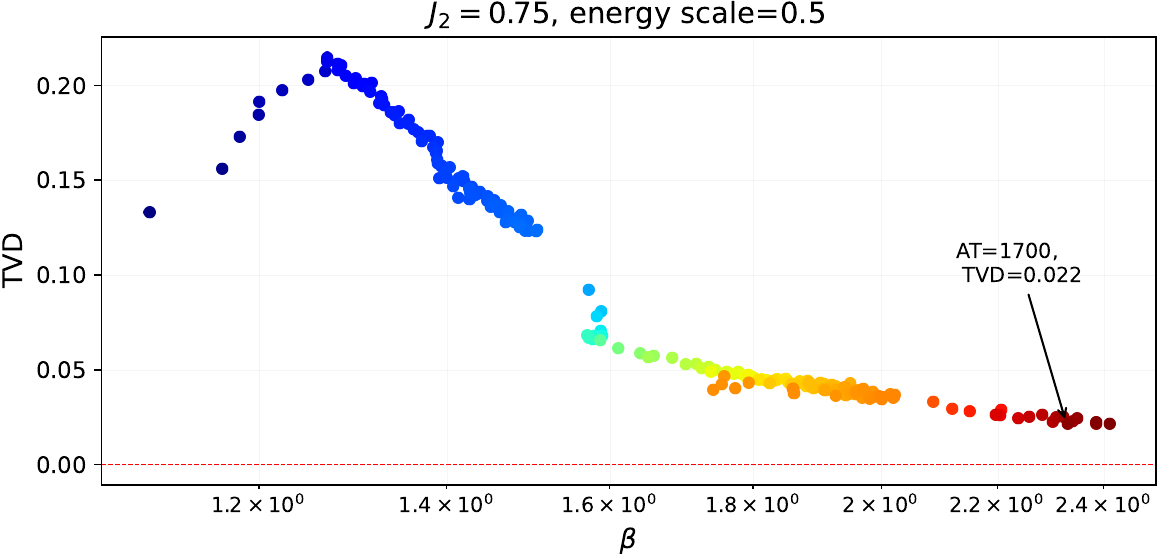}
    \includegraphics[width=0.495\linewidth]{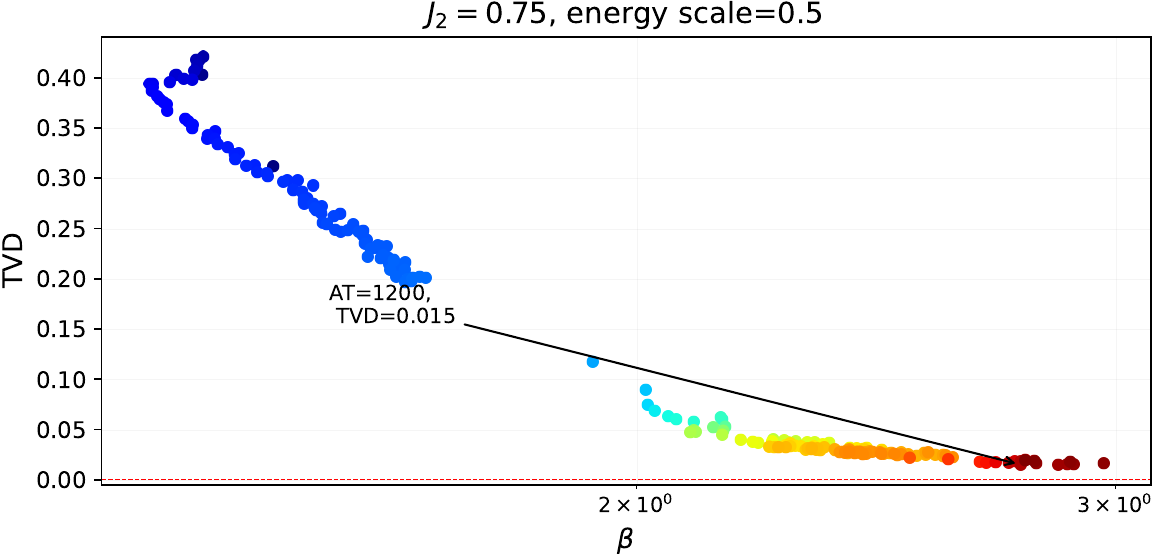}
    \includegraphics[width=0.495\linewidth]{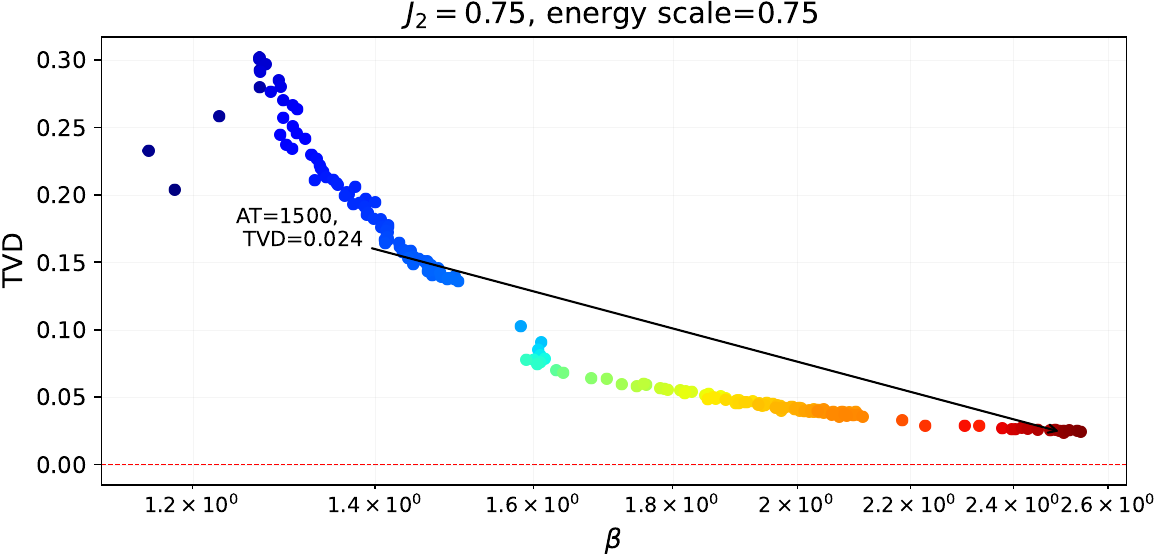}
    \includegraphics[width=0.495\linewidth]{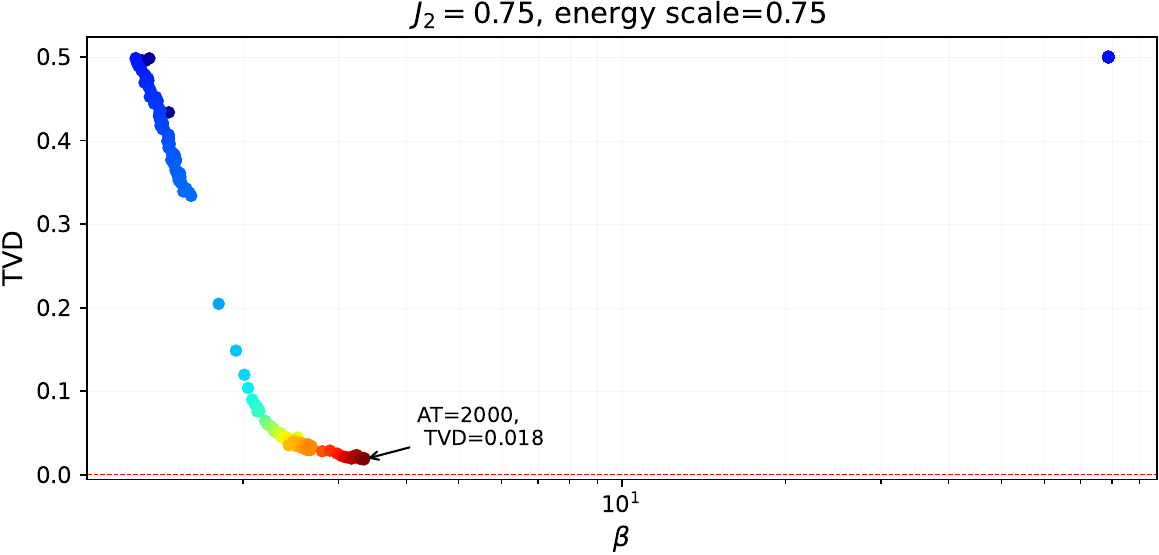}
    \includegraphics[width=0.495\linewidth]{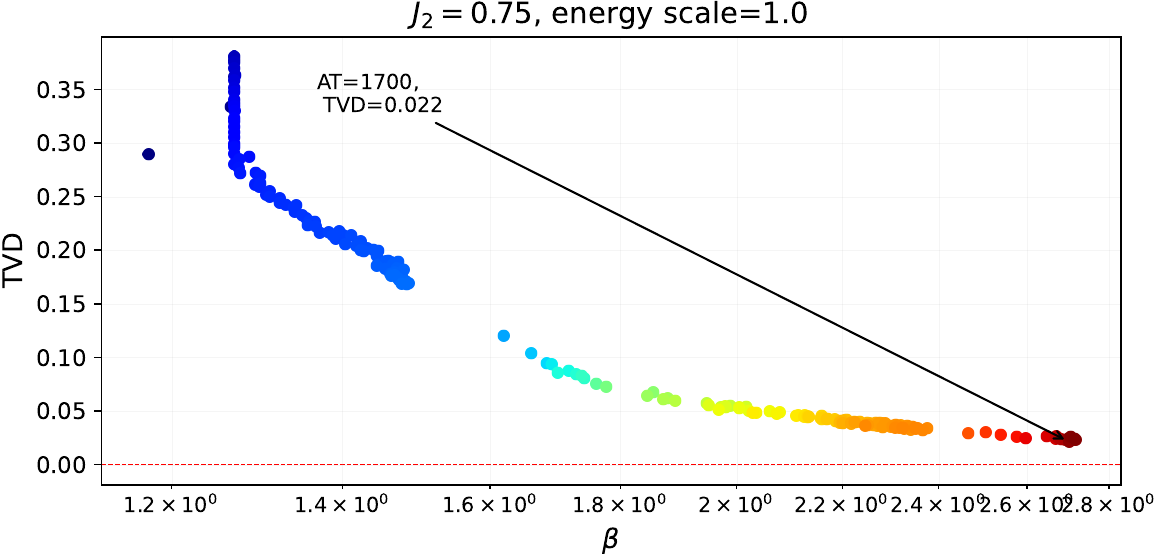}
    \includegraphics[width=0.495\linewidth]{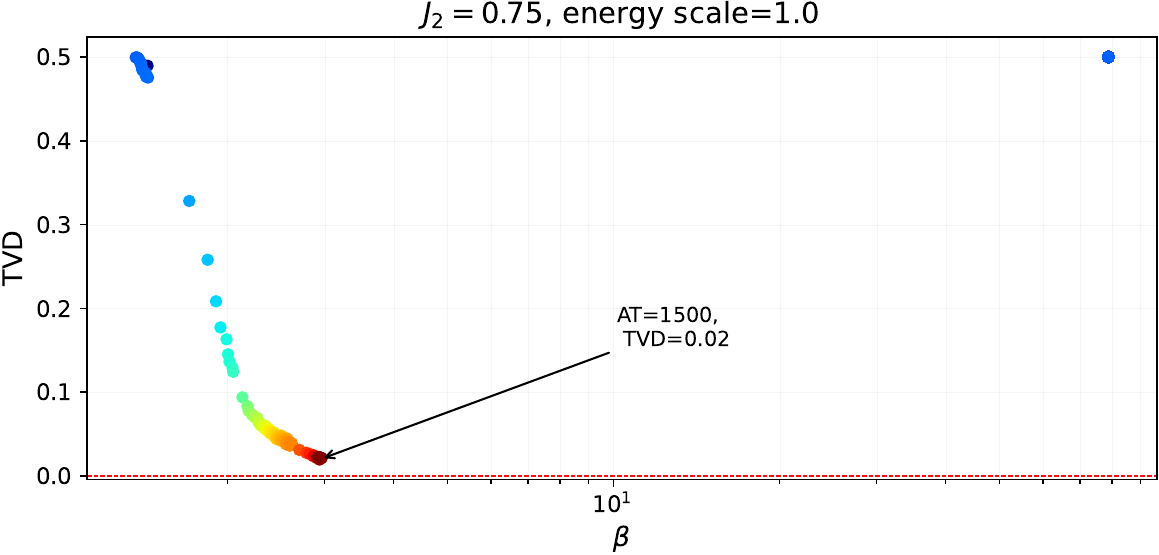}
    \includegraphics[width=0.6\linewidth]{figures/scatter_beta_vs_TVD/AT_colorbar.pdf}
    \caption{ $J_2=0.75$ ANNNI frustration parameter, run on \texttt{Advantage2\_system1.4} (right column) and \texttt{Advantage\_system4.1} (left column).  }
    \label{fig:J2_0.75_QPU_comparison}
\end{figure*}

\begin{figure*}[ht!]
    \centering
    \includegraphics[width=0.495\linewidth]{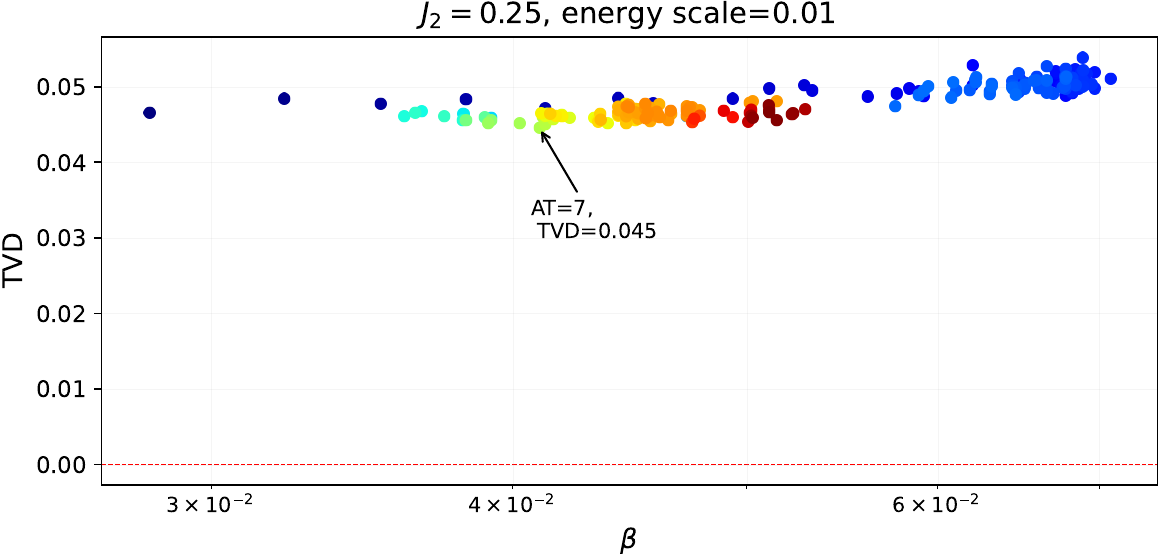}
    \includegraphics[width=0.495\linewidth]{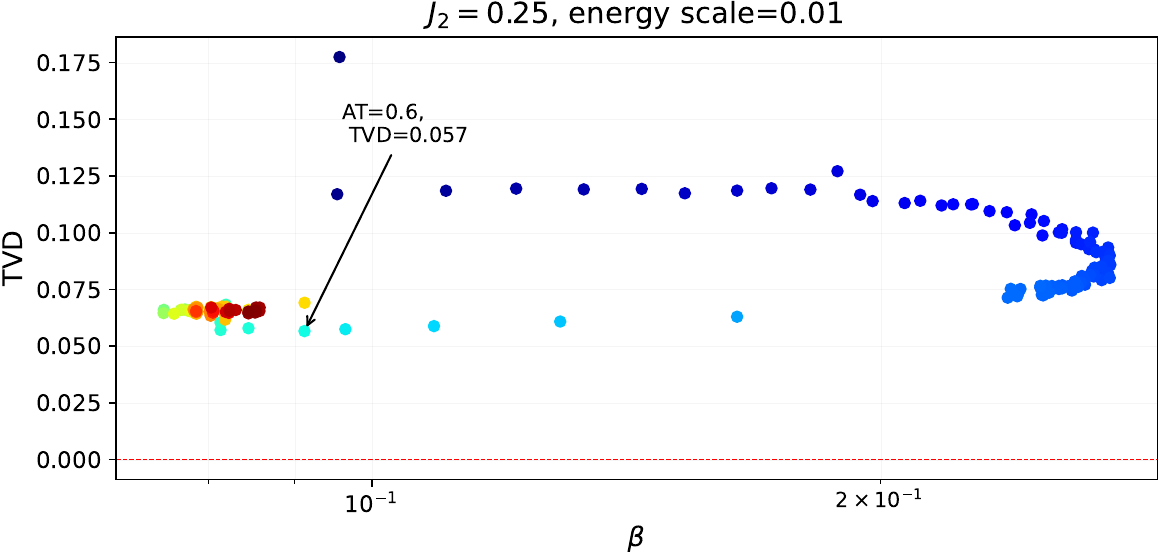}
    \includegraphics[width=0.495\linewidth]{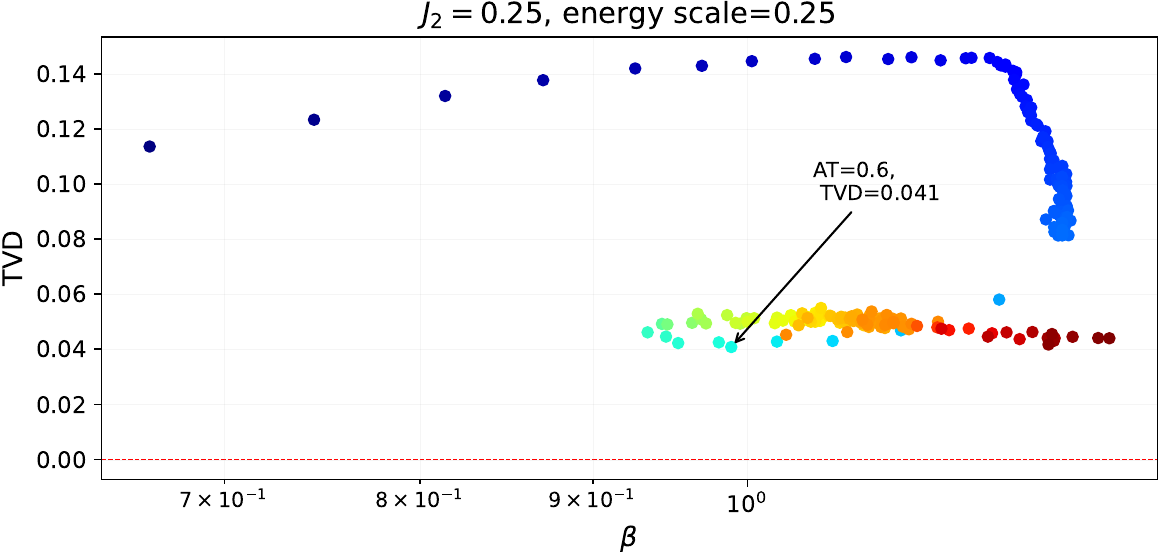}
    \includegraphics[width=0.495\linewidth]{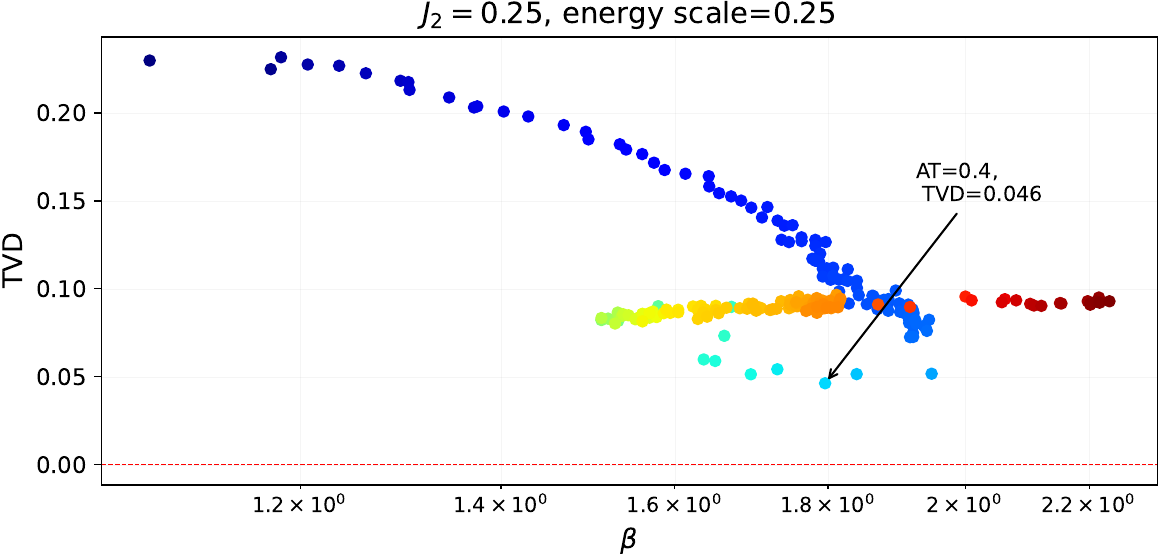}
    \includegraphics[width=0.495\linewidth]{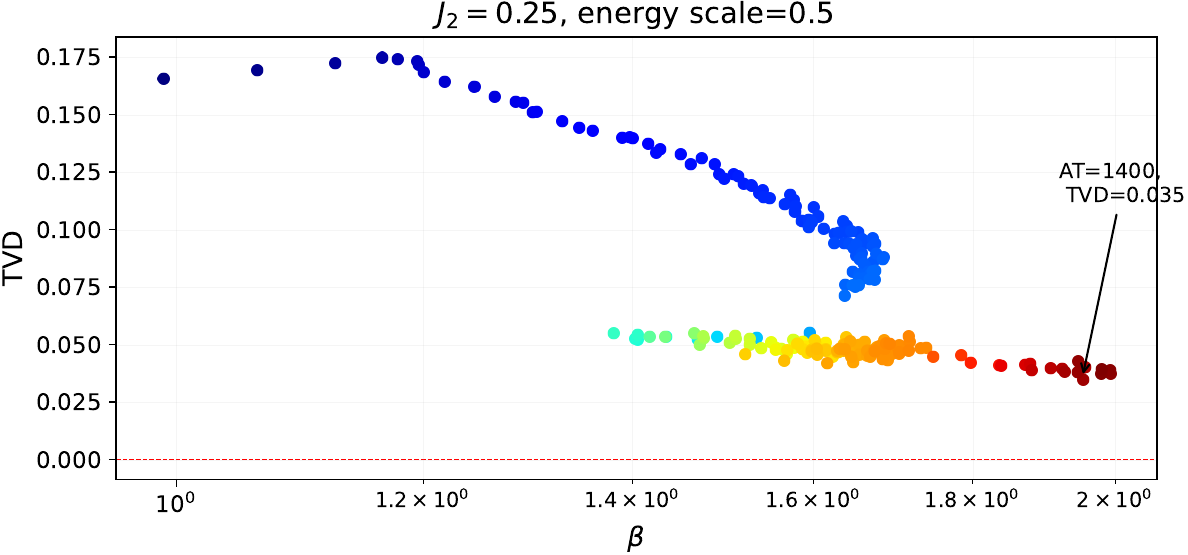}
    \includegraphics[width=0.495\linewidth]{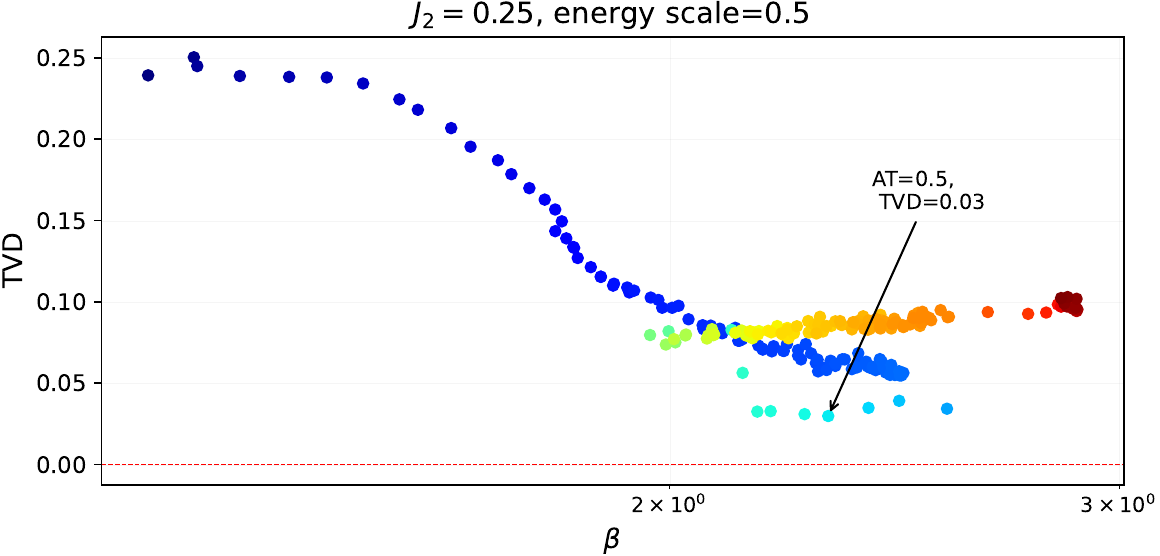}
    \includegraphics[width=0.495\linewidth]{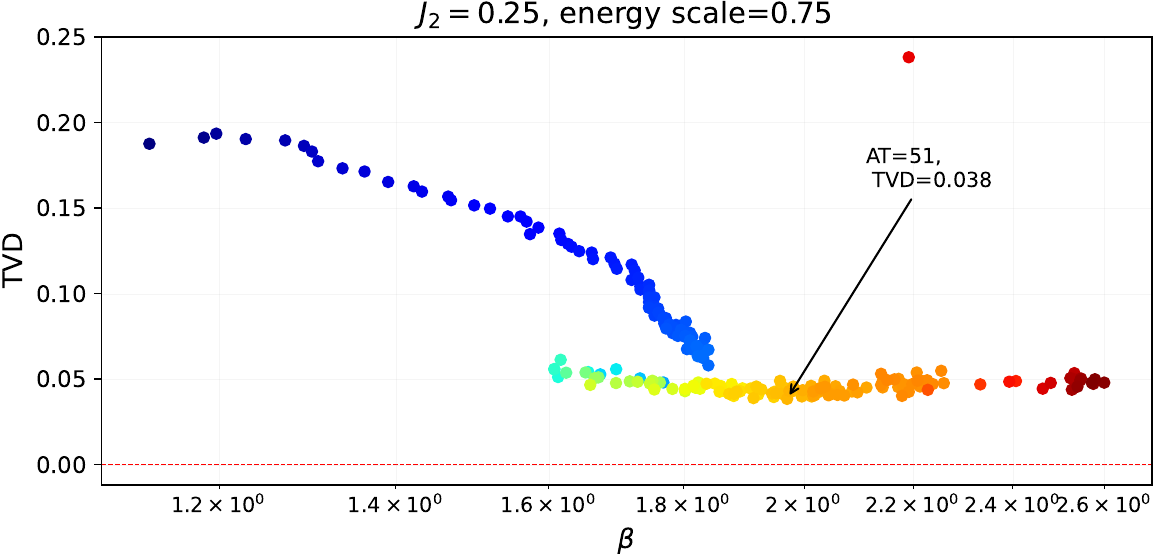}
    \includegraphics[width=0.495\linewidth]{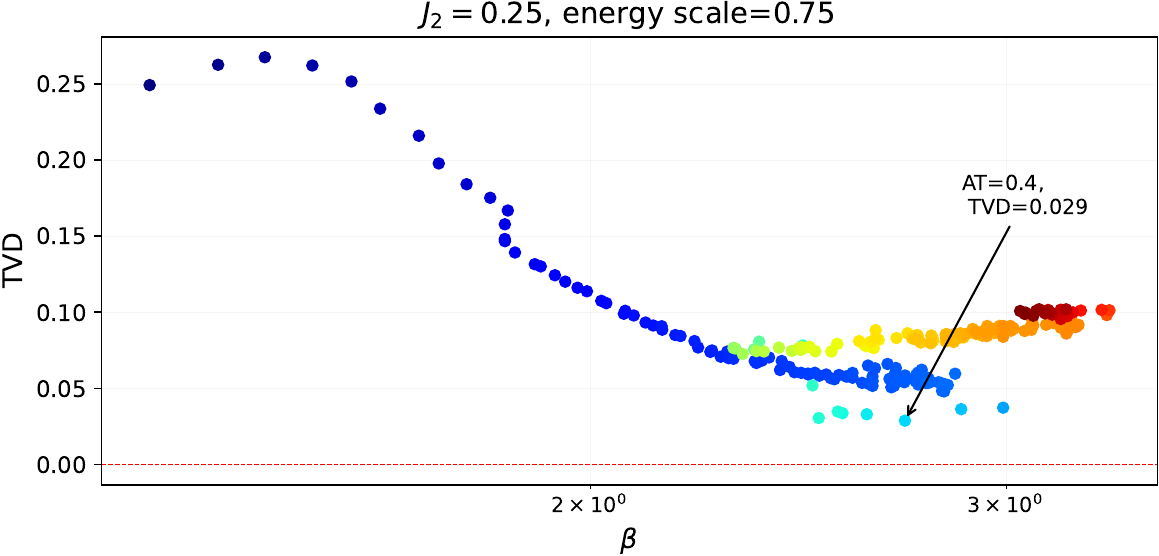}
    \includegraphics[width=0.495\linewidth]{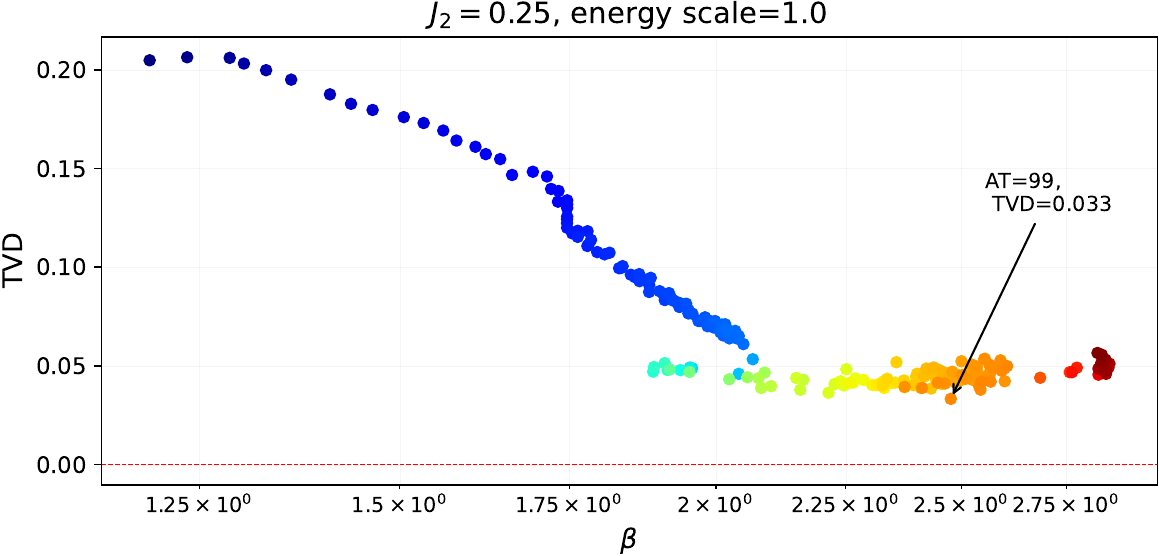}
    \includegraphics[width=0.495\linewidth]{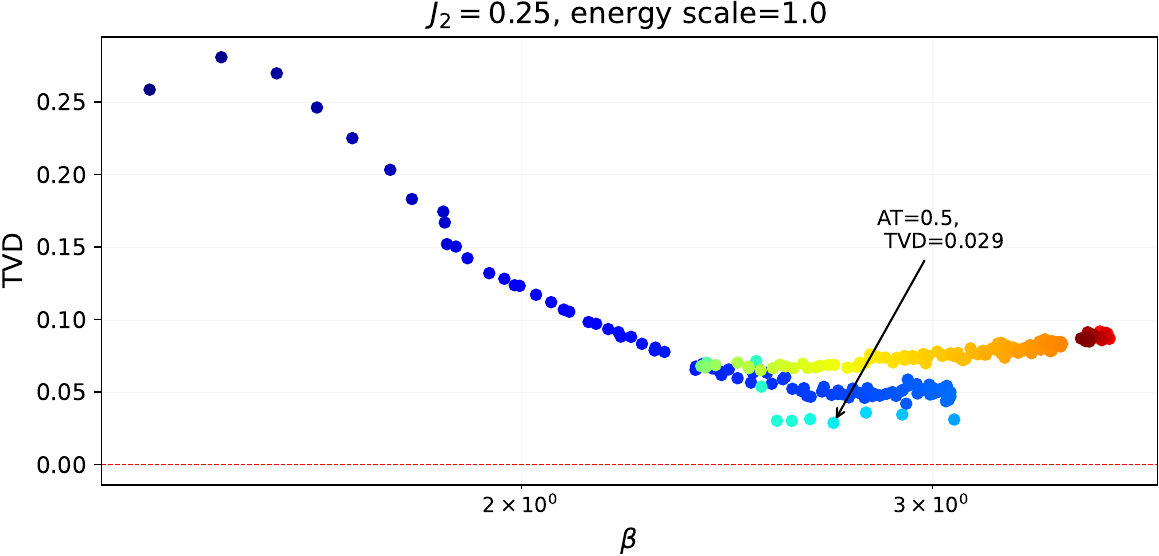}
    \includegraphics[width=0.6\linewidth]{figures/scatter_beta_vs_TVD/AT_colorbar.pdf}
    \caption{ $J_2=0.25$ ANNNI frustration parameter, run on \texttt{Advantage2\_system1.4} (right column) and \texttt{Advantage\_system4.1} (left column).  }
    \label{fig:J2_0.25_QPU_comparison}
\end{figure*}

\begin{figure*}[ht!]
    \centering
    \includegraphics[width=0.495\linewidth]{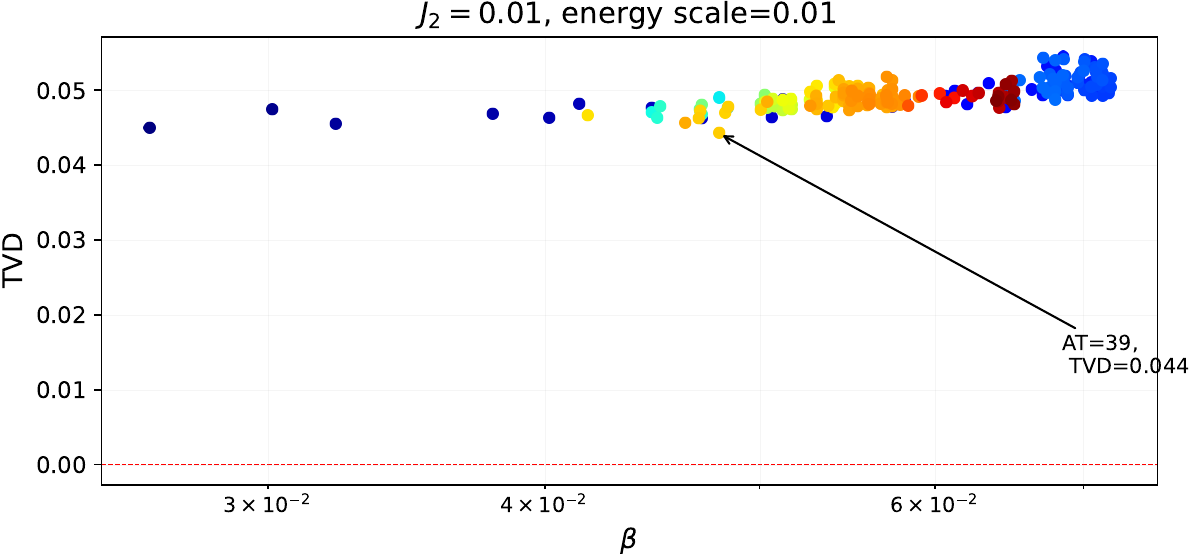}
    \includegraphics[width=0.495\linewidth]{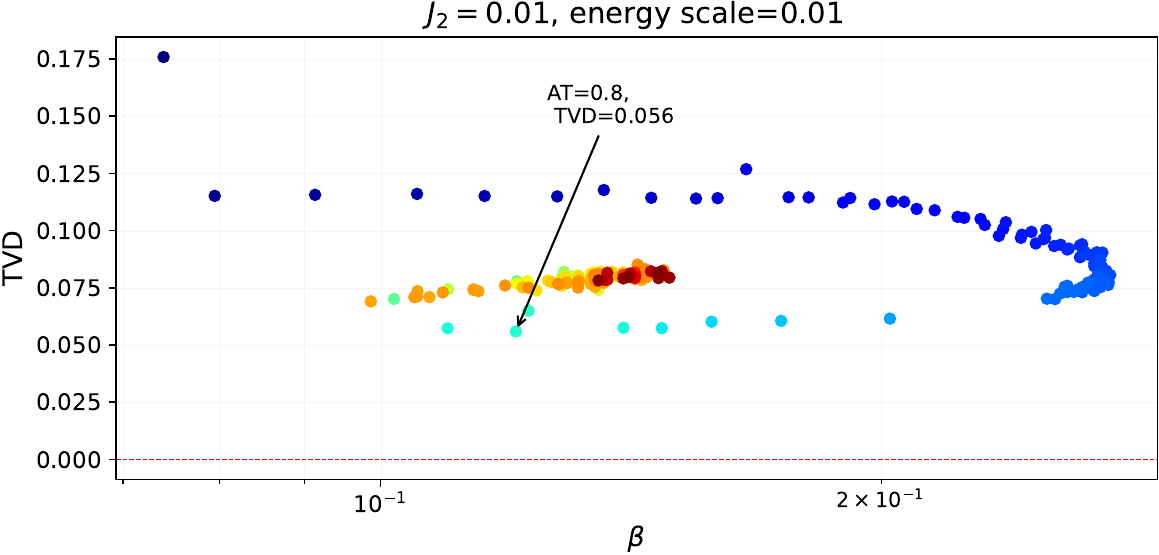}
    \includegraphics[width=0.495\linewidth]{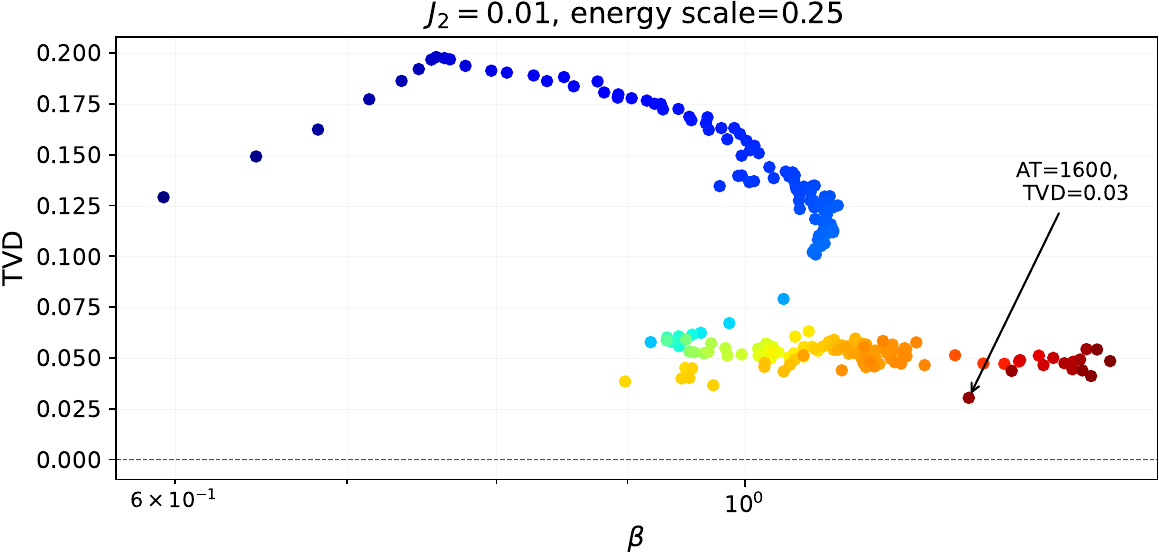}
    \includegraphics[width=0.495\linewidth]{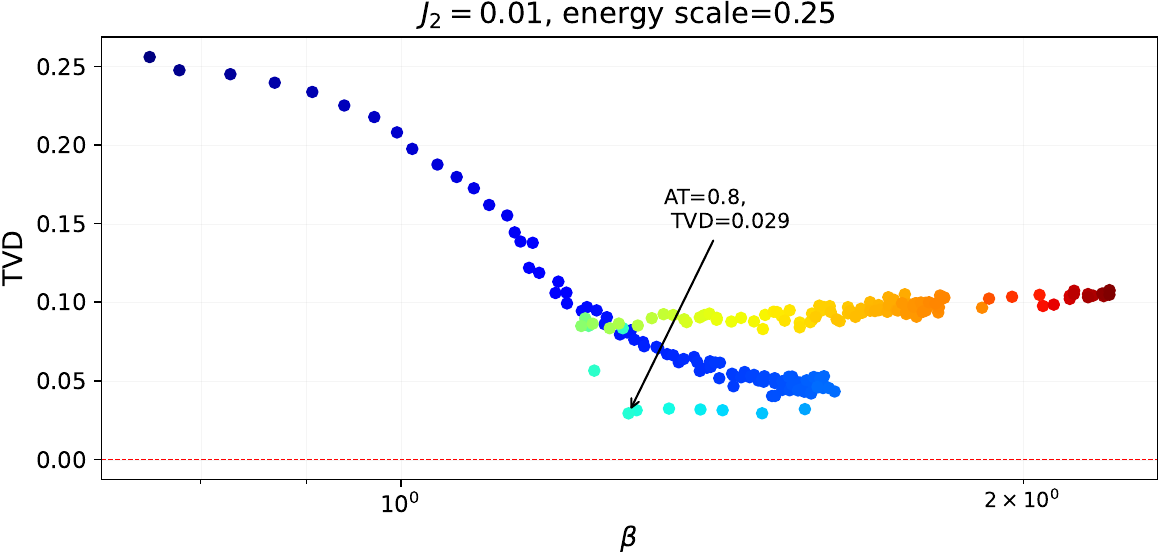}
    \includegraphics[width=0.495\linewidth]{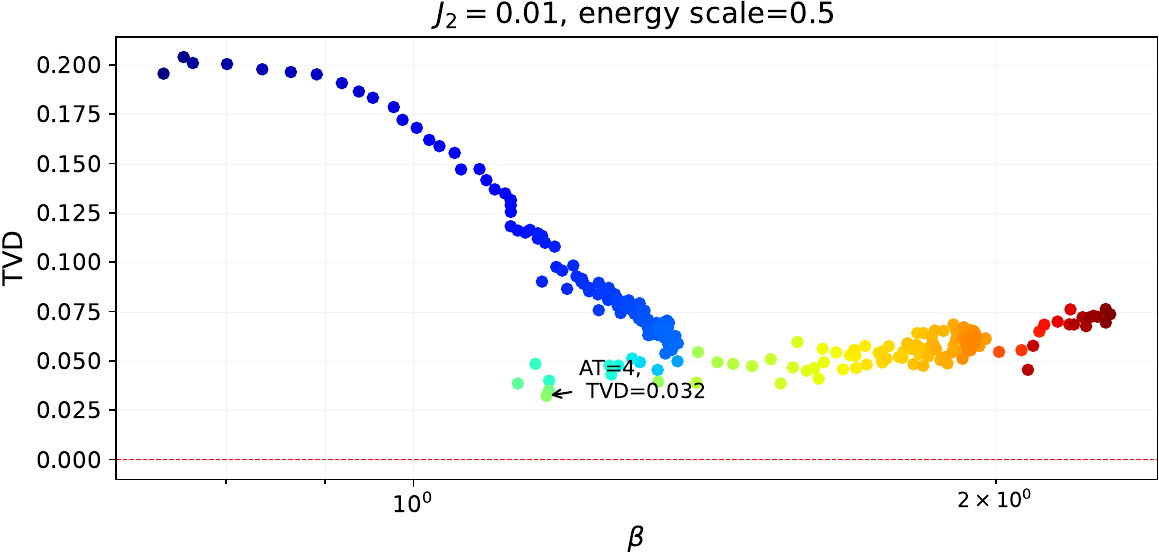}
    \includegraphics[width=0.495\linewidth]{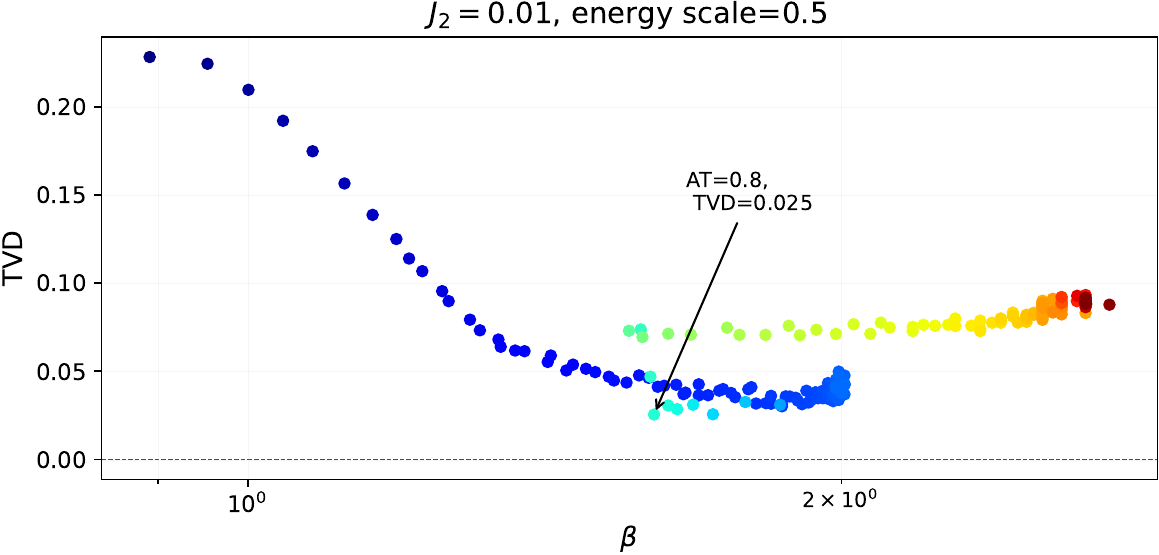}
    \includegraphics[width=0.495\linewidth]{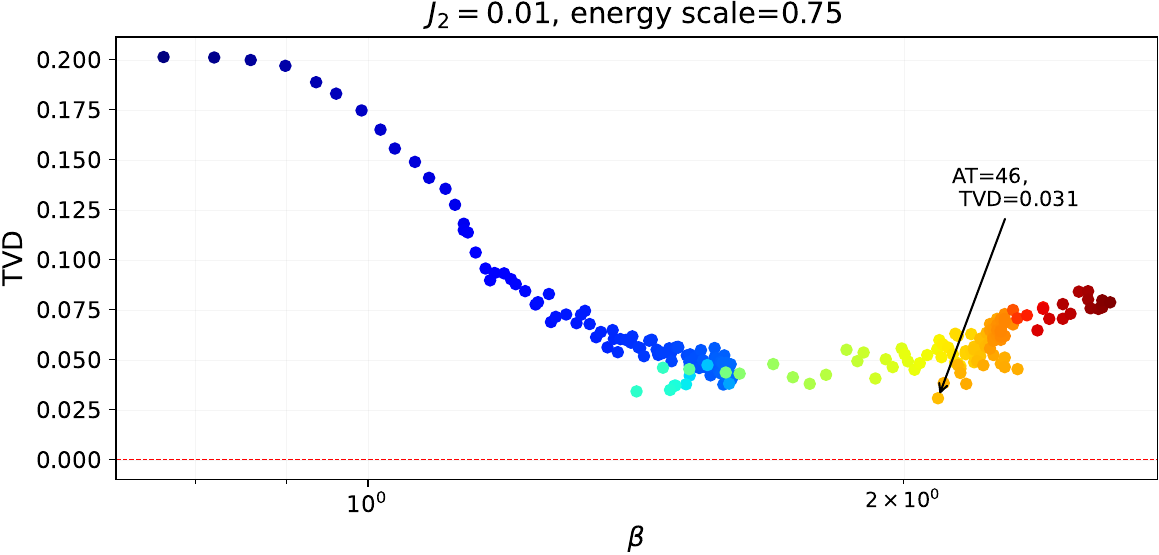}
    \includegraphics[width=0.495\linewidth]{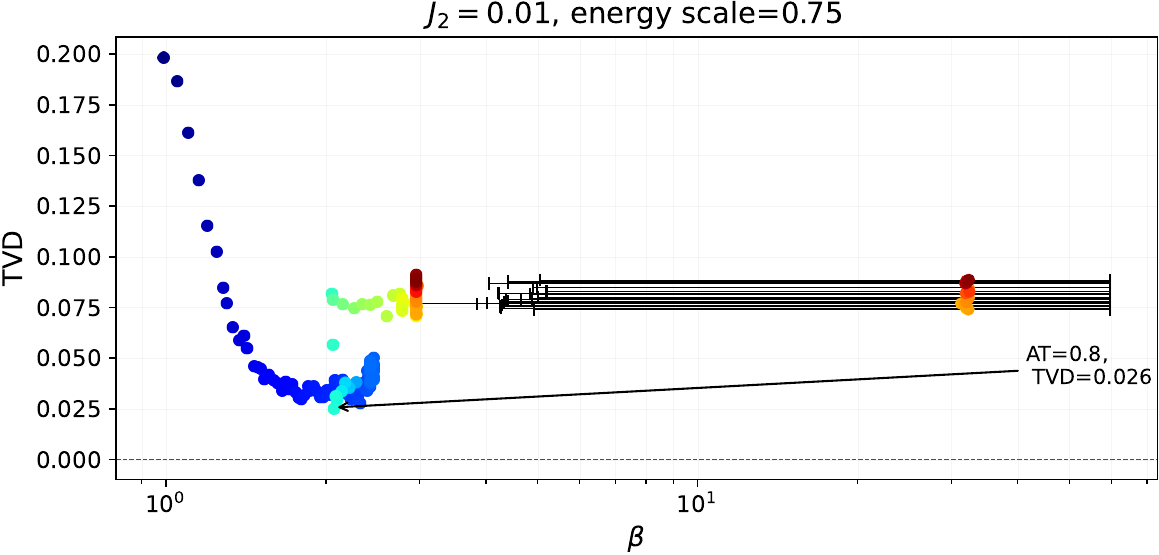}
    \includegraphics[width=0.495\linewidth]{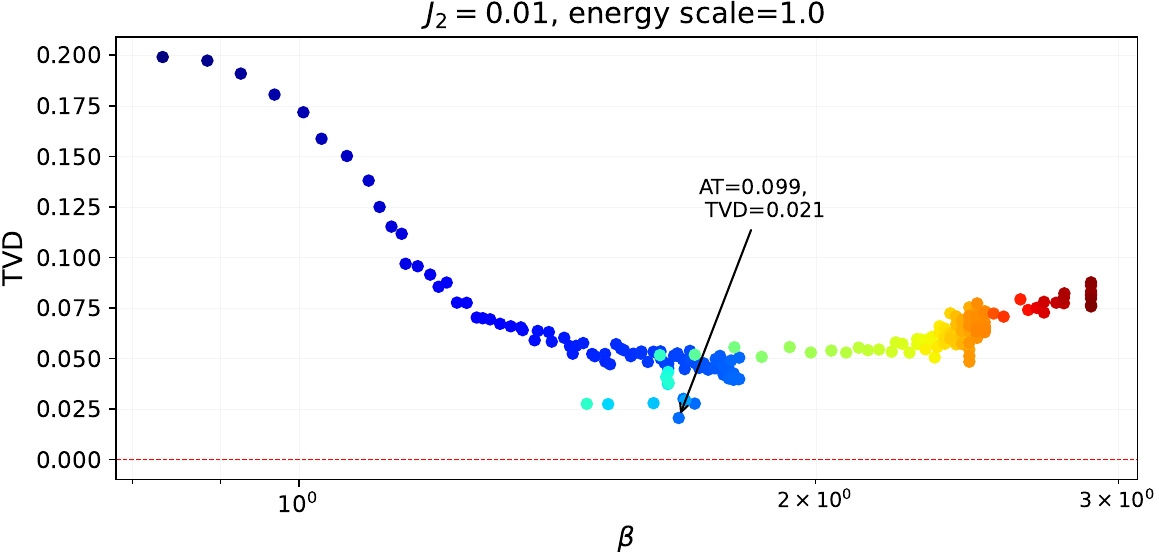}
    \includegraphics[width=0.495\linewidth]{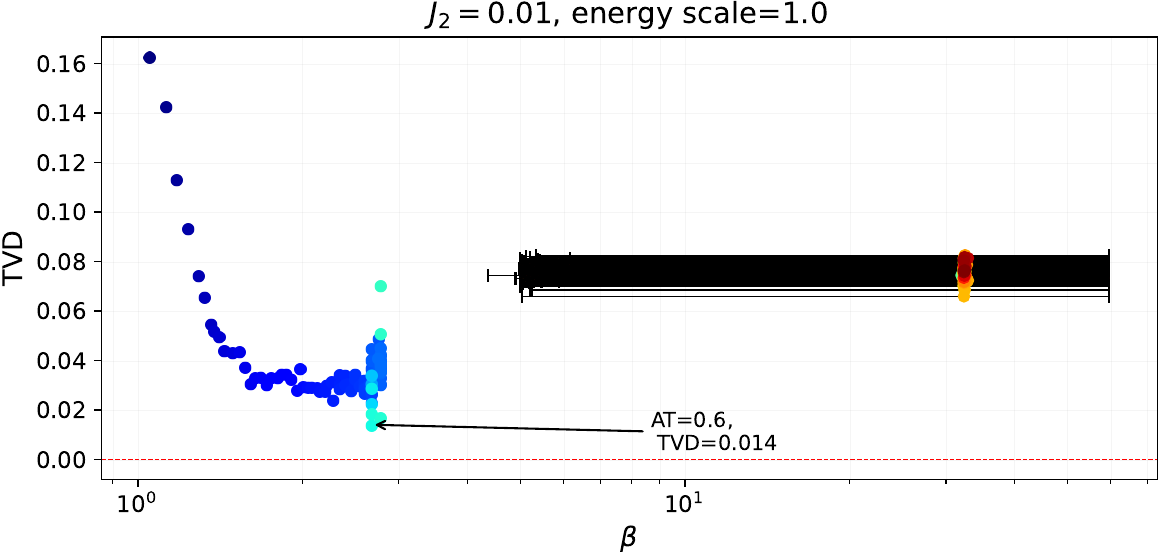}
    \includegraphics[width=0.6\linewidth]{figures/scatter_beta_vs_TVD/AT_colorbar.pdf}
    \caption{ $J_2=0.01$ ANNNI frustration parameter, run on \texttt{Advantage2\_system1.4} (right column) and \texttt{Advantage\_system4.1} (left column).     }
    \label{fig:J2_0.01_QPU_comparison}
\end{figure*}

Figure~\ref{fig:J2_1_QPU_comparison} shows complete error rate versus $\beta$ distributions for the full range of annealing times, at the $J_2=1$ ANNNI frustration parameter. Similarly, Figure~\ref{fig:J2_0.75_QPU_comparison} shows the same for $J_2=0.75$, Figure~\ref{fig:J2_0.25_QPU_comparison} shows the same for $J_2=0.25$, and Figure~\ref{fig:J2_0.01_QPU_comparison} shows the same for $J_2=0.01$. Interestingly, similar to the $J_2=0.5$ results shown in the main text, Figure~\ref{fig:J2_1_QPU_comparison} and Figure~\ref{fig:J2_0.01_QPU_comparison} show us that $J_2=0.01$ and $J_2=1$ also result in some uncertainty in the sampled effective temperature, albeit only at some of the annealing times.

\begin{figure*}[ht!]
    \centering
    \includegraphics[width=0.495\linewidth]{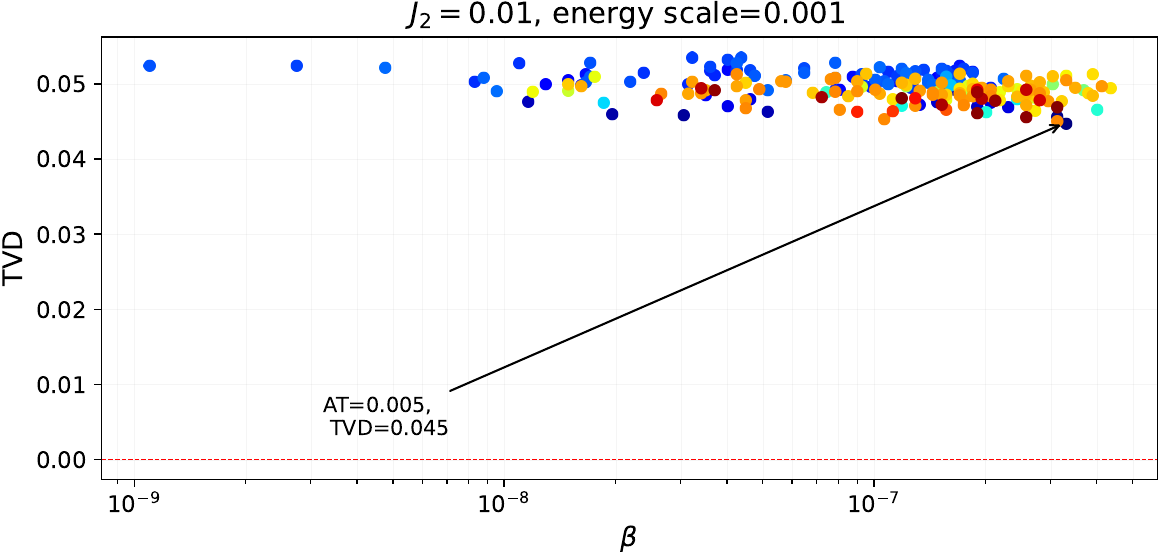}
    \includegraphics[width=0.495\linewidth]{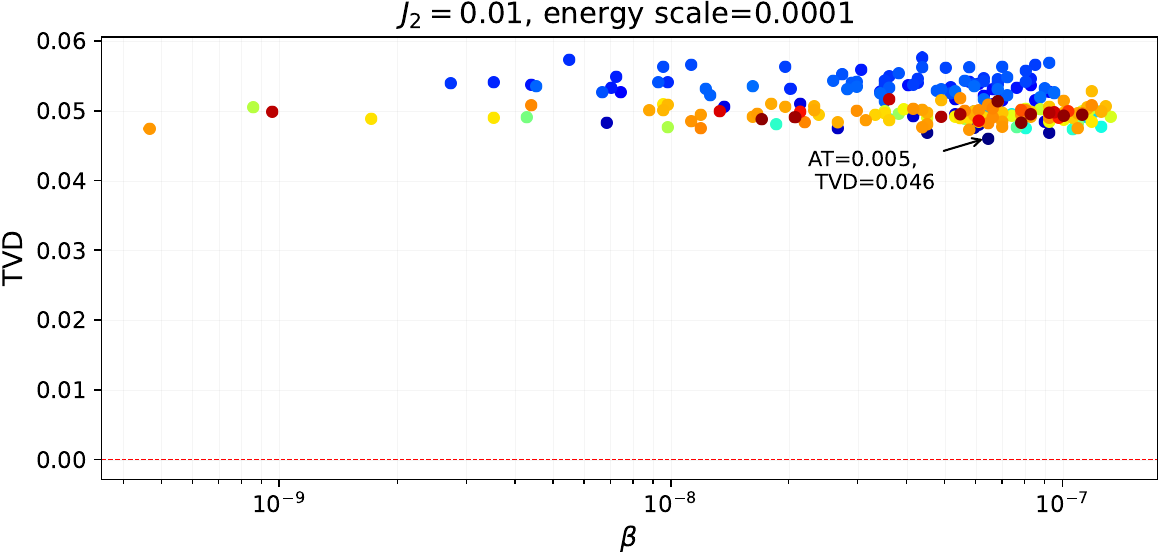}
    \includegraphics[width=0.495\linewidth]{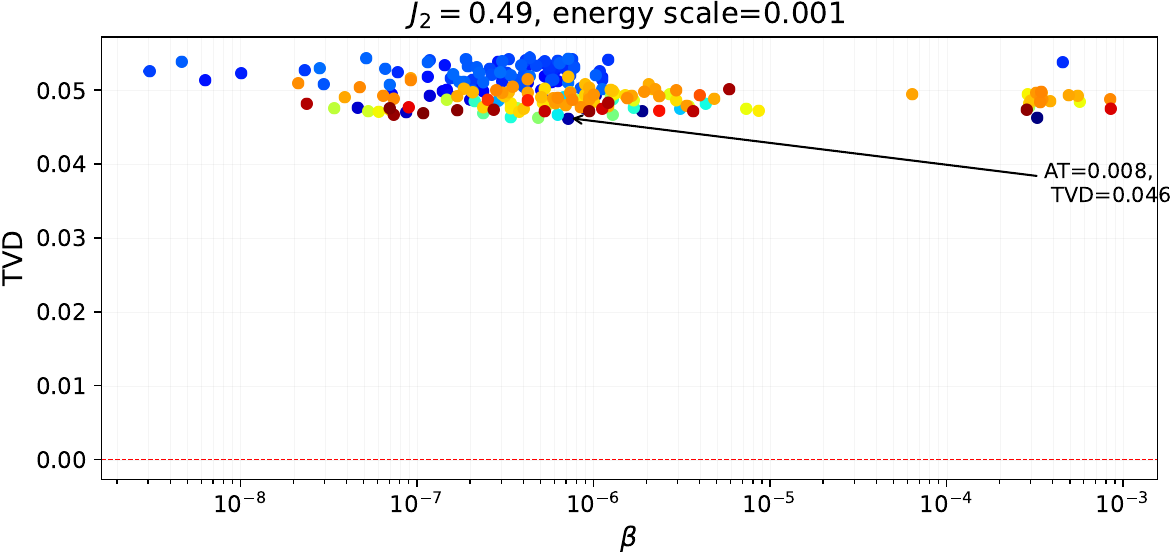}
    \includegraphics[width=0.495\linewidth]{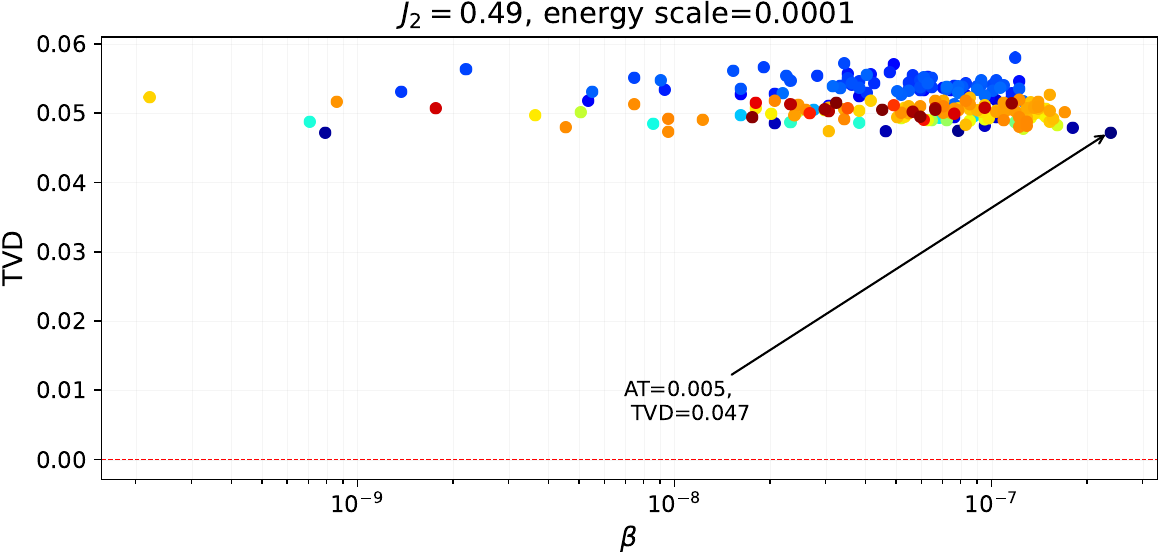}
    \includegraphics[width=0.495\linewidth]{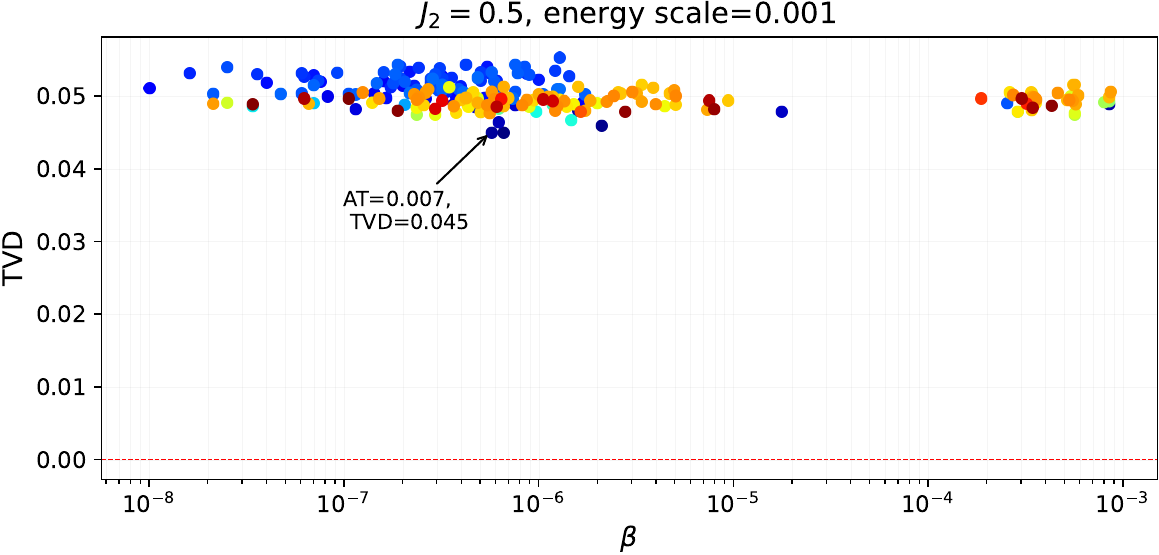}
    \includegraphics[width=0.495\linewidth]{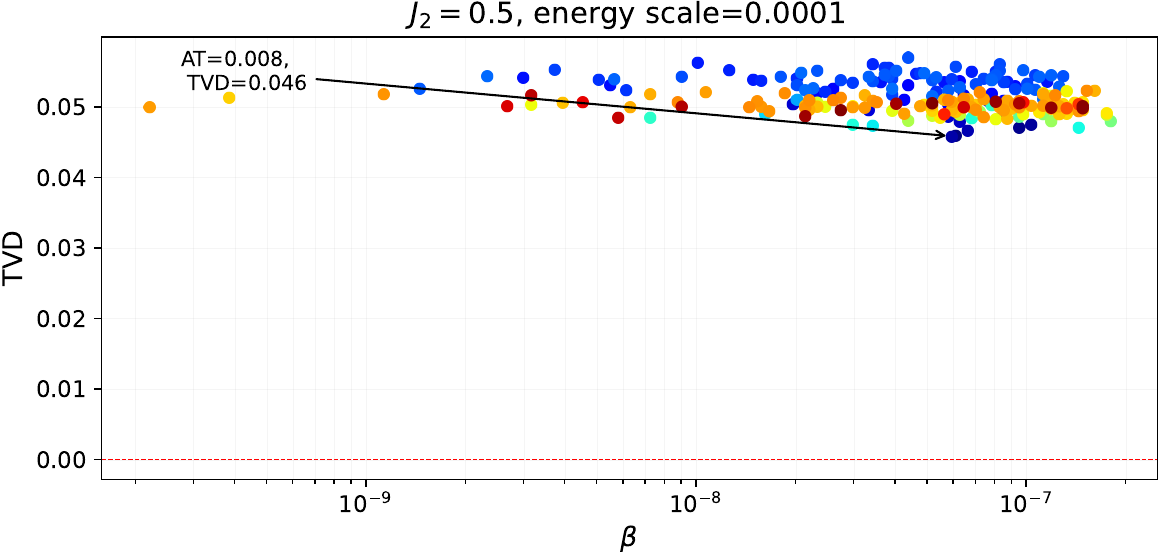}
    \includegraphics[width=0.495\linewidth]{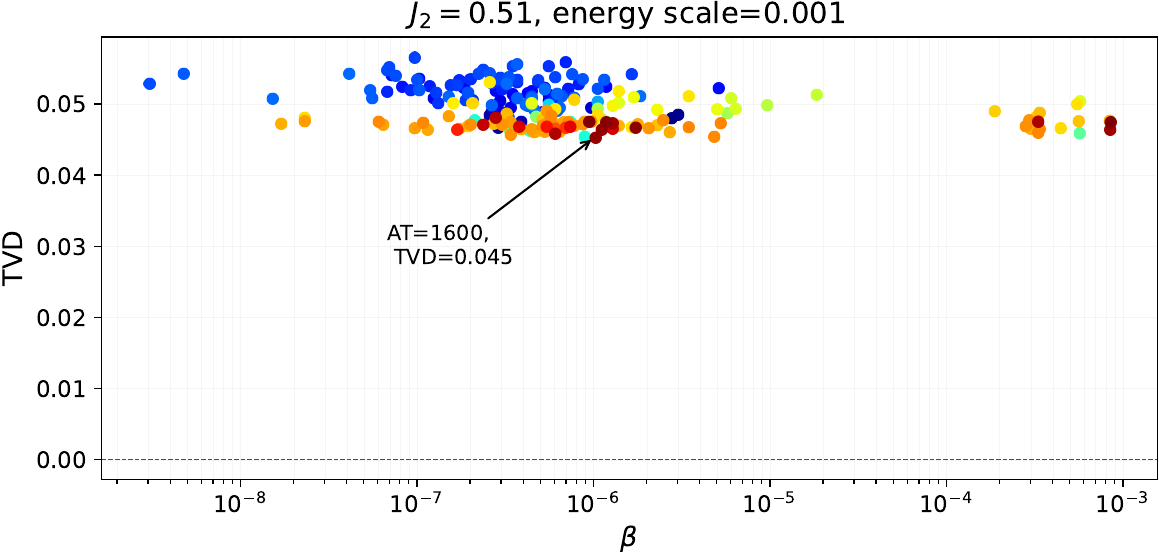}
    \includegraphics[width=0.495\linewidth]{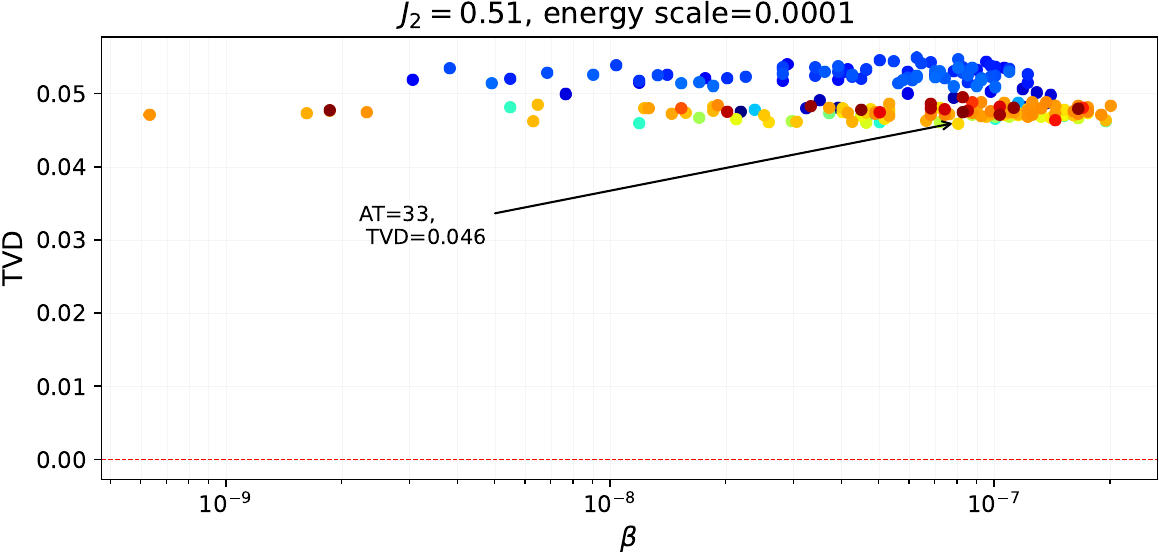}
    \includegraphics[width=0.495\linewidth]{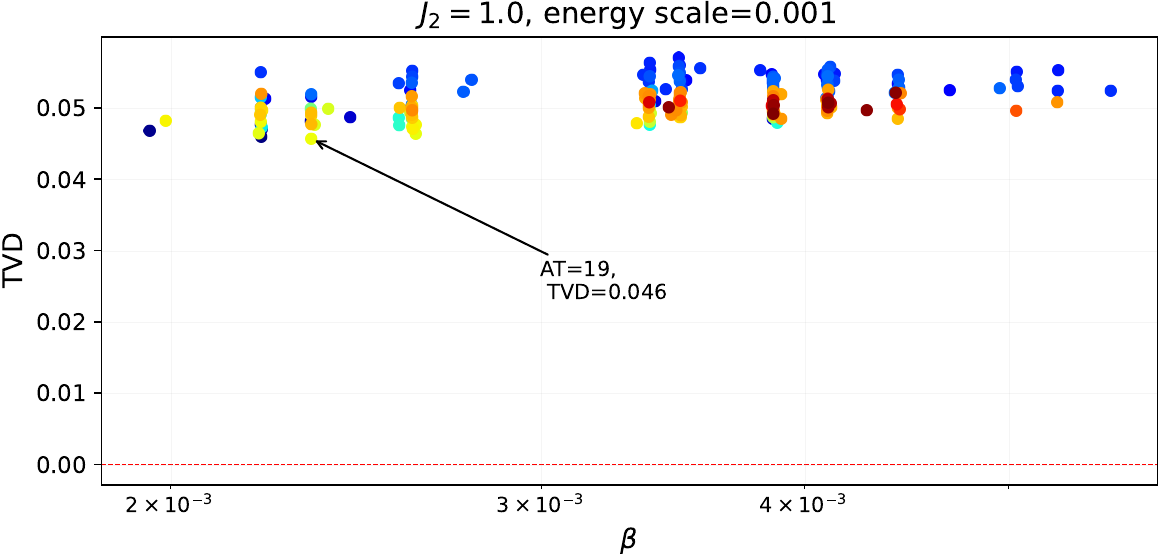}
    \includegraphics[width=0.495\linewidth]{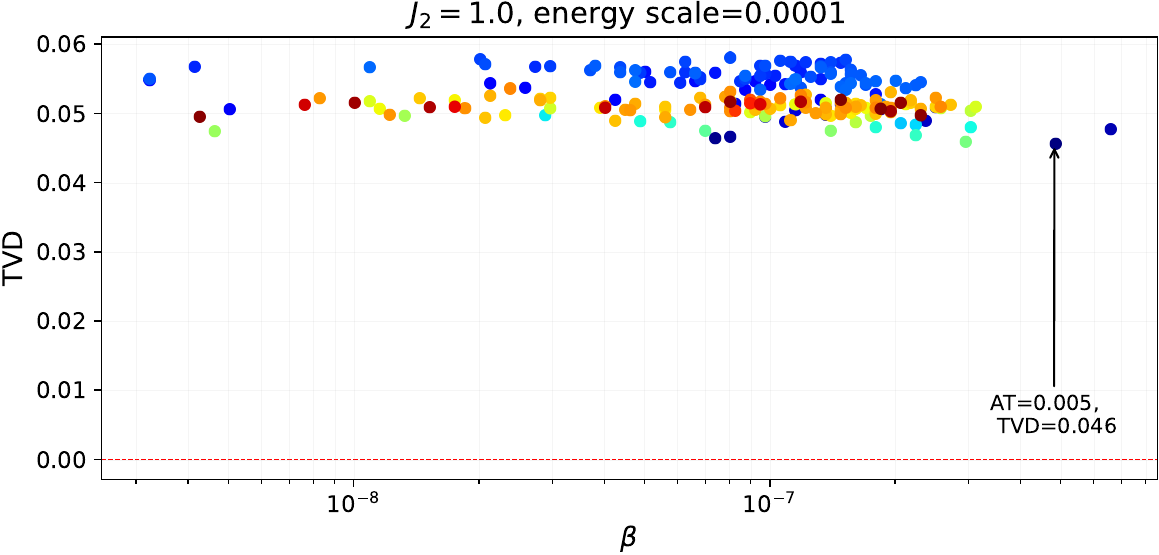}
    \includegraphics[width=0.6\linewidth]{figures/scatter_beta_vs_TVD/AT_colorbar.pdf}
    \caption{ Comparing small energy scales on the analog hardware, with an overall J coupler energy scale of $0.001$ (left column) and $0.0001$ (right column). Results from \texttt{Advantage\_system4.1}.   }
    \label{fig:small_J_coeff_precision_comparison_Pegasus}
\end{figure*}

\begin{figure*}[ht!]
    \centering
    \includegraphics[width=0.495\linewidth]{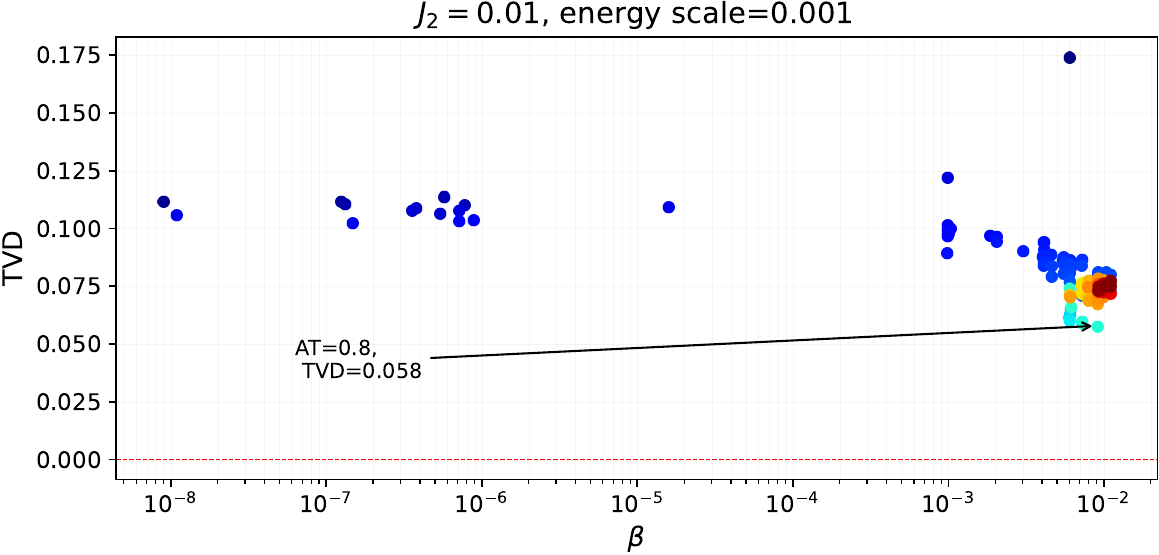}
    \includegraphics[width=0.495\linewidth]{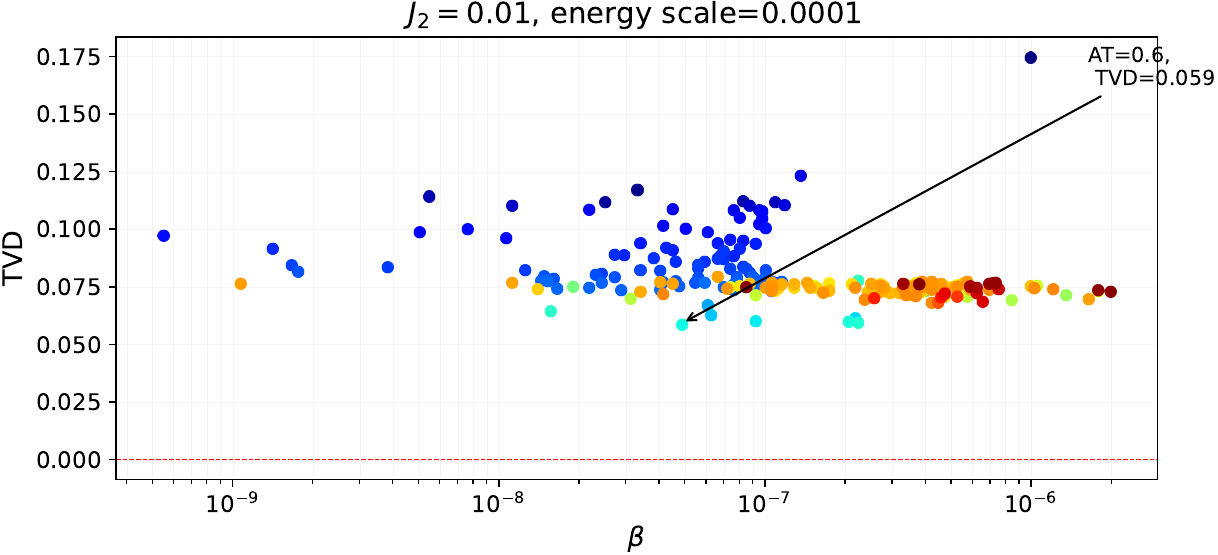}
    \includegraphics[width=0.495\linewidth]{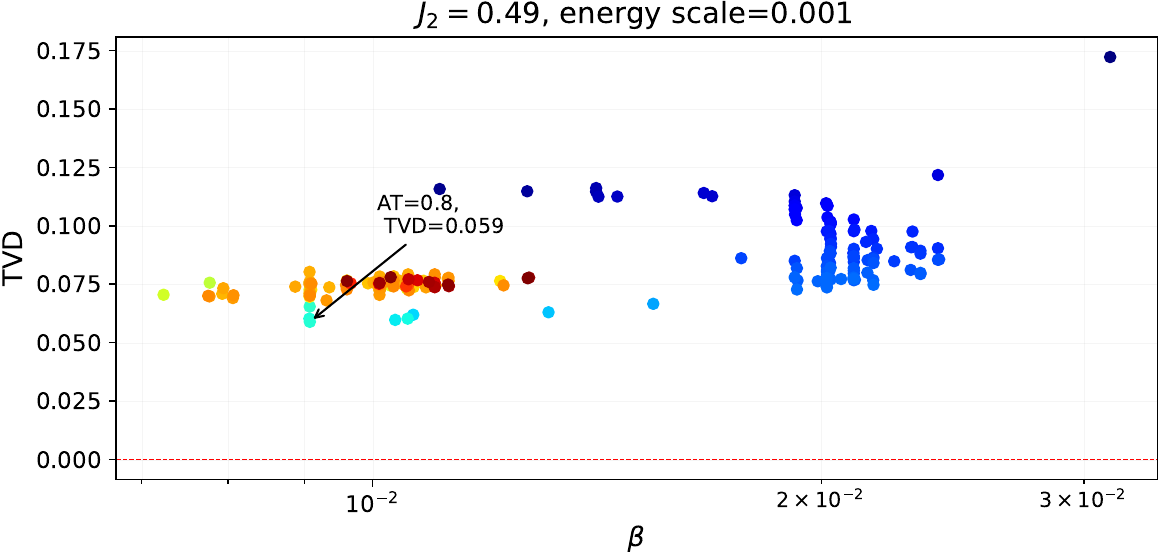}
    \includegraphics[width=0.495\linewidth]{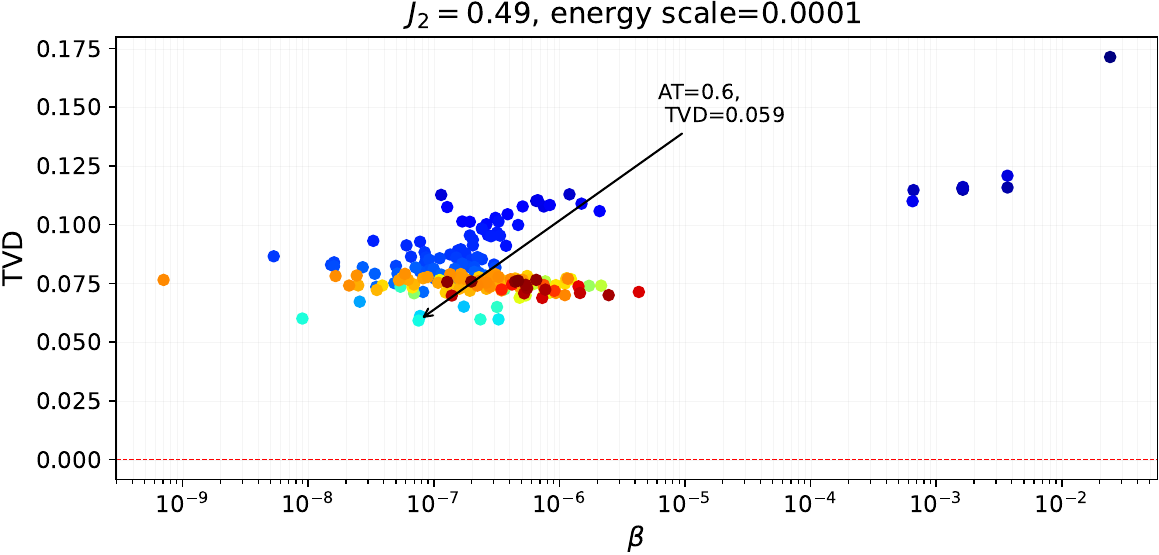}
    \includegraphics[width=0.495\linewidth]{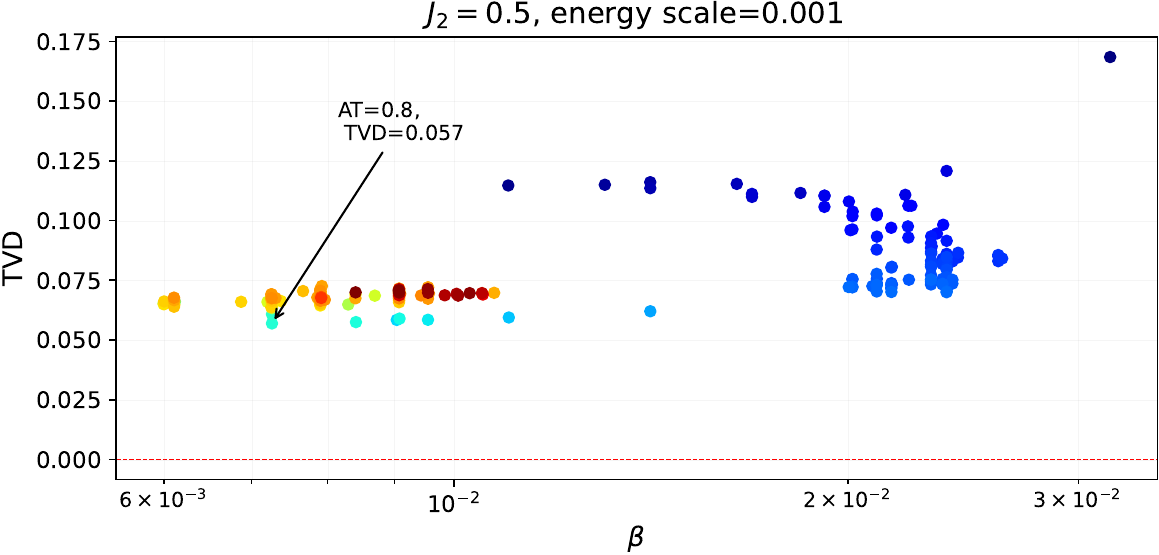}
    \includegraphics[width=0.495\linewidth]{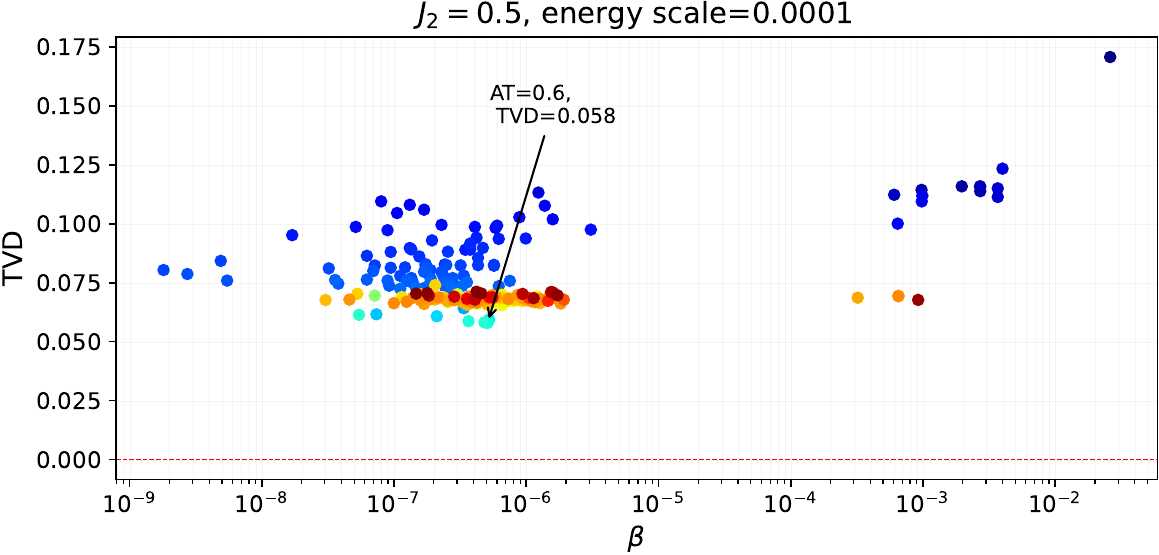}
    \includegraphics[width=0.495\linewidth]{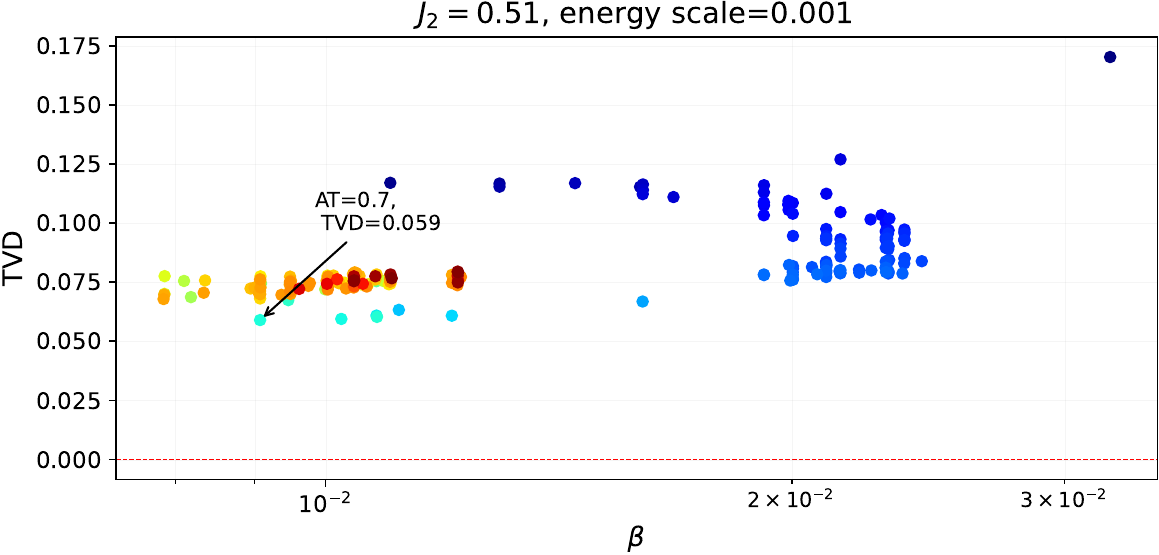}
    \includegraphics[width=0.495\linewidth]{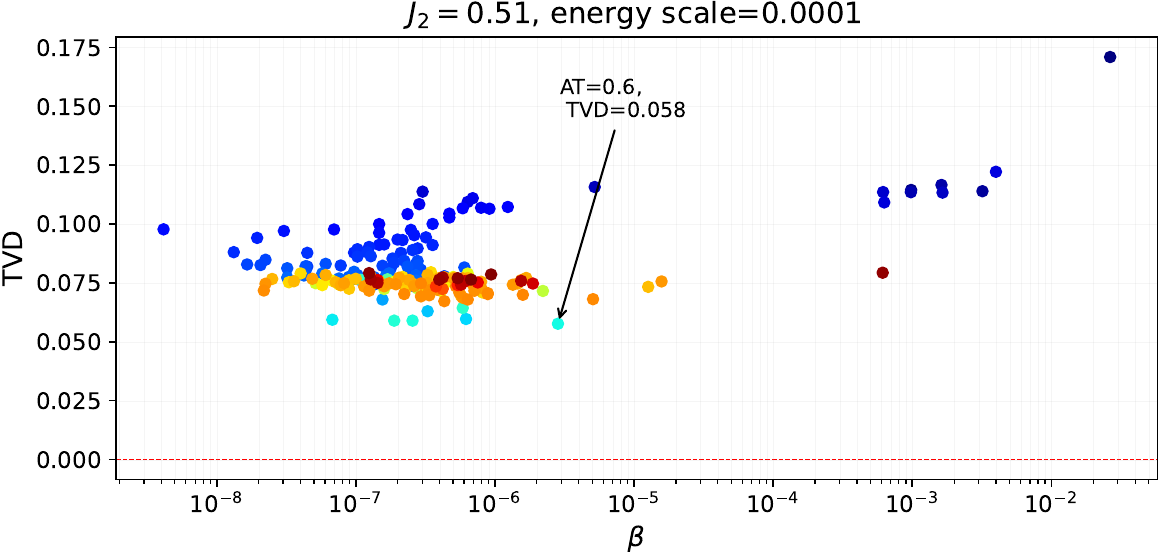}
    \includegraphics[width=0.495\linewidth]{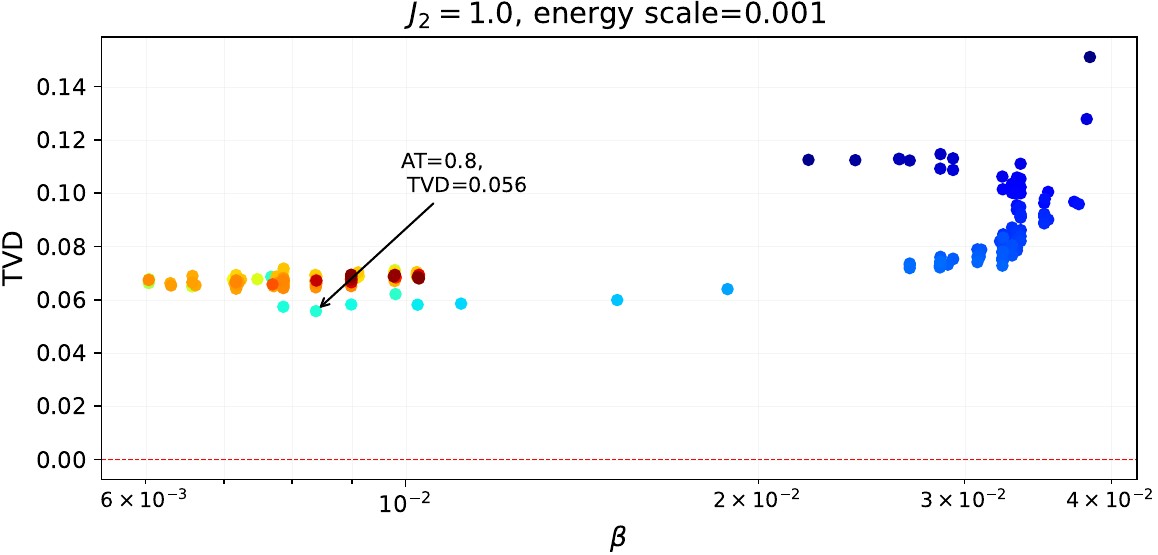}
    \includegraphics[width=0.495\linewidth]{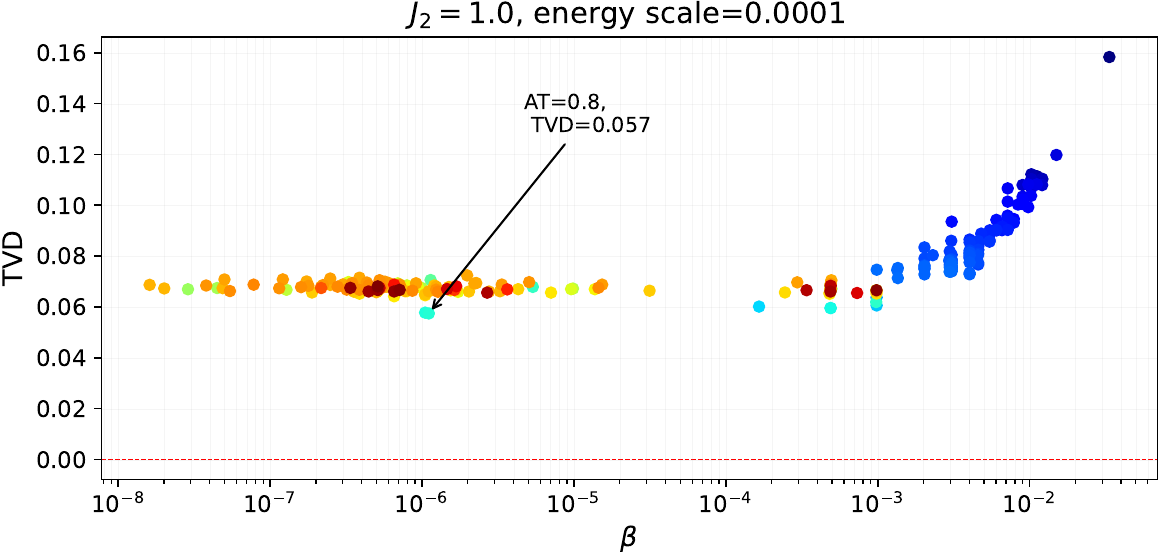}
    \includegraphics[width=0.6\linewidth]{figures/scatter_beta_vs_TVD/AT_colorbar.pdf}
    \caption{ Comparing small energy scales on the analog hardware, with an overall J scale of $0.001$ (left column) and $0.0001$ (right column). Results from \texttt{Advantage2\_system1.4}.   }
    \label{fig:small_J_coeff_precision_comparison_Zephyr2}
\end{figure*}

Figure~\ref{fig:small_J_coeff_precision_comparison_Pegasus} and Figure~\ref{fig:small_J_coeff_precision_comparison_Zephyr2} both show error rate as a function of inverse temperature $\beta$ distributions when the overall J energy scale is very small -- $0.001$. This means that the effective temperature of the sampling on the hardware is quite high. Importantly, this shows that this coupler energy scale is close to the overall analog precision limit on the hardware, and in particular when the frustration parameter $J_2$ is small in these plots, the effective antiferromagnetic couplers on the hardware are very likely effectively zero.

\clearpage

\bibliographystyle{apsrev4-2-titles}
\bibliography{references}

\begin{thebibliography}{99}%
\makeatletter
\providecommand \@ifxundefined [1]{%
 \@ifx{#1\undefined}
}%
\providecommand \@ifnum [1]{%
 \ifnum #1\expandafter \@firstoftwo
 \else \expandafter \@secondoftwo
 \fi
}%
\providecommand \@ifx [1]{%
 \ifx #1\expandafter \@firstoftwo
 \else \expandafter \@secondoftwo
 \fi
}%
\providecommand \natexlab [1]{#1}%
\providecommand \enquote  [1]{``#1''}%
\providecommand \bibnamefont  [1]{#1}%
\providecommand \bibfnamefont [1]{#1}%
\providecommand \citenamefont [1]{#1}%
\providecommand \href@noop [0]{\@secondoftwo}%
\providecommand \href [0]{\begingroup \@sanitize@url \@href}%
\providecommand \@href[1]{\@@startlink{#1}\@@href}%
\providecommand \@@href[1]{\endgroup#1\@@endlink}%
\providecommand \@sanitize@url [0]{\catcode `\\12\catcode `\$12\catcode `\&12\catcode `\#12\catcode `\^12\catcode `\_12\catcode `\%12\relax}%
\providecommand \@@startlink[1]{}%
\providecommand \@@endlink[0]{}%
\providecommand \url  [0]{\begingroup\@sanitize@url \@url }%
\providecommand \@url [1]{\endgroup\@href {#1}{\urlprefix }}%
\providecommand \urlprefix  [0]{URL }%
\providecommand \Eprint [0]{\href }%
\providecommand \doibase [0]{https://doi.org/}%
\providecommand \selectlanguage [0]{\@gobble}%
\providecommand \bibinfo  [0]{\@secondoftwo}%
\providecommand \bibfield  [0]{\@secondoftwo}%
\providecommand \translation [1]{[#1]}%
\providecommand \BibitemOpen [0]{}%
\providecommand \bibitemStop [0]{}%
\providecommand \bibitemNoStop [0]{.\EOS\space}%
\providecommand \EOS [0]{\spacefactor3000\relax}%
\providecommand \BibitemShut  [1]{\csname bibitem#1\endcsname}%
\let\auto@bib@innerbib\@empty
\bibitem [{\citenamefont {Elliott}(1961)}]{PhysRev.124.346}%
  \BibitemOpen
  \bibfield  {author} {\bibinfo {author} {\bibfnamefont {R.~J.}\ \bibnamefont {Elliott}},\ }\bibfield  {title} {\emph {\bibinfo {title} {Phenomenological discussion of magnetic ordering in the heavy rare-earth metals}},\ }\href {https://doi.org/10.1103/PhysRev.124.346} {\bibfield  {journal} {\bibinfo  {journal} {Phys. Rev.}\ }\textbf {\bibinfo {volume} {124}},\ \bibinfo {pages} {346--353} (\bibinfo {year} {1961})}\BibitemShut {NoStop}%
\bibitem [{\citenamefont {Fisher}\ and\ \citenamefont {Selke}(1980)}]{PhysRevLett.44.1502}%
  \BibitemOpen
  \bibfield  {author} {\bibinfo {author} {\bibfnamefont {M.~E.}\ \bibnamefont {Fisher}}\ and\ \bibinfo {author} {\bibfnamefont {W.}~\bibnamefont {Selke}},\ }\bibfield  {title} {\emph {\bibinfo {title} {{Infinitely Many Commensurate Phases in a Simple Ising Model}}},\ }\href {https://doi.org/10.1103/PhysRevLett.44.1502} {\bibfield  {journal} {\bibinfo  {journal} {Phys. Rev. Lett.}\ }\textbf {\bibinfo {volume} {44}},\ \bibinfo {pages} {1502--1505} (\bibinfo {year} {1980})}\BibitemShut {NoStop}%
\bibitem [{\citenamefont {Selke}(1988)}]{selke1988annni}%
  \BibitemOpen
  \bibfield  {author} {\bibinfo {author} {\bibfnamefont {W.}~\bibnamefont {Selke}},\ }\bibfield  {title} {\emph {\bibinfo {title} {{The ANNNI model—theoretical analysis and experimental application}}},\ }\href@noop {} {\bibfield  {journal} {\bibinfo  {journal} {Physics Reports}\ }\textbf {\bibinfo {volume} {170}},\ \bibinfo {pages} {213--264} (\bibinfo {year} {1988})}\BibitemShut {NoStop}%
\bibitem [{\citenamefont {Fisher}\ and\ \citenamefont {Selke}(1981)}]{fisher1981low}%
  \BibitemOpen
  \bibfield  {author} {\bibinfo {author} {\bibfnamefont {M.~E.}\ \bibnamefont {Fisher}}\ and\ \bibinfo {author} {\bibfnamefont {W.}~\bibnamefont {Selke}},\ }\bibfield  {title} {\emph {\bibinfo {title} {{Low temperature analysis of the axial next-nearest neighbour Ising model near its multiphase point}}},\ }\href@noop {} {\bibfield  {journal} {\bibinfo  {journal} {Philosophical Transactions of the Royal Society of London. Series A, Mathematical and Physical Sciences}\ }\textbf {\bibinfo {volume} {302}},\ \bibinfo {pages} {1--44} (\bibinfo {year} {1981})}\BibitemShut {NoStop}%
\bibitem [{\citenamefont {Sciortino}\ \emph {et~al.}(2004)\citenamefont {Sciortino}, \citenamefont {Mossa}, \citenamefont {Zaccarelli},\ and\ \citenamefont {Tartaglia}}]{PhysRevLett.93.055701}%
  \BibitemOpen
  \bibfield  {author} {\bibinfo {author} {\bibfnamefont {F.}~\bibnamefont {Sciortino}}, \bibinfo {author} {\bibfnamefont {S.}~\bibnamefont {Mossa}}, \bibinfo {author} {\bibfnamefont {E.}~\bibnamefont {Zaccarelli}},\ and\ \bibinfo {author} {\bibfnamefont {P.}~\bibnamefont {Tartaglia}},\ }\bibfield  {title} {\emph {\bibinfo {title} {Equilibrium cluster phases and low-density arrested disordered states: The role of short-range attraction and long-range repulsion}},\ }\href {https://doi.org/10.1103/PhysRevLett.93.055701} {\bibfield  {journal} {\bibinfo  {journal} {Phys. Rev. Lett.}\ }\textbf {\bibinfo {volume} {93}},\ \bibinfo {pages} {055701} (\bibinfo {year} {2004})}\BibitemShut {NoStop}%
\bibitem [{\citenamefont {Rieger}\ and\ \citenamefont {Uimin}(1996)}]{Rieger_1996}%
  \BibitemOpen
  \bibfield  {author} {\bibinfo {author} {\bibfnamefont {H.}~\bibnamefont {Rieger}}\ and\ \bibinfo {author} {\bibfnamefont {G.}~\bibnamefont {Uimin}},\ }\bibfield  {title} {\emph {\bibinfo {title} {{The one-dimensional ANNNI model in a transverse field: analytic and numerical study of effective Hamiltonians}}},\ }\href {https://doi.org/10.1007/s002570050252} {\bibfield  {journal} {\bibinfo  {journal} {Zeitschrift für Physik B Condensed Matter}\ }\textbf {\bibinfo {volume} {101}},\ \bibinfo {pages} {597–611} (\bibinfo {year} {1996})}\BibitemShut {NoStop}%
\bibitem [{\citenamefont {Rastelli}\ \emph {et~al.}(2010)\citenamefont {Rastelli}, \citenamefont {Regina},\ and\ \citenamefont {Tassi}}]{PhysRevB.81.094425}%
  \BibitemOpen
  \bibfield  {author} {\bibinfo {author} {\bibfnamefont {E.}~\bibnamefont {Rastelli}}, \bibinfo {author} {\bibfnamefont {S.}~\bibnamefont {Regina}},\ and\ \bibinfo {author} {\bibfnamefont {A.}~\bibnamefont {Tassi}},\ }\bibfield  {title} {\emph {\bibinfo {title} {{Specific heat and structure factor in the square ANNNI model by Monte Carlo simulation}}},\ }\href {https://doi.org/10.1103/PhysRevB.81.094425} {\bibfield  {journal} {\bibinfo  {journal} {Phys. Rev. B}\ }\textbf {\bibinfo {volume} {81}},\ \bibinfo {pages} {094425} (\bibinfo {year} {2010})}\BibitemShut {NoStop}%
\bibitem [{\citenamefont {Grynberg}\ and\ \citenamefont {Tanatar}(1992)}]{PhysRevB.45.2876}%
  \BibitemOpen
  \bibfield  {author} {\bibinfo {author} {\bibfnamefont {M.~D.}\ \bibnamefont {Grynberg}}\ and\ \bibinfo {author} {\bibfnamefont {B.}~\bibnamefont {Tanatar}},\ }\bibfield  {title} {\emph {\bibinfo {title} {{Square Ising model with second-neighbor interactions and the Ising chain in a transverse field}}},\ }\href {https://doi.org/10.1103/PhysRevB.45.2876} {\bibfield  {journal} {\bibinfo  {journal} {Phys. Rev. B}\ }\textbf {\bibinfo {volume} {45}},\ \bibinfo {pages} {2876--2882} (\bibinfo {year} {1992})}\BibitemShut {NoStop}%
\bibitem [{\citenamefont {Dutta}\ \emph {et~al.}(2015)\citenamefont {Dutta}, \citenamefont {Aeppli}, \citenamefont {Chakrabarti}, \citenamefont {Divakaran}, \citenamefont {Rosenbaum},\ and\ \citenamefont {Sen}}]{dutta2015quantumphasetransitionstransverse}%
  \BibitemOpen
  \bibfield  {author} {\bibinfo {author} {\bibfnamefont {A.}~\bibnamefont {Dutta}}, \bibinfo {author} {\bibfnamefont {G.}~\bibnamefont {Aeppli}}, \bibinfo {author} {\bibfnamefont {B.~K.}\ \bibnamefont {Chakrabarti}}, \bibinfo {author} {\bibfnamefont {U.}~\bibnamefont {Divakaran}}, \bibinfo {author} {\bibfnamefont {T.~F.}\ \bibnamefont {Rosenbaum}},\ and\ \bibinfo {author} {\bibfnamefont {D.}~\bibnamefont {Sen}},\ }\href {https://arxiv.org/abs/1012.0653} {\bibinfo {title} {Quantum phase transitions in transverse field spin models: from statistical physics to quantum information}} (\bibinfo {year} {2015}),\ \Eprint {https://arxiv.org/abs/1012.0653} {arXiv:1012.0653 [cond-mat.stat-mech]} \BibitemShut {NoStop}%
\bibitem [{\citenamefont {Chandra}\ and\ \citenamefont {Dasgupta}(2007)}]{Chandra_2007}%
  \BibitemOpen
  \bibfield  {author} {\bibinfo {author} {\bibfnamefont {A.~K.}\ \bibnamefont {Chandra}}\ and\ \bibinfo {author} {\bibfnamefont {S.}~\bibnamefont {Dasgupta}},\ }\bibfield  {title} {\emph {\bibinfo {title} {{Floating phase in a 2D ANNNI model}}},\ }\href {https://doi.org/10.1088/1751-8113/40/24/001} {\bibfield  {journal} {\bibinfo  {journal} {Journal of Physics A: Mathematical and Theoretical}\ }\textbf {\bibinfo {volume} {40}},\ \bibinfo {pages} {6251–6265} (\bibinfo {year} {2007})}\BibitemShut {NoStop}%
\bibitem [{\citenamefont {Kadowaki}\ and\ \citenamefont {Nishimori}(1998)}]{Kadowaki_1998}%
  \BibitemOpen
  \bibfield  {author} {\bibinfo {author} {\bibfnamefont {T.}~\bibnamefont {Kadowaki}}\ and\ \bibinfo {author} {\bibfnamefont {H.}~\bibnamefont {Nishimori}},\ }\bibfield  {title} {\emph {\bibinfo {title} {Quantum annealing in the transverse ising model}},\ }\href {https://doi.org/10.1103/physreve.58.5355} {\bibfield  {journal} {\bibinfo  {journal} {Physical Review E}\ }\textbf {\bibinfo {volume} {58}},\ \bibinfo {pages} {5355--5363} (\bibinfo {year} {1998})},\ \Eprint {https://arxiv.org/abs/cond-mat/9804280} {arXiv:cond-mat/9804280} \BibitemShut {NoStop}%
\bibitem [{\citenamefont {Santoro}\ and\ \citenamefont {Tosatti}(2006)}]{santoro2006optimization}%
  \BibitemOpen
  \bibfield  {author} {\bibinfo {author} {\bibfnamefont {G.~E.}\ \bibnamefont {Santoro}}\ and\ \bibinfo {author} {\bibfnamefont {E.}~\bibnamefont {Tosatti}},\ }\bibfield  {title} {\emph {\bibinfo {title} {Optimization using quantum mechanics: quantum annealing through adiabatic evolution}},\ }\href@noop {} {\bibfield  {journal} {\bibinfo  {journal} {Journal of Physics A: Mathematical and General}\ }\textbf {\bibinfo {volume} {39}},\ \bibinfo {pages} {R393} (\bibinfo {year} {2006})}\BibitemShut {NoStop}%
\bibitem [{\citenamefont {Santoro}\ \emph {et~al.}(2002)\citenamefont {Santoro}, \citenamefont {Marto{\v{n}}\'{a}k}, \citenamefont {Tosatti},\ and\ \citenamefont {Car}}]{Santoro_2002}%
  \BibitemOpen
  \bibfield  {author} {\bibinfo {author} {\bibfnamefont {G.~E.}\ \bibnamefont {Santoro}}, \bibinfo {author} {\bibfnamefont {R.}~\bibnamefont {Marto{\v{n}}\'{a}k}}, \bibinfo {author} {\bibfnamefont {E.}~\bibnamefont {Tosatti}},\ and\ \bibinfo {author} {\bibfnamefont {R.}~\bibnamefont {Car}},\ }\bibfield  {title} {\emph {\bibinfo {title} {{Theory of Quantum Annealing of an Ising Spin Glass}}},\ }\href {https://doi.org/10.1126/science.1068774} {\bibfield  {journal} {\bibinfo  {journal} {Science}\ }\textbf {\bibinfo {volume} {295}},\ \bibinfo {pages} {2427–2430} (\bibinfo {year} {2002})}\BibitemShut {NoStop}%
\bibitem [{\citenamefont {Morita}\ and\ \citenamefont {Nishimori}(2008)}]{Morita_2008}%
  \BibitemOpen
  \bibfield  {author} {\bibinfo {author} {\bibfnamefont {S.}~\bibnamefont {Morita}}\ and\ \bibinfo {author} {\bibfnamefont {H.}~\bibnamefont {Nishimori}},\ }\bibfield  {title} {\emph {\bibinfo {title} {Mathematical foundation of quantum annealing}},\ }\bibfield  {journal} {\bibinfo  {journal} {Journal of Mathematical Physics}\ }\textbf {\bibinfo {volume} {49}},\ \href {https://doi.org/10.1063/1.2995837} {10.1063/1.2995837} (\bibinfo {year} {2008})\BibitemShut {NoStop}%
\bibitem [{\citenamefont {Farhi}\ \emph {et~al.}(2000)\citenamefont {Farhi}, \citenamefont {Goldstone}, \citenamefont {Gutmann},\ and\ \citenamefont {Sipser}}]{farhi2000quantumcomputationadiabaticevolution}%
  \BibitemOpen
  \bibfield  {author} {\bibinfo {author} {\bibfnamefont {E.}~\bibnamefont {Farhi}}, \bibinfo {author} {\bibfnamefont {J.}~\bibnamefont {Goldstone}}, \bibinfo {author} {\bibfnamefont {S.}~\bibnamefont {Gutmann}},\ and\ \bibinfo {author} {\bibfnamefont {M.}~\bibnamefont {Sipser}},\ }\href {https://arxiv.org/abs/quant-ph/0001106} {\bibinfo {title} {{Quantum Computation by Adiabatic Evolution}}} (\bibinfo {year} {2000}),\ \Eprint {https://arxiv.org/abs/quant-ph/0001106} {arXiv:quant-ph/0001106 [quant-ph]} \BibitemShut {NoStop}%
\bibitem [{\citenamefont {Harris}\ \emph {et~al.}(2018)\citenamefont {Harris}, \citenamefont {Sato}, \citenamefont {Berkley}, \citenamefont {Reis}, \citenamefont {Altomare}, \citenamefont {Amin}, \citenamefont {Boothby}, \citenamefont {Bunyk}, \citenamefont {Deng}, \citenamefont {Enderud} \emph {et~al.}}]{harris2018phase}%
  \BibitemOpen
  \bibfield  {author} {\bibinfo {author} {\bibfnamefont {R.}~\bibnamefont {Harris}}, \bibinfo {author} {\bibfnamefont {Y.}~\bibnamefont {Sato}}, \bibinfo {author} {\bibfnamefont {A.~J.}\ \bibnamefont {Berkley}}, \bibinfo {author} {\bibfnamefont {M.}~\bibnamefont {Reis}}, \bibinfo {author} {\bibfnamefont {F.}~\bibnamefont {Altomare}}, \bibinfo {author} {\bibfnamefont {M.}~\bibnamefont {Amin}}, \bibinfo {author} {\bibfnamefont {K.}~\bibnamefont {Boothby}}, \bibinfo {author} {\bibfnamefont {P.}~\bibnamefont {Bunyk}}, \bibinfo {author} {\bibfnamefont {C.}~\bibnamefont {Deng}}, \bibinfo {author} {\bibfnamefont {C.}~\bibnamefont {Enderud}}, \emph {et~al.},\ }\bibfield  {title} {\emph {\bibinfo {title} {Phase transitions in a programmable quantum spin glass simulator}},\ }\href@noop {} {\bibfield  {journal} {\bibinfo  {journal} {Science}\ }\textbf {\bibinfo {volume} {361}},\ \bibinfo {pages} {162--165} (\bibinfo {year} {2018})}\BibitemShut {NoStop}%
\bibitem [{\citenamefont {King}\ \emph {et~al.}(2025)\citenamefont {King}, \citenamefont {Nocera}, \citenamefont {Rams}, \citenamefont {Dziarmaga}, \citenamefont {Wiersema}, \citenamefont {Bernoudy}, \citenamefont {Raymond}, \citenamefont {Kaushal}, \citenamefont {Heinsdorf}, \citenamefont {Harris} \emph {et~al.}}]{king2025beyond}%
  \BibitemOpen
  \bibfield  {author} {\bibinfo {author} {\bibfnamefont {A.~D.}\ \bibnamefont {King}}, \bibinfo {author} {\bibfnamefont {A.}~\bibnamefont {Nocera}}, \bibinfo {author} {\bibfnamefont {M.~M.}\ \bibnamefont {Rams}}, \bibinfo {author} {\bibfnamefont {J.}~\bibnamefont {Dziarmaga}}, \bibinfo {author} {\bibfnamefont {R.}~\bibnamefont {Wiersema}}, \bibinfo {author} {\bibfnamefont {W.}~\bibnamefont {Bernoudy}}, \bibinfo {author} {\bibfnamefont {J.}~\bibnamefont {Raymond}}, \bibinfo {author} {\bibfnamefont {N.}~\bibnamefont {Kaushal}}, \bibinfo {author} {\bibfnamefont {N.}~\bibnamefont {Heinsdorf}}, \bibinfo {author} {\bibfnamefont {R.}~\bibnamefont {Harris}}, \emph {et~al.},\ }\bibfield  {title} {\emph {\bibinfo {title} {Beyond-classical computation in quantum simulation}},\ }\href@noop {} {\bibfield  {journal} {\bibinfo  {journal} {Science}\ }\textbf {\bibinfo {volume} {388}},\ \bibinfo {pages} {199--204} (\bibinfo {year} {2025})}\BibitemShut {NoStop}%
\bibitem [{\citenamefont {King}\ \emph {et~al.}(2021{\natexlab{a}})\citenamefont {King}, \citenamefont {Batista}, \citenamefont {Raymond}, \citenamefont {Lanting}, \citenamefont {Ozfidan}, \citenamefont {Poulin-Lamarre}, \citenamefont {Zhang},\ and\ \citenamefont {Amin}}]{PRXQuantum.2.030317}%
  \BibitemOpen
  \bibfield  {author} {\bibinfo {author} {\bibfnamefont {A.~D.}\ \bibnamefont {King}}, \bibinfo {author} {\bibfnamefont {C.~D.}\ \bibnamefont {Batista}}, \bibinfo {author} {\bibfnamefont {J.}~\bibnamefont {Raymond}}, \bibinfo {author} {\bibfnamefont {T.}~\bibnamefont {Lanting}}, \bibinfo {author} {\bibfnamefont {I.}~\bibnamefont {Ozfidan}}, \bibinfo {author} {\bibfnamefont {G.}~\bibnamefont {Poulin-Lamarre}}, \bibinfo {author} {\bibfnamefont {H.}~\bibnamefont {Zhang}},\ and\ \bibinfo {author} {\bibfnamefont {M.~H.}\ \bibnamefont {Amin}},\ }\bibfield  {title} {\emph {\bibinfo {title} {{Quantum Annealing Simulation of Out-of-Equilibrium Magnetization in a Spin-Chain Compound}}},\ }\href {https://doi.org/10.1103/PRXQuantum.2.030317} {\bibfield  {journal} {\bibinfo  {journal} {PRX Quantum}\ }\textbf {\bibinfo {volume} {2}},\ \bibinfo {pages} {030317} (\bibinfo {year} {2021}{\natexlab{a}})}\BibitemShut {NoStop}%
\bibitem [{\citenamefont {King}\ \emph {et~al.}(2021{\natexlab{b}})\citenamefont {King}, \citenamefont {Nisoli}, \citenamefont {Dahl}, \citenamefont {Poulin-Lamarre},\ and\ \citenamefont {Lopez-Bezanilla}}]{qubit_spin_ice}%
  \BibitemOpen
  \bibfield  {author} {\bibinfo {author} {\bibfnamefont {A.~D.}\ \bibnamefont {King}}, \bibinfo {author} {\bibfnamefont {C.}~\bibnamefont {Nisoli}}, \bibinfo {author} {\bibfnamefont {E.~D.}\ \bibnamefont {Dahl}}, \bibinfo {author} {\bibfnamefont {G.}~\bibnamefont {Poulin-Lamarre}},\ and\ \bibinfo {author} {\bibfnamefont {A.}~\bibnamefont {Lopez-Bezanilla}},\ }\bibfield  {title} {\emph {\bibinfo {title} {Qubit spin ice}},\ }\href {https://doi.org/10.1126/science.abe2824} {\bibfield  {journal} {\bibinfo  {journal} {Science}\ }\textbf {\bibinfo {volume} {373}},\ \bibinfo {pages} {576–580} (\bibinfo {year} {2021}{\natexlab{b}})}\BibitemShut {NoStop}%
\bibitem [{\citenamefont {Kairys}\ \emph {et~al.}(2020)\citenamefont {Kairys}, \citenamefont {King}, \citenamefont {Ozfidan}, \citenamefont {Boothby}, \citenamefont {Raymond}, \citenamefont {Banerjee},\ and\ \citenamefont {Humble}}]{PRXQuantum.1.020320}%
  \BibitemOpen
  \bibfield  {author} {\bibinfo {author} {\bibfnamefont {P.}~\bibnamefont {Kairys}}, \bibinfo {author} {\bibfnamefont {A.~D.}\ \bibnamefont {King}}, \bibinfo {author} {\bibfnamefont {I.}~\bibnamefont {Ozfidan}}, \bibinfo {author} {\bibfnamefont {K.}~\bibnamefont {Boothby}}, \bibinfo {author} {\bibfnamefont {J.}~\bibnamefont {Raymond}}, \bibinfo {author} {\bibfnamefont {A.}~\bibnamefont {Banerjee}},\ and\ \bibinfo {author} {\bibfnamefont {T.~S.}\ \bibnamefont {Humble}},\ }\bibfield  {title} {\emph {\bibinfo {title} {{Simulating the Shastry-Sutherland Ising Model Using Quantum Annealing}}},\ }\href {https://doi.org/10.1103/PRXQuantum.1.020320} {\bibfield  {journal} {\bibinfo  {journal} {PRX Quantum}\ }\textbf {\bibinfo {volume} {1}},\ \bibinfo {pages} {020320} (\bibinfo {year} {2020})}\BibitemShut {NoStop}%
\bibitem [{\citenamefont {Pelofske}\ \emph {et~al.}(2024)\citenamefont {Pelofske}, \citenamefont {Bärtschi},\ and\ \citenamefont {Eidenbenz}}]{pelofske2024simulatingheavyhextransversefield}%
  \BibitemOpen
  \bibfield  {author} {\bibinfo {author} {\bibfnamefont {E.}~\bibnamefont {Pelofske}}, \bibinfo {author} {\bibfnamefont {A.}~\bibnamefont {Bärtschi}},\ and\ \bibinfo {author} {\bibfnamefont {S.}~\bibnamefont {Eidenbenz}},\ }\href {https://arxiv.org/abs/2311.01657} {\bibinfo {title} {{Simulating Heavy-Hex Transverse Field Ising Model Magnetization Dynamics Using Programmable Quantum Annealers}}} (\bibinfo {year} {2024}),\ \Eprint {https://arxiv.org/abs/2311.01657} {arXiv:2311.01657 [quant-ph]} \BibitemShut {NoStop}%
\bibitem [{\citenamefont {Narasimhan}\ \emph {et~al.}(2024)\citenamefont {Narasimhan}, \citenamefont {Humeniuk}, \citenamefont {Roy},\ and\ \citenamefont {Drouin-Touchette}}]{Narasimhan_2024}%
  \BibitemOpen
  \bibfield  {author} {\bibinfo {author} {\bibfnamefont {P.}~\bibnamefont {Narasimhan}}, \bibinfo {author} {\bibfnamefont {S.}~\bibnamefont {Humeniuk}}, \bibinfo {author} {\bibfnamefont {A.}~\bibnamefont {Roy}},\ and\ \bibinfo {author} {\bibfnamefont {V.}~\bibnamefont {Drouin-Touchette}},\ }\bibfield  {title} {\emph {\bibinfo {title} {{Simulating the transverse-field Ising model on the kagome lattice using a programmable quantum annealer}}},\ }\bibfield  {journal} {\bibinfo  {journal} {Physical Review B}\ }\textbf {\bibinfo {volume} {110}},\ \href {https://doi.org/10.1103/physrevb.110.054432} {10.1103/physrevb.110.054432} (\bibinfo {year} {2024})\BibitemShut {NoStop}%
\bibitem [{\citenamefont {Marshall}\ \emph {et~al.}(2019)\citenamefont {Marshall}, \citenamefont {Venturelli}, \citenamefont {Hen},\ and\ \citenamefont {Rieffel}}]{PhysRevApplied.11.044083}%
  \BibitemOpen
  \bibfield  {author} {\bibinfo {author} {\bibfnamefont {J.}~\bibnamefont {Marshall}}, \bibinfo {author} {\bibfnamefont {D.}~\bibnamefont {Venturelli}}, \bibinfo {author} {\bibfnamefont {I.}~\bibnamefont {Hen}},\ and\ \bibinfo {author} {\bibfnamefont {E.~G.}\ \bibnamefont {Rieffel}},\ }\bibfield  {title} {\emph {\bibinfo {title} {{Power of Pausing: Advancing Understanding of Thermalization in Experimental Quantum Annealers}}},\ }\href {https://doi.org/10.1103/PhysRevApplied.11.044083} {\bibfield  {journal} {\bibinfo  {journal} {Phys. Rev. Appl.}\ }\textbf {\bibinfo {volume} {11}},\ \bibinfo {pages} {044083} (\bibinfo {year} {2019})}\BibitemShut {NoStop}%
\bibitem [{\citenamefont {Izquierdo}\ \emph {et~al.}(2021)\citenamefont {Izquierdo}, \citenamefont {Hen},\ and\ \citenamefont {Albash}}]{Izquierdo_2021}%
  \BibitemOpen
  \bibfield  {author} {\bibinfo {author} {\bibfnamefont {Z.~G.}\ \bibnamefont {Izquierdo}}, \bibinfo {author} {\bibfnamefont {I.}~\bibnamefont {Hen}},\ and\ \bibinfo {author} {\bibfnamefont {T.}~\bibnamefont {Albash}},\ }\bibfield  {title} {\emph {\bibinfo {title} {{Testing a Quantum Annealer as a Quantum Thermal Sampler}}},\ }\href {https://doi.org/10.1145/3464456} {\bibfield  {journal} {\bibinfo  {journal} {{ACM} Transactions on Quantum Computing}\ }\textbf {\bibinfo {volume} {2}},\ \bibinfo {pages} {1--20} (\bibinfo {year} {2021})},\ \Eprint {https://arxiv.org/abs/2003.00361} {arXiv:2003.00361} \BibitemShut {NoStop}%
\bibitem [{\citenamefont {Marshall}\ \emph {et~al.}(2017{\natexlab{a}})\citenamefont {Marshall}, \citenamefont {Rieffel},\ and\ \citenamefont {Hen}}]{Marshall_2017}%
  \BibitemOpen
  \bibfield  {author} {\bibinfo {author} {\bibfnamefont {J.}~\bibnamefont {Marshall}}, \bibinfo {author} {\bibfnamefont {E.~G.}\ \bibnamefont {Rieffel}},\ and\ \bibinfo {author} {\bibfnamefont {I.}~\bibnamefont {Hen}},\ }\bibfield  {title} {\emph {\bibinfo {title} {{Thermalization, Freeze-out, and Noise: Deciphering Experimental Quantum Annealers}}},\ }\bibfield  {journal} {\bibinfo  {journal} {Physical Review Applied}\ }\textbf {\bibinfo {volume} {8}},\ \href {https://doi.org/10.1103/physrevapplied.8.064025} {10.1103/physrevapplied.8.064025} (\bibinfo {year} {2017}{\natexlab{a}})\BibitemShut {NoStop}%
\bibitem [{\citenamefont {Nelson}\ \emph {et~al.}(2021)\citenamefont {Nelson}, \citenamefont {Vuffray}, \citenamefont {Lokhov},\ and\ \citenamefont {Coffrin}}]{nelson2021single}%
  \BibitemOpen
  \bibfield  {author} {\bibinfo {author} {\bibfnamefont {J.}~\bibnamefont {Nelson}}, \bibinfo {author} {\bibfnamefont {M.}~\bibnamefont {Vuffray}}, \bibinfo {author} {\bibfnamefont {A.~Y.}\ \bibnamefont {Lokhov}},\ and\ \bibinfo {author} {\bibfnamefont {C.}~\bibnamefont {Coffrin}},\ }\bibfield  {title} {\emph {\bibinfo {title} {Single-qubit fidelity assessment of quantum annealing hardware}},\ }\href@noop {} {\bibfield  {journal} {\bibinfo  {journal} {IEEE Transactions on Quantum Engineering}\ }\textbf {\bibinfo {volume} {2}},\ \bibinfo {pages} {1--10} (\bibinfo {year} {2021})}\BibitemShut {NoStop}%
\bibitem [{\citenamefont {Vuffray}\ \emph {et~al.}(2022)\citenamefont {Vuffray}, \citenamefont {Coffrin}, \citenamefont {Kharkov},\ and\ \citenamefont {Lokhov}}]{PRXQuantum.3.020317}%
  \BibitemOpen
  \bibfield  {author} {\bibinfo {author} {\bibfnamefont {M.}~\bibnamefont {Vuffray}}, \bibinfo {author} {\bibfnamefont {C.}~\bibnamefont {Coffrin}}, \bibinfo {author} {\bibfnamefont {Y.~A.}\ \bibnamefont {Kharkov}},\ and\ \bibinfo {author} {\bibfnamefont {A.~Y.}\ \bibnamefont {Lokhov}},\ }\bibfield  {title} {\emph {\bibinfo {title} {{Programmable Quantum Annealers as Noisy Gibbs Samplers}}},\ }\href {https://doi.org/10.1103/PRXQuantum.3.020317} {\bibfield  {journal} {\bibinfo  {journal} {PRX Quantum}\ }\textbf {\bibinfo {volume} {3}},\ \bibinfo {pages} {020317} (\bibinfo {year} {2022})}\BibitemShut {NoStop}%
\bibitem [{\citenamefont {Nelson}\ \emph {et~al.}(2022)\citenamefont {Nelson}, \citenamefont {Vuffray}, \citenamefont {Lokhov}, \citenamefont {Albash},\ and\ \citenamefont {Coffrin}}]{PhysRevApplied.17.044046}%
  \BibitemOpen
  \bibfield  {author} {\bibinfo {author} {\bibfnamefont {J.}~\bibnamefont {Nelson}}, \bibinfo {author} {\bibfnamefont {M.}~\bibnamefont {Vuffray}}, \bibinfo {author} {\bibfnamefont {A.~Y.}\ \bibnamefont {Lokhov}}, \bibinfo {author} {\bibfnamefont {T.}~\bibnamefont {Albash}},\ and\ \bibinfo {author} {\bibfnamefont {C.}~\bibnamefont {Coffrin}},\ }\bibfield  {title} {\emph {\bibinfo {title} {{High-Quality Thermal Gibbs Sampling with Quantum Annealing Hardware}}},\ }\href {https://doi.org/10.1103/PhysRevApplied.17.044046} {\bibfield  {journal} {\bibinfo  {journal} {Phys. Rev. Appl.}\ }\textbf {\bibinfo {volume} {17}},\ \bibinfo {pages} {044046} (\bibinfo {year} {2022})}\BibitemShut {NoStop}%
\bibitem [{\citenamefont {Buffoni}\ and\ \citenamefont {Campisi}(2020)}]{buffoni2020thermodynamics}%
  \BibitemOpen
  \bibfield  {author} {\bibinfo {author} {\bibfnamefont {L.}~\bibnamefont {Buffoni}}\ and\ \bibinfo {author} {\bibfnamefont {M.}~\bibnamefont {Campisi}},\ }\bibfield  {title} {\emph {\bibinfo {title} {Thermodynamics of a quantum annealer}},\ }\href@noop {} {\bibfield  {journal} {\bibinfo  {journal} {Quantum Science and Technology}\ }\textbf {\bibinfo {volume} {5}},\ \bibinfo {pages} {035013} (\bibinfo {year} {2020})}\BibitemShut {NoStop}%
\bibitem [{\citenamefont {Sathe}\ \emph {et~al.}(2025)\citenamefont {Sathe}, \citenamefont {King}, \citenamefont {Mniszewski}, \citenamefont {Coffrin}, \citenamefont {Nisoli},\ and\ \citenamefont {Caravelli}}]{sathe2025classicalcriticalityquantumannealing}%
  \BibitemOpen
  \bibfield  {author} {\bibinfo {author} {\bibfnamefont {P.}~\bibnamefont {Sathe}}, \bibinfo {author} {\bibfnamefont {A.~D.}\ \bibnamefont {King}}, \bibinfo {author} {\bibfnamefont {S.~M.}\ \bibnamefont {Mniszewski}}, \bibinfo {author} {\bibfnamefont {C.}~\bibnamefont {Coffrin}}, \bibinfo {author} {\bibfnamefont {C.}~\bibnamefont {Nisoli}},\ and\ \bibinfo {author} {\bibfnamefont {F.}~\bibnamefont {Caravelli}},\ }\href {https://arxiv.org/abs/2505.13625} {\bibinfo {title} {{Classical Criticality via Quantum Annealing}}} (\bibinfo {year} {2025}),\ \Eprint {https://arxiv.org/abs/2505.13625} {arXiv:2505.13625 [cond-mat.stat-mech]} \BibitemShut {NoStop}%
\bibitem [{\citenamefont {Raymond}\ \emph {et~al.}(2016)\citenamefont {Raymond}, \citenamefont {Yarkoni},\ and\ \citenamefont {Andriyash}}]{Raymond_2016}%
  \BibitemOpen
  \bibfield  {author} {\bibinfo {author} {\bibfnamefont {J.}~\bibnamefont {Raymond}}, \bibinfo {author} {\bibfnamefont {S.}~\bibnamefont {Yarkoni}},\ and\ \bibinfo {author} {\bibfnamefont {E.}~\bibnamefont {Andriyash}},\ }\bibfield  {title} {\emph {\bibinfo {title} {{Global Warming: Temperature Estimation in Annealers}}},\ }\bibfield  {journal} {\bibinfo  {journal} {Frontiers in ICT}\ }\textbf {\bibinfo {volume} {3}},\ \href {https://doi.org/10.3389/fict.2016.00023} {10.3389/fict.2016.00023} (\bibinfo {year} {2016})\BibitemShut {NoStop}%
\bibitem [{\citenamefont {Sandt}\ and\ \citenamefont {Spatschek}(2023)}]{sandt2023efficient}%
  \BibitemOpen
  \bibfield  {author} {\bibinfo {author} {\bibfnamefont {R.}~\bibnamefont {Sandt}}\ and\ \bibinfo {author} {\bibfnamefont {R.}~\bibnamefont {Spatschek}},\ }\bibfield  {title} {\emph {\bibinfo {title} {{Efficient low temperature Monte Carlo sampling using quantum annealing}}},\ }\href@noop {} {\bibfield  {journal} {\bibinfo  {journal} {Scientific Reports}\ }\textbf {\bibinfo {volume} {13}},\ \bibinfo {pages} {6754} (\bibinfo {year} {2023})}\BibitemShut {NoStop}%
\bibitem [{\citenamefont {Shibukawa}\ \emph {et~al.}(2024)\citenamefont {Shibukawa}, \citenamefont {Tamura},\ and\ \citenamefont {Tsuda}}]{PhysRevResearch.6.043050}%
  \BibitemOpen
  \bibfield  {author} {\bibinfo {author} {\bibfnamefont {R.}~\bibnamefont {Shibukawa}}, \bibinfo {author} {\bibfnamefont {R.}~\bibnamefont {Tamura}},\ and\ \bibinfo {author} {\bibfnamefont {K.}~\bibnamefont {Tsuda}},\ }\bibfield  {title} {\emph {\bibinfo {title} {{Boltzmann sampling with quantum annealers via fast Stein correction}}},\ }\href {https://doi.org/10.1103/PhysRevResearch.6.043050} {\bibfield  {journal} {\bibinfo  {journal} {Phys. Rev. Res.}\ }\textbf {\bibinfo {volume} {6}},\ \bibinfo {pages} {043050} (\bibinfo {year} {2024})}\BibitemShut {NoStop}%
\bibitem [{\citenamefont {Chancellor}\ \emph {et~al.}(2016)\citenamefont {Chancellor}, \citenamefont {Szoke}, \citenamefont {Vinci}, \citenamefont {Aeppli},\ and\ \citenamefont {Warburton}}]{Chancellor_2016}%
  \BibitemOpen
  \bibfield  {author} {\bibinfo {author} {\bibfnamefont {N.}~\bibnamefont {Chancellor}}, \bibinfo {author} {\bibfnamefont {S.}~\bibnamefont {Szoke}}, \bibinfo {author} {\bibfnamefont {W.}~\bibnamefont {Vinci}}, \bibinfo {author} {\bibfnamefont {G.}~\bibnamefont {Aeppli}},\ and\ \bibinfo {author} {\bibfnamefont {P.~A.}\ \bibnamefont {Warburton}},\ }\bibfield  {title} {\emph {\bibinfo {title} {{Maximum-Entropy Inference with a Programmable Annealer}}},\ }\bibfield  {journal} {\bibinfo  {journal} {Scientific Reports}\ }\textbf {\bibinfo {volume} {6}},\ \href {https://doi.org/10.1038/srep22318} {10.1038/srep22318} (\bibinfo {year} {2016})\BibitemShut {NoStop}%
\bibitem [{\citenamefont {Tan}\ \emph {et~al.}(2010)\citenamefont {Tan}, \citenamefont {Li},\ and\ \citenamefont {Stoica}}]{5495896}%
  \BibitemOpen
  \bibfield  {author} {\bibinfo {author} {\bibfnamefont {X.}~\bibnamefont {Tan}}, \bibinfo {author} {\bibfnamefont {J.}~\bibnamefont {Li}},\ and\ \bibinfo {author} {\bibfnamefont {P.}~\bibnamefont {Stoica}},\ }in\ \href {https://doi.org/10.1109/ICASSP.2010.5495896} {\emph {\bibinfo {booktitle} {2010 IEEE International Conference on Acoustics, Speech and Signal Processing}}}\ (\bibinfo {year} {2010})\ pp.\ \bibinfo {pages} {3634--3637}\BibitemShut {NoStop}%
\bibitem [{\citenamefont {Metropolis}\ \emph {et~al.}(1953)\citenamefont {Metropolis}, \citenamefont {Rosenbluth}, \citenamefont {Rosenbluth}, \citenamefont {Teller},\ and\ \citenamefont {Teller}}]{metropolis1953equation}%
  \BibitemOpen
  \bibfield  {author} {\bibinfo {author} {\bibfnamefont {N.}~\bibnamefont {Metropolis}}, \bibinfo {author} {\bibfnamefont {A.~W.}\ \bibnamefont {Rosenbluth}}, \bibinfo {author} {\bibfnamefont {M.~N.}\ \bibnamefont {Rosenbluth}}, \bibinfo {author} {\bibfnamefont {A.~H.}\ \bibnamefont {Teller}},\ and\ \bibinfo {author} {\bibfnamefont {E.}~\bibnamefont {Teller}},\ }\bibfield  {title} {\emph {\bibinfo {title} {Equation of state calculations by fast computing machines}},\ }\href@noop {} {\bibfield  {journal} {\bibinfo  {journal} {The journal of chemical physics}\ }\textbf {\bibinfo {volume} {21}},\ \bibinfo {pages} {1087--1092} (\bibinfo {year} {1953})}\BibitemShut {NoStop}%
\bibitem [{\citenamefont {Andrews}\ \emph {et~al.}(2009)\citenamefont {Andrews}, \citenamefont {De~Sterck}, \citenamefont {Inglis},\ and\ \citenamefont {Melko}}]{PhysRevE.79.041127}%
  \BibitemOpen
  \bibfield  {author} {\bibinfo {author} {\bibfnamefont {S.}~\bibnamefont {Andrews}}, \bibinfo {author} {\bibfnamefont {H.}~\bibnamefont {De~Sterck}}, \bibinfo {author} {\bibfnamefont {S.}~\bibnamefont {Inglis}},\ and\ \bibinfo {author} {\bibfnamefont {R.~G.}\ \bibnamefont {Melko}},\ }\bibfield  {title} {\emph {\bibinfo {title} {{Monte Carlo study of degenerate ground states and residual entropy in a frustrated honeycomb lattice Ising model}}},\ }\href {https://doi.org/10.1103/PhysRevE.79.041127} {\bibfield  {journal} {\bibinfo  {journal} {Phys. Rev. E}\ }\textbf {\bibinfo {volume} {79}},\ \bibinfo {pages} {041127} (\bibinfo {year} {2009})}\BibitemShut {NoStop}%
\bibitem [{\citenamefont {Huang}\ \emph {et~al.}(2024)\citenamefont {Huang}, \citenamefont {Perkins},\ and\ \citenamefont {Potechin}}]{huang2024hardnesssamplingantiferromagneticising}%
  \BibitemOpen
  \bibfield  {author} {\bibinfo {author} {\bibfnamefont {N.}~\bibnamefont {Huang}}, \bibinfo {author} {\bibfnamefont {W.}~\bibnamefont {Perkins}},\ and\ \bibinfo {author} {\bibfnamefont {A.}~\bibnamefont {Potechin}},\ }\href {https://arxiv.org/abs/2409.03974} {\bibinfo {title} {{Hardness of sampling for the anti-ferromagnetic Ising model on random graphs}}} (\bibinfo {year} {2024}),\ \Eprint {https://arxiv.org/abs/2409.03974} {arXiv:2409.03974 [math.PR]} \BibitemShut {NoStop}%
\bibitem [{\citenamefont {Spiridon}\ and\ \citenamefont {Minh}(2017)}]{spiridon2017hamiltonian}%
  \BibitemOpen
  \bibfield  {author} {\bibinfo {author} {\bibfnamefont {L.}~\bibnamefont {Spiridon}}\ and\ \bibinfo {author} {\bibfnamefont {D.~D.}\ \bibnamefont {Minh}},\ }\bibfield  {title} {\emph {\bibinfo {title} {Hamiltonian monte carlo with constrained molecular dynamics as gibbs sampling}},\ }\href@noop {} {\bibfield  {journal} {\bibinfo  {journal} {Journal of chemical theory and computation}\ }\textbf {\bibinfo {volume} {13}},\ \bibinfo {pages} {4649--4659} (\bibinfo {year} {2017})}\BibitemShut {NoStop}%
\bibitem [{\citenamefont {Bodini}\ and\ \citenamefont {Ponty}(2010)}]{bodini2010multi}%
  \BibitemOpen
  \bibfield  {author} {\bibinfo {author} {\bibfnamefont {O.}~\bibnamefont {Bodini}}\ and\ \bibinfo {author} {\bibfnamefont {Y.}~\bibnamefont {Ponty}},\ }\bibfield  {title} {\emph {\bibinfo {title} {{Multi-dimensional Boltzmann sampling of languages}}},\ }\href@noop {} {\bibfield  {journal} {\bibinfo  {journal} {Discrete Mathematics \& Theoretical Computer Science}\ } (\bibinfo {year} {2010})}\BibitemShut {NoStop}%
\bibitem [{\citenamefont {No{\'e}}\ \emph {et~al.}(2019)\citenamefont {No{\'e}}, \citenamefont {Olsson}, \citenamefont {K{\"o}hler},\ and\ \citenamefont {Wu}}]{noe2019boltzmann}%
  \BibitemOpen
  \bibfield  {author} {\bibinfo {author} {\bibfnamefont {F.}~\bibnamefont {No{\'e}}}, \bibinfo {author} {\bibfnamefont {S.}~\bibnamefont {Olsson}}, \bibinfo {author} {\bibfnamefont {J.}~\bibnamefont {K{\"o}hler}},\ and\ \bibinfo {author} {\bibfnamefont {H.}~\bibnamefont {Wu}},\ }\bibfield  {title} {\emph {\bibinfo {title} {Boltzmann generators: Sampling equilibrium states of many-body systems with deep learning}},\ }\href@noop {} {\bibfield  {journal} {\bibinfo  {journal} {Science}\ }\textbf {\bibinfo {volume} {365}},\ \bibinfo {pages} {eaaw1147} (\bibinfo {year} {2019})}\BibitemShut {NoStop}%
\bibitem [{\citenamefont {Goto}\ \emph {et~al.}(2018)\citenamefont {Goto}, \citenamefont {Lin},\ and\ \citenamefont {Nakamura}}]{goto2018boltzmann}%
  \BibitemOpen
  \bibfield  {author} {\bibinfo {author} {\bibfnamefont {H.}~\bibnamefont {Goto}}, \bibinfo {author} {\bibfnamefont {Z.}~\bibnamefont {Lin}},\ and\ \bibinfo {author} {\bibfnamefont {Y.}~\bibnamefont {Nakamura}},\ }\bibfield  {title} {\emph {\bibinfo {title} {{Boltzmann sampling from the Ising model using quantum heating of coupled nonlinear oscillators}}},\ }\href@noop {} {\bibfield  {journal} {\bibinfo  {journal} {Scientific reports}\ }\textbf {\bibinfo {volume} {8}},\ \bibinfo {pages} {7154} (\bibinfo {year} {2018})}\BibitemShut {NoStop}%
\bibitem [{\citenamefont {Margiani}\ \emph {et~al.}(2025)\citenamefont {Margiani}, \citenamefont {Ameye}, \citenamefont {Zilberberg},\ and\ \citenamefont {Eichler}}]{npp4-b1xb}%
  \BibitemOpen
  \bibfield  {author} {\bibinfo {author} {\bibfnamefont {G.}~\bibnamefont {Margiani}}, \bibinfo {author} {\bibfnamefont {O.}~\bibnamefont {Ameye}}, \bibinfo {author} {\bibfnamefont {O.}~\bibnamefont {Zilberberg}},\ and\ \bibinfo {author} {\bibfnamefont {A.}~\bibnamefont {Eichler}},\ }\bibfield  {title} {\emph {\bibinfo {title} {Three strongly coupled kerr parametric oscillators forming a boltzmann machine}},\ }\href {https://doi.org/10.1103/npp4-b1xb} {\bibfield  {journal} {\bibinfo  {journal} {Phys. Rev. Lett.}\ }\textbf {\bibinfo {volume} {135}},\ \bibinfo {pages} {097201} (\bibinfo {year} {2025})}\BibitemShut {NoStop}%
\bibitem [{\citenamefont {Margiani}\ \emph {et~al.}(2023)\citenamefont {Margiani}, \citenamefont {del Pino}, \citenamefont {Heugel}, \citenamefont {Bousse}, \citenamefont {Guerrero}, \citenamefont {Kenny}, \citenamefont {Zilberberg}, \citenamefont {Sabonis},\ and\ \citenamefont {Eichler}}]{PhysRevResearch.5.L012029}%
  \BibitemOpen
  \bibfield  {author} {\bibinfo {author} {\bibfnamefont {G.}~\bibnamefont {Margiani}}, \bibinfo {author} {\bibfnamefont {J.}~\bibnamefont {del Pino}}, \bibinfo {author} {\bibfnamefont {T.~L.}\ \bibnamefont {Heugel}}, \bibinfo {author} {\bibfnamefont {N.~E.}\ \bibnamefont {Bousse}}, \bibinfo {author} {\bibfnamefont {S.}~\bibnamefont {Guerrero}}, \bibinfo {author} {\bibfnamefont {T.~W.}\ \bibnamefont {Kenny}}, \bibinfo {author} {\bibfnamefont {O.}~\bibnamefont {Zilberberg}}, \bibinfo {author} {\bibfnamefont {D.}~\bibnamefont {Sabonis}},\ and\ \bibinfo {author} {\bibfnamefont {A.}~\bibnamefont {Eichler}},\ }\bibfield  {title} {\emph {\bibinfo {title} {Deterministic and stochastic sampling of two coupled kerr parametric oscillators}},\ }\href {https://doi.org/10.1103/PhysRevResearch.5.L012029} {\bibfield  {journal} {\bibinfo  {journal} {Phys. Rev. Res.}\ }\textbf {\bibinfo {volume} {5}},\ \bibinfo {pages} {L012029} (\bibinfo {year} {2023})}\BibitemShut {NoStop}%
\bibitem [{\citenamefont {Hasegawa}\ and\ \citenamefont {Shibata}(2025)}]{hasegawa2025residual}%
  \BibitemOpen
  \bibfield  {author} {\bibinfo {author} {\bibfnamefont {H.}~\bibnamefont {Hasegawa}}\ and\ \bibinfo {author} {\bibfnamefont {N.}~\bibnamefont {Shibata}},\ }\bibfield  {title} {\emph {\bibinfo {title} {Residual errors after quantum annealing in the axial next nearest neighbor ising model: Impact of critical points and modulated correlation}},\ }\href@noop {} {\bibfield  {journal} {\bibinfo  {journal} {Journal of the Physical Society of Japan}\ }\textbf {\bibinfo {volume} {94}},\ \bibinfo {pages} {094004} (\bibinfo {year} {2025})}\BibitemShut {NoStop}%
\bibitem [{\citenamefont {Pexe}\ \emph {et~al.}(2024)\citenamefont {Pexe}, \citenamefont {Rattighieri}, \citenamefont {Malvezzi},\ and\ \citenamefont {Fanchini}}]{Pexe_2024}%
  \BibitemOpen
  \bibfield  {author} {\bibinfo {author} {\bibfnamefont {G.~E.~L.}\ \bibnamefont {Pexe}}, \bibinfo {author} {\bibfnamefont {L.~A.~M.}\ \bibnamefont {Rattighieri}}, \bibinfo {author} {\bibfnamefont {A.~L.}\ \bibnamefont {Malvezzi}},\ and\ \bibinfo {author} {\bibfnamefont {F.~F.}\ \bibnamefont {Fanchini}},\ }\bibfield  {title} {\emph {\bibinfo {title} {{Using a feedback-based quantum algorithm to analyze the critical properties of the ANNNI model without classical optimization}}},\ }\bibfield  {journal} {\bibinfo  {journal} {Physical Review B}\ }\textbf {\bibinfo {volume} {110}},\ \href {https://doi.org/10.1103/physrevb.110.224422} {10.1103/physrevb.110.224422} (\bibinfo {year} {2024})\BibitemShut {NoStop}%
\bibitem [{\citenamefont {Cea}\ \emph {et~al.}(2024)\citenamefont {Cea}, \citenamefont {Grossi}, \citenamefont {Monaco}, \citenamefont {Rico}, \citenamefont {Tagliacozzo},\ and\ \citenamefont {Vallecorsa}}]{cea2024exploringphasediagramquantum}%
  \BibitemOpen
  \bibfield  {author} {\bibinfo {author} {\bibfnamefont {M.}~\bibnamefont {Cea}}, \bibinfo {author} {\bibfnamefont {M.}~\bibnamefont {Grossi}}, \bibinfo {author} {\bibfnamefont {S.}~\bibnamefont {Monaco}}, \bibinfo {author} {\bibfnamefont {E.}~\bibnamefont {Rico}}, \bibinfo {author} {\bibfnamefont {L.}~\bibnamefont {Tagliacozzo}},\ and\ \bibinfo {author} {\bibfnamefont {S.}~\bibnamefont {Vallecorsa}},\ }\href {https://arxiv.org/abs/2402.11022} {\bibinfo {title} {{Exploring the Phase Diagram of the quantum one-dimensional ANNNI model}}} (\bibinfo {year} {2024}),\ \Eprint {https://arxiv.org/abs/2402.11022} {arXiv:2402.11022 [cond-mat.str-el]} \BibitemShut {NoStop}%
\bibitem [{\citenamefont {Marin}\ \emph {et~al.}(2024)\citenamefont {Marin}, \citenamefont {Fontana}, \citenamefont {Bellani}, \citenamefont {Pederiva}, \citenamefont {Quaranta}, \citenamefont {Rossella}, \citenamefont {Salamon},\ and\ \citenamefont {Salina}}]{marin2024modelingfrustratedisingsquare}%
  \BibitemOpen
  \bibfield  {author} {\bibinfo {author} {\bibfnamefont {C.}~\bibnamefont {Marin}}, \bibinfo {author} {\bibfnamefont {A.}~\bibnamefont {Fontana}}, \bibinfo {author} {\bibfnamefont {V.}~\bibnamefont {Bellani}}, \bibinfo {author} {\bibfnamefont {F.}~\bibnamefont {Pederiva}}, \bibinfo {author} {\bibfnamefont {A.}~\bibnamefont {Quaranta}}, \bibinfo {author} {\bibfnamefont {F.}~\bibnamefont {Rossella}}, \bibinfo {author} {\bibfnamefont {A.}~\bibnamefont {Salamon}},\ and\ \bibinfo {author} {\bibfnamefont {G.}~\bibnamefont {Salina}},\ }\href {https://arxiv.org/abs/2409.11259} {\bibinfo {title} {{Modeling a frustrated Ising square lattice with the D-Wave Quantum Annealer}}} (\bibinfo {year} {2024}),\ \Eprint {https://arxiv.org/abs/2409.11259} {arXiv:2409.11259 [physics.comp-ph]} \BibitemShut {NoStop}%
\bibitem [{\citenamefont {Johnson}\ \emph {et~al.}(2011)\citenamefont {Johnson}, \citenamefont {Amin}, \citenamefont {Gildert}, \citenamefont {Lanting}, \citenamefont {Hamze}, \citenamefont {Dickson}, \citenamefont {Harris}, \citenamefont {Berkley}, \citenamefont {Johansson}, \citenamefont {Bunyk} \emph {et~al.}}]{johnson2011quantum}%
  \BibitemOpen
  \bibfield  {author} {\bibinfo {author} {\bibfnamefont {M.~W.}\ \bibnamefont {Johnson}}, \bibinfo {author} {\bibfnamefont {M.~H.}\ \bibnamefont {Amin}}, \bibinfo {author} {\bibfnamefont {S.}~\bibnamefont {Gildert}}, \bibinfo {author} {\bibfnamefont {T.}~\bibnamefont {Lanting}}, \bibinfo {author} {\bibfnamefont {F.}~\bibnamefont {Hamze}}, \bibinfo {author} {\bibfnamefont {N.}~\bibnamefont {Dickson}}, \bibinfo {author} {\bibfnamefont {R.}~\bibnamefont {Harris}}, \bibinfo {author} {\bibfnamefont {A.~J.}\ \bibnamefont {Berkley}}, \bibinfo {author} {\bibfnamefont {J.}~\bibnamefont {Johansson}}, \bibinfo {author} {\bibfnamefont {P.}~\bibnamefont {Bunyk}}, \emph {et~al.},\ }\bibfield  {title} {\emph {\bibinfo {title} {Quantum annealing with manufactured spins}},\ }\href {https://doi.org/10.1038/nature10012} {\bibfield  {journal} {\bibinfo  {journal} {Nature}\ }\textbf {\bibinfo {volume} {473}},\ \bibinfo {pages} {194--198} (\bibinfo {year} {2011})}\BibitemShut {NoStop}%
\bibitem [{\citenamefont {Bunyk}\ \emph {et~al.}(2014)\citenamefont {Bunyk}, \citenamefont {Hoskinson}, \citenamefont {Johnson}, \citenamefont {Tolkacheva}, \citenamefont {Altomare}, \citenamefont {Berkley}, \citenamefont {Harris}, \citenamefont {Hilton}, \citenamefont {Lanting}, \citenamefont {Przybysz},\ and\ \citenamefont {Whittaker}}]{Bunyk_2014}%
  \BibitemOpen
  \bibfield  {author} {\bibinfo {author} {\bibfnamefont {P.~I.}\ \bibnamefont {Bunyk}}, \bibinfo {author} {\bibfnamefont {E.~M.}\ \bibnamefont {Hoskinson}}, \bibinfo {author} {\bibfnamefont {M.~W.}\ \bibnamefont {Johnson}}, \bibinfo {author} {\bibfnamefont {E.}~\bibnamefont {Tolkacheva}}, \bibinfo {author} {\bibfnamefont {F.}~\bibnamefont {Altomare}}, \bibinfo {author} {\bibfnamefont {A.~J.}\ \bibnamefont {Berkley}}, \bibinfo {author} {\bibfnamefont {R.}~\bibnamefont {Harris}}, \bibinfo {author} {\bibfnamefont {J.~P.}\ \bibnamefont {Hilton}}, \bibinfo {author} {\bibfnamefont {T.}~\bibnamefont {Lanting}}, \bibinfo {author} {\bibfnamefont {A.~J.}\ \bibnamefont {Przybysz}},\ and\ \bibinfo {author} {\bibfnamefont {J.}~\bibnamefont {Whittaker}},\ }\bibfield  {title} {\emph {\bibinfo {title} {{Architectural Considerations in the Design of a Superconducting Quantum Annealing Processor}}},\ }\href {https://doi.org/10.1109/tasc.2014.2318294} {\bibfield  {journal} {\bibinfo  {journal} {IEEE Transactions on Applied
  Superconductivity}\ }\textbf {\bibinfo {volume} {24}},\ \bibinfo {pages} {1–10} (\bibinfo {year} {2014})}\BibitemShut {NoStop}%
\bibitem [{\citenamefont {Dickson}\ \emph {et~al.}(2013)\citenamefont {Dickson}, \citenamefont {Johnson}, \citenamefont {Amin}, \citenamefont {Harris}, \citenamefont {Altomare}, \citenamefont {Berkley}, \citenamefont {Bunyk}, \citenamefont {Cai}, \citenamefont {Chapple}, \citenamefont {Chavez} \emph {et~al.}}]{dickson2013thermally}%
  \BibitemOpen
  \bibfield  {author} {\bibinfo {author} {\bibfnamefont {N.~G.}\ \bibnamefont {Dickson}}, \bibinfo {author} {\bibfnamefont {M.~W.}\ \bibnamefont {Johnson}}, \bibinfo {author} {\bibfnamefont {M.}~\bibnamefont {Amin}}, \bibinfo {author} {\bibfnamefont {R.}~\bibnamefont {Harris}}, \bibinfo {author} {\bibfnamefont {F.}~\bibnamefont {Altomare}}, \bibinfo {author} {\bibfnamefont {A.~J.}\ \bibnamefont {Berkley}}, \bibinfo {author} {\bibfnamefont {P.}~\bibnamefont {Bunyk}}, \bibinfo {author} {\bibfnamefont {J.}~\bibnamefont {Cai}}, \bibinfo {author} {\bibfnamefont {E.}~\bibnamefont {Chapple}}, \bibinfo {author} {\bibfnamefont {P.}~\bibnamefont {Chavez}}, \emph {et~al.},\ }\bibfield  {title} {\emph {\bibinfo {title} {Thermally assisted quantum annealing of a 16-qubit problem}},\ }\href@noop {} {\bibfield  {journal} {\bibinfo  {journal} {Nature communications}\ }\textbf {\bibinfo {volume} {4}},\ \bibinfo {pages} {1903} (\bibinfo {year} {2013})}\BibitemShut {NoStop}%
\bibitem [{\citenamefont {Harris}\ \emph {et~al.}(2010)\citenamefont {Harris}, \citenamefont {Johnson}, \citenamefont {Lanting}, \citenamefont {Berkley}, \citenamefont {Johansson}, \citenamefont {Bunyk}, \citenamefont {Tolkacheva}, \citenamefont {Ladizinsky}, \citenamefont {Ladizinsky}, \citenamefont {Oh}, \citenamefont {Cioata}, \citenamefont {Perminov}, \citenamefont {Spear}, \citenamefont {Enderud}, \citenamefont {Rich}, \citenamefont {Uchaikin}, \citenamefont {Thom}, \citenamefont {Chapple}, \citenamefont {Wang}, \citenamefont {Wilson}, \citenamefont {Amin}, \citenamefont {Dickson}, \citenamefont {Karimi}, \citenamefont {Macready}, \citenamefont {Truncik},\ and\ \citenamefont {Rose}}]{PhysRevB.82.024511}%
  \BibitemOpen
  \bibfield  {author} {\bibinfo {author} {\bibfnamefont {R.}~\bibnamefont {Harris}}, \bibinfo {author} {\bibfnamefont {M.~W.}\ \bibnamefont {Johnson}}, \bibinfo {author} {\bibfnamefont {T.}~\bibnamefont {Lanting}}, \bibinfo {author} {\bibfnamefont {A.~J.}\ \bibnamefont {Berkley}}, \bibinfo {author} {\bibfnamefont {J.}~\bibnamefont {Johansson}}, \bibinfo {author} {\bibfnamefont {P.}~\bibnamefont {Bunyk}}, \bibinfo {author} {\bibfnamefont {E.}~\bibnamefont {Tolkacheva}}, \bibinfo {author} {\bibfnamefont {E.}~\bibnamefont {Ladizinsky}}, \bibinfo {author} {\bibfnamefont {N.}~\bibnamefont {Ladizinsky}}, \bibinfo {author} {\bibfnamefont {T.}~\bibnamefont {Oh}}, \bibinfo {author} {\bibfnamefont {F.}~\bibnamefont {Cioata}}, \bibinfo {author} {\bibfnamefont {I.}~\bibnamefont {Perminov}}, \bibinfo {author} {\bibfnamefont {P.}~\bibnamefont {Spear}}, \bibinfo {author} {\bibfnamefont {C.}~\bibnamefont {Enderud}}, \bibinfo {author} {\bibfnamefont {C.}~\bibnamefont {Rich}}, \bibinfo {author} {\bibfnamefont {S.}~\bibnamefont
  {Uchaikin}}, \bibinfo {author} {\bibfnamefont {M.~C.}\ \bibnamefont {Thom}}, \bibinfo {author} {\bibfnamefont {E.~M.}\ \bibnamefont {Chapple}}, \bibinfo {author} {\bibfnamefont {J.}~\bibnamefont {Wang}}, \bibinfo {author} {\bibfnamefont {B.}~\bibnamefont {Wilson}}, \bibinfo {author} {\bibfnamefont {M.~H.~S.}\ \bibnamefont {Amin}}, \bibinfo {author} {\bibfnamefont {N.}~\bibnamefont {Dickson}}, \bibinfo {author} {\bibfnamefont {K.}~\bibnamefont {Karimi}}, \bibinfo {author} {\bibfnamefont {B.}~\bibnamefont {Macready}}, \bibinfo {author} {\bibfnamefont {C.~J.~S.}\ \bibnamefont {Truncik}},\ and\ \bibinfo {author} {\bibfnamefont {G.}~\bibnamefont {Rose}},\ }\bibfield  {title} {\emph {\bibinfo {title} {Experimental investigation of an eight-qubit unit cell in a superconducting optimization processor}},\ }\href {https://doi.org/10.1103/PhysRevB.82.024511} {\bibfield  {journal} {\bibinfo  {journal} {Phys. Rev. B}\ }\textbf {\bibinfo {volume} {82}},\ \bibinfo {pages} {024511} (\bibinfo {year} {2010})}\BibitemShut
  {NoStop}%
\bibitem [{\citenamefont {Johnson}\ \emph {et~al.}(2010)\citenamefont {Johnson}, \citenamefont {Bunyk}, \citenamefont {Maibaum}, \citenamefont {Tolkacheva}, \citenamefont {Berkley}, \citenamefont {Chapple}, \citenamefont {Harris}, \citenamefont {Johansson}, \citenamefont {Lanting}, \citenamefont {Perminov}, \citenamefont {Ladizinsky}, \citenamefont {Oh},\ and\ \citenamefont {Rose}}]{Johnson_2010}%
  \BibitemOpen
  \bibfield  {author} {\bibinfo {author} {\bibfnamefont {M.~W.}\ \bibnamefont {Johnson}}, \bibinfo {author} {\bibfnamefont {P.}~\bibnamefont {Bunyk}}, \bibinfo {author} {\bibfnamefont {F.}~\bibnamefont {Maibaum}}, \bibinfo {author} {\bibfnamefont {E.}~\bibnamefont {Tolkacheva}}, \bibinfo {author} {\bibfnamefont {A.~J.}\ \bibnamefont {Berkley}}, \bibinfo {author} {\bibfnamefont {E.~M.}\ \bibnamefont {Chapple}}, \bibinfo {author} {\bibfnamefont {R.}~\bibnamefont {Harris}}, \bibinfo {author} {\bibfnamefont {J.}~\bibnamefont {Johansson}}, \bibinfo {author} {\bibfnamefont {T.}~\bibnamefont {Lanting}}, \bibinfo {author} {\bibfnamefont {I.}~\bibnamefont {Perminov}}, \bibinfo {author} {\bibfnamefont {E.}~\bibnamefont {Ladizinsky}}, \bibinfo {author} {\bibfnamefont {T.}~\bibnamefont {Oh}},\ and\ \bibinfo {author} {\bibfnamefont {G.}~\bibnamefont {Rose}},\ }\bibfield  {title} {\emph {\bibinfo {title} {A scalable control system for a superconducting adiabatic quantum optimization processor}},\ }\href
  {https://doi.org/10.1088/0953-2048/23/6/065004} {\bibfield  {journal} {\bibinfo  {journal} {Superconductor Science and Technology}\ }\textbf {\bibinfo {volume} {23}},\ \bibinfo {pages} {065004} (\bibinfo {year} {2010})}\BibitemShut {NoStop}%
\bibitem [{\citenamefont {Matsuda}\ \emph {et~al.}(2009)\citenamefont {Matsuda}, \citenamefont {Nishimori},\ and\ \citenamefont {Katzgraber}}]{matsuda2009quantum}%
  \BibitemOpen
  \bibfield  {author} {\bibinfo {author} {\bibfnamefont {Y.}~\bibnamefont {Matsuda}}, \bibinfo {author} {\bibfnamefont {H.}~\bibnamefont {Nishimori}},\ and\ \bibinfo {author} {\bibfnamefont {H.~G.}\ \bibnamefont {Katzgraber}},\ }in\ \href {https://doi.org/10.1088/1742-6596/143/1/012003} {\emph {\bibinfo {booktitle} {Journal of Physics: Conference Series}}},\ Vol.\ \bibinfo {volume} {143}\ (\bibinfo {organization} {IOP Publishing},\ \bibinfo {year} {2009})\ p.\ \bibinfo {pages} {012003}\BibitemShut {NoStop}%
\bibitem [{\citenamefont {Pelofske}\ \emph {et~al.}(2021)\citenamefont {Pelofske}, \citenamefont {Golden}, \citenamefont {Bärtschi}, \citenamefont {O’Malley},\ and\ \citenamefont {Eidenbenz}}]{9605329}%
  \BibitemOpen
  \bibfield  {author} {\bibinfo {author} {\bibfnamefont {E.}~\bibnamefont {Pelofske}}, \bibinfo {author} {\bibfnamefont {J.}~\bibnamefont {Golden}}, \bibinfo {author} {\bibfnamefont {A.}~\bibnamefont {Bärtschi}}, \bibinfo {author} {\bibfnamefont {D.}~\bibnamefont {O’Malley}},\ and\ \bibinfo {author} {\bibfnamefont {S.}~\bibnamefont {Eidenbenz}},\ }in\ \href {https://doi.org/10.1109/QCE52317.2021.00038} {\emph {\bibinfo {booktitle} {2021 IEEE International Conference on Quantum Computing and Engineering (QCE)}}}\ (\bibinfo {year} {2021})\ pp.\ \bibinfo {pages} {207--217}\BibitemShut {NoStop}%
\bibitem [{\citenamefont {K\"onz}\ \emph {et~al.}(2019)\citenamefont {K\"onz}, \citenamefont {Mazzola}, \citenamefont {Ochoa}, \citenamefont {Katzgraber},\ and\ \citenamefont {Troyer}}]{PhysRevA.100.030303}%
  \BibitemOpen
  \bibfield  {author} {\bibinfo {author} {\bibfnamefont {M.~S.}\ \bibnamefont {K\"onz}}, \bibinfo {author} {\bibfnamefont {G.}~\bibnamefont {Mazzola}}, \bibinfo {author} {\bibfnamefont {A.~J.}\ \bibnamefont {Ochoa}}, \bibinfo {author} {\bibfnamefont {H.~G.}\ \bibnamefont {Katzgraber}},\ and\ \bibinfo {author} {\bibfnamefont {M.}~\bibnamefont {Troyer}},\ }\bibfield  {title} {\emph {\bibinfo {title} {Uncertain fate of fair sampling in quantum annealing}},\ }\href {https://doi.org/10.1103/PhysRevA.100.030303} {\bibfield  {journal} {\bibinfo  {journal} {Phys. Rev. A}\ }\textbf {\bibinfo {volume} {100}},\ \bibinfo {pages} {030303} (\bibinfo {year} {2019})}\BibitemShut {NoStop}%
\bibitem [{\citenamefont {Mandr\`a}\ \emph {et~al.}(2017)\citenamefont {Mandr\`a}, \citenamefont {Zhu},\ and\ \citenamefont {Katzgraber}}]{PhysRevLett.118.070502}%
  \BibitemOpen
  \bibfield  {author} {\bibinfo {author} {\bibfnamefont {S.}~\bibnamefont {Mandr\`a}}, \bibinfo {author} {\bibfnamefont {Z.}~\bibnamefont {Zhu}},\ and\ \bibinfo {author} {\bibfnamefont {H.~G.}\ \bibnamefont {Katzgraber}},\ }\bibfield  {title} {\emph {\bibinfo {title} {Exponentially biased ground-state sampling of quantum annealing machines with transverse-field driving hamiltonians}},\ }\href {https://doi.org/10.1103/PhysRevLett.118.070502} {\bibfield  {journal} {\bibinfo  {journal} {Phys. Rev. Lett.}\ }\textbf {\bibinfo {volume} {118}},\ \bibinfo {pages} {070502} (\bibinfo {year} {2017})}\BibitemShut {NoStop}%
\bibitem [{\citenamefont {Zhu}\ \emph {et~al.}(2019)\citenamefont {Zhu}, \citenamefont {Ochoa},\ and\ \citenamefont {Katzgraber}}]{PhysRevE.99.063314}%
  \BibitemOpen
  \bibfield  {author} {\bibinfo {author} {\bibfnamefont {Z.}~\bibnamefont {Zhu}}, \bibinfo {author} {\bibfnamefont {A.~J.}\ \bibnamefont {Ochoa}},\ and\ \bibinfo {author} {\bibfnamefont {H.~G.}\ \bibnamefont {Katzgraber}},\ }\bibfield  {title} {\emph {\bibinfo {title} {Fair sampling of ground-state configurations of binary optimization problems}},\ }\href {https://doi.org/10.1103/PhysRevE.99.063314} {\bibfield  {journal} {\bibinfo  {journal} {Phys. Rev. E}\ }\textbf {\bibinfo {volume} {99}},\ \bibinfo {pages} {063314} (\bibinfo {year} {2019})}\BibitemShut {NoStop}%
\bibitem [{\citenamefont {Job}\ and\ \citenamefont {Lidar}(2018)}]{job2018test}%
  \BibitemOpen
  \bibfield  {author} {\bibinfo {author} {\bibfnamefont {J.}~\bibnamefont {Job}}\ and\ \bibinfo {author} {\bibfnamefont {D.}~\bibnamefont {Lidar}},\ }\bibfield  {title} {\emph {\bibinfo {title} {Test-driving 1000 qubits}},\ }\href@noop {} {\bibfield  {journal} {\bibinfo  {journal} {Quantum Science and Technology}\ }\textbf {\bibinfo {volume} {3}},\ \bibinfo {pages} {030501} (\bibinfo {year} {2018})}\BibitemShut {NoStop}%
\bibitem [{\citenamefont {Albash}\ \emph {et~al.}(2015)\citenamefont {Albash}, \citenamefont {Vinci}, \citenamefont {Mishra}, \citenamefont {Warburton},\ and\ \citenamefont {Lidar}}]{PhysRevA.91.042314}%
  \BibitemOpen
  \bibfield  {author} {\bibinfo {author} {\bibfnamefont {T.}~\bibnamefont {Albash}}, \bibinfo {author} {\bibfnamefont {W.}~\bibnamefont {Vinci}}, \bibinfo {author} {\bibfnamefont {A.}~\bibnamefont {Mishra}}, \bibinfo {author} {\bibfnamefont {P.~A.}\ \bibnamefont {Warburton}},\ and\ \bibinfo {author} {\bibfnamefont {D.~A.}\ \bibnamefont {Lidar}},\ }\bibfield  {title} {\emph {\bibinfo {title} {Consistency tests of classical and quantum models for a quantum annealer}},\ }\href {https://doi.org/10.1103/PhysRevA.91.042314} {\bibfield  {journal} {\bibinfo  {journal} {Phys. Rev. A}\ }\textbf {\bibinfo {volume} {91}},\ \bibinfo {pages} {042314} (\bibinfo {year} {2015})}\BibitemShut {NoStop}%
\bibitem [{\citenamefont {Albash}\ and\ \citenamefont {Lidar}(2015)}]{Albash_2015_decoherence}%
  \BibitemOpen
  \bibfield  {author} {\bibinfo {author} {\bibfnamefont {T.}~\bibnamefont {Albash}}\ and\ \bibinfo {author} {\bibfnamefont {D.~A.}\ \bibnamefont {Lidar}},\ }\bibfield  {title} {\emph {\bibinfo {title} {Decoherence in adiabatic quantum computation}},\ }\bibfield  {journal} {\bibinfo  {journal} {Physical Review A}\ }\textbf {\bibinfo {volume} {91}},\ \href {https://doi.org/10.1103/physreva.91.062320} {10.1103/physreva.91.062320} (\bibinfo {year} {2015})\BibitemShut {NoStop}%
\bibitem [{\citenamefont {Boixo}\ \emph {et~al.}(2013)\citenamefont {Boixo}, \citenamefont {Albash}, \citenamefont {Spedalieri}, \citenamefont {Chancellor},\ and\ \citenamefont {Lidar}}]{boixo2013experimental}%
  \BibitemOpen
  \bibfield  {author} {\bibinfo {author} {\bibfnamefont {S.}~\bibnamefont {Boixo}}, \bibinfo {author} {\bibfnamefont {T.}~\bibnamefont {Albash}}, \bibinfo {author} {\bibfnamefont {F.~M.}\ \bibnamefont {Spedalieri}}, \bibinfo {author} {\bibfnamefont {N.}~\bibnamefont {Chancellor}},\ and\ \bibinfo {author} {\bibfnamefont {D.~A.}\ \bibnamefont {Lidar}},\ }\bibfield  {title} {\emph {\bibinfo {title} {Experimental signature of programmable quantum annealing}},\ }\href@noop {} {\bibfield  {journal} {\bibinfo  {journal} {Nature communications}\ }\textbf {\bibinfo {volume} {4}},\ \bibinfo {pages} {2067} (\bibinfo {year} {2013})}\BibitemShut {NoStop}%
\bibitem [{\citenamefont {Redner}(1981)}]{redner1981one}%
  \BibitemOpen
  \bibfield  {author} {\bibinfo {author} {\bibfnamefont {S.}~\bibnamefont {Redner}},\ }\bibfield  {title} {\emph {\bibinfo {title} {{One-dimensional Ising chain with competing interactions: Exact results and connection with other statistical models}}},\ }\href@noop {} {\bibfield  {journal} {\bibinfo  {journal} {Journal of Statistical Physics}\ }\textbf {\bibinfo {volume} {25}},\ \bibinfo {pages} {15--23} (\bibinfo {year} {1981})}\BibitemShut {NoStop}%
\bibitem [{\citenamefont {Hu}\ and\ \citenamefont {Charbonneau}(2021{\natexlab{a}})}]{PhysRevB.103.094441}%
  \BibitemOpen
  \bibfield  {author} {\bibinfo {author} {\bibfnamefont {Y.}~\bibnamefont {Hu}}\ and\ \bibinfo {author} {\bibfnamefont {P.}~\bibnamefont {Charbonneau}},\ }\bibfield  {title} {\emph {\bibinfo {title} {Resolving the two-dimensional axial next-nearest-neighbor ising model using transfer matrices}},\ }\href {https://doi.org/10.1103/PhysRevB.103.094441} {\bibfield  {journal} {\bibinfo  {journal} {Phys. Rev. B}\ }\textbf {\bibinfo {volume} {103}},\ \bibinfo {pages} {094441} (\bibinfo {year} {2021}{\natexlab{a}})}\BibitemShut {NoStop}%
\bibitem [{\citenamefont {Stephenson}(1970)}]{PhysRevB.1.4405}%
  \BibitemOpen
  \bibfield  {author} {\bibinfo {author} {\bibfnamefont {J.}~\bibnamefont {Stephenson}},\ }\bibfield  {title} {\emph {\bibinfo {title} {{Ising Model with Antiferromagnetic Next-Nearest-Neighbor Coupling: Spin Correlations and Disorder Points}}},\ }\href {https://doi.org/10.1103/PhysRevB.1.4405} {\bibfield  {journal} {\bibinfo  {journal} {Phys. Rev. B}\ }\textbf {\bibinfo {volume} {1}},\ \bibinfo {pages} {4405--4409} (\bibinfo {year} {1970})}\BibitemShut {NoStop}%
\bibitem [{\citenamefont {Hu}\ and\ \citenamefont {Charbonneau}(2021{\natexlab{b}})}]{PhysRevB.104.144429}%
  \BibitemOpen
  \bibfield  {author} {\bibinfo {author} {\bibfnamefont {Y.}~\bibnamefont {Hu}}\ and\ \bibinfo {author} {\bibfnamefont {P.}~\bibnamefont {Charbonneau}},\ }\bibfield  {title} {\emph {\bibinfo {title} {{Numerical transfer matrix study of frustrated next-nearest-neighbor Ising models on square lattices}}},\ }\href {https://doi.org/10.1103/PhysRevB.104.144429} {\bibfield  {journal} {\bibinfo  {journal} {Phys. Rev. B}\ }\textbf {\bibinfo {volume} {104}},\ \bibinfo {pages} {144429} (\bibinfo {year} {2021}{\natexlab{b}})}\BibitemShut {NoStop}%
\bibitem [{\citenamefont {Shirakura}\ \emph {et~al.}(2014)\citenamefont {Shirakura}, \citenamefont {Matsubara},\ and\ \citenamefont {Suzuki}}]{PhysRevB.90.144410}%
  \BibitemOpen
  \bibfield  {author} {\bibinfo {author} {\bibfnamefont {T.}~\bibnamefont {Shirakura}}, \bibinfo {author} {\bibfnamefont {F.}~\bibnamefont {Matsubara}},\ and\ \bibinfo {author} {\bibfnamefont {N.}~\bibnamefont {Suzuki}},\ }\bibfield  {title} {\emph {\bibinfo {title} {{Kosterlitz-Thouless phase transition of the axial next-nearest-neighbor Ising model in two dimensions}}},\ }\href {https://doi.org/10.1103/PhysRevB.90.144410} {\bibfield  {journal} {\bibinfo  {journal} {Phys. Rev. B}\ }\textbf {\bibinfo {volume} {90}},\ \bibinfo {pages} {144410} (\bibinfo {year} {2014})}\BibitemShut {NoStop}%
\bibitem [{\citenamefont {Dhar}\ \emph {et~al.}(2000)\citenamefont {Dhar}, \citenamefont {Shastry},\ and\ \citenamefont {Dasgupta}}]{Dhar_2000}%
  \BibitemOpen
  \bibfield  {author} {\bibinfo {author} {\bibfnamefont {A.}~\bibnamefont {Dhar}}, \bibinfo {author} {\bibfnamefont {B.~S.}\ \bibnamefont {Shastry}},\ and\ \bibinfo {author} {\bibfnamefont {C.}~\bibnamefont {Dasgupta}},\ }\bibfield  {title} {\emph {\bibinfo {title} {{Equilibrium and dynamical properties of the axial next-nearest-neighbor Ising chain at the multiphase point}}},\ }\href {https://doi.org/10.1103/physreve.62.1592} {\bibfield  {journal} {\bibinfo  {journal} {Physical Review E}\ }\textbf {\bibinfo {volume} {62}},\ \bibinfo {pages} {1592–1600} (\bibinfo {year} {2000})}\BibitemShut {NoStop}%
\bibitem [{\citenamefont {Yeomans}(1995)}]{yeomans1995splittingmultiphasepoint}%
  \BibitemOpen
  \bibfield  {author} {\bibinfo {author} {\bibfnamefont {J.~M.}\ \bibnamefont {Yeomans}},\ }\href {https://arxiv.org/abs/cond-mat/9506129} {\bibinfo {title} {Splitting the multiphase point}} (\bibinfo {year} {1995}),\ \Eprint {https://arxiv.org/abs/cond-mat/9506129} {arXiv:cond-mat/9506129 [cond-mat]} \BibitemShut {NoStop}%
\bibitem [{\citenamefont {Torre}\ \emph {et~al.}(2025)\citenamefont {Torre}, \citenamefont {Catalano}, \citenamefont {Kožić}, \citenamefont {Franchini},\ and\ \citenamefont {Giampaolo}}]{Torre_2025}%
  \BibitemOpen
  \bibfield  {author} {\bibinfo {author} {\bibfnamefont {G.}~\bibnamefont {Torre}}, \bibinfo {author} {\bibfnamefont {A.~G.}\ \bibnamefont {Catalano}}, \bibinfo {author} {\bibfnamefont {S.~B.}\ \bibnamefont {Kožić}}, \bibinfo {author} {\bibfnamefont {F.}~\bibnamefont {Franchini}},\ and\ \bibinfo {author} {\bibfnamefont {S.~M.}\ \bibnamefont {Giampaolo}},\ }\bibfield  {title} {\emph {\bibinfo {title} {{Interplay between local and non-local frustration in the 1D ANNNI chain: I. The even case}}},\ }\href {https://doi.org/10.1088/1742-5468/adc897} {\bibfield  {journal} {\bibinfo  {journal} {Journal of Statistical Mechanics: Theory and Experiment}\ }\textbf {\bibinfo {volume} {2025}},\ \bibinfo {pages} {053101} (\bibinfo {year} {2025})}\BibitemShut {NoStop}%
\bibitem [{\citenamefont {McCreesh}\ \emph {et~al.}(2020)\citenamefont {McCreesh}, \citenamefont {Prosser},\ and\ \citenamefont {Trimble}}]{mccreesh2020glasgow}%
  \BibitemOpen
  \bibfield  {author} {\bibinfo {author} {\bibfnamefont {C.}~\bibnamefont {McCreesh}}, \bibinfo {author} {\bibfnamefont {P.}~\bibnamefont {Prosser}},\ and\ \bibinfo {author} {\bibfnamefont {J.}~\bibnamefont {Trimble}},\ }in\ \href@noop {} {\emph {\bibinfo {booktitle} {International Conference on Graph Transformation}}}\ (\bibinfo {organization} {Springer},\ \bibinfo {year} {2020})\ pp.\ \bibinfo {pages} {316--324}\BibitemShut {NoStop}%
\bibitem [{\citenamefont {Chern}\ \emph {et~al.}(2023)\citenamefont {Chern}, \citenamefont {Boothby}, \citenamefont {Raymond}, \citenamefont {Farré},\ and\ \citenamefont {King}}]{Chern_2023}%
  \BibitemOpen
  \bibfield  {author} {\bibinfo {author} {\bibfnamefont {K.}~\bibnamefont {Chern}}, \bibinfo {author} {\bibfnamefont {K.}~\bibnamefont {Boothby}}, \bibinfo {author} {\bibfnamefont {J.}~\bibnamefont {Raymond}}, \bibinfo {author} {\bibfnamefont {P.}~\bibnamefont {Farré}},\ and\ \bibinfo {author} {\bibfnamefont {A.~D.}\ \bibnamefont {King}},\ }\bibfield  {title} {\emph {\bibinfo {title} {Tutorial: calibration refinement in quantum annealing}},\ }\bibfield  {journal} {\bibinfo  {journal} {Frontiers in Computer Science}\ }\textbf {\bibinfo {volume} {5}},\ \href {https://doi.org/10.3389/fcomp.2023.1238988} {10.3389/fcomp.2023.1238988} (\bibinfo {year} {2023})\BibitemShut {NoStop}%
\bibitem [{\citenamefont {Cai}\ \emph {et~al.}(2014)\citenamefont {Cai}, \citenamefont {Macready},\ and\ \citenamefont {Roy}}]{cai2014practicalheuristicfindinggraph}%
  \BibitemOpen
  \bibfield  {author} {\bibinfo {author} {\bibfnamefont {J.}~\bibnamefont {Cai}}, \bibinfo {author} {\bibfnamefont {W.~G.}\ \bibnamefont {Macready}},\ and\ \bibinfo {author} {\bibfnamefont {A.}~\bibnamefont {Roy}},\ }\href {https://arxiv.org/abs/1406.2741} {\bibinfo {title} {A practical heuristic for finding graph minors}} (\bibinfo {year} {2014}),\ \Eprint {https://arxiv.org/abs/1406.2741} {arXiv:1406.2741 [quant-ph]} \BibitemShut {NoStop}%
\bibitem [{\citenamefont {Pelofske}\ \emph {et~al.}(2022{\natexlab{a}})\citenamefont {Pelofske}, \citenamefont {Hahn},\ and\ \citenamefont {Djidjev}}]{parallel_QA}%
  \BibitemOpen
  \bibfield  {author} {\bibinfo {author} {\bibfnamefont {E.}~\bibnamefont {Pelofske}}, \bibinfo {author} {\bibfnamefont {G.}~\bibnamefont {Hahn}},\ and\ \bibinfo {author} {\bibfnamefont {H.~N.}\ \bibnamefont {Djidjev}},\ }\bibfield  {title} {\emph {\bibinfo {title} {Parallel quantum annealing}},\ }\bibfield  {journal} {\bibinfo  {journal} {Scientific Reports}\ }\textbf {\bibinfo {volume} {12}},\ \href {https://doi.org/10.1038/s41598-022-08394-8} {10.1038/s41598-022-08394-8} (\bibinfo {year} {2022}{\natexlab{a}})\BibitemShut {NoStop}%
\bibitem [{\citenamefont {Pelofske}\ \emph {et~al.}(2022{\natexlab{b}})\citenamefont {Pelofske}, \citenamefont {Hahn}, \citenamefont {O’Malley}, \citenamefont {Djidjev},\ and\ \citenamefont {Alexandrov}}]{Pelofske_2022_boolean}%
  \BibitemOpen
  \bibfield  {author} {\bibinfo {author} {\bibfnamefont {E.}~\bibnamefont {Pelofske}}, \bibinfo {author} {\bibfnamefont {G.}~\bibnamefont {Hahn}}, \bibinfo {author} {\bibfnamefont {D.}~\bibnamefont {O’Malley}}, \bibinfo {author} {\bibfnamefont {H.~N.}\ \bibnamefont {Djidjev}},\ and\ \bibinfo {author} {\bibfnamefont {B.~S.}\ \bibnamefont {Alexandrov}},\ }\bibfield  {title} {\emph {\bibinfo {title} {{Quantum annealing algorithms for Boolean tensor networks}}},\ }\bibfield  {journal} {\bibinfo  {journal} {Scientific Reports}\ }\textbf {\bibinfo {volume} {12}},\ \href {https://doi.org/10.1038/s41598-022-12611-9} {10.1038/s41598-022-12611-9} (\bibinfo {year} {2022}{\natexlab{b}})\BibitemShut {NoStop}%
\bibitem [{\citenamefont {Dattani}\ \emph {et~al.}(2019)\citenamefont {Dattani}, \citenamefont {Szalay},\ and\ \citenamefont {Chancellor}}]{dattani2019pegasussecondconnectivitygraph}%
  \BibitemOpen
  \bibfield  {author} {\bibinfo {author} {\bibfnamefont {N.}~\bibnamefont {Dattani}}, \bibinfo {author} {\bibfnamefont {S.}~\bibnamefont {Szalay}},\ and\ \bibinfo {author} {\bibfnamefont {N.}~\bibnamefont {Chancellor}},\ }\href {https://arxiv.org/abs/1901.07636} {\bibinfo {title} {{Pegasus: The second connectivity graph for large-scale quantum annealing hardware}}} (\bibinfo {year} {2019}),\ \Eprint {https://arxiv.org/abs/1901.07636} {arXiv:1901.07636 [quant-ph]} \BibitemShut {NoStop}%
\bibitem [{\citenamefont {Boothby}\ \emph {et~al.}(2020)\citenamefont {Boothby}, \citenamefont {Bunyk}, \citenamefont {Raymond},\ and\ \citenamefont {Roy}}]{boothby2020nextgenerationtopologydwavequantum}%
  \BibitemOpen
  \bibfield  {author} {\bibinfo {author} {\bibfnamefont {K.}~\bibnamefont {Boothby}}, \bibinfo {author} {\bibfnamefont {P.}~\bibnamefont {Bunyk}}, \bibinfo {author} {\bibfnamefont {J.}~\bibnamefont {Raymond}},\ and\ \bibinfo {author} {\bibfnamefont {A.}~\bibnamefont {Roy}},\ }\href {https://arxiv.org/abs/2003.00133} {\bibinfo {title} {{Next-Generation Topology of D-Wave Quantum Processors}}} (\bibinfo {year} {2020}),\ \Eprint {https://arxiv.org/abs/2003.00133} {arXiv:2003.00133 [quant-ph]} \BibitemShut {NoStop}%
\bibitem [{\citenamefont {Boothby}\ \emph {et~al.}(2021)\citenamefont {Boothby}, \citenamefont {King},\ and\ \citenamefont {Raymond}}]{zephyr}%
  \BibitemOpen
  \bibfield  {author} {\bibinfo {author} {\bibfnamefont {K.}~\bibnamefont {Boothby}}, \bibinfo {author} {\bibfnamefont {A.~D.}\ \bibnamefont {King}},\ and\ \bibinfo {author} {\bibfnamefont {J.}~\bibnamefont {Raymond}},\ }\href {https://www.dwavesys.com/media/2uznec4s/14-1056a-a_zephyr_topology_of_d-wave_quantum_processors.pdf} {\bibinfo {title} {Zephyr topology of d-wave quantum processors}} (\bibinfo {year} {2021})\BibitemShut {NoStop}%
\bibitem [{\citenamefont {Marshall}\ \emph {et~al.}(2020)\citenamefont {Marshall}, \citenamefont {Di~Gioacchino},\ and\ \citenamefont {Rieffel}}]{PhysRevResearch.2.023020}%
  \BibitemOpen
  \bibfield  {author} {\bibinfo {author} {\bibfnamefont {J.}~\bibnamefont {Marshall}}, \bibinfo {author} {\bibfnamefont {A.}~\bibnamefont {Di~Gioacchino}},\ and\ \bibinfo {author} {\bibfnamefont {E.~G.}\ \bibnamefont {Rieffel}},\ }\bibfield  {title} {\emph {\bibinfo {title} {Perils of embedding for sampling problems}},\ }\href {https://doi.org/10.1103/PhysRevResearch.2.023020} {\bibfield  {journal} {\bibinfo  {journal} {Phys. Rev. Res.}\ }\textbf {\bibinfo {volume} {2}},\ \bibinfo {pages} {023020} (\bibinfo {year} {2020})}\BibitemShut {NoStop}%
\bibitem [{\citenamefont {Kibble}(1976)}]{kibble1976topology}%
  \BibitemOpen
  \bibfield  {author} {\bibinfo {author} {\bibfnamefont {T.~W.}\ \bibnamefont {Kibble}},\ }\bibfield  {title} {\emph {\bibinfo {title} {Topology of cosmic domains and strings}},\ }\href@noop {} {\bibfield  {journal} {\bibinfo  {journal} {Journal of Physics A: Mathematical and General}\ }\textbf {\bibinfo {volume} {9}},\ \bibinfo {pages} {1387} (\bibinfo {year} {1976})}\BibitemShut {NoStop}%
\bibitem [{\citenamefont {Zurek}(1985)}]{zurek1985cosmological}%
  \BibitemOpen
  \bibfield  {author} {\bibinfo {author} {\bibfnamefont {W.~H.}\ \bibnamefont {Zurek}},\ }\bibfield  {title} {\emph {\bibinfo {title} {Cosmological experiments in superfluid helium?}},\ }\href@noop {} {\bibfield  {journal} {\bibinfo  {journal} {Nature}\ }\textbf {\bibinfo {volume} {317}},\ \bibinfo {pages} {505--508} (\bibinfo {year} {1985})}\BibitemShut {NoStop}%
\bibitem [{\citenamefont {Bando}\ \emph {et~al.}(2020)\citenamefont {Bando}, \citenamefont {Susa}, \citenamefont {Oshiyama}, \citenamefont {Shibata}, \citenamefont {Ohzeki}, \citenamefont {G{\'o}mez-Ruiz}, \citenamefont {Lidar}, \citenamefont {Suzuki}, \citenamefont {Del~Campo},\ and\ \citenamefont {Nishimori}}]{bando2020probing}%
  \BibitemOpen
  \bibfield  {author} {\bibinfo {author} {\bibfnamefont {Y.}~\bibnamefont {Bando}}, \bibinfo {author} {\bibfnamefont {Y.}~\bibnamefont {Susa}}, \bibinfo {author} {\bibfnamefont {H.}~\bibnamefont {Oshiyama}}, \bibinfo {author} {\bibfnamefont {N.}~\bibnamefont {Shibata}}, \bibinfo {author} {\bibfnamefont {M.}~\bibnamefont {Ohzeki}}, \bibinfo {author} {\bibfnamefont {F.~J.}\ \bibnamefont {G{\'o}mez-Ruiz}}, \bibinfo {author} {\bibfnamefont {D.~A.}\ \bibnamefont {Lidar}}, \bibinfo {author} {\bibfnamefont {S.}~\bibnamefont {Suzuki}}, \bibinfo {author} {\bibfnamefont {A.}~\bibnamefont {Del~Campo}},\ and\ \bibinfo {author} {\bibfnamefont {H.}~\bibnamefont {Nishimori}},\ }\bibfield  {title} {\emph {\bibinfo {title} {{Probing the universality of topological defect formation in a quantum annealer: Kibble-Zurek mechanism and beyond}}},\ }\href@noop {} {\bibfield  {journal} {\bibinfo  {journal} {Physical Review Research}\ }\textbf {\bibinfo {volume} {2}},\ \bibinfo {pages} {033369} (\bibinfo {year} {2020})}\BibitemShut
  {NoStop}%
\bibitem [{\citenamefont {King}\ \emph {et~al.}(2023)\citenamefont {King}, \citenamefont {Raymond}, \citenamefont {Lanting}, \citenamefont {Harris}, \citenamefont {Zucca}, \citenamefont {Altomare}, \citenamefont {Berkley}, \citenamefont {Boothby}, \citenamefont {Ejtemaee}, \citenamefont {Enderud} \emph {et~al.}}]{king2023quantum}%
  \BibitemOpen
  \bibfield  {author} {\bibinfo {author} {\bibfnamefont {A.~D.}\ \bibnamefont {King}}, \bibinfo {author} {\bibfnamefont {J.}~\bibnamefont {Raymond}}, \bibinfo {author} {\bibfnamefont {T.}~\bibnamefont {Lanting}}, \bibinfo {author} {\bibfnamefont {R.}~\bibnamefont {Harris}}, \bibinfo {author} {\bibfnamefont {A.}~\bibnamefont {Zucca}}, \bibinfo {author} {\bibfnamefont {F.}~\bibnamefont {Altomare}}, \bibinfo {author} {\bibfnamefont {A.~J.}\ \bibnamefont {Berkley}}, \bibinfo {author} {\bibfnamefont {K.}~\bibnamefont {Boothby}}, \bibinfo {author} {\bibfnamefont {S.}~\bibnamefont {Ejtemaee}}, \bibinfo {author} {\bibfnamefont {C.}~\bibnamefont {Enderud}}, \emph {et~al.},\ }\bibfield  {title} {\emph {\bibinfo {title} {Quantum critical dynamics in a 5,000-qubit programmable spin glass}},\ }\href@noop {} {\bibfield  {journal} {\bibinfo  {journal} {Nature}\ }\textbf {\bibinfo {volume} {617}},\ \bibinfo {pages} {61--66} (\bibinfo {year} {2023})}\BibitemShut {NoStop}%
\bibitem [{\citenamefont {King}\ \emph {et~al.}(2022)\citenamefont {King}, \citenamefont {Suzuki}, \citenamefont {Raymond}, \citenamefont {Zucca}, \citenamefont {Lanting}, \citenamefont {Altomare}, \citenamefont {Berkley}, \citenamefont {Ejtemaee}, \citenamefont {Hoskinson}, \citenamefont {Huang}, \citenamefont {Ladizinsky}, \citenamefont {MacDonald}, \citenamefont {Marsden}, \citenamefont {Oh}, \citenamefont {Poulin-Lamarre}, \citenamefont {Reis}, \citenamefont {Rich}, \citenamefont {Sato}, \citenamefont {Whittaker}, \citenamefont {Yao}, \citenamefont {Harris}, \citenamefont {Lidar}, \citenamefont {Nishimori},\ and\ \citenamefont {Amin}}]{King_2022}%
  \BibitemOpen
  \bibfield  {author} {\bibinfo {author} {\bibfnamefont {A.~D.}\ \bibnamefont {King}}, \bibinfo {author} {\bibfnamefont {S.}~\bibnamefont {Suzuki}}, \bibinfo {author} {\bibfnamefont {J.}~\bibnamefont {Raymond}}, \bibinfo {author} {\bibfnamefont {A.}~\bibnamefont {Zucca}}, \bibinfo {author} {\bibfnamefont {T.}~\bibnamefont {Lanting}}, \bibinfo {author} {\bibfnamefont {F.}~\bibnamefont {Altomare}}, \bibinfo {author} {\bibfnamefont {A.~J.}\ \bibnamefont {Berkley}}, \bibinfo {author} {\bibfnamefont {S.}~\bibnamefont {Ejtemaee}}, \bibinfo {author} {\bibfnamefont {E.}~\bibnamefont {Hoskinson}}, \bibinfo {author} {\bibfnamefont {S.}~\bibnamefont {Huang}}, \bibinfo {author} {\bibfnamefont {E.}~\bibnamefont {Ladizinsky}}, \bibinfo {author} {\bibfnamefont {A.~J.~R.}\ \bibnamefont {MacDonald}}, \bibinfo {author} {\bibfnamefont {G.}~\bibnamefont {Marsden}}, \bibinfo {author} {\bibfnamefont {T.}~\bibnamefont {Oh}}, \bibinfo {author} {\bibfnamefont {G.}~\bibnamefont {Poulin-Lamarre}}, \bibinfo {author} {\bibfnamefont
  {M.}~\bibnamefont {Reis}}, \bibinfo {author} {\bibfnamefont {C.}~\bibnamefont {Rich}}, \bibinfo {author} {\bibfnamefont {Y.}~\bibnamefont {Sato}}, \bibinfo {author} {\bibfnamefont {J.~D.}\ \bibnamefont {Whittaker}}, \bibinfo {author} {\bibfnamefont {J.}~\bibnamefont {Yao}}, \bibinfo {author} {\bibfnamefont {R.}~\bibnamefont {Harris}}, \bibinfo {author} {\bibfnamefont {D.~A.}\ \bibnamefont {Lidar}}, \bibinfo {author} {\bibfnamefont {H.}~\bibnamefont {Nishimori}},\ and\ \bibinfo {author} {\bibfnamefont {M.~H.}\ \bibnamefont {Amin}},\ }\bibfield  {title} {\emph {\bibinfo {title} {Coherent quantum annealing in a programmable 2,000 qubit ising chain}},\ }\href {https://doi.org/10.1038/s41567-022-01741-6} {\bibfield  {journal} {\bibinfo  {journal} {Nature Physics}\ }\textbf {\bibinfo {volume} {18}},\ \bibinfo {pages} {1324–1328} (\bibinfo {year} {2022})}\BibitemShut {NoStop}%
\bibitem [{\citenamefont {Amin}(2015)}]{Amin_2015}%
  \BibitemOpen
  \bibfield  {author} {\bibinfo {author} {\bibfnamefont {M.~H.}\ \bibnamefont {Amin}},\ }\bibfield  {title} {\emph {\bibinfo {title} {Searching for quantum speedup in quasistatic quantum annealers}},\ }\bibfield  {journal} {\bibinfo  {journal} {Physical Review A}\ }\textbf {\bibinfo {volume} {92}},\ \href {https://doi.org/10.1103/physreva.92.052323} {10.1103/physreva.92.052323} (\bibinfo {year} {2015})\BibitemShut {NoStop}%
\bibitem [{\citenamefont {Pelofske}\ \emph {et~al.}(2023)\citenamefont {Pelofske}, \citenamefont {Hahn},\ and\ \citenamefont {Djidjev}}]{Pelofske_2023_noise}%
  \BibitemOpen
  \bibfield  {author} {\bibinfo {author} {\bibfnamefont {E.}~\bibnamefont {Pelofske}}, \bibinfo {author} {\bibfnamefont {G.}~\bibnamefont {Hahn}},\ and\ \bibinfo {author} {\bibfnamefont {H.~N.}\ \bibnamefont {Djidjev}},\ }\bibfield  {title} {\emph {\bibinfo {title} {Noise dynamics of quantum annealers: estimating the effective noise using idle qubits}},\ }\href {https://doi.org/10.1088/2058-9565/accbe6} {\bibfield  {journal} {\bibinfo  {journal} {Quantum Science and Technology}\ }\textbf {\bibinfo {volume} {8}},\ \bibinfo {pages} {035005} (\bibinfo {year} {2023})}\BibitemShut {NoStop}%
\bibitem [{\citenamefont {Morrell}\ \emph {et~al.}(2023)\citenamefont {Morrell}, \citenamefont {Vuffray}, \citenamefont {Lokhov}, \citenamefont {B\"artschi}, \citenamefont {Albash},\ and\ \citenamefont {Coffrin}}]{PhysRevApplied.19.034053}%
  \BibitemOpen
  \bibfield  {author} {\bibinfo {author} {\bibfnamefont {Z.}~\bibnamefont {Morrell}}, \bibinfo {author} {\bibfnamefont {M.}~\bibnamefont {Vuffray}}, \bibinfo {author} {\bibfnamefont {A.~Y.}\ \bibnamefont {Lokhov}}, \bibinfo {author} {\bibfnamefont {A.}~\bibnamefont {B\"artschi}}, \bibinfo {author} {\bibfnamefont {T.}~\bibnamefont {Albash}},\ and\ \bibinfo {author} {\bibfnamefont {C.}~\bibnamefont {Coffrin}},\ }\bibfield  {title} {\emph {\bibinfo {title} {{Signatures of Open and Noisy Quantum Systems in Single-Qubit Quantum Annealing}}},\ }\href {https://doi.org/10.1103/PhysRevApplied.19.034053} {\bibfield  {journal} {\bibinfo  {journal} {Phys. Rev. Appl.}\ }\textbf {\bibinfo {volume} {19}},\ \bibinfo {pages} {034053} (\bibinfo {year} {2023})}\BibitemShut {NoStop}%
\bibitem [{\citenamefont {Zaborniak}\ and\ \citenamefont {de~Sousa}(2021)}]{Zaborniak_2021}%
  \BibitemOpen
  \bibfield  {author} {\bibinfo {author} {\bibfnamefont {T.}~\bibnamefont {Zaborniak}}\ and\ \bibinfo {author} {\bibfnamefont {R.}~\bibnamefont {de~Sousa}},\ }\bibfield  {title} {\emph {\bibinfo {title} {{Benchmarking Hamiltonian Noise in the D-Wave Quantum Annealer}}},\ }\href {https://doi.org/10.1109/tqe.2021.3050449} {\bibfield  {journal} {\bibinfo  {journal} {IEEE Transactions on Quantum Engineering}\ }\textbf {\bibinfo {volume} {2}},\ \bibinfo {pages} {1–6} (\bibinfo {year} {2021})}\BibitemShut {NoStop}%
\bibitem [{\citenamefont {Marshall}\ \emph {et~al.}(2017{\natexlab{b}})\citenamefont {Marshall}, \citenamefont {Rieffel},\ and\ \citenamefont {Hen}}]{PhysRevApplied.8.064025}%
  \BibitemOpen
  \bibfield  {author} {\bibinfo {author} {\bibfnamefont {J.}~\bibnamefont {Marshall}}, \bibinfo {author} {\bibfnamefont {E.~G.}\ \bibnamefont {Rieffel}},\ and\ \bibinfo {author} {\bibfnamefont {I.}~\bibnamefont {Hen}},\ }\bibfield  {title} {\emph {\bibinfo {title} {{Thermalization, Freeze-out, and Noise: Deciphering Experimental Quantum Annealers}}},\ }\href {https://doi.org/10.1103/PhysRevApplied.8.064025} {\bibfield  {journal} {\bibinfo  {journal} {Phys. Rev. Appl.}\ }\textbf {\bibinfo {volume} {8}},\ \bibinfo {pages} {064025} (\bibinfo {year} {2017}{\natexlab{b}})}\BibitemShut {NoStop}%
\bibitem [{\citenamefont {Tüysüz}\ \emph {et~al.}(2025)\citenamefont {Tüysüz}, \citenamefont {Jayakumar}, \citenamefont {Coffrin}, \citenamefont {Vuffray},\ and\ \citenamefont {Lokhov}}]{tüysüz2025learningresponsefunctionsanalog}%
  \BibitemOpen
  \bibfield  {author} {\bibinfo {author} {\bibfnamefont {C.}~\bibnamefont {Tüysüz}}, \bibinfo {author} {\bibfnamefont {A.}~\bibnamefont {Jayakumar}}, \bibinfo {author} {\bibfnamefont {C.}~\bibnamefont {Coffrin}}, \bibinfo {author} {\bibfnamefont {M.}~\bibnamefont {Vuffray}},\ and\ \bibinfo {author} {\bibfnamefont {A.~Y.}\ \bibnamefont {Lokhov}},\ }\href {https://arxiv.org/abs/2503.12520} {\bibinfo {title} {Learning response functions of analog quantum computers: analysis of neutral-atom and superconducting platforms}} (\bibinfo {year} {2025}),\ \Eprint {https://arxiv.org/abs/2503.12520} {arXiv:2503.12520 [quant-ph]} \BibitemShut {NoStop}%
\bibitem [{\citenamefont {Lanting}\ \emph {et~al.}(2020)\citenamefont {Lanting}, \citenamefont {Amin}, \citenamefont {Baron}, \citenamefont {Babcock}, \citenamefont {Boschee}, \citenamefont {Boixo}, \citenamefont {Smelyanskiy}, \citenamefont {Foygel},\ and\ \citenamefont {Petukhov}}]{lanting2020probingenvironmentalspinpolarization}%
  \BibitemOpen
  \bibfield  {author} {\bibinfo {author} {\bibfnamefont {T.}~\bibnamefont {Lanting}}, \bibinfo {author} {\bibfnamefont {M.~H.}\ \bibnamefont {Amin}}, \bibinfo {author} {\bibfnamefont {C.}~\bibnamefont {Baron}}, \bibinfo {author} {\bibfnamefont {M.}~\bibnamefont {Babcock}}, \bibinfo {author} {\bibfnamefont {J.}~\bibnamefont {Boschee}}, \bibinfo {author} {\bibfnamefont {S.}~\bibnamefont {Boixo}}, \bibinfo {author} {\bibfnamefont {V.~N.}\ \bibnamefont {Smelyanskiy}}, \bibinfo {author} {\bibfnamefont {M.}~\bibnamefont {Foygel}},\ and\ \bibinfo {author} {\bibfnamefont {A.~G.}\ \bibnamefont {Petukhov}},\ }\href {https://arxiv.org/abs/2003.14244} {\bibinfo {title} {{Probing Environmental Spin Polarization with Superconducting Flux Qubits}}} (\bibinfo {year} {2020}),\ \Eprint {https://arxiv.org/abs/2003.14244} {arXiv:2003.14244 [quant-ph]} \BibitemShut {NoStop}%
\bibitem [{\citenamefont {Virtanen}\ \emph {et~al.}(2020)\citenamefont {Virtanen}, \citenamefont {Gommers}, \citenamefont {Oliphant}, \citenamefont {Haberland}, \citenamefont {Reddy}, \citenamefont {Cournapeau}, \citenamefont {Burovski}, \citenamefont {Peterson}, \citenamefont {Weckesser}, \citenamefont {Bright}, \citenamefont {{van der Walt}}, \citenamefont {Brett}, \citenamefont {Wilson}, \citenamefont {Millman}, \citenamefont {Mayorov}, \citenamefont {Nelson}, \citenamefont {Jones}, \citenamefont {Kern}, \citenamefont {Larson}, \citenamefont {Carey}, \citenamefont {Polat}, \citenamefont {Feng}, \citenamefont {Moore}, \citenamefont {{VanderPlas}}, \citenamefont {Laxalde}, \citenamefont {Perktold}, \citenamefont {Cimrman}, \citenamefont {Henriksen}, \citenamefont {Quintero}, \citenamefont {Harris}, \citenamefont {Archibald}, \citenamefont {Ribeiro}, \citenamefont {Pedregosa}, \citenamefont {{van Mulbregt}},\ and\ \citenamefont {{SciPy 1.0 Contributors}}}]{2020SciPy-NMeth}%
  \BibitemOpen
  \bibfield  {author} {\bibinfo {author} {\bibfnamefont {P.}~\bibnamefont {Virtanen}}, \bibinfo {author} {\bibfnamefont {R.}~\bibnamefont {Gommers}}, \bibinfo {author} {\bibfnamefont {T.~E.}\ \bibnamefont {Oliphant}}, \bibinfo {author} {\bibfnamefont {M.}~\bibnamefont {Haberland}}, \bibinfo {author} {\bibfnamefont {T.}~\bibnamefont {Reddy}}, \bibinfo {author} {\bibfnamefont {D.}~\bibnamefont {Cournapeau}}, \bibinfo {author} {\bibfnamefont {E.}~\bibnamefont {Burovski}}, \bibinfo {author} {\bibfnamefont {P.}~\bibnamefont {Peterson}}, \bibinfo {author} {\bibfnamefont {W.}~\bibnamefont {Weckesser}}, \bibinfo {author} {\bibfnamefont {J.}~\bibnamefont {Bright}}, \bibinfo {author} {\bibfnamefont {S.~J.}\ \bibnamefont {{van der Walt}}}, \bibinfo {author} {\bibfnamefont {M.}~\bibnamefont {Brett}}, \bibinfo {author} {\bibfnamefont {J.}~\bibnamefont {Wilson}}, \bibinfo {author} {\bibfnamefont {K.~J.}\ \bibnamefont {Millman}}, \bibinfo {author} {\bibfnamefont {N.}~\bibnamefont {Mayorov}}, \bibinfo {author} {\bibfnamefont
  {A.~R.~J.}\ \bibnamefont {Nelson}}, \bibinfo {author} {\bibfnamefont {E.}~\bibnamefont {Jones}}, \bibinfo {author} {\bibfnamefont {R.}~\bibnamefont {Kern}}, \bibinfo {author} {\bibfnamefont {E.}~\bibnamefont {Larson}}, \bibinfo {author} {\bibfnamefont {C.~J.}\ \bibnamefont {Carey}}, \bibinfo {author} {\bibfnamefont {{\.I}.}~\bibnamefont {Polat}}, \bibinfo {author} {\bibfnamefont {Y.}~\bibnamefont {Feng}}, \bibinfo {author} {\bibfnamefont {E.~W.}\ \bibnamefont {Moore}}, \bibinfo {author} {\bibfnamefont {J.}~\bibnamefont {{VanderPlas}}}, \bibinfo {author} {\bibfnamefont {D.}~\bibnamefont {Laxalde}}, \bibinfo {author} {\bibfnamefont {J.}~\bibnamefont {Perktold}}, \bibinfo {author} {\bibfnamefont {R.}~\bibnamefont {Cimrman}}, \bibinfo {author} {\bibfnamefont {I.}~\bibnamefont {Henriksen}}, \bibinfo {author} {\bibfnamefont {E.~A.}\ \bibnamefont {Quintero}}, \bibinfo {author} {\bibfnamefont {C.~R.}\ \bibnamefont {Harris}}, \bibinfo {author} {\bibfnamefont {A.~M.}\ \bibnamefont {Archibald}}, \bibinfo {author}
  {\bibfnamefont {A.~H.}\ \bibnamefont {Ribeiro}}, \bibinfo {author} {\bibfnamefont {F.}~\bibnamefont {Pedregosa}}, \bibinfo {author} {\bibfnamefont {P.}~\bibnamefont {{van Mulbregt}}},\ and\ \bibinfo {author} {\bibnamefont {{SciPy 1.0 Contributors}}},\ }\bibfield  {title} {\emph {\bibinfo {title} {{{SciPy} 1.0: Fundamental Algorithms for Scientific Computing in Python}}},\ }\href {https://doi.org/10.1038/s41592-019-0686-2} {\bibfield  {journal} {\bibinfo  {journal} {Nature Methods}\ }\textbf {\bibinfo {volume} {17}},\ \bibinfo {pages} {261--272} (\bibinfo {year} {2020})}\BibitemShut {NoStop}%
\bibitem [{\citenamefont {Gao}\ and\ \citenamefont {Han}(2012)}]{10.1007/s10589-010-9329-3}%
  \BibitemOpen
  \bibfield  {author} {\bibinfo {author} {\bibfnamefont {F.}~\bibnamefont {Gao}}\ and\ \bibinfo {author} {\bibfnamefont {L.}~\bibnamefont {Han}},\ }\bibfield  {title} {\emph {\bibinfo {title} {Implementing the nelder-mead simplex algorithm with adaptive parameters}},\ }\href {https://doi.org/10.1007/s10589-010-9329-3} {\bibfield  {journal} {\bibinfo  {journal} {Comput. Optim. Appl.}\ }\textbf {\bibinfo {volume} {51}},\ \bibinfo {pages} {259–277} (\bibinfo {year} {2012})}\BibitemShut {NoStop}%
\bibitem [{\citenamefont {Nocedal}\ and\ \citenamefont {Wright}(2006)}]{nocedal2006numerical}%
  \BibitemOpen
  \bibfield  {author} {\bibinfo {author} {\bibfnamefont {J.}~\bibnamefont {Nocedal}}\ and\ \bibinfo {author} {\bibfnamefont {S.~J.}\ \bibnamefont {Wright}},\ }\href@noop {} {\emph {\bibinfo {title} {Numerical optimization}}}\ (\bibinfo  {publisher} {Springer},\ \bibinfo {year} {2006})\BibitemShut {NoStop}%
\bibitem [{\citenamefont {Conn}\ \emph {et~al.}(2000)\citenamefont {Conn}, \citenamefont {Gould},\ and\ \citenamefont {Toint}}]{conn2000trust}%
  \BibitemOpen
  \bibfield  {author} {\bibinfo {author} {\bibfnamefont {A.~R.}\ \bibnamefont {Conn}}, \bibinfo {author} {\bibfnamefont {N.~I.}\ \bibnamefont {Gould}},\ and\ \bibinfo {author} {\bibfnamefont {P.~L.}\ \bibnamefont {Toint}},\ }\href@noop {} {\emph {\bibinfo {title} {Trust region methods}}}\ (\bibinfo  {publisher} {SIAM},\ \bibinfo {year} {2000})\BibitemShut {NoStop}%
\bibitem [{\citenamefont {Lam}\ \emph {et~al.}(2015)\citenamefont {Lam}, \citenamefont {Pitrou},\ and\ \citenamefont {Seibert}}]{10.1145/2833157.2833162}%
  \BibitemOpen
  \bibfield  {author} {\bibinfo {author} {\bibfnamefont {S.~K.}\ \bibnamefont {Lam}}, \bibinfo {author} {\bibfnamefont {A.}~\bibnamefont {Pitrou}},\ and\ \bibinfo {author} {\bibfnamefont {S.}~\bibnamefont {Seibert}},\ }in\ \href {https://doi.org/10.1145/2833157.2833162} {\emph {\bibinfo {booktitle} {Proceedings of the Second Workshop on the LLVM Compiler Infrastructure in HPC}}},\ \bibinfo {series and number} {LLVM '15}\ (\bibinfo  {publisher} {Association for Computing Machinery},\ \bibinfo {address} {New York, NY, USA},\ \bibinfo {year} {2015})\BibitemShut {NoStop}%
\bibitem [{\citenamefont {Harris}\ \emph {et~al.}(2020)\citenamefont {Harris}, \citenamefont {Millman}, \citenamefont {van~der Walt}, \citenamefont {Gommers}, \citenamefont {Virtanen}, \citenamefont {Cournapeau}, \citenamefont {Wieser}, \citenamefont {Taylor}, \citenamefont {Berg}, \citenamefont {Smith}, \citenamefont {Kern}, \citenamefont {Picus}, \citenamefont {Hoyer}, \citenamefont {van Kerkwijk}, \citenamefont {Brett}, \citenamefont {Haldane}, \citenamefont {del R{\'{i}}o}, \citenamefont {Wiebe}, \citenamefont {Peterson}, \citenamefont {G{\'{e}}rard-Marchant}, \citenamefont {Sheppard}, \citenamefont {Reddy}, \citenamefont {Weckesser}, \citenamefont {Abbasi}, \citenamefont {Gohlke},\ and\ \citenamefont {Oliphant}}]{harris2020array}%
  \BibitemOpen
  \bibfield  {author} {\bibinfo {author} {\bibfnamefont {C.~R.}\ \bibnamefont {Harris}}, \bibinfo {author} {\bibfnamefont {K.~J.}\ \bibnamefont {Millman}}, \bibinfo {author} {\bibfnamefont {S.~J.}\ \bibnamefont {van~der Walt}}, \bibinfo {author} {\bibfnamefont {R.}~\bibnamefont {Gommers}}, \bibinfo {author} {\bibfnamefont {P.}~\bibnamefont {Virtanen}}, \bibinfo {author} {\bibfnamefont {D.}~\bibnamefont {Cournapeau}}, \bibinfo {author} {\bibfnamefont {E.}~\bibnamefont {Wieser}}, \bibinfo {author} {\bibfnamefont {J.}~\bibnamefont {Taylor}}, \bibinfo {author} {\bibfnamefont {S.}~\bibnamefont {Berg}}, \bibinfo {author} {\bibfnamefont {N.~J.}\ \bibnamefont {Smith}}, \bibinfo {author} {\bibfnamefont {R.}~\bibnamefont {Kern}}, \bibinfo {author} {\bibfnamefont {M.}~\bibnamefont {Picus}}, \bibinfo {author} {\bibfnamefont {S.}~\bibnamefont {Hoyer}}, \bibinfo {author} {\bibfnamefont {M.~H.}\ \bibnamefont {van Kerkwijk}}, \bibinfo {author} {\bibfnamefont {M.}~\bibnamefont {Brett}}, \bibinfo {author} {\bibfnamefont
  {A.}~\bibnamefont {Haldane}}, \bibinfo {author} {\bibfnamefont {J.~F.}\ \bibnamefont {del R{\'{i}}o}}, \bibinfo {author} {\bibfnamefont {M.}~\bibnamefont {Wiebe}}, \bibinfo {author} {\bibfnamefont {P.}~\bibnamefont {Peterson}}, \bibinfo {author} {\bibfnamefont {P.}~\bibnamefont {G{\'{e}}rard-Marchant}}, \bibinfo {author} {\bibfnamefont {K.}~\bibnamefont {Sheppard}}, \bibinfo {author} {\bibfnamefont {T.}~\bibnamefont {Reddy}}, \bibinfo {author} {\bibfnamefont {W.}~\bibnamefont {Weckesser}}, \bibinfo {author} {\bibfnamefont {H.}~\bibnamefont {Abbasi}}, \bibinfo {author} {\bibfnamefont {C.}~\bibnamefont {Gohlke}},\ and\ \bibinfo {author} {\bibfnamefont {T.~E.}\ \bibnamefont {Oliphant}},\ }\bibfield  {title} {\emph {\bibinfo {title} {Array programming with {NumPy}}},\ }\href {https://doi.org/10.1038/s41586-020-2649-2} {\bibfield  {journal} {\bibinfo  {journal} {Nature}\ }\textbf {\bibinfo {volume} {585}},\ \bibinfo {pages} {357--362} (\bibinfo {year} {2020})}\BibitemShut {NoStop}%
\bibitem [{\citenamefont {Suzuki}\ \emph {et~al.}(2012)\citenamefont {Suzuki}, \citenamefont {Inoue},\ and\ \citenamefont {Chakrabarti}}]{suzuki2012quantum}%
  \BibitemOpen
  \bibfield  {author} {\bibinfo {author} {\bibfnamefont {S.}~\bibnamefont {Suzuki}}, \bibinfo {author} {\bibfnamefont {J.-i.}\ \bibnamefont {Inoue}},\ and\ \bibinfo {author} {\bibfnamefont {B.~K.}\ \bibnamefont {Chakrabarti}},\ }\href@noop {} {\emph {\bibinfo {title} {{Quantum Ising phases and transitions in transverse Ising models}}}},\ Vol.\ \bibinfo {volume} {862}\ (\bibinfo  {publisher} {Springer},\ \bibinfo {year} {2012})\BibitemShut {NoStop}%
\bibitem [{\citenamefont {Haldar}\ \emph {et~al.}(2021)\citenamefont {Haldar}, \citenamefont {Mallayya}, \citenamefont {Heyl}, \citenamefont {Pollmann}, \citenamefont {Rigol},\ and\ \citenamefont {Das}}]{PhysRevX.11.031062}%
  \BibitemOpen
  \bibfield  {author} {\bibinfo {author} {\bibfnamefont {A.}~\bibnamefont {Haldar}}, \bibinfo {author} {\bibfnamefont {K.}~\bibnamefont {Mallayya}}, \bibinfo {author} {\bibfnamefont {M.}~\bibnamefont {Heyl}}, \bibinfo {author} {\bibfnamefont {F.}~\bibnamefont {Pollmann}}, \bibinfo {author} {\bibfnamefont {M.}~\bibnamefont {Rigol}},\ and\ \bibinfo {author} {\bibfnamefont {A.}~\bibnamefont {Das}},\ }\bibfield  {title} {\emph {\bibinfo {title} {{Signatures of Quantum Phase Transitions after Quenches in Quantum Chaotic One-Dimensional Systems}}},\ }\href {https://doi.org/10.1103/PhysRevX.11.031062} {\bibfield  {journal} {\bibinfo  {journal} {Phys. Rev. X}\ }\textbf {\bibinfo {volume} {11}},\ \bibinfo {pages} {031062} (\bibinfo {year} {2021})}\BibitemShut {NoStop}%
\end{thebibliography}%
\end{document}